\documentclass[12pt,thmsa,a4paper]{article}

\usepackage{amsmath}
\usepackage{amssymb}
\usepackage{epsfig}
\usepackage[latin1]{inputenc}

\newcommand{\p}{\partial}
\newcommand{\n}{\nabla}
\newcommand{\Ss}{\sqrt{1-\frac{2M}{r}}}
\newcommand{\Sss}{\left(1-\frac{2M}{r}\right)}
\newcommand{\Th}{T^{\theta}{}_{\theta}}
\newcommand{\E}[1]{\left< #1 \right>}
\newcommand{\V}[1]{\left<0\left| #1 \right|0\right>}
\newcommand{\St}[1]{\left| #1 \right>}
\newcommand{\U}[1]{\left<U\left| #1 \right|U\right>}
\newcommand{\HH}[1]{\left<H\left| #1 \right|H\right>}
\newcommand{\B}[1]{\left<B\left| #1 \right|B\right>}
\newcommand{\Lc}{{\cal L}}
\newcommand{\vp}{\varphi}
\newcommand{\ve}{\varepsilon}
\newcommand{\1}{1\hspace{-0.243em}\text{l}}

\begin{document}

\begin{titlepage}
  \newcommand{\mytitle}{\textbf{\huge Quantum Radiation  blablabla}} 
  \newlength{\titlepagewidth}
  \settowidth{\titlepagewidth}{\mytitle}
  \newcommand{\amytitle}{\textbf{\huge Quantum Radiation\\
from\vspace*{0.4cm}\\
Black Holes}}
  \centering
  \begin{minipage}[b][\textheight][t]{\titlepagewidth}
    \centering \large \vspace*{1.5cm}
  
    \textbf{\huge DISSERTATION} \\
        
    \vspace{1.8cm}
        
    \amytitle
    
    \vspace{2.2cm}
    
    ausgeführt zum Zwecke der Erlangung des akademischen Grades eines
    Doktors der technischen Wissenschaften unter der Leitung von\\
    
    \vspace{0.8cm}
        
    O.\,Univ.\,Prof.\,DI.\ 
    Dr.\,techn.\ Wolfgang Kummer\\E136 Institut für Theoretische Physik
    
    \vspace{0.8cm}
    eingereicht an der
    Technischen Universität Wien\\ Fakultät für Technische Naturwissenschaften
    und Informatik\\[0.8cm]
    von
    
    \vspace{0.8cm}
    
    \textbf{Daniele Hofmann} e9326914 \\ Nordwestbahnstrasse 35A/5 A-1020
    Wien
  
    \vspace{\fill}
  
    \leftline{Wien, im Mai 2002}
 \end{minipage}

\end{titlepage}

\newpage
\vspace*{-2.2truecm}

\enlargethispage{0cm}
\tableofcontents

\newpage
\section{Introduction}

Already before I started to study physics I was fascinated
by the milestones of modern theoretical physics which
are supposed to explain the whole universe: the theory of
General Relativity (GR), which governs the macrocosm, the world
of galaxies, Black Holes (BH), and the universe as a whole,
all being parts of a single curved manifold which evolves
according to a deterministic law.
And Quantum Field Theory (QFT) which describes the world
on the microscopic scale as a steady succession of
unpredictable interactions between the immense multitude
of fundamental particles.

Until today these two theories have remained the best verified
and most fundamental theories in physics, although
it has become clear very soon that both break down
within a certain range of their parameters.
Einstein has shown that the gravitational force is
mediated by the geometry through the gravitational field which propagates
information and energy with a finite velocity (the finiteness
is already a consequence of Special Relativity).
The dynamics of the geometry is described by the
Einstein equations
\begin{equation}
G_{\mu\nu}=R_{\mu\nu}-\frac{g_{\mu\nu}}{2}=-\frac{8\pi G}{c^2}T_{\mu\nu},
\label{Einstein-eq.}
\end{equation}
which are the basis of GR. $G_{\mu\nu}$ is the Einstein tensor that
represents the geometric content of the theory. On the r.h.s. stands
the Energy-Momentum (EM) tensor (also called stress-energy tensor)
which contains the matter part.
The sign in (\ref{Einstein-eq.}) is chosen such that a positive energy
density $T_{tt}>0$ leads to an attractive gravitational
potential in the Newtonian approximation. 
The theory of GR already incorporates various concepts:
the equivalence principle
(the inertial mass of a body equals its gravitational mass)
which suggests a geometrical description; the self-interaction
of the gravitational field (information is transported by energy that
produces information etc.); the equivalence of accelerated
systems (introducing partly Mach's principle into GR).
However, it does not include the particle properties of the gravitational
field. It is generally assumed that the gravitational interaction
is governed by the rules of quantum mechanics as soon as
the exchange of gravitons becomes comparable with that of other
particles. The underlying idea is that the coupling constants of
the known forces approach each other with increasing particle energy
and decrease until they finally meet with the gravitational coupling
constant $G=6.7\cdot10^{-8}\frac{cm^3}{g\cdot s}$.
The energy scale at which quantum effects play a significant
role in gravity is given by the Planck scale $m_{Pl}=\sqrt{\frac{\hbar c}{G}}
\approx2.2\cdot10^{-5}g\hat{=}10^{19}\,GeV$. The characteristic length
scale at which spacetime is expected to deviate from the classical
description (of GR) is $l_{Pl}=\sqrt{\frac{\hbar G}{c^3}}=1.6\cdot10^{-33}cm$.
At this point the known theories are assumed to fail and
correct physical predictions could only be made by a theory
of Quantum Gravity.

The lack of knowledge beyond the Planck scale manifests itself
already in ordinary QFT and GR: the virtual particles exchanged
in loop graphs may carry arbitrary energies which generally leads
to ultraviolet (UV) divergences. They indicate that the
separation into a fixed spacetime and point-particles which
propagate on it does not describe accurately a physical process
at extremely high energy densities. A step into this direction
has been taken by string theory, where the point particles
are replaced by extended objects which themselves reveal
an inner structure (by their geometry and topology); the spacetime
structure, however, remains classical.
Further, the problem of the particle masses and couplings of the fundamental
particles in the Standard Model has not been resolved.
They enter as parameters, although a fundamental theory, including
the gravitational interaction, should allow to calculate them
from basic principles. As in the classical theory of self-interacting
point particles also in GR the theory itself predicts its
own breakdown as it contains singular solutions, where the spacetime curvature,
and thus the energy density, diverges. Singularities appear in the
most accepted and common applications such as the Friedmann solutions,
describing the evolving universe until the Big Bang, and the
Schwarzschild solution, when it describes a BH.
All in all, there is convincing evidence that the present picture
one has of the spacetime structure and particle interactions
is limited and that a new theory, which is capable to answer
the open questions, may require fundamentally different concepts
to all current approaches.

The aim of a theory of Quantum Gravity is to unify the principles
of quantum mechanics with GR which can only succeed if the
the main conceptual problem is overcome, namely, to deal with
a quantum field (e.g. the metric or the connection) that describes
the spacetime on which it lives. It is generally believed
that the solution lies in a non-perturbative treatment because
there is no natural background that serves as a starting point
for the perturbation; one can show by simple arguments that
the perturbation series of the self-interacting gravitational
field is non-renormalisable.

In my thesis I will consider a problem which, within a certain
range of the parameters, can be resolved by the ``classical''
methods of GR and QFT.
I will investigate the interaction of a quantum field with a strong
(but still classical, i.e. below the Planck scale) gravitational
field, or in other words, \emph{the quantisation of free scalar particles
on a curved spacetime}. I consider scalar particles because they
involve the characteristic features of such calculations but
are easier to handle than higher-spin fields; the extension
to arbitrary spin is tedious but not fundamentally different.
More specifically, I will take a Schwarzschild spacetime (describing
a BH) as background manifold, whereby I restrict myself to 
sufficiently large BH masses $M\gg m_{Pl}$ so that
the gravitational field outside the horizon lies far below the
Planck scale. In this setting the latter can be treated as
a classical field which acts on the quantum field as an external
source. I will not examine the interior of the horizon where
the fields become arbitrarily large, independently of the BH
mass. For an external observer a huge BH, from the
gravitational point of view, is a classical system whose
field is perfectly described by GR.

Nevertheless, already this ``external'' combination of QFT and GR
(as compared to the internal combination when quantising the
gravitational field itself) reveals fundamentally new physical
phenomena which are hidden in the classical theory.
A particularly nice example is Hawking radiation
from a BH. Although classically no particles
can leave the BH (by definition), quantum theory
predicts the spontaneous production of particles close
to the event horizon which can leave the BH to
infinity and decrease thereby the BH mass.
The effect of particle production always occurs in the
presence of strong fields which ``feed'' the vacuum
of some quantum field (a similar effect can be observed
in electrodynamics where the photon field produces
electron-positron pairs). What makes the situation particular
are the global properties of the BH spacetime
which are characterised by the event horizon.
As it prevents information and particles from leaving the
interior region of the BH, it gives the whole
spacetime some causal structure; the latter is not
affected by the Hawking radiation as it is a thermal
distribution which is not related to the inner structure
of a BH. Although the Hawking effect
clarifies the question of the BH evolution,
the problem of its final fate, and the predicted transition
between different spacetime topologies, requires a
theory of Quantum Gravity. In my thesis the calculation
of the Hawking flux will be a ``by-product'' of the
quantisation procedure -- on the other hand it serves
as a motivation because it provides a direct interpretation
of the results and application to cosmological models.

The starting point of my computations is the Einstein-Hilbert
action 
\begin{equation}
L_{EH}=\int_M\left[\frac{c^2}{16\pi G}R+\frac{(\p S)^2}{2}
-\frac{m^2S^2}{2}\right]\sqrt{-g}d^4x\label{scalar action}
\end{equation}
which contains the Lagrangian of a massive scalar field.
It describes the classical propagation of a scalar particle
$S$ with mass $m$ on a spacetime $M$ (I use
the same symbol for the manifold as for the BH mass)
with metric $g_{\mu\nu}$, where $R$ is the scalar curvature.
If the gravitational effect of the scalar particle is small, compared
to that of the BH, the r.h.s. of (\ref{Einstein-eq.})
can be set to $0$. The geometry is then that of a vacuum
spacetime. This approximation is maintained when the scalar
field is quantised, as long as the gravitational field on the
horizon does not reach the Planck scale (see the estimate
in Section 1.2).

Throughout this thesis I will use natural (Planck) units:
I set $c=\hbar=G=k_B=1$. This means that times, distances, energies
and temperatures become dimensionless quantities which
are measured in multiples of the Planck unit $m_{Pl}=l_{Pl}=1$.
The value of a quantity in arbitrary units is obtained by multiplying
it with the corresponding Planck quantity in these units.
For estimates of loop orders etc. I reinsert $\hbar$
in the respective equation.

The plan of my thesis is the following:
in the present Chapter I introduce the basic concepts
of GR and QFT. Most importantly I show how a scalar
field can be quantised in the semi-classical approximation
by the path integral formalism.

In the second Chapter I present the Christensen-Fulling (CF) approach
\cite{chf77}, by which the problem of computing the expectation value
of the EM tensor can be reduced to the computation of
two basic components. Then I introduce the
two-dimensional dilaton model. It allows to describe a Schwarzschild
spacetime by a two-dimensional Einstein-Hilbert action, whereby the additional
structure enters by the appearance of the dilaton field.
Further, I discuss the boundary conditions and their relation
to the quantum states of the scalar field, and show
how they can be fixed by the choice of integration constants
in the CF approach. In this course I can show the state-independence
of the basic components.

In Chapter three I discuss the effective action from which
I can derive all expectation values of the quantised
free scalar field. Thereby I will distinguish between
massive and massless scalar fields and develop a local,
respectively non-local, perturbation theory.
The starting point in both cases is the so-called
heat kernel which is regularised by the zeta-function
regularisation. In particular I examine the convergence condition
of the local Seeley-DeWitt expansion \cite{sch51,dew63}
in Section 3.1.3. Further, I will adopt the covariant perturbation
theory \cite{bav87} (presented in Section 3.2) for arbitrary 
two-dimensional models.

The following two Chapters include the main parts
of my thesis: in Chapter four I consider massive
scalar particles in four dimensions. I calculate
the expectation value of the EM tensor by the local
expansion of the effective action and the CF
method and discuss the relevance of the obtained results.

In the fifth Chapter I repeat this procedure for
massless particles in the dilaton model.
Most importantly I can show that the corresponding
non-local effective action can be derived directly from the
heat kernel by the covariant perturbation theory.
A major part of this Chapter is devoted to the examination
of the two-dimensional Green functions by an appropriate
perturbation series. In this context I can show the
consistency of the effective action with the assumed
boundary conditions and the infrared (IR) renormalisation.
Finally, I reproduce the results for the Hawking flux
obtained in the literature \cite{kuv99} which have now been derived
by a closed and rigorous calculation.

Figure 1 shows the logical flow of my thesis. The left side
of the diagram corresponds to the dilaton model, the right
side to the four-dimensional theory. The objects in the center
are associated to both, respectively. 
\\
\\
\begin{figure}[h]

\hspace{0cm}\epsfig{file=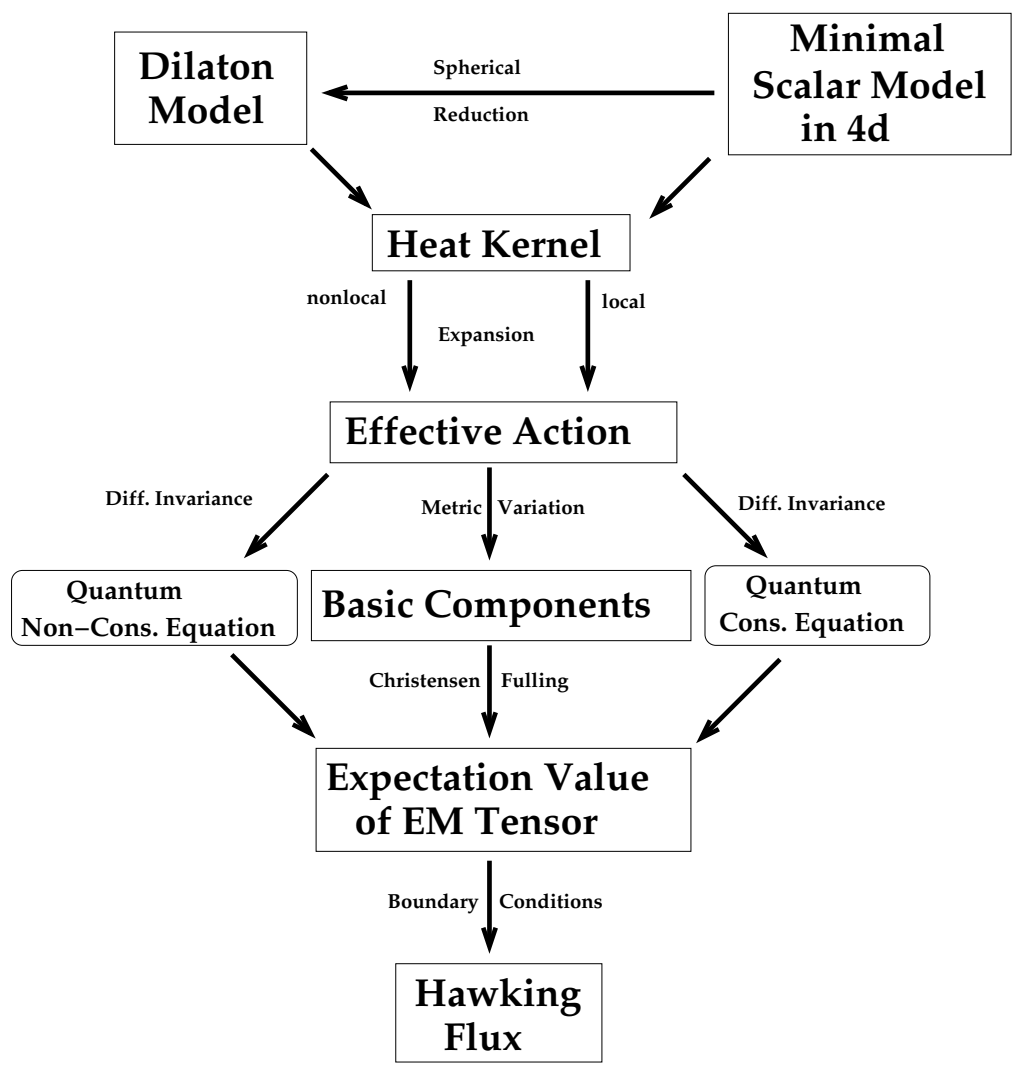,height=14cm}

\caption[fig1]{Logical Flow of my Thesis}
\end{figure}

\newpage

\subsection{Black Holes}

A BH is defined as a region in spacetime where
the gravitational field is strong enough to prevent
even light from escaping to infinity. It is formed by a body
of mass $M$ when it is contracted to a size smaller than
$2M$. There is ample evidence that such objects indeed
exist, especially at the centre of galaxies, although
clearly a BH has never been observed directly, but
only by its gravitational interaction.
The classical picture of BH formation
is that very heavy stars, such as neutron stars or white dwarfs,
may have sufficiently high energy densities so that
no known force can stop it from collapsing to a
point-like object, the singularity. The Chandrasekhar
limit for a white dwarf
is $1.2-1.4$ masses of the sun until it exceeds the necessary
energy density (for neutron stars there are other limits).
As long as there exists no satisfying theory of Quantum Gravity,
one can only argue about the inner structure of a BH --
quantum theoretical arguments suggest that the mass
is not concentrated at a single point but merely distributed
over highly excited gravitons. Anyway, the existence
of BHs as cosmological objects is rather well established
and during the last years they have become a playground
for astronomers as well as theorists.

In this Section I will give a short introduction on the
basic concepts of BHs, whereby I concentrate
on the aspects and methods needed in this work.

\subsubsection{The Schwarzschild Black Hole}

A Schwarzschild metric describes a spacetime that is characterised
only by a mass which is concentrated at its origin.
The BH is the region of spacetime that is hidden
by the event horizon and from which matter and even light cannot
escape. Because it has no angular momentum
the spacetime is spherically symmetric. Further, the whole
spacetime is static -- the space outside of the BH is empty,
no radiation or matter falls in or out, and the mass
remains constant. Also it cannot be decreased by
gravitational radiation since there are no spherically symmetric
gravitational waves (s-waves), according to the Birkhoff
theorem \cite{wal84}. For this reason the Schwarzschild BH is also 
called ``eternal''.

The Schwarzschild spacetime is an exact solution of the vacuum Einstein
equations\footnote{The BH mass can be introduced in the Einstein equations
by a delta-function-like EM tensor \cite{ban93,ban94}.
As I am only interested in the region outside of the BH I can
neglect this term.} $G_{\mu\nu}=0$, see (\ref{Einstein-eq.}) for $T_{\mu\nu}=0$,
under the additional condition of spherical symmetry. The solution can be
expressed in several coordinate systems which are valid for different
patches of the spacetime. For the exterior of the event horizon the
Schwarzschild gauge\footnote{Throughout this work I will use the 
sign convention $(1,-1,-1,-1)$ for the Lorentzian spacetime metric.}
\begin{equation}
ds^2=\Sss dt^2-\frac{1}{\Sss}dr^2-r^2[d\theta^2+\sin^2\theta\,d\vp^2]
\label{SS-metric}
\end{equation}
is appropriate. One immediately observes that the metric
is singular on the event horizon $r=2M$. This is only a coordinate
singularity which in this gauge prevents us from calculating
beyond the horizon. Of course there are other gauges that allow calculations 
inside the BH, see Appendix A.3. The physical singularity
of course lies at $r=0$. Here the spacetime curvature becomes singular.

By a conformal transformation, see Appendix C, with infinite
conformal factor in the asymptotic region, the Schwarzschild spacetime can
be mapped onto a Penrose diagram (Figure 1).

\begin{figure}[h]

\hspace{1.4cm}\epsfig{file=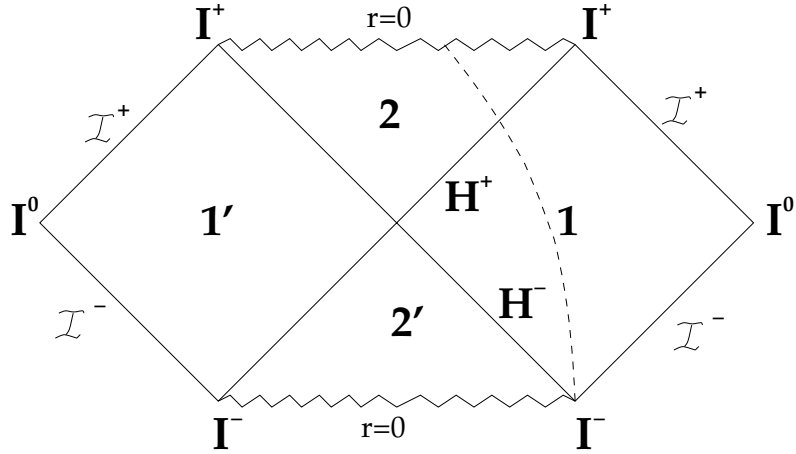,height=6cm}

\caption[fig1]{Penrose Diagram of a Schwarzschild Spacetime}
\end{figure}

In this diagram the asymptotic region is mapped onto five distinct
parts: spacelike infinity $I^0$, future null infinity ${\cal I}^+$,
past null infinity ${\cal I}^-$, future timelike infinity $I^+$,
and past timelike infinity $I^-$. They correspond to the points
(or regions), where spacelike, lightlike and timelike geodesics
end if they are infinitely continued into the future or past.
For instance, incoming massless particles are emitted somewhere
on ${\cal I}^-$ in region $1$ and travel on a lightlike geodesic
until they cross the future horizon $H^+$. Note that the conformal
transformation preserves the angles between geodesics and hence
the conformal structure of the spacetime -- lightlike geodesics
always have $45$ degree inclination with respect to a horizontal line.

Region $1$ corresponds to the patch of the spacetime which is covered
by Schwarzschild coordinates. Region $2$ represents the BH. 
One can see from the diagram that no causal (i.e. timelike or light-like)
geodesic starting from this region can escape to ${\cal I}^+$.
The event horizon $H^+$ separates these two regions. 
The (future and past) singularity is drawn as a jagged line.
Surprisingly, it is identified with a spacelike part of the
spacetime (and not with a timelike as one might expect).
One can see indeed from the metric (\ref{SS-metric}) that
the notions of spacelike and timelike change their role on the horizon.
Note that there are regions in the diagram, namely $1'$ and $2'$, which do
not represent regions in a realistic BH spacetime. Region
$2'$ is called the White Hole and it is the necessary mathematical
continuation of the BH if the latter shall be static. 
Region $1'$ is not even causally connected with the visible part
of the spacetime $1$ and it can be considered as a hidden parallel
universe. The problem of these unphysical parts of the diagram
is resolved when describing the BH as a cosmological object
that was formed by a collapsing body. The dashed line in the diagram,
which starts from $I^-$ and ends somewhere inside the BH,
shows the surface of a such a body. Obviously regions $1'$ and
$2'$ now are found in the interior of this collapsing body which
is no more correctly described by the Schwarzschild solution.

Although the Schwarzschild spacetime cannot describe correctly
the evolution of a cosmological BH it is a good
approximation in the quasi-static phase, where the BH
has formed completely and the surrounding spacetime is almost
empty (region $1$ and $2$ to the right of the dashed line).
If the BH is very massive, its surface temperature
is low and it evolves slowly by continuously radiating away
massless particles (see Section 1.1.4).
The geometry outside of the horizon then is nicely described
by the Schwarzschild metric (\ref{SS-metric}).
In the calculation of quantum mechanical expectation values
I will employ the \emph{static approximation} by
inserting for the spacetime geometry the Schwarzschild solution.

\subsubsection{Isometries}

An active diffeomorphism from a manifold to itself that preserves the 
metric is called an isometry. If the isometry is generated by
a vector field $V^{\mu}$ that gives the displacement at each spacetime point
this vector field is called a \emph{Killing field}.
On a given manifold Killing fields can be found by solving
the Killing equation
\begin{equation}
{\cal L}_Vg_{\mu\nu}=V^{\rho}\n_{\rho}g_{\mu\nu}
+(\n_{\mu}V^{\rho})g_{\rho\nu}+(\n_{\nu}V^{\rho})g_{\mu\rho}
=2\n_{(\mu}V_{\nu)}=0,
\end{equation}
where ${\cal L}_V$ is the Lie-derivative into the direction of
the vector field $V$.
This reflects the fact that the metric and hence all geometrical
properties of the spacetime do not change by moving into the
direction of a Killing field.
The Schwarzschild spacetime has four globally independent Killing fields,
one of which is timelike. From the time-independence of the metric
components of the Schwarzschild metric we immediately observe that
$(1,0,0,0)$ is a timelike Killing vector field. It expresses the
time-translation symmetry of the Schwarzschild spacetime that characterises
all stationary spacetimes. In this specific case the timelike Killing
field is also hypersurface orthogonal which means that Schwarzschild
is static.

The other three Killing fields are spacelike and generate the
spherical symmetry. The orbits they form are two-spheres $S^2$.
Locally one only has two independent Killing fields, e.g. the
basis vectors $\p_{\theta},\p_{\vp}$ of a tangent basis on $S^2$.
Because the two-sphere cannot be covered by a single coordinate
patch one needs a third Killing field to form a global
basis of the tangent space. In practical calculations it is
sufficient to work with $\p_{\theta},\p_{\vp}$, because the
basis vector $\p_{\theta}$ becomes singular only at isolated
points, namely the poles.

A general tensor field is said to be invariant under translations
into the direction of a vector field if the Lie-derivative
of the tensor field vanishes. The Lie-derivative into
a coordinate direction is simply given by the partial
derivative for this coordinate (in the corresponding coordinate basis).
The invariance condition can thus be written as
\begin{equation}
{\cal L}_{\p_{\mu}}T^{\alpha\beta\dots}=\p_{\mu}T^{\alpha\beta\dots}=0.
\end{equation}
In a generally relativistic system the existence of Killing fields
of the manifold normally implies the invariance of the matter
fields under translations into the direction of the Killing fields.
This is the case if the inhomogeneous
Einstein equations (\ref{Einstein-eq.}) can be solved exactly.
If the Lie-derivative into the symmetry-direction is then applied
to the whole equation the result follows immediately.

In this work I will describe the spacetime geometry by that of the
Schwarz\-schild solution, although the r.h.s. of the Einstein equations
is not exactly zero. Therefore it is possible that the
matter, produced by the quantum radiation from the BH, the Hawking effect,
does not obey the symmetry conditions of the spacetime (a spherically symmetric
potential may exhibit asymmetric solutions).
Nevertheless, it is believed that the spherically symmetric part of
the Hawking radiation, i.e. the s-waves, give the major contribution
to the total flux and thus one imposes the symmetry conditions
\begin{equation}
{\cal L}_{\p_{\theta}}S=0\,\,,\,\,{\cal L}_{\p_{\vp}}S=0
\end{equation}
on the scalar field $S$. With respect to quantum mechanical
expectation values it is obvious that the symmetry conditions
must be imposed on the observables. In Section 2.1.1 I show
how the form of the EM tensor is restricted by the s-wave
condition.

Note that a realistic BH spacetime exhibits no time-translation
symmetry! Hence, also the matter fields are not invariant
under time-translations
\begin{equation}
{\cal L}_{\p_t}S\neq0.
\end{equation}

\subsubsection{Energy and Flux}

The interesting physical observables of the scalar field are the
energy density and the local fluxes. They are given by the timelike
components $T_{t\mu}$ of the EM tensor
\begin{equation}
T_{\mu\nu}=\frac{2}{\sqrt{-g}}\frac{\delta L_m}{\delta g^{\mu\nu}},
\label{EM tensor}
\end{equation}
where $L_m$ is the matter action, i.e. the part of (\ref{scalar action})
that contains the scalar field $S$.
The prefactor is chosen such that $T_{\mu\nu}$ is indeed a tensorial
object (see below its derivation).
The EM tensor can be defined as the Noether current which is
conserved under a translation from one spacetime point with
coordinates $x^{\mu}$ to some other $(x')^{\mu}=x^{\mu}+\xi^{\mu}$.
The change of the metric and the scalar field $S$ under this
transformation is given by the Lie-derivative into the direction of $\xi$:
\begin{equation}
\delta_{\xi}g^{\mu\nu}={\cal L}_{\xi}g^{\mu\nu}=
-\xi^{\mu;\nu}-\xi^{\nu;\mu}\,\,,\,\,
\delta_{\xi}S={\cal L}_{\xi}S=\xi^{\mu}\p_{\mu}S.
\end{equation}
The matter part of the action changes as
\begin{eqnarray}
\delta_{\xi} L_m[x\to x']&=&\int\delta_{\xi}g^{\mu\nu}\frac{\delta L_m}
{\delta g^{\mu\nu}}d^4x+\int\delta_{\xi}S\frac{\delta L_m}{\delta S}
d^4x\nonumber\\
&\stackrel{EOM}{=}&\int-(\xi^{\mu;\nu}+\xi^{\nu;\mu})T_{\mu\nu}
\frac{\sqrt{-g}}{2}d^4x\nonumber\\
&\stackrel{EOM}{=}&\int\xi^{\mu}\n^{\nu}T_{\mu\nu}\sqrt{-g}d^4x.
\end{eqnarray}
In the first line I have used $\frac{\delta L_m}{\delta S}=0$,
which means that $S$ fulfils the classical equations of motion (EOM).
If the action functional shall be invariant
under spacetime translations (diffeomorphism invariance)
the EM tensor must be locally conserved:
\begin{equation}
\n^{\mu}T_{\mu\nu}=0.\label{cons.equation}
\end{equation}
This is the \emph{conservation equation of the EM tensor} which
is one of the basic concepts of this work. Namely, it
establishes relations between the components of the EM
tensor and thereby reduces significantly the number of
independent components at a certain spacetime point.
Most importantly, I will show that \emph{the conservation of the
EM tensor extends to quantum mechanical expectation values}
(Section 2.1 and Section 2.2.3). This is not trivial since
I have used the classical EOM in the classical derivation.

The EM tensor, and its conservation law, includes all kinds
of matter fields but does not define a notion of gravitational
energy. Unfortunately, there is no similar local expression in the metric
such that a general energy conservation law could be formulated,
including gravitational waves (except in two-dimensional gravity \cite{grk00}).
For globally hyperbolic\footnote{A globally hyperbolic spacetime
is one that contains a Cauchy surface. A Cauchy surface is a
three-dimensional spacelike submanifold (of a four-dimensional
time-orientable spacetime) which is intersected by all
future-directed and past-directed causal geodesics -- hence
a Cauchy surface contains the information of the whole spacetime.
It is believed that all physically meaningful spacetimes are globally
hyperbolic \cite{adm62}.}, and asymptotically flat spacetimes there
exists at least a global conservation law. Namely, one can define
the energy and momentum of the whole spacetime at spatial infinity
$I^0$ by an integral over some expression in the metric and
the intrinsic curvature on a given Cauchy surface.
The so defined energy and momentum are independent of that
surface by which they are calculated and can be interpreted
as the total available energy and momentum of the considered spacetime.
In particular, I introduce the term ADM mass for the total energy
on a BH spacetime with the demanded properties.
For a Schwarzschild spacetime it is identical to the mass
of the BH $m_{ADM}=M$.

The fact that some energy and flux is ``hidden'' in the gravitational
degrees of freedom has another intriguing consequence: the classical
assumption that the energies and fluxes of the matter fields
are strictly positive may be violated on a curved spacetime.
There exist several independent energy conditions
corresponding to different physical requirements.
The \emph{weak energy condition}
\begin{equation}
T_{\mu\nu}\xi^{\mu}\xi^{\nu}\ge0
\end{equation}
demands that the energy density, as measured by any observer, is positive, 
$\xi^{\mu}$ being a timelike vector field which is tangent to the
geodesic of the observer. This is strictly
valid for all classical fields whose EM tensor is given by the
variation of a classical action as in (\ref{EM tensor}) (see
at the end of this Section). \emph{For the expectation values of
quantum fields the weak energy condition may be violated}.
In particular, in the evaporation process of a BH
a flux of virtual particles with negative energy goes into the BH
and thereby decreases its mass.
\\
\\
In order to illustrate the physical meaning of the components
of the EM tensor and to identify the correct signs, I will integrate
the conservation equation
over a small spatial volume $\triangle V$ with surface $\p V$.
I will do this for a flat spacetime,
keeping in mind the peculiarities of curved spacetime:
\begin{equation}
\int_{\triangle V}\p_tT^t{}_td^3x
=\p_t\int_{\triangle V}T^t{}_td^3x=-\int_{\triangle V}\p_{\kappa}
T^{\kappa}{}_td^3x=-\oint_{\p V}T_{t\kappa}d^2f^{\kappa}.
\end{equation}
The measure of the surface $\p V$ is $d^2f^{\kappa}
=\ve^{\kappa}{}_{\mu\nu}dx^{\mu}\wedge dx^{\nu}\frac{\sqrt{-g}}{d!}$
and $d=3$ is the space dimension.
The change in time of the energy in $\triangle V$ is minus the
flux through the surface $\p V$. Thus, \emph{$T_{tt}$ can be interpreted
as the energy density and $T_{t\kappa}\,,\kappa={1,2,3}$ as the
fluxes into the corresponding coordinate directions}.
Further, one can derive the relation
\begin{equation}
\p_t\int_{\triangle V}T^t{}_{\kappa}d^3x
=-\oint_{\p V}T_{\kappa\lambda}d^2f^{\lambda}
\end{equation}
which shows that the change of momentum of $\triangle V$
is determined by the total pressure acting on its surface.
The pressure is given by the integral over the stresses $T_{ij}$.
On a spherically symmetric spacetime the only nonzero stresses
are $T_{rr},T_{\theta\theta}$ and $T_{\vp\vp}$. If I set
$\kappa=r$ in the last equation I get $\p_t\int_{\triangle V}T^t{}_rd^3x
=-\frac{1}{3}\oint_{\p V}T^r{}_{r}\sqrt{-g}d\theta\wedge d\vp$.
Thus, the change of radial flux in time is given by the change of
the stress in the r-direction. For $\kappa=\theta,\vp$, the r.h.s.
becomes zero because $T_{\theta\theta},T_{\vp\vp}$
are independent of $\theta,\vp$. Accordingly the fluxes
into the $\theta,\vp$-directions are constant on a spherically symmetric
spacetime (in Section 2.1.1 I will show that they are even zero).

On a flat spacetime one can introduce the four-momentum vector
of a system
\begin{equation}
P^{\mu}=\oint_{\p V}T^{\mu}{}_{\nu}d^2f^{\nu}.
\end{equation}
This concept cannot be generalised easily to curved spacetimes
because there the tangent spaces differ at each spacetime
point. Nevertheless, for globally hyperbolic and asymptotically flat 
spacetimes one can define the ADM mass of the spacetime
by a similar integral, where $V$ is the whole spacetime \cite{adm62}.

In the discussion of boundary conditions a light-cone coordinate system
(\ref{light-cone}) will be particularly convenient.
In this gauge the components $T_{--},T_{++}$ (\ref{EM-light2},\ref{EM-light3})
are the outgoing and incoming fluxes. On a Schwarzschild spacetime
the two can be combined to the total flux by the
relation
\begin{equation}
T^{r_{\ast}}{}_t=-T^t{}_{r_{\ast}}=\frac{1}{\Sss}(T_{--}-T_{++}),
\label{total flux}
\end{equation}
where $r_{\ast}$ is the Regge-Wheeler coordinate, see Appendix A.3.
In my sign-convention $T^{r_{\ast}}{}_t$ is positive for matter
moving into the positive r-direction. Note that in Wald's convention
(see Appendix A.1) the positive flux is given by $T^t{}_{r_{\ast}}$. 
In both conventions $T_{tr}$ is \emph{negative} for an outgoing flux
of particles with positive energy!

A scalar particle $S$ with action (\ref{scalar action}) has an EM tensor
\begin{equation}
T_{\mu\nu}=\p_{\mu}S\p_{\nu}S-\frac{g_{\mu\nu}}{2}[(\p S)^2-m^2S^2],
\label{EM-scalar}
\end{equation}
where I have used the relation\footnote{The $\delta$-function on a
general manifold is defined by $\int_M\delta(x-x')d^4x=1$.}
(\ref{var.measure})
\begin{equation}
\frac{\delta\sqrt{-g}(x)}{\delta g^{\mu\nu}(x')}=-g_{\mu\nu}
\frac{\sqrt{-g}}{2}\delta(x-x').
\end{equation}
The trace of the EM tensor then is $T=2m^2S^2-(\p S)^2$. The signs
in the scalar action (\ref{scalar action}) and the defining
equation of the EM tensor (\ref{EM tensor})
are chosen such that, in my sign convention, \emph{the energy density of 
a classical massive scalar field is strictly positive}:
\begin{equation}
T_{tt}=\frac{(\p_tS)^2}{2}-g_{tt}\frac{g^{rr}(\p_rS)^2+g^{\theta\theta}
(\p_{\theta}S)^2+g^{\vp\vp}(\p_{\vp}S)^2}{2}
+g_{tt}\frac{m^2S^2}{2}\ge0.
\end{equation}
As already mentioned this is no more valid for the expectation
values of quantum fields. 

In Appendix A.5 I discuss some further properties of the
EM tensor like the effect of non-minimal coupling to the curvature.

\subsubsection{Hawking Radiation}

The idea that BHs radiate when quantum theory is incorporated
to describe the matter fields was introduced by Stephen Hawking in the middle
of the seventies \cite{haw75,hah76}, building upon previous work
by Unruh \cite{unr76}.
Before that, it was considered as a fact that the event horizon
of a BH cannot decrease which would mean that a BH, once
produced, could never disappear from spacetime. The final state of a 
BH has been seen as a stationary state, completely described by 
the mass, the angular momentum, and the charge (if any).
With Hawking's discovery this scenario changed dramatically. It was soon
realized that a continuously radiating BH looses its mass and
finally may disappear completely. This fact and the evaporation process itself
raised lots of new interesting questions, many of which are still not
answered. On the phenomenological side one may ask if the Hawking radiation
of a BH may be observed directly. This seems to be rather delicate
since the known BHs have been found by their extremely
high-energetic X-rays, produced by the accretion of mass from nearby neutron 
stars. In comparison to this high amount of ``classical'' radiation
the Hawking flux is almost negligible. Because the lifetime of small
BHs may be less than the age of the universe, one might conjecture that
we are surrounded by a large amount of small BHs formed at the
time of the Big Bang, the so-called primordial BHs \cite{Car75}.
Their possible existence and investigation could give further hints on the
inhomogeneity of the universe at very early times.
On the theoretical side the open questions are linked with the lack
of a theory of Quantum Gravity. In the final period of the evaporation
process the quantum fluctuations of the metric become dominant.
Thus one needs exact control of the backreaction of the metric and its further
action on the particle vacuum and so on. It is assumed that some feedback
between the radiation and the gravitational field at this stage settles
the Hawking flux, which in the semi-classical approximation would tend
to infinity as the BH mass decreases. Unfortunately, the exact solution is
still unknown. This lack of knowledge about the final BH evolution
prevents us from understanding some fundamental problems such as the
information loss puzzle. Namely, the information once swallowed by
the BH (say a system of pure quantum states) is lost
forever if it disappears at the end of the
evaporation process. There is no evidence that the Hawking radiation
(as a thermal mixture of quantum states) has somehow encoded
the information of the matter which has passed the
event horizon. The only possibility that the information be released
could be at the very final stage which is still unknown ground.
If this is not the case the unitarity of quantum theory would be
violated (the probability to find the particle which had fallen
into the BH somewhere in the universe might become zero)!
Figure 2 shows a Penrose diagram of a realistic BH.
The dashed line again marks the surface of the collapsing body that
forms the BH. The information loss problem is illustrated
by the ``Cauchy surfaces'' $\Sigma_1,\Sigma_2$: information that leaves
$\Sigma_1$ on causal geodesics into the future may either reach
$\Sigma_2$ or fall into the spacelike singularity (indicated by a
jagged line). The spacetime is no more globally hyperbolic.

\begin{figure}[h]

\hspace{4cm}\epsfig{file=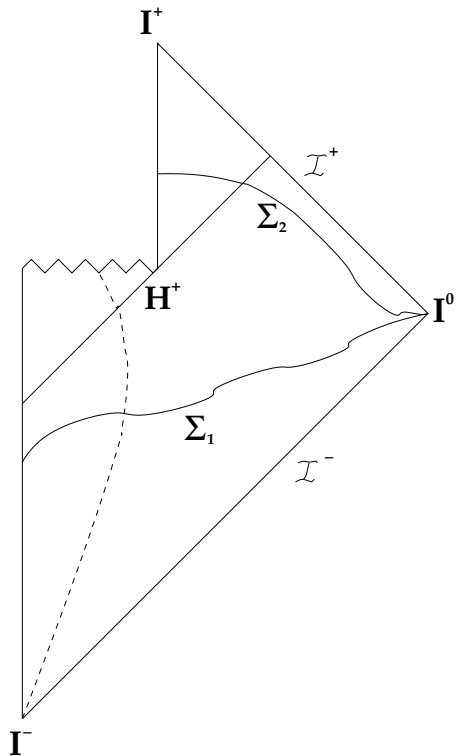,height=10cm}

\caption[fig1]{Black Hole Evolution}
\end{figure}

The quantum mechanical effect that enables BHs to
radiate away their mass is known as \emph{particle production}.
It always takes place when a quantum vacuum of some particle species
interacts with an external field. The vacuum is assumed
to be filled with virtual particle-antiparticle pairs whose
total energy is zero. 
Thus, one of the particles of a pair carries negative energy (violating
the weak energy condition), while the other particle may carry
sufficient positive energy to be on the mass-shell. 
If the particles interact with some
external field they may acquire some additional energy.
If it is sufficient, both particles become real and can be
measured in a detector. This physical process is well-known
for strong electromagnetic fields.
Near a BH the situation is more subtle. In principle
there is enough gravitational energy to produce real particles
but the gravitational radiation has to tunnel through the
event horizon. Alternatively we can think of a particle
with negative energy, produced in the pair-production process, that
falls into the BH and thereby decreases its mass. As it
is trapped in the BH we are not confronted with the
situation that a particle of negative energy might be measured.
The other particle (which is real) can probably escape to infinity.

The main result of Stephen Hawking's famous calculation on BH
radiance was \cite{hah76}: BHs emit radiation at a characteristic
temperature
\begin{equation}
T_H=\frac{1}{8\pi M},\label{Hawking temp.}
\end{equation}
where $M$ is the BH mass.
Hawking speaks of a temperature (instead of energy or frequency)
to emphasize the relation to thermodynamics.

Hawking did not compute explicitly the expectation value of the
EM tensor, while this is one of the aims of my thesis.
Instead he circumvented
this problem by relating the amplitude of a particle
(with a certain energy $E$) emitted by the
BH to the one of a particle absorbed by the BH. The ratio of these
two probabilities already implies that the radiation corresponds to the one
of a Black Body at a certain temperature. It has the form of the
\emph{Boltzmann distribution}
\begin{equation}
\frac{\text{P(emit particle with E)}}{\text{P(absorb particle with E)}}
=e^{-\frac{E}{T_H}}
\end{equation}
and shows that the probability to emit particles with an energy higher
than $T_H$ is exponentially damped. Surprisingly, this result was obtained
without explicitly calculating the probabilities!
By this and some statistical physics one can already calculate
the Hawking radiation, i.e. the amount of energy radiated away
through a unit surface per unit time.

In the following I calculate the Hawking flux of massless
particles with spin $0$ starting from the Black Body hypothesis.
I consider the surface of a BH as a perfect
Black Body which is described by infinitely many
oscillators with energies $E_r$. The probability that an oscillator
is in the state $E_r$ is $e^{-\frac{E_r}{T_H}}$. Because
the total number of particles is not fixed, the partition function
is given by a sum over all possible occupation numbers in all
possible states:
\begin{multline}
Z[T_H]=\sum_{n_1=0}^{\infty}\sum_{n_2=0}^{\infty}
\cdots e^{-\frac{(n_1E_1+n_2E_2+\dots)}{T_H}}\\
=\left(\sum_{n_1=0}^{\infty}e^{-\frac{n_1E_1}{T_H}}\right)
\cdot \left(\sum_{n_2=0}^{\infty}e^{-\frac{n_2E_2}{T_H}}\right)\cdots
=\prod_{E_r}\frac{1}{1-e^{-\frac{E_r}{T_H}}}.
\end{multline}
The separation of the summations into products of sums is possible
because all sums go to $\infty$. The average occupation number
of some energy-mode is given by
\begin{equation}
\E{n_r}=-T_H\frac{\p\ln Z}{\p E_r}=\frac{e^{-\frac{E_r}{T_H}}}
{1-e^{-\frac{E_r}{T_H}}}=\frac{1}{e^{\frac{E_r}{T_H}}-1}.
\end{equation}
What is still missing is the number and distribution of energy states
in a unit spacetime volume. The energy states are identified with
the states of translational kinetic energy, characterised by a
momentum three-vector (the energy clearly only depends on the
absolute value of the momentum $p$). In a unit volume these states are
counted as $f(p)dp=1\cdot\frac{p^2}{2\pi^2}dp=\frac{\omega^2}{2\pi^2}d\omega
=\frac{E^2}{2\pi^2}dE$,
where I have written a $1$ for the unit volume and $\nu$ is the frequency
corresponding to the momentum $p$ (in ordinary units one has the relation
$p=\frac{\hbar\omega}{c}$). Combining the measure in
the state-space with the average state-occupation number, one obtains
the distribution of particles in the state-space in a unit volume:
\begin{equation}
dN_E=\frac{1}{2\pi^2}\frac{E^2}{e^{\frac{E}{T_H}}-1}dE.
\end{equation}
Planck's law of Black Body radiation is finally obtained by
multiplying by the energy in the given state:
\begin{equation}
E\cdot dN_E=\frac{1}{2\pi^2}\frac{E^3}{e^{\frac{E}{T_H}}-1}dE.
\end{equation}
The total flux, i.e. the energy radiated away per unit time, of a BH
is now simply given by integrating over the whole range of energy
and multiplying with the surface $A$ of the BH and the speed
of light $c=1$:
\begin{equation}
\text{Flux}_{tot}=\frac{1}{4}\frac{A}{2\pi^2}\int_0^{\infty}
\frac{E^3}{e^{\frac{E}{T_H}}-1}dE=\frac{\pi^2A(T_H)^4}{120}.
\label{Flux.tot}
\end{equation}
The factor $\frac{1}{4}$ comes in because only the energy
radiated into the half-plane from some infinitesimal hole
in the Black Body contributes: $\text{Flux}_{tot}\propto\\ \int_{\text{hp}}
\cos\theta\frac{d^2\Omega}{4\pi}=\frac{1}{4}$; the factor $\cos\theta$
enters because the radiation leaves the BH under an angle
$\theta$.
If we insert the Hawking temperature $T_H=\frac{1}{8\pi M}$,
and the area of the event horizon\footnote{As can be seen from
the qualitative behaviour of the flux and energy density
(Figures 6,12,13,14) the region near the horizon
($r=\gamma\cdot M\,,\,1\le\gamma\le100$) exhibits special properties,
differing from the ones of a usual Black Body. Therefore, the
area $A$ perhaps should be replaced by $A_{eff}=\gamma^2\cdot A$.}
$A=16\pi M^2$, we obtain
\begin{equation}
\text{Flux}_{tot}=\frac{1}{30720\pi M^2}.
\label{flux.massless.blackbody}
\end{equation}
The local flux is obtained by dividing by $\frac{1}{4\pi r^2}$.

Although Hawking's result was revolutionary since it showed that
BHs are dynamical objects that may evaporate and finally disappear,
it was just the trigger for subsequent calculations on quantised fields
in curved spacetime.
For many physical considerations the explicit form of the quantum
EM tensor is needed.
Even more, when the final state of an evaporating BH is investigated,
one has to deal with the full interaction between the metric and the quantised 
fields. The latter problem, because of its nonlinearity, goes far beyond the
quantisation of non-interacting fields on a curved background and is not within
the scope of this work. Nevertheless, such calculations cannot be avoided when
seeking answers to the information loss puzzle or other fundamental questions 
that arise in the extreme regimes of singularities.

\subsection{Semi-Classical Quantum Gravity}

A complete theory of Quantum Gravity should describe gravitational effects
at very high energy densities or very small distances. In these
regions the classical deterministic description of the gravitational
field by the Einstein equations breaks down and quantum effects,
like the uncertainty principle, become important. One expects that this happens
at the Planck scale, when the spacetime curvature becomes comparable
to the Planck curvature
\begin{equation}
R_{\text{Planck}}=\frac{c^3}{\hbar\cdot G}\approx 3.829\cdot 10^{65}\text{cm}^{-2}
\,\hat{=}1\,\text{(Planck units)}.
\end{equation}
Such high energy densities only exist near singularities
like the one at the centre of BHs. Unfortunately, such
a theory does not yet exist. Conceptual problems, like
the dual role of the metric as a dynamical field and the background,
have not yet been overcome. Further, there is no experimental evidence
for quantum gravitational effects because they take place only under
extreme conditions.

In this work the metric always remains a classical external field that
interacts with the quantum fields which live on the curved background.
This means that the metric still obeys classical,
deterministic EOM, while the matter is 
described by a quantum mechanical probability function.
The interaction is then described by the semi-classical Einstein 
equations which I will ``derive'' in the following from
fundamental considerations.

I assume that there exists some generating functional $Z$
that contains the whole information on all physical observables.
It can be written as a path integral over all physical
variables, weighted by the corresponding action functionals:
\begin{equation}
Z=\int{\cal D}g\det{\cal F}[g]{\cal D}S\cdots
e^{i(L_{EH}[g,S]+L_{GF}[g]+\dots)}.\label{path-integral}
\end{equation}
Here $\det{\cal F}[g]$ is the Fadeev-Popov determinant
of the metric field and $L_{GF}$ is the gauge-fixing part
of the action (gravity is a non-Abelian gauge theory \cite{het92}).
Clearly, the expectation values are independent of the choice
of gauge (corresponding to a choice of coordinate system).
The following steps only have a formal character, thus I will
neglect the peculiarities of gravity as a gauge theory and
discard the Fadeev-Popov determinant and the gauge-fixing term.
In principle one could add all known matter fields but I will
only consider the case of a scalar field with action (\ref{scalar action}).
The generating functional depends on the sources $j$ that
are coupled to the matter fields and whose variations lead
to the expectation values. The path integral must be invariant
under (local) translations $g_{\mu\nu}\to g_{\mu\nu}+\delta g_{\mu\nu}$
($\delta g$ is \emph{independent} of the metric):
\begin{multline}
0=\delta_g Z=\delta_g\int{\cal D}g{\cal D}S\,e^{iL_{EH}[g,S]}\\
=i\int\delta g_{\mu\nu}\int{\cal D}g{\cal D}S
\left(\frac{\delta L_{EH}[g,S]}{\delta g_{\mu\nu}}\right)e^{iL_{EH}[g,S]}d^4x.
\end{multline}
Note that any reasonable path integral measure is invariant
under translations
${\cal D}g={\cal D}(g+\delta g)$. If we add some normalising
factor (see below), we recover the Einstein equations for the
expectation values of the quantum operators of the fields
(denoted by a hat on top):
\begin{equation}
\E{\hat{G}_{\mu\nu}}=-\frac{1}{2}\E{\hat{T}_{\mu\nu}}.\label{Einstein-eq.qu.}
\end{equation}
In the same way one can derive the classical EOM
for the expectation value of the scalar field. The fact that
the expectation values of quantum fields obey the classical EOM 
is known as Ehrenfest theorem.

Equation (\ref{Einstein-eq.qu.}) holds exactly. Now I introduce
the semi-classical approximation by replacing 
$\E{\hat{G}_{\mu\nu}}$ by $G_{\mu\nu}[\E{\hat{g}}]$, where 
$\E{\hat{g}_{\mu\nu}}=g^0_{\mu\nu}+\hbar g^1_{\mu\nu}+\dots$ is the
expectation value of the metric field expanded in orders of $\hbar$.
Clearly the two expressions differ in general. Nevertheless, in
situations where
the classical metric $g^0$ is dominant, the approximation
$\E{\hat{G}_{\mu\nu}}\approx G_{\mu\nu}[\E{\hat{g}}]$ is justified.
The spacetime geometry is then described by the
semi-classical Einstein equations:
\begin{equation}
G_{\mu\nu}[\E{g}]=-\frac{1}{2}\E{\hat{T}_{\mu\nu}}.\label{Einstein-eq.sc.}
\end{equation}
In particular, the semi-classical approximation can be applied
in the exterior region of heavy Schwarzschild BHs $M\gg m_{Pl}=1$:
the curvature
$R^{\mu}{}_{\nu\sigma\tau}$ behaves like $M r^{-3}$ (\ref{Riemann.SS})
and is of the order $M^{-2}\ll1$ at the
horizon. The radiative components of the vacuum expectation value of the
EM tensor behave like $r^{-2}$ and are of the order $cM^{-4}$,
where $c=\frac{1}{10^6\pi^2}$ by (\ref{flux.massless.blackbody}).
Thus, the classically induced spacetime
curvature is dominant near the horizon which is the region
where the physically interesting processes take place.
The quantum fields dominate far away from the BH,
but their energy density still falls off sufficiently rapidly so that
the spacetime is considered asymptotically flat.

Now I can expand both sides of (\ref{Einstein-eq.sc.}) in orders
of $\hbar$:
\begin{equation}
G^0_{\mu\nu}[g_0]+\Delta^1_{\mu\nu}[g_0,g_1]+\dots=-\frac{1}{2}
\left(T_{\mu\nu}^0+\E{\hat{T}_{\mu\nu}}^1[g_0]+\dots\right).
\label{Einstein-eq.sc.2}
\end{equation}
The terms of zeroth order in $\hbar$ correspond to the classical
expressions. Because of the non-linearity in the metric
the higher order terms on the l.h.s. such as $\Delta^1_{\mu\nu}$
do not have the analytical form of the Einstein tensor $G_{\mu\nu}$.
Note that the first quantum order of the matter fields $\E{\hat{T}_{\mu\nu}}^1$,
calculated by field quantisation on a given background, only depends
on the classical metric $g_0$.
The first quantum correction $g_1$ of the metric often is called
the \emph{backreaction} (see Section 1.2.3) of the spacetime on
the quantum field. It is of particular interest if one starts with a static,
classical metric $g_0$, because it encodes the evolution of
the BH (in a range where the semi-classical approximation holds).
In this thesis I only consider the zeroth order of the geometry
$g_0$, which is determined by the vacuum Einstein
equations $G^0_{\mu\nu}[g_0]=-\frac{1}{2}T_{\mu\nu}^0=0$ as the
Schwarzschild solution (\ref{SS-metric}). Then I
compute the first order of the r.h.s. of (\ref{Einstein-eq.sc.2})
$\E{\hat{T}_{\mu\nu}}^1[g_0]$ by quantising the scalar field $S$ on
this background.

\subsubsection{Expectation Values}

The main subject of this work will be the computation of
the quantum mechanical expectation values of the EM tensor
\begin{equation}
\E{\hat{T}_{\mu\nu}}
\end{equation}
in a general quantum state which shall not be specified for the moment.
$\hat{T}_{\mu\nu}(x)$ is the local operator that corresponds to
a measurement of the EM tensor\footnote{In the following I omit
the hat on top of quantum operators.}.
The classical EM tensor of a scalar field (\ref{EM tensor}) is a quadratic
expression in the fields. Thus one might consider the process of measuring
energy and momentum as the production of some test-particle
at a certain spacetime point $x$ that propagates in a closed loop and is
then annihilated at the same point, Figure 3.
Such loops without external scalar field legs are responsible for the
infinite vacuum energy in ordinary QFT (in flat spacetime).
The difference of the respective values in curved spacetime
and flat spacetime is the amount of energy supplied by
strong gravitational fields for spontaneous particle production.
From perturbation theory we know that closed loops correspond
to orders in $\hbar$. This means that the vacuum expectation
value of the EM tensor is a pure quantum effect that contributes
solely to the order $\hbar^1$ if there is no self-interaction or interaction
with other particles\footnote{If there were some (self-)interaction
one could make a perturbation around the free scalar field. This
would lead to higher scalar-loop interaction graphs.}
and if the metric is classical.
The interaction of the scalar particle with the gravitational
field is considered as a classical process (which in Feynman diagrams
are represented by tree-graphs). It can be visualized
by an external line that intersects the scalar loop at the point
of measurement. If the scalar particles are massless
the interaction with gravity is a non-local process because the
scalar loops can then become infinitely large and the interaction
may occur arbitrarily apart from the point of measurement $x$.
Note that higher quantum orders in the metric (like the backreaction)
could be represented by graviton loops and therefore would
contribute only to the order $\hbar^2$ to the expectation value of the EM
tensor.

\begin{figure}[h]

\hspace{1.7cm}\epsfig{file=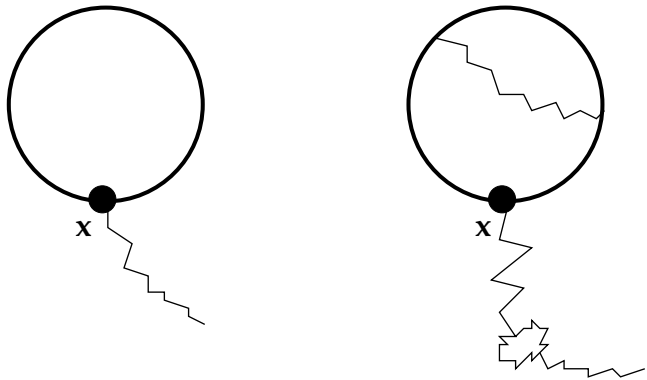,height=5cm}

\caption[fig1]{Vacuum Loop without and with Gravitons: the scalar loops
are drawn as straight lines, the gravitational field as jagged lines.}
\end{figure}

In this work I will use the path integral method to quantise
the scalar field $S$, while the metric is considered as a classical field.
The basic object in this approach is the generating functional
\begin{equation}
Z[g]={\cal N}\int_{\left|\bullet\right>_g}{\cal D}S\cdot e^{i L_m[g,S]}
\label{gen.functional}
\end{equation}
which contains the whole information on the quantum system (such as
eigenstates). $L_m$ is the matter action of the scalar field and
${\cal N}$ is some (infinite, but field-independent)
normalisation constant. Symbolically I have marked the path integral
by a quantum state $\left|\bullet\right>_g$ (which is not yet specified)
to emphasize the dependence of the generating functional on the
boundary conditions. The transition from
the full path integral (\ref{path-integral}) over all variables
to (\ref{gen.functional}) can be accomplished by the introduction
of a delta-function $\delta(g-g_0)$ into (\ref{path-integral})
which restricts the geometry to the classical value.

From $Z$ one can already derive the expectation
value of the EM tensor.
The metric which enters (\ref{gen.functional}) is the
expectation value\footnote{For simplicity I just write $g$.} $\E{g}$
in the actual state of the system and may in general contain the backreaction
and hence orders in $\hbar$.
Clearly, as I do not know the full quantum metric from the
beginning, I insert an approximate (static) metric (which will be the
classical Schwarzschild one) to obtain a first order solution of
$\E{T_{\mu\nu}}$. From this one may calculate the backreaction which
reinserted into the path integral gives the next order of the EM
tensor and so forth (which is not within the scope of this work).
It is convenient to introduce the generating functional of the
connected graphs $W$ by
\begin{equation}
Z[g]=e^{i W[g]}\,\,\to\,\,W[g]=-i\ln\left({\cal N}\int{\cal D}S
\cdot e^{i L_m[g,S]}\right)\label{eff.action}.
\end{equation}
It produces the expectation value of the EM tensor by variation
for the spacetime metric as in (\ref{EM tensor})
\begin{equation}
\E{T_{\mu\nu}(x)}=\frac{2}{\sqrt{-g}}\frac{\delta W[g]}
{\delta g^{\mu\nu}(x)}=\frac{\int{\cal D}S\cdot {\cal T}_{\mu\nu}[S](x)
\cdot e^{i L_m[g,S]}}{\int{\cal D}S\cdot e^{i L_m[g,S]}}.\label{exp.value}
\end{equation}
The expectation value of an arbitrary observable\footnote{I
denote observables by italics. Their arguments are not operators
but classical functions (which does not mean that they obey the classical
EOM).} of the scalar field $S$ can be obtained by the introduction
of an external source $j(x)$, coupled to the observable, into the classical action:
\begin{equation}
L_m[g,S,j]=L_m[g,S]+\int j(x){\cal O}[S](x)\sqrt{-g}d^4x.
\end{equation}
It is then obtained by variation of $W[g,j]$ for this
source and subsequently setting it to zero:
\begin{equation}
\E{O}=\frac{1}{\sqrt{-g}}\frac{\delta W[g,j]}{\delta j(x)}\biggm|_{j=0}.
\label{exp.value2}
\end{equation}
Normally one introduces the generating functional of the
one-particle irreducible (1PI) graphs $\Gamma[-\!\!\!\!\! S]$ (where
$-\!\!\!\!\! S$ is the mean field defined by $-\!\!\!\!\! S=\frac{\delta W}
{\delta j}$ if ${\cal O}=S$), also called the \emph{effective action},
in the course of the renormalisation procedure.
It is related to the connected functional $W$ by a Legendre transform
and differs from it (among other things) by the fact that
the generated 1PI graphs do not possess external legs (as compared
to the connected graphs).
For \emph{non-interacting} fields the only possible loop graphs
are single loops without external legs as there exists no interaction
vertex
to connect a propagator to the loop, hence $W$ and $\Gamma$
are equivalent. Therefore, I will call $W$ the effective action
as it is common in the literature.

If the complete matter action $L_m[g,S,j]$ is a \emph{quadratic expression}
in the scalar field, the effective action is a Gaussian path integral
that can be integrated out. In particular, the observable ${\cal O}$
must also be quadratic in $S$ which is the case for the EM tensor.
In this thesis I will not introduce a source term but
calculate the expectation values as in (\ref{exp.value}).

The scalar field in the path integral of (\ref{gen.functional})
can be separated into a classical and a quantum part: $S=S_0+S_q$.
Accordingly, the classical action can be expanded around the classical
solution
\begin{multline}
L_m=L_m[g,S_0]+\int S_q(x)\frac{\delta L_m}{\delta S}\biggm|_{S=S_0}d^4x\\
+\int\int S_q(x)S_q(y)\frac{\delta^2L_m}{(\delta S)^2}\biggm|_{S=S_0}d^4xd^4y,
\end{multline}
where the second term vanishes $\frac{\delta L_m}{\delta S}|_{S=S_0}=0$.
As $S_0$ is some fixed classical solution the path integral measure
becomes ${\cal D}S_q$. The first term $L_m[g,S_0]$
can be pulled out of the path integral, but nevertheless, it
contributes to the expectation values, namely by the classical
value. The quadratic term simply reproduces the matter action,
whereby the total field $S$ is replaced by the quantum part.
Since there are no higher orders, as I consider free fields,
the perturbation can be written as
\begin{equation}
L_m[g,S]=L_m[g,S_0]+L_m[g,S_q].
\end{equation}
Analogously the observable of the EM tensor ${\cal T}_{\mu\nu}$
can be expanded around its classical value ${\cal T}_{\mu\nu}[S]=
{\cal T}_{\mu\nu}[S_0]+{\cal T}_{\mu\nu}[S_0,S_q]+{\cal T}_{\mu\nu}[S_q]$.
If a classical solution $S_0$ is inserted which mimics a collapsing body forming
a BH, one obtains contributions from the mixed term
${\cal T}_{\mu\nu}[S_0,S_q]$ as one calculates the expectation value.
This might lead to the grey-body factors which modify the expected
Hawking flux.

In this thesis I set $S_0=0$ and thus $S=S_q$. This choice of
classical solution fixes the boundary conditions (see below)
and hence the quantum state of the effective action.
This state will turn out to be the so-called \emph{Boulware state}
which I denote with $\left|B\right>$ (I define it properly in Section 2.4.1).
It is not the vacuum state of the theory, although it corresponds
to the state of lowest asymptotic energy density because it exhibits
unphysical properties on the horizon (see Section 2.4)!
Throughout this work I will always assume that the effective
action is in the Boulware state and that all expectation values
calculated from it thus correspond to this state. Namely, beside
the fact that one does not have to care about boundary terms etc.,
another advantage of this state is that it fits best to the
static approximation of the spacetime metric. The latter implies
asymptotic flatness\footnote{The effective action turns out to be a purely
geometric expression (see below) which in the static approximation
exhibits the necessary fall-off conditions. A different asymptotic
behaviour would require a more complicated geometric representation.}
and this is only consistent if the scalar field vanishes
asymptotically. Further, I will show that any quantum state can
be recovered easily at the level of expectation values.

Of particular importance is the vacuum state $\left|0\right>$.
It is defined as the quantum state which yields the minimum
value for the expectation value of the energy density
$\V{T_{tt}}:=\text{min}\E{T_{tt}}$ in a given background geometry.
Classically this would be realized by a field configuration
where the field vanishes on the whole spacetime. If pair-production
is possible there are two contributions to the vacuum energy: first,
by the particle production, as described in Section 1.1.4. Second, by the
\emph{vacuum polarisation} which I will discuss now.
It is given by contribution of the disconnected scalar loops that
are always present in the vacuum state. In the free theory, there
is only one loop (namely the one representing the measurement
of the EM tensor), whereas in models with interacting fields
there are disconnected graphs at each loop order (as they do not
depend on the point of measurement, they only contribute to
the infinitesimal normalisation and are formally eliminated
by the denominator in (\ref{exp.value})). The contribution
of the vacuum loop is divergent in general and needs to be renormalised,
see next Section. Both effects, the particle production and
the vacuum polarisation, contribute to the vacuum energy --
the former mainly by real and the latter by virtual particles.
As the particle production near the horizon of a BH
involves virtual particles, namely the ones swallowed by it,
it is not possible to rigorously distinguish between
the two contributions.

The vacuum state is not only characterised by some minimal finite
energy density, but also by some outgoing flux (otherwise
there would be no particle production).
However, the incoming flux is zero:
a finite incoming flux would increase the energy density beyond
the minimum value. Because the outgoing particles carry away
energy from the BH the vacuum state changes continuously.
This suggests to define a vacuum state $\left|0\right>_{\Sigma}$
on each time-slice $\Sigma$ as the state of lowest
total energy of the spacelike submanifold, orthogonal to the timelike
curve that defines the time-slicing\footnote{Such a time-slicing
exists for all globally hyperbolic spacetimes \cite{wal84}.}.

One further observation is of interest: in flat spacetime a vacuum expectation
value is defined by the so-called in-vacuum $\left<\text{in}\right|O\left|
\text{in}\right>$. The state $\left|\text{in}\right>$ corresponds to the vacuum
state in the remote past which has been propagated forward in time to the
spacetime point of measurement. If we trace back the evolution of the BH
before the time of the collapse we find that our ``vacuum state'' is occupied
by the particles that have formed the BH. In this respect one
cannot speak of a vacuum state in the common sense -- such a state can only
be defined on a spacetime with zero ADM mass, i.e. without BH.
Nevertheless, this notion is sensible in the static approximation
and bearing in mind that we have a multi-particle system.
Thus the vacuum state is only defined for some time-slice by the actual
mass of the BH and the emptiness of the states in the remote past of 
the corresponding static solution.

\subsubsection{Renormalisation}

The mathematical expression of the scalar loop, and thus the
vacuum energy, in general is
divergent. This fundamental problem emerges in every QFT
and reflects the fact that the physics at very
high energies (in Quantum Gravity at least the Planck scale)
or small distances is not
yet understood. Accordingly one speaks of an UV divergence.
UV divergences generally appear in loop graphs to all orders,
as there one includes infinitesimally small loops in the path integral
which lead to infinite energy densities. By restricting
the range of the momenta by the introduction of some cut-off
one can regularise the expectation values. Then one relates
the measured observable to some reference (renormalisation) point
(e.g. by simply substracting the value at this point) and thereby obtains
a finite value when removing the cut-off. This procedure
is known as \emph{renormalisation} and it guarantees that
the fundamental QFTs yield sensible
results in the range where the physics is well-understood.

The ultimate basis of the renormalisation is that one knows
``experimentally''
the value of some observables at some significant
point and then extrapolates within some range that is
within the scope of the theory (e.g. below the Planck scale
where Quantum Gravity is supposed to play a role).
In flat spacetime the vacuum energy is simply renormalised
to zero. This is in nice agreement with
the observations which suggest that the spacetime
is almost perfectly flat (the problem of reproducing
the finite but extremely small cosmological constant
by the vacuum energy of the known fundamental particles
is still unresolved). However, as QFT
mainly deals with microscopic systems that do not
significantly affect the spacetime curvature, the
vacuum energy can be considered constant
and can thus be set to an arbitrary value -- if gravity
is neglected energy has no absolute meaning and one
only measures differences of energies.

In the present context gravitational effects clearly play
a crucial role and the vacuum energy depends on the
renormalisation point. Thus it is necessary to fix its
absolute value at some reference point where the
vacuum energy is known. I define the renormalised
vacuum expectation value of the EM tensor by substracting
the flat spacetime value:
\begin{equation}
\V{T_{\mu\nu}}_{ren;g}:=\V{T_{\mu\nu}}_g-\V{T_{\mu\nu}}_{flat}.
\end{equation}
This can be generalised to expectation values in arbitrary
quantum states. Unfortunately, this definition of the
renormalised EM tensor does not eliminate all divergences.
However, it demonstrates the basic concept of the renormalisation
on a curved manifold. The remaining problems shall be clarified
as soon as they emerge.

\subsubsection{Backreaction}

The backreaction of the quantum field on the spacetime
geometry is given by the higher order terms $g^1,g^2\dots$ 
of the metric in $\hbar$ which are produced by the loop
contributions of the EM tensor. In principle it
can be calculated iteratively by equation (\ref{Einstein-eq.sc.2}).
The classical metric $g^0$ alone determines the
one-loop order of the EM tensor $\E{T_{\mu\nu}}^1[g^0]$.
Thus, we get a system of coupled second order differential equations
for the components of $g^1$, namely
\begin{equation}
\Delta^1_{\mu\nu}=\frac{\hbar}{2}\left[2\n^{\kappa}\n_{(\nu}
\bar{g}^1_{\mu)\kappa}-\square\bar{g}^1_{\mu\nu}-g_{\mu\nu}\n^{\kappa}
\n^{\lambda}\bar{g}^1_{\kappa\lambda}\right]=-\frac{1}{2}\E{T_{\mu\nu}}^1.
\label{backreaction}
\end{equation}
Here I have introduced the auxiliary metric $\bar{g}^1_{\mu\nu}
=g^1_{\mu\nu}-\frac{g_{\mu\nu}}{2}g^1\,,\,\bar{g}^1=-g^1
=-g^{\mu\nu}g^1_{\mu\nu}$. If $g^1$ is known one can
calculate the next order of the EM tensor and so on.
As I consider a free scalar field the computation of the EM
tensor to all orders involves calculating a single
one-loop graph, where the metric includes increasing orders
in $\hbar$. The crucial point is thus to solve the differential
equation (\ref{backreaction}).

In the problem of Hawking radiation the control over
the backreaction is necessary to study the Hawking
radiation in an evolving BH spacetime,
i.e. when the static approximation is no more justified.
Thereby one has to bear in mind that the perturbational
expansion of the Einstein equations (\ref{Einstein-eq.sc.2})
breaks down, together with the semi-classical approximation,
for $M\approx m_{Pl}=1$. If one wants to calculate beyond this, one
has to include the full backreaction by some non-perturbative
method, e.g. by integrating out exactly the gravitational
degrees of freedom (in the dilaton model, Section 2.2, such a
calculation already exists \cite{klv97a} -- note, however, that there
is no dynamical degree of freedom in two-dimensional models).
In the slowly evolving phase one expects a damping of
the Hawking flux by the backreaction -- by extrapolation
to the late-time evolution one could probably avoid
the infinite temperature of infinitely small BHs
$T_H\propto\frac{1}{M}$ predicted by the semi-classical
calculation. Further, backreaction effects might change
significantly the estimates on the lifetime of BHs
which could have far-reaching consequences for many cosmological
models.

\newpage
\section{Christensen-Fulling Approach in $4d$ and $2d$}

The main task in calculating Hawking radiation is to find an
expression for the vacuum expectation value of the EM tensor. Clearly
not all components are of direct interest. Some components
are already eliminated by symmetry conditions as I will only
consider the s-waves of the radiation.
Christensen and Fulling \cite{chf77} 
have shown that \emph{by the use of the conservation equation
(\ref{cons.equation}) on a Schwarzschild
spacetime the number of independent components of the EM tensor reduces
to two}. The remaining two non-vanishing components,
which in the following I will call the \emph{basic components},
are obtained by integrating
the conservation equation of the EM tensor. Thereby enter two integration
constants which determine the quantum state of the system.
The latter is related to the boundary conditions, that e.g. fix the
incoming flux on ${\cal I}^-$,  and it will be an important part
of this work to clarify this relation and the problem of the
correct quantum state in general.

The method of Christensen and Fulling is neither the only way to
compute the EM tensor nor does it provide a means to obtain
the vacuum expectation values of the basic components. The computation of the latter
will be the most difficult part of the whole problem
and one must rely on elaborate methods to calculate
expectation values of quantum fields on a curved spacetime.

What makes the CF approach so appealing is that
it allows to control easily the boundary conditions and hence
the quantum state of the system. Most importantly, it separates
those components of the EM tensor that are independent of the
quantum state (the basic ones) and others that are not (these are the ones
containing real particle states).
It will turn out that the effective action in the static approximation
only produces expectation values in the unphysical $\St{B}$-state,
where no asymptotic particle states are occupied. By the CF
method one can add the missing terms to reconstruct the physically
correct quantum state.

Generally, this method is only applicable in the \emph{static approximation}
because it is based on the conservation equation in a Schwarzschild
geometry. This means that when backreaction
effects become important, the CF representation does
not provide the correct relation between the components of the EM tensor!

I start with the original derivation in four spacetime dimensions.
Then I shortly present the two-dimensional dilaton model that
describes the dynamics of a classical field on a four-dimensional,
spherically symmetric spacetime and show that the CF
method can be established also in this model.
Finally, I discuss the boundary conditions and quantum states
of the expectation values and how they fit into the framework
derived in this Chapter.

\subsection{Christensen-Fulling Representation in $4d$}

The basic principle of the CF approach is to use
the energy-momentum conservation equation for the expectation
value of the EM tensor\footnote{In the following I will sometimes
omit the expectation value brackets for simplicity.}.
Formally, the conservation equation
at the quantum level can be derived by demanding general coordinate (diffeomorphism)
invariance of the full path integral (\ref{gen.functional}).
The metric and the scalar field transform under a diffeomorphism
$(x')^{\rho}=x^{\rho}+\xi^{\rho}$ as
\begin{eqnarray}
\delta_{\xi} g^{\rho\sigma}&=&{\cal L}_{\xi}g^{\rho\sigma}=-\n^{\rho}\xi^{\sigma}
-\n^{\sigma}\xi^{\rho}\\
\delta_{\xi} S&=&{\cal L}_{\xi}S=\xi^{\rho}\p_{\rho}S,
\end{eqnarray}
where ${\cal L}_{\xi}$ is the Lie-derivative into the direction of $\xi$.
The variation of the generating functional $Z$ under a diffeomorphism
transformation shall vanish:
\begin{eqnarray}
0&=&\delta_{\xi}Z[g]\nonumber\\
&=&i{\cal N}\int{\cal D}S\int_M\delta_{\xi}g^{\rho\sigma}\frac{\delta L_m}
{\delta g^{\rho\sigma}}d^4x\cdot e^{iL_m[S]}
+\lim_{y\to x}\int_M\delta_{\xi}S(x)\frac{\delta Z[g]}{\delta S(y)}
d^4x\nonumber\\
&=&i{\cal N}\int{\cal D}S\int_M(-\n^{\rho}\xi^{\sigma}-\n^{\sigma}\xi^{\rho})
\frac{1}{2}T_{\rho\sigma}\sqrt{-g}d^4x\cdot e^{iL_m[S]}\nonumber\\
&&+\lim_{y\to x}{\cal N}\int{\cal D}S\int_M\xi^{\rho}\p_{\rho}S(x)
\frac{\delta}{\delta S(y)}e^{iL_m[S]}d^4x\nonumber\\
&=&i{\cal N}\int{\cal D}S\int_M\xi^{\rho}\n^{\sigma}T_{\rho\sigma}\sqrt{-g}d^4x\cdot
e^{iL_m[S]}\nonumber\\
&&-\lim_{y\to x}{\cal N}\int{\cal D}S\int_Me^{iL_m[S]}
\frac{\delta}{\delta S(y)}[\xi^{\rho}\p_{\rho}S(x)]d^4x\nonumber\\
&=&i\int_M\xi^{\rho}\n^{\sigma}\left(\E{T_{\rho\sigma}}+\frac{ig_{\rho\sigma}}{\sqrt{-g}}
\lim_{y\to x}\delta(x-y)\right)\sqrt{-g}d^4x.
\end{eqnarray}
From the third to the forth equality I have dropped a ``surface term''
\newline $\int{\cal D}S\frac{\delta}{\delta S}\dots$. The delta-function in the
last line represents the divergent part of the zero-point energy. 
The substraction of this term corresponds to the normal ordering in
the operator approach. The result is a finite renormalised EM tensor.
It obeys
the conservation equation
\begin{equation}
\n_{\rho}\E{T^{\rho}{}_{\sigma}}_{ren}=0\label{EM.cons.quant}.
\end{equation}

\subsubsection{Symmetries of the Energy-Momentum Tensor}

Before writing down the conservation equation for a Schwarzschild BH 
it proves useful to find the most general form of the EM tensor
on a spherically symmetric spacetime (I do not assume staticity at this
stage). 
It is restricted by the existence
of the three Killing vector fields that characterise a spherically symmetric
spacetime and which have two-spheres $S^2$ as orbits.
The symmetry condition is that the Lie-derivatives of the EM tensor
into the directions of the Killing fields have to vanish.
For consistency with the conservation equation I must
impose the same condition for the divergence of the EM tensor.
Locally the three Killing vector fields that form the $so(3)$ algebra
are linearly dependent. In particular, when using spherical coordinates
$\theta,\vp$, the tangent vectors $\p_{\theta},\p_{\vp}$
form a complete basis of the isometry algebra except at the poles of the sphere. 
Thus, bearing in mind that the poles are isolated, regular points
of the manifold, it suffices to demand
\begin{equation}
\Lc_{\p_{\theta}}T^{\rho}{}_{\sigma}=\Lc_
{\p_{\vp}}T^{\rho}{}_{\sigma}=0.
\end{equation}
In a coordinate basis the Lie-derivative
along a basis vector coincides with the partial derivative into the same
direction. Thus the necessary condition is that $T^{\rho}{}_{\sigma}$ does not
depend on $\theta,\vp$: $\p_{\theta,\vp}T^{\rho}{}_{\sigma}=0$.

Now I come to the conservation equation that has to obey
\begin{equation}
\Lc_{\p_{\theta}}(\n_{\rho}T^{\rho}{}_{\sigma})=\Lc_
{\p_{\vp}}(\n_{\rho}T^{\rho}{}_{\sigma})=0.
\end{equation}
For convenience I use a vielbein
frame, see Appendix A.4. In this formalism
the above condition reads $E_{2,3}(e_m{}^{\rho}e^n{}_{\sigma}T^m{}_n)=0$
and therefore $E_{2,3}T^m{}_n=0$ for all $m,n$
except $E_2T^2{}_3=-\frac{\cot\theta}{r}\,T^2{}_3$.
I start with writing down the symmetry conditions for the conservation
equation in a coordinate basis and then change to a vielbein frame
to calculate the covariant derivatives (the connection one-form on
a Schwarzschild spacetime is (\ref{con.4d})). The first new condition is
\begin{multline}
\Lc_{\theta}\left(\n_{\rho}T^{\rho}{}_{t}\right)
=\p_{\theta}\left(\n_{\vp}T^{\vp}{}_{t}\right)=\p_{\theta}
\left(\Ss\n_3T^3{}_0\right)\\
=\p_{\theta}\left(\frac{\Sss}{r}T^1{}_0+\frac{\cot\theta
\Ss}{r}T^2{}_0\right)=-\frac{1}{\sin^2\theta\cdot r}T^2{}_0=0.
\end{multline}
This component of the EM tensor must be identically
zero on a spherically symmetric spacetime. By the symmetry
of the coordinate directions $\theta$ and $\vp$ the component
$T^3{}_0=0$ must also vanish. In the same way I get
\begin{multline}
\Lc_{\theta}\left(\n_{\rho}T^{\rho}{}_{r}\right)
=\p_{\theta}\left(\n_{\vp}T^{\vp}{}_{r}\right)=\p_{\theta}
\left(\Ss\n_3T^3{}_1\right)\\
=\p_{\theta}\left(\frac{\Sss}{r}T^1{}_1+\frac{\cot\theta
\Ss}{r}T^2{}_1\right)=-\frac{1}{\sin^2\theta\cdot r}T^2{}_1=0,
\end{multline}
and hence $T^2{}_1=0$. From the symmetry between the $\theta$- and
$\vp$-coordinate also $T^3{}_1=0$ follows.
The next condition is
\begin{multline}
\Lc_{\theta}(\n_{\rho}T^{\rho}{}_{\theta})
=\p_{\theta}(\n_{\vp}T^{\vp}{}_{\theta})
=\p_{\theta}(r\n_3T^3{}_2)\\
=\p_{\theta}\left[\cot\theta(T^2{}_2-T^3{}_3)+\Ss T^1{}_2\right]
=-\frac{1}{\sin^2\theta}(T^2{}_2-T^3{}_3)=0,
\end{multline}
i.e. $T^2{}_2=T^3{}_3$ or in a coordinate basis
$\Th=T^{\vp}{}_{\vp}$. Finally, we have
\begin{multline}
\Lc_{\theta}(\n_{\rho}T^{\rho}{}_{\vp})=\Lc_{\theta}(\n_{\theta}
T^{\theta}{}_{\vp}+\n_{\vp}T^{\vp}{}_{\vp})
=\p_{\theta}\left[r\sin\theta(\n_2T^2{}_3+\n_3T^3{}_3)\right]\\
=\p_{\theta}\left(2\sin\theta\Ss T^1{}_3
+\cos\theta\,T^2{}_3\right)\\
=2\cos\theta\Ss T^1{}_3-\sin\theta\,T^2{}_3=0.
\end{multline}
If I bring the term in $T^2{}_3$ to the r.h.s., divide
the equation by the prefactors of the l.h.s. (if $\theta\neq\frac{\pi}{2}$),
and let a derivative $E_2$ act on it I obtain
\begin{multline}
2E_2T^1{}_3=0=\frac{1}{r\Ss\cos^2\theta}T^2{}_3+\frac{\tan\theta}{\Ss}E_2T^2{}_3\\
=\frac{1}{r\Ss\cos^2\theta}T^2{}_3-\frac{\tan\theta\cot\theta}{\Ss}T^2{}_3
=\frac{\tan^2\theta}{r\Ss}T^2{}_3.
\end{multline}
This means that I get two conditions, namely $T^2{}_3=0$
and $T^1{}_3=0$, whereby the latter has been guessed already
by symmetry considerations. Alternatively, this could have been seen already
from the above equation by setting $\theta=0$, respectively
$\theta=\frac{\pi}{2}$, because the components of the EM tensor are independent
of $\theta$.

To sum it up, by demanding $\Lc_{\p_{\theta},
\p_{\vp}}T^{\rho}{}_{\sigma}=0$ and for consistency\newline
$\Lc_{\p_{\theta},\p_{\vp}}\n_{\rho}T^{\rho}{}_{\sigma}=0$, the form of the 
EM tensor is constrained to \cite{chf77}
\begin{equation}
T^m{}_n=\left(\begin{array}{cccc}
T^0{}_0 & T^0{}_1 & 0 & 0 \\
-T^0{}_1 & T^1{}_1 & 0 & 0 \\
0 & 0 & T^2{}_2 & 0 \\
0 & 0 & 0 & T^2{}_2 \end{array}\right)\label{s-wave-1}.
\end{equation}
From now on I will always assume implicitly that the EM tensor has this
form! Because of $T^2{}_3=0$, the relations
\begin{equation}
E_iT^m{}_n=0\label{s-wave-2}
\end{equation}
now hold for all $m,n$ and $i=2,3$.

In the beginning of Section 2.1 I have shown that the EM tensor is still
conserved at the quantum level (\ref{EM.cons.quant}).
Its explicit form does not enter the symmetry
considerations of the current Section. Hence, I can simply
replace it by the quantum mechanical expectation
value of the EM tensor operator and obtain the same result:
the vacuum expectation value of the EM tensor on a spherically
symmetric spacetime has the non-vanishing components
\begin{equation}
\E{T^m{}_n}=\left(\begin{array}{cccc}
\E{T^0{}_0} & \E{T^0{}_1} & 0 & 0 \\
-\E{T^0{}_1} & \E{T^1{}_1} & 0 & 0 \\
0 & 0 & \E{T^2{}_2} & 0 \\
0 & 0 & 0 & \E{T^2{}_2} \end{array}\right)\label{s-wave-quant}.
\end{equation}
This follows directly from the fact that the geometry
is described as a classical physical system. The EM
tensor \emph{operator} clearly may break the spherical
symmetry and is constrained in no way as long as it is not
applied to physical states.

Again I emphasize that (\ref{s-wave-1}) is the form of the EM
tensor on a \emph{general, spherically symmetric spacetime}. I have shown
this for a Schwarzschild spacetime, but the results remain the
same if the metric is changed in the first block
$g_{\alpha\beta};\alpha,\beta\in\{t,r\}$, e.g. for a non-static
metric. At no point I have used the staticity condition
${\cal L}_{\p_t}T^{\rho}{}_{\sigma}=0$! The physical manifold
describing an evolving BH in fact \emph{does not possess
a timelike Killing field}. If such a symmetry were present
the radiation components of the EM tensor $T_{01}=T_{10}$
also would have to vanish.

In the following I will give some physical picture
to clarify the significance of the calculations in this
Section. In GR matter and geometry are intimately related
and the existence of some spacetime symmetry (that
is always accompanied by some Killing field) means that also
the matter-distribution has the same symmetry structure.
For instance, a spherically symmetric spacetime implies that
the fields on this spacetime are invariant under translations
into the direction of the spherical Killing fields.
On the other hand, one can scatter plane waves on a large BH
without significantly disturbing the spherical symmetry,
although strictly speaking the symmetry conditions are violated.
In this respect one might consider the part of the EM tensor
of the form (\ref{s-wave-1}) as the spherically symmetric (s-wave)
contribution of the radiation which possibly possesses modes
with higher angular momentum though the s-waves represent
the main contribution.

\subsubsection{Conservation Equation}

The conservation equation $\n_{\rho}T^{\rho}{}_{\sigma}=0$ consists of four 
independent equations. In a vielbein frame the first two of them read
\begin{eqnarray}
\n_mT^m{}_0&=&E_1 T^1{}_0+\left(\frac{2M}{r^2\Ss}+\frac{2\Ss}{r}\right)
T^1{}_0=0\\
\n_mT^m{}_1&=&E_1T^1{}_1+\left(\frac{M}{r^2\Ss}+\frac{2\Ss}{r}\right)T^1{}_1
\nonumber\\
&&-\frac{\Ss}{r}(T^2{}_2+T^3{}_3)-\frac{M}{r^2\Ss}T^0{}_0=0.
\end{eqnarray}
The third equation $\n_mT^m{}_2=0$ again gives $T^2{}_2=T^3{}_3$
which is already included by the representation (\ref{s-wave-1})
of the EM tensor, while the last equation $\n_mT^m{}_3=0$ is trivially
fulfilled. In a coordinate basis the two new equations have the form
\begin{eqnarray}
\p_rT^r{}_t&=&-\frac{2}{r}T^r{}_t\label{cons1}\\
\p_rT^r{}_r&=&-\left(\frac{2}{r}+\frac{M}{r^2\Sss}\right)T^r{}_r+\frac{2}{r}
\Th+\frac{M}{r^2\Sss}T^t{}_t,\label{cons2}
\end{eqnarray}
where I have set $\Th=T^{\vp}{}_{\vp}$.
(\ref{cons1}) has the exact solution
\begin{equation}
T^r{}_t=-\frac{K}{M^2r^2},\label{sol1}
\end{equation}
where $K$ is an integration constant. The solution of (\ref{cons2}) 
can formally be written as \cite{chf77}
\begin{equation}
T^r{}_r=\frac{1}{r^2\Sss}\left\{\frac{Q-K}{M^2}+
\int_{2M}^r\Bigm[MT(r')+2(r'-3M)\Th(r')\Bigm]dr'\right\}.
\label{sol2}
\end{equation}
Here $Q$ is another integration constant and $T$ is the trace of the 
EM tensor. The component $T^t{}_{t}$ has been eliminated
by the relation $T^t{}_t=T-T^r{}_r-2\Th$.
Therefore, the complete EM tensor only depends on \emph{two
independent constants $Q,K$ and two independent and unknown
functions  $T,\Th$ which I call the basic components}. The integration constants
will be determined by the boundary conditions imposed on the EM tensor.
Thereby different choices of boundary conditions will lead
to different quantum states.
The main difficulty lies in finding the quantum mechanical expectation values 
of the basic components $\E{T},\E{\Th}$.

% EDDINGTON-FINKELSTEIN gauge

% $T^r{}_t=T^{r'}{}_{x^-}=-\frac{K}{M^2r^2}$. From $T^r{}_r=T^{r'}{}_{r'}
% -\Sss^{-1}T^{r'}{}_{x^-}$ we get $T^{r'}{}_{r'}=1/r^2\Sss\cdot[(Q-2K)/M^2+f(r)]$.
% Accordingly, in the Unruh state $K=M^2f(\infty)/2,Q=0$ we have
% $T^{r'}{}_{r'}(\infty)=T^{x^-}{}_{x^-}(\infty)=0$ (if $T(\infty)=\Th(\infty)=0$).

\subsection{Dilaton Model}

The dilaton model has been invented to describe the
dynamics of spherically symmetric (scalar) matter on
a spherically symmetric four-dimensional spacetime
by a modified two-dimensional Einstein-Hilbert action.
If the matter does not exhibit spherical symmetry in $4d$
the dilaton model only describes the s-waves of an
angular-momentum decomposition, i.e. the spherically symmetric
part (this, certainly, only makes sense if the remaining
part causes negligible perturbations of the geometry).
Its name is due to a scalar field, the dilaton field, which
is part of the original four-dimensional metric and
appears as a scalar field in the $2d$ action.
In fact it represents no dynamical degree of freedom
(i.e. it is pure gauge) as already the
four-dimensional spherically symmetric action
possesses none: the Birkhoff theorem states that
there is no spherically symmetric gravitational radiation,
i.e. such systems (without matter) are static!
The four-dimensional manifold $M$ can be imagined as a two-dimensional
submanifold $L$ with Minkowski signature, spanned by the
coordinates\footnote{This can be any pair of coordinates describing $L$.
Symbolically I write a time- and radius-coordinate.} $t,r$,
where at each point two-spheres $S^2$ of varying size are
attached. If necessary I will mark the geometrical objects by an
index according to their associated (sub-)manifold (objects belonging
to the two-sphere $S^2$ are marked by an index $S$).
Sometimes I will only use an index $4$ or $2$ to emphasize
association to $M$ respectively $L$.
The physics shall not depend on the value of
the sphere-coordinates $\theta,\vp$, it \emph{does} instead depend
on the size of the sphere which by its intrinsic curvature
contributes to the total spacetime curvature.
This picture suggests that the four-dimensional
theory can indeed be described by the dilaton model
if the intrinsic curvature of the two-sphere, depending only
on the position of the sphere on $L$ and hence on $t,r$,
is added to the curvature of $L$ (more precisely, one also must
add an embedding term).
The procedure to compute the two-dimensional action
of the dilaton model and, in particular, its scalar curvature
is called \emph{spherical reduction} and carried out
in detail in Appendix D.

Before going into detail I mention some general
aspects of dimensional reduction. Although the gravitational
part of the model exhibits no dynamical degrees of freedom the
metric in general is non-static because radiation may occur
by the matter fields (and those certainly can radiate
by s-waves) -- this is exactly the situation to be described
in this work. Any four-dimensional spacetime that possesses
at least two independent Killing-fields can be reduced
to a two-dimensional model. There are even pure gravity models
that possess dynamical degrees of freedoms which are represented
by dilaton models in a two-dimensional action (for instance,
the Gowdy model \cite{gow71} describes a vacuum spacetime with cylindrical
symmetry and has two independent gravitational degrees of
freedom which are inherited by two dilaton fields in the $2d$ action).
The Schwarzschild spacetime, which is employed in the static
approximation, has (beside the spacelike Killing-fields)
a timelike Killing field and could thus be further reduced
to a one-dimensional model. Clearly, such a model would be trivial
and could not describe the evolution of a realistic BH.
Finally, the dilaton model is not restricted to scalar fields.
It can also be used to describe fermions on a spherically symmetric
spacetime.

The most general spherically symmetric four-dimensional
line-element can be written as\footnote{If confusion is possible
I will use different indices for coordinates and tensors
belonging to different (sub-)manifolds, see Appendix A.2.}:
\begin{equation}
ds^2=g^M_{\alpha\beta}(t,r)dx^{\alpha}dx^{\beta}-X(t,r)\left(d\theta^2
+\sin^2\theta\,d\vp^2\right).
\end{equation}
\emph{$X$ is the dilaton field}. It is a function of the coordinates
$x^{\alpha}$ on $L$ and its value gives the area of the two-sphere $S^2$
at the actual point. The explicit form of the dilaton field depends
on the choice of coordinate system on $L$. For a Schwarzschild spacetime
in Schwarzschild coordinates it becomes $X=r^2$. Generally (for non-static
spacetimes) some time-dependence may appear; however, as the dilaton
represents a gauge degree of freedom, the choice $X=r^2$ is always admissible.
The two-dimensional line-element of $L$ is simply given by
\begin{equation}
ds^2=g^L_{\alpha\beta}(t,r)dx^{\alpha}dx^{\beta},
\end{equation}
which means that the first block of the four-dimensional metric
$g_{\rho\sigma}$ can be identified with the metric on $L$:
$g^M_{\alpha\beta}\equiv g^L_{\alpha\beta}$.
In the following I will sketch the spherical reduction procedure
(the tedious part of the calculation is done in Appendix D).
I start from the four-dimensional scalar field action functional
(\ref{scalar action}). Then I replace all quantities by their reduced
expressions.
For the kinetic term $(\p S)^2$ the s-mode condition
${\cal L}_{\p_{\theta,\p_{\vp}}}S
=\p_{\theta,\vp}S=0$ ($S=S(t,r)$) globally leads to
$g^{\rho\sigma}(\p_{\rho}S)(\p_{\sigma}S)=g^{\alpha\beta}(\p_{\alpha}S)(\p_{\beta}S)$
which means it is sufficient to replace the four-dimensional metric by the
two-dimensional one. The spacetime measure transforms like
$\sqrt{-g_M}=X\sqrt{-g_L}$. The transformation of the scalar 
curvature is more complicated and carried out explicitly in Appendix D,
(\ref{R-sph.red.4d}).
After having replaced all expressions in the $4d$ action I can
integrate over the isometric coordinates $\theta,\vp$ and obtain
the two-dimensional action of the dilaton model
\begin{equation}
L_{dil}=\int_L\left\{XR+\frac{(\p X)^2}{2X}-2+X\left[\frac{(\p S)^2}{2}
-\frac{m^2S^2}{2}\right]\right\}\sqrt{-g_L}d^2x\label{2daction}.
\end{equation}
For convenience I have divided by a factor $4\pi$ that has to be
recast if the physical four-dimensional observables are considered.
The dilaton field has a kinetic term and is coupled non-minimally
to the scalar curvature. Of particular importance is the \emph{non-minimal
coupling of the scalar field $S$ to the dilaton}. If this coupling
were absent, the dilaton field would decouple from its dynamics
and the scalar field would only ``feel'' the intrinsic geometry of the
two-dimensional manifold $L$ described by the metric $g_L$.
The dilaton would then become superfluous and hence I will call
the minimally coupled model also \emph{intrinsic two-dimensional theory}.

Here I will not show explicitly the classical equivalence of the
action (\ref{2daction}) with (\ref{scalar action}) for its s-mode solutions,
for this see e.g. \cite{grk00}. 
Just for completeness I state the existence of a first order formalism
by the introduction of auxiliary fields that facilitates path integrals
of the geometric variables \cite{grk00}. In this work the geometry remains
classical, hence the form (\ref{2daction}) is sufficient.

\subsubsection{Reconstruction of the $4d$ Energy-Momentum Tensor}

We have seen that the EM tensor describing s-waves on a spherically
symmetric, four-dimensional spacetime has only $4$ independent
components (\ref{s-wave-1},\ref{s-wave-quant}). In the $4d$ theory
the whole EM tensor is obtained by variation of the matter
action for the metric. In the dilaton model the variation for
the metric yields a $2\times2$-matrix that, up to a factor $X$,
corresponds to the first quadrant of the four-dimensional EM
tensor:
\begin{equation}
T^2_{\alpha\beta}=\frac{2}{\sqrt{-g_L}}\frac{\delta L_{dil}^m}
{\delta g^{\alpha\beta}}=X\cdot\left\{(\p_{\alpha}S)(\p_{\beta}S)
-\frac{g_{\alpha\beta}}{2}\left[(\p S)^2-m^2S^2\right]\right\}.
\end{equation}
Note that the scalar $(\p S)^2$ is contracted by the $2d$ metric
and hence differs by a term $g^{\theta\theta}(\p_{\theta}S)^2
+g^{\vp\vp}(\p_{\vp}S)^2$ from the corresponding expression in
(\ref{EM-scalar}). For s-waves this term vanishes
because of the symmetry condition ${\cal L}_{\theta,\vp}S=0$
and thus the relation
\begin{equation}
T_{\alpha\beta}:=\frac{1}{4\pi X}T^2_{\alpha\beta}\label{EM-tensor.4d-2d}
\end{equation}
between the components of the EM tensor in the two-dimensional
dilaton model and the first quadrant of the four-dimensional EM tensor
can be established (I have not marked the $4d$ EM tensor by an index
$4$ because it is computed here in the dilaton model).
Equation (\ref{EM-tensor.4d-2d})
can be interpreted as the s-wave approximation to the
four-dimensional EM tensor $T^4_{\rho\sigma}$.

The remaining non-vanishing component of the EM tensor is
$\Th$. In $4d$ it is given by
\begin{equation}
(T^{\theta}{}_{\theta})_4=\frac{1}{2}\left[m^2S^2-g^{\alpha\beta}
(\p_{\alpha}S)(\p_{\beta}S)\right]=\frac{1}{2}\left[m^2S^2
-(\p S)^2_{(2)}\right].
\end{equation}
The index $(2)$ in the last expression indicates that
the contraction may be performed by the two-dimensional
metric of the $2d$ manifold $L$. This shows that
$\Th$ only consists of quantities accessible in the
dilaton model, \emph{even if the s-wave condition is not
fulfilled by $S$}. It can be seen easily that this
component is obtained by varying the dilaton action
(more precisely its matter part $L_{dil}^m$) for the dilaton field:
\begin{equation}
\Th:=-\frac{1}{4\pi\sqrt{-g_L}}\frac{\delta L_{dil}^m}{\delta X}.\label{Th}
\end{equation}
At the classical level this relations holds for fields
with arbitrary angular momentum (as long as the geometry
is almost perfectly spherically symmetric, i.e. if the spacetime
is almost vacuum). The component $\Th$ has no intrinsic
meaning in the two-dimensional dilaton model.
For convenience I define a quantity that differs from
(\ref{Th}) by a factor $4\pi$
\begin{equation}
(\Th)_2:=-\frac{1}{\sqrt{-g_L}}\frac{\delta L_{dil}^m}{\delta X},
\label{Th-2}
\end{equation}
as the formal analogue of $(\Th)_4$ in the dilaton model.

At the classical level one strictly has
\begin{equation}
T_{\alpha\beta}^4\bigm|_{s-modes}=T_{\alpha\beta}\,\,,\,\,
(\Th)_4=\Th.
\end{equation}
This follows directly from the equivalence of the classical
EOM. However, it is not yet obvious that this equivalence extends
to the level of quantum mechanical expectation values.
Anyway, I will use equations (\ref{EM-tensor.4d-2d},\ref{Th})
to reconstruct the four-dimensional EM tensor from the expectation
values $\E{T_{\alpha\beta}}_2,\E{\Th}_2$ computed in the dilaton model.

\subsubsection{Non-Conservation Equation in $2d$}

As we have seen in Section 2.1
the basis of the CF approach in $4d$ is the energy-momentum
conservation equation. It allows to calculate all components
of the EM tensor on a Schwarzschild spacetime from the
four-dimensional trace $T$ and the $\Th$-component.

It is a nice feature of the dilaton model that it
reveals an analogue of that.
If one extends the diffeomorphism invariance
to the dilaton field $X$, which is necessary in $2d$ because there the dilaton
is just a scalar field like $S$,
the conservation of the two-dimensional EM tensor is spoilt by
the appearance of an extra-term; the latter turns out to be
nothing but the $\Th$-component of the EM tensor. If the
components of the EM tensor are transformed into four-dimensional
components (see last Section) one recovers the $4d$ conservation
equation. In the present Section the EM tensor is always understood
as the two-dimensional one if not indicated otherwise.

Under a diffeomorphism transformation $(x')^{\alpha}=x^{\alpha}+\xi^{\alpha}$
the metric changes as $\delta_{\xi}g^{\alpha\beta}=\Lc_{\xi}g^{\alpha\beta}=
-\n^{\alpha}\xi^{\beta}-\n^{\beta}\xi^{\alpha}$. Because the
dilaton field is an ordinary scalar field in the two-dimensional action
it transforms as $\delta_{\xi}X=\Lc_{\xi}X=\xi^{\alpha}\p_{\alpha}X$.
The matter part of the $2d$ action changes as
\begin{multline}
\delta_{\xi} L_{dil}^m[x\to x']=\int_L\delta_{\xi}g^{\alpha\beta}
\frac{\delta L_{dil}^m}{\delta g^{\alpha\beta}}d^2x+\int_L\delta_{\xi}X
\frac{\delta L_{dil}^m}{\delta X}d^2x+\int_L\delta_{\xi}S
\frac{\delta L_{dil}^m}{\delta S}d^2x\\
\stackrel{EOM}{=}\int_L\left\{-(\xi^{\alpha;\beta}+\xi^{\beta;\alpha})
T_{\alpha\beta}
+\xi^{\alpha}\p_{\alpha}X[(\p S)^2-m^2S^2]\right\}\frac{\sqrt{-g_L}}{2}d^2x\\
\stackrel{EOM}{=}\int_L\xi^{\alpha}\left\{2\n^{\beta}T
_{\alpha\beta}+\p_{\alpha}X[(\p S)^2-m^2S^2]\right\}
\frac{\sqrt{-g_L}}{2}d^2x.\label{diffinv}
\end{multline}
The result is the ``non-conservation equation'' \cite{kuv99}
\begin{equation}
\n_{\alpha}T^{\alpha}{}_{\beta}=\frac{\p_{\beta}X}{2}[m^2S^2-(\p S)^2]
=-\frac{\p_{\beta}X}{\sqrt{-g_L}}\frac{\delta L_{dil}^m}{\delta X}
\label{non-cons-eq.},
\end{equation}
whereas in an intrinsically two-dimensional model (without dilaton field)
the r.h.s. would be zero. (\ref{non-cons-eq.}) can be checked by
considering s-mode solutions in the four-dimensional conservation equation
($T^m{}_n$ is now the four-dimensional EM tensor, the index $m$ runs from $0$
to $d$):
\begin{multline}
0=\n_mT^m{}_c=\n_aT^a{}_c+\n_iT^i{}_c\\
=E_aT^a{}_c+\omega^a{}_m(E_a)T^m{}_c-\omega^m{}_c(E_a)T^a{}_m
+\omega^i{}_m(E_i)T^m{}_c-\omega^m{}_c(E_i)T^i{}_m\\
=[E_aT^a{}_c+\omega^a{}_b(E_a)T^b{}_c-\omega^b{}_c(E_a)T^a{}_b]
+\omega^i{}_a(E_i)T^a{}_c-\omega^i{}_c(E_k)T^k{}_i\\
=\n_a\frac{(T^a{}_c)_2}{4\pi X}+\frac{E_aX}{X}T^a{}_c-\frac{E_cX}{2X}T^i{}_i
=\frac{1}{X}\left[\frac{1}{4\pi}\n_a(T^a{}_c)_2-(E_cX)\Th\right].
\end{multline}
Here I have used the s-wave conditions (\ref{s-wave-1},\ref{s-wave-2}) and
the relations between the connections on $M$ and $L$ Appendix D
(\ref{con.sph.red.}). By inserting (\ref{Th}) one indeed arrives at
(\ref{non-cons-eq.}).

The CF representation in the dilaton model is obtained
by solving the non-conservation equation for a (four-dimensional)
Schwarz\-schild spacetime. In particular, this means that the gauge for
the dilaton field has to be fixed as $X=r^2$. To keep track of
the dilaton field I leave it as $X$ in the equations as long as possible.
In contrast to the four-dimensional conservation equation its
two-dimensional analogue only produces two independent equations that,
however, contain the whole information (remember that the two additional
equations in $4d$ were redundant):
\begin{eqnarray}
\p_rT^r{}_t&=&-\frac{\p_tX}{\sqrt{-g_L}}\frac{\delta L_{dil}^m}{\delta X}=0\\
\p_rT^r{}_r&=&\frac{M}{r^2\Sss}(T^t{}_t-T^r{}_r)-\frac{\p_rX}{\sqrt{-g_L}}
\frac{\delta L_{dil}^m}{\delta X}.
\end{eqnarray}
The corresponding solutions can be written as
\begin{eqnarray}
T^r{}_t&=&-\frac{K_2}{M^2}\label{2dsol1}\\
T^r{}_r&=&\frac{1}{\Sss}\biggm\{\frac{Q_2-K_2}{M^2}\nonumber\\
&&+\int_{2M}^{r}\left[\frac{MT}{(r')^2}-\left(1-\frac{2M}{r'}\right)
\frac{\p_rX}{\sqrt{-g_L}}\frac{\delta L_{dil}^m}{\delta X}\right]dr'\biggm\},
\label{2dsol2}
\end{eqnarray}
where I have substituted $T^t{}_t=T-T^r{}_r$.

They are classically equivalent to the four-dimensional solutions
(\ref{sol1},\ref{sol2}). To see this I first show the relation between
the trace of the four-dimensional EM tensor for s-waves and the
trace of the two-dimensional EM tensor:
\begin{equation}
T_4=g^{\rho\sigma}T^4_{\rho\sigma}=g^{\alpha\beta}T^4_{\alpha\beta}+2T^{\theta}
{}_{\theta}\stackrel{s-modes}{=}\frac{T_2}{4\pi X}
-\frac{1}{2\pi\sqrt{-g_L}}\frac{\delta L_{dil}}{\delta X}.
\label{T4-T2}
\end{equation}
If I put this into the solution of the four-dimensional conservation
equation (\ref{sol1},\ref{sol2}) and replace the components of the EM tensor
by the relations (\ref{EM-tensor.4d-2d},\ref{Th}), I obtain
the solutions (\ref{2dsol1},\ref{2dsol2}) with the constants multiplied
by a factor $4\pi$:
\begin{equation}
K_4=4\pi K_2\,\,,\,\,Q_4=4\pi Q_2.
\end{equation}
As I will fix these constants in the particular model in which
I work I will never use this relation explicitly, hence I can
omit the dimension index.
Thus the CF approach is recovered in the dilaton
model by means of the non-conservation equation.
The whole EM tensor is still determined by two constants
and two functions (the basic components), which in $4d$ were $T_4,(\Th)_4$
and in the dilaton model are $T_2,(\Th)_2$. Note that the indices $2,4$
emphasize by which model the marked object has been calculated (this
difference becomes important at the quantum level).
The reconstructed components carry no index as they live
in $4d$ but are calculated by the dilaton model.
If I work with the dilaton
model I will compute all components of the EM tensor in $2d$ 
(by the non-conservation equation) and
finally reconstruct the four-dimensional EM tensor by
the relations (\ref{EM-tensor.4d-2d},\ref{Th}).

In two dimensions the light-cone gauge (\ref{light-cone}), Appendix A.5.1
is particularly useful. The non-conservation equation there becomes
\begin{eqnarray}
\p_+T_{--}&=&-\p_-T_{-+}+2(\p_-\rho)T_{-+}-
\frac{e^{-2\rho}}{4}(\p_-X)(\p S)^2\\
\p_-T_{++}&=&-\p_+T_{-+}+2(\p_+\rho)T_{-+}-
\frac{e^{-2\rho}}{4}(\p_+X)(\p S)^2.
\end{eqnarray}
In the static approximation $\p_+=-\p_-=\frac{1}{2}\p_{r_{\ast}}$ both
equations become identical! Therefore, the components $T_{--},T_{++}$
only differ by an integration constant, namely $K$. It is this constant,
being the difference of incoming and outgoing flux, that determines
the total flux, see (\ref{total flux},\ref{2dsol1}).

\subsubsection{Quantum Equivalence}

What fundamentally distinguishes quantum theory from classical
theory is the fact that one can only make statements on the
average behaviour of a particle or system. The expectation value
of the EM tensor tells us something about the average energy
density and flux of a sufficiently large number of measured
particles. A single particle's properties may differ infinitely much
from this average value, though quantum theory can at least
give the probability for this to occur (which may be infinitely small).

In particular, the probability wave function might describe
a ``classical'' particle (if interpreted in this ``wrong'' way)
that violates the spherical symmetry condition if applied
directly to the wave function (and not to the expectation
values) as it would suggest the classical interpretation.
From the classical point of view a spherically symmetric
scalar field configuration (s-waves) has to fulfil
${\cal L}_{\theta}S=\p_{\theta}S=0$ (and further ${\cal L}_{\vp}S=0$).
In the quantum mechanical interpretation the physical
observables are the expectation values of the quantum operators
and the symmetry condition must therefore hold only when
applied to the expectation values in physical states.

This means that the classical condition $\p_{\theta}S=0$
must be replaced by $\p_{\theta}\E{S}=0$ when
going over to the quantum picture. Note that for a classical
field $\p_{\theta}S=0$ implies that $(\p_{\theta}S)^2=0$.
The same is true for an operator $\hat{O}:=\p_{\theta}S$:
if $\hat{O}\left|\text{phys}\right>=0$ for all physical states
$\left|\text{phys}\right>$, it then follows automatically that
$\hat{O}^2\left|\text{phys}\right>=0$. However, the condition
$\p_{\theta}\E{S}=0$ does not necessarily imply $\E{(\p_{\theta}S)^2}=0$.
This means that the expectation value $\E{(\p_{\theta}S)^2}$
might contribute to the EM tensor though the mean field $\E{S}$ fulfils
the symmetry condition. If this is indeed the case
the equivalence between the dilaton model and the $4d$ theory
is broken because the dilaton model cannot produce such a term.

In particular this term appears in the trace of the four-dimensional
EM tensor
\begin{equation}
T_4=2m^2S^2-(\p S)^2=2m^2S^2-g^{\alpha\beta}\p_{\alpha}S\p_{\beta}S
-2g^{\theta\theta}(\p_{\theta}S)^2.
\end{equation}
Here I have used the relation $g^{\theta\theta}(\p_{\theta}S)^2
=g^{\vp\vp}(\p_{\vp}S)^2$ that follows from the spherical symmetry
condition of the EM tensor $T^{\theta}{}_{\theta}=T^{\vp}{}_{\vp}$:
\begin{multline}
T^{\theta}{}_{\theta}=g^{\theta\theta}(\p_{\theta}S)^2
+\frac{1}{2}\left[m^2S^2-(\p S)^2\right]\\
=g^{\vp\vp}(\p_{\vp}S)^2+\frac{1}{2}\left[m^2S^2-(\p S)^2\right]
=T^{\vp}{}_{\vp}.
\end{multline}
Because it is derived purely from symmetry considerations this relation
also holds for expectation values:
$\E{g^{\theta\theta}(\p_{\theta}S)^2}_4=\E{g^{\vp\vp}(\p_{\vp}S)^2}_4$.

For classical s-waves the condition $(\p_{\theta}S)^2=0$
guarantees that relation (\ref{T4-T2}) holds and that
the solutions (\ref{sol1},\ref{sol2}) of the four-dimensional
conservation equation are equivalent to those of the
two-dimensional one (\ref{2dsol1},\ref{2dsol2}).

At the quantum level this equivalence might be broken
even if the symmetry conditions for the mean field $\p_{\theta}\E{S}=0$
and the EM tensor (\ref{s-wave-quant}) hold, because
the l.h.s. and r.h.s. of (\ref{T4-T2}) differ by the
expectation value $\E{g^{\theta\theta}(\p_{\theta}S)^2}_4$.

Also other expectation values like $\E{S^2}$
may differ because in the four-dimensional path integral
functions $S$, having an angular-dependence, may contribute
to the expectation values, but there the situation is less transparent.
\\
\\
Although the expectation values entering the
non-conservation equation might deviate from those
calculated by the $4d$ theory the equation itself
is still valid at the quantum level. In other words,
\emph{the diffeomorphism invariance is not broken
in the quantised theory}. 

To show this I introduce the non-minimal coupling of the scalar field
to the dilaton field into the path integral.
Further, I must add the complete geometric part $L_g$ of the dilaton action
because it encodes the dynamics of the dilaton field.
A $2d$ diffeomorphism transformation applied to the path integral
yields the quantum non-conservation equation:
\small
\begin{eqnarray}
0&=&\delta_{\xi}Z_{2d}[g]={\cal N}\delta_{\xi}\int{\cal D}S\cdot
e^{i\int_L\left[XR+\frac{(\p X)^2}{2}+X\frac{(\p S)^2}{2}
-X\frac{m^2S^2}{2}\right]\sqrt{-g_L}d^2x}\nonumber\\
&=&{\cal N}\int{\cal D}S\int_L\biggm\{\delta_{\xi}g^{\alpha\beta}
\frac{\delta L_g+L_{dil}^m}{\delta g^{\alpha\beta}}+\delta_{\xi}X
\frac{\delta L_g+L_{dil}^m}{\delta X}+\delta_{\xi}S\frac{\delta L_{dil}^m}{\delta S}
\biggm\}d^2x\cdot e^{i(L_g+L_{dil}^m)}\nonumber\\
&=&\E{\1}i\int_L2\xi^{\alpha}\n^{\beta}\biggm\{X\left(R_{\alpha\beta}
-g_{\alpha\beta}\frac{R}{2}\right)+g_{\alpha\beta}\square X-\n_{\alpha}
\n_{\beta}X\nonumber\\
&&+\frac{(\n_{\alpha}X)(\n_{\beta}X)}{2}-g_{\alpha\beta}\frac{(\p X)^2}{4}
\biggm\}\sqrt{-g_L}d^2x\nonumber+i\int_L\xi^{\alpha}\n^{\beta}
\E{T_{\alpha\beta}}\sqrt{-g_L}d^2x\nonumber\\
&&+i\int_L\xi^{\alpha}(\p_{\alpha}X)\biggm\{\E{\1}[R-\square X]
+\frac{1}{2}\E{(\p S)^2-m^2S^2}\biggm\}\sqrt{-g_L}d^2x
\nonumber\\
&&+\lim_{y\to x}{\cal N}\int{\cal D}S\int_L\xi^{\alpha}
\left[\p_{\alpha}S\frac{\delta}{\delta S(y)}\right]e^{i\int_L\frac{X}{2}
[(\p S)^2-m^2S^2]\sqrt{-g_L}d^2x}d^2x\nonumber\\
&=&\E{\1}i\int_L\xi^{\alpha}\biggm\{2R_{\alpha}{}^{\beta}\n_{\beta}X
-R\n_{\alpha}X+2\n_{\alpha}\square X-2R_{\alpha}{}^{\beta}\n_{\beta}X
-2\n_{\alpha}\square X\nonumber\\
&&+(\n^{\beta}\n_{\alpha}X)\n_{\beta}X+(\n_{\alpha}X)\square X
-(\n_{\alpha}\n^{\beta}X)\n_{\beta}X\biggm\}\sqrt{-g_L}d^2x+\dots\nonumber\\
&=&i\int_L\xi^{\alpha}\biggm\{\n^{\beta}\Bigm(\E{T_{\alpha\beta}}
+\frac{ig_{\alpha\beta}}{\sqrt{-g_L}}\lim_{y\to x}\delta(x-y)\Bigm)
\nonumber\\
&&+\frac{\p_{\alpha}X}{2}\E{(\p S)^2-m^2S^2}\biggm\}\sqrt{-g_{L}}d^2x.
\end{eqnarray}
\normalsize
Again, I interpret the contribution of the delta-function
as the infinite zero-point energy of the quantised scalar field $S$.
Upon substraction of this term, the renormalised EM tensor
fulfils
\begin{equation}
\n^{\beta}\E{T_{\alpha\beta}}_{ren}=\frac{\p_{\alpha}X}{2}
\E{m^2S^2-(\p S)^2}\label{EM.cons.quant.dil}
\end{equation}
which is indeed the expectation value of the non-conservation 
equation (\ref{non-cons-eq.}).

\subsection{Basic Components - Trace Anomaly}

By the CF method one can calculate
all components of the EM tensor starting from two \emph{basic
components} which are the trace of the EM tensor $T$
and the $\Th$-component (and in the dilaton model the formal analogue
of the latter).
This will turn out to be a great advantage at the
quantum level because the basic components are independent
of the quantum state (see Section 2.4.2).
For convenience I collect their expectation
values in Table 1 ($W$ is the effective dilaton action).
\\
\\ \Large
\hspace*{3.2cm}\begin{tabular}{|c|c|c|}
\hline
\large 4d & \large $\E{T}_4$ & \large $\E{\Th}_4$ \\ \hline
\large 2d & \large $\E{T}_2$ & \large $\E{\Th}_2=-\frac{1}{\sqrt{-g_L}}
\frac{\delta W}{\delta X}$ \\ 
\hline\end{tabular}\\ \\ \small\hspace*{6cm}Table 1\\
\normalsize
According to my opinion this state-independence of the basic
components is the main reason why the CF approach
is this useful. Namely, to fix the quantum state directly in the
effective action is much trickier than to do this simply by
adjusting the constants $Q,K$. Besides, the explicit use
of the conservation equation guarantees that the energy-momentum
conservation, which has been shown to be manifest at the level
of expectation values, is always fulfilled.
Christensen and Fulling themselves pointed out that the
main advantage of their approach lies in the knowledge of the
trace anomaly for general (even) dimensions. According to them
this reduces the problem to finding the expectation value $\E{\Th}$.
Indeed, in the dilaton model (which was not examined by
CF) one can refer to the trace-anomaly to obtain
the expectation value $\E{T}_2$. But then one still misses the
other basic component which has to be derived in a quite different
manner. In $4d$ the trace anomaly does not enter at all if one
considers minimally coupled scalars (as I do). To clarify my
point of view in these matters I will shortly discuss the notion
of general couplings and how they affect the basic components.

In a non-dilatonic gravitational action (i.e. there is no dilaton field,
$S$ is the only field apart from the metric) a scalar field is
said to be coupled non-minimally to gravity if the Lagrangian
contains a term $\xi S^2R$, where $\xi\neq0$ (see (\ref{non-min.action})).
For dimensional reasons the scalar field appears quadratically and
$\xi$ is a dimensionless constant. If the scalar field is massless
one can find in any dimension a value for $\xi$ such that the trace
of the EM tensor vanishes on-shell. In this case one says that
the scalar field is \emph{conformally coupled}. Two dimensions are
somewhat particular as there $\xi=0$ corresponds to conformal coupling;
further, the trace then already vanishes off-shell.

During the last years
theories with non-minimally coupled scalar fields have become fashionable
again in different areas of gravitational physics. Already Kaluza
discovered that the compactification of an empty five-dimen\-sional
spacetime produces a non-minimally coupled scalar field
in four dimensions, but he avoided an interpretation by
setting it constant \cite{kal21}. Jordan adopted his calculations and interpreted
this scalar field as a local gravitational coupling constant \cite{jor55},
following ideas of Mach \cite{dic62} and Dirac.
In order to find a theoretical description of the recently discovered
far-distant accelerating galaxies this
idea was resurrected and the non-minimally coupled scalar field
was named ``quintessence'' \cite{zws99}. In this course various
potentials were invented to mimic the late-time evolution of
the expanding universe, carried by an increasing cosmological constant.
Because they are also plagued with compactification problems,
string theorists and SuperGravitationalists occupy themselves
by examining non-minimally coupled scalar fields and deriving scalar
potentials that serve the cosmologists to fit their data.
A special role thereby play conformally coupled scalar fields:
the string coordinates are ``conformally coupled'' scalar fields
because they live on a two-dimensional manifold, the string, and
their conformal coupling coincides with minimal (i.e. no) coupling.
What makes the conformal coupling interesting (and gives it its name)
is that it introduces a new symmetry into the theory: conformal invariance.
If the metric is conformally transformed in an active
way\footnote{A conformal transformation may also be a (passive!)
coordinate transformation that leaves the metric \emph{components} invariant
up to multiplication by a constant, i.e. angles do not change but
distances do! In contrast to that, if the metric (not only its components)
is multiplied by some function one speaks of an active conformal transformation.
In this case the invariant line-element of the spacetime is changed
and one has in fact a new manifold. All EOM in GR
can be written in generally covariant form which means
that arbitrary (passive) coordinate transformations leave these equations
form-invariant. The physically motivated invariances, like local
Lorentz invariance, are essentially valid as \emph{active} transformations,
consider e.g. a physically performed boost at a certain spacetime point.}
as $g_{\rho\sigma}\to\tilde{g}_{\rho\sigma}=\Omega^2(x)g_{\rho\sigma}$
the action changes as
\begin{equation}
\delta_{g}L_m[g,S]=(\Omega^2-1)g_{\rho\sigma}\frac{\delta L_m[g,S]}
{\delta g_{\rho\sigma}}\propto T.
\end{equation}
This shows that the action is \emph{form-invariant} under a
conformal transformation if the scalar field is conformally coupled.
Note that GR is \emph{not invariant under
conformal transformations} \cite{ghk00b}!
Namely, what characterises a general relativistic spacetime
is its geometry which is described by the metric. If the
metric $g$ is conformally transformed to a metric $\tilde{g}$
it does not describe the same spacetime anymore. By a singular conformal
transformation one could even map a Schwarzschild spacetime
onto flat spacetime \cite{qui99}. This is not so for a the string worldsheet --
the physical content is purely described by its topology and is
therefore conformally invariant. Beside string theory
conformal invariance appears in solid state physics in certain
phenomena like ferromagnetism. This in the past suggested that the conformal
group, including the Lorentz-group, might be a fundamental symmetry
group in nature. A crucial test is whether the conformal
symmetry of a system survives the quantisation procedure.
If a classical system is invariant under conformal transformations
but the quantised system is not one speaks of a \emph{conformal
anomaly}, see Appendix C.2. Because the breaking of the conformal invariance
is accompanied by an acquisition of a non-vanishing trace of the EM tensor one
also speaks of a \emph{trace anomaly}. Indeed, one can show
that a conformally coupled scalar field possesses a non-vanishing
trace of the quantised EM tensor (\ref{trace-anomaly4d},\ref{trace-anomaly2d}).
Generically the renormalisation procedure in QFT breaks the
conformal invariance (only string theory seems to escape this problem).
Also this, and the non-invariance of GR, make it highly unlikely
that conformally coupled matter correctly describes Nature.

In this thesis I will always assume \emph{minimal coupling\footnote{In the following
I will use the notion of minimal or non-minimal coupling exclusively for the
coupling of the scalar field to the dilaton in $2d$.}
of the scalar fields in the $4d$ action}. Nevertheless, there is
one reason why in the context of Hawking radiation matter has
often been coupled conformally in the $4d$ action: by coincidence
the quantum mechanical expectation value of $T$ turns out
to be computable in a surprisingly simple manner if the
classical theory was conformally invariant, i.e. if $\E{T}$ is
a trace anomaly (in Appendix C.2 I calculate the trace anomaly
in $4d$ and $2d$). Unfortunately, this does
not help in finding the other basic component $\E{\Th}$.
Further, one has to decide whether one couples the scalar
field conformally in $4d$ or in $2d$ because the spherical
reduction destroys this property of the $4d$ action!
With my choice of minimal coupling in $4d$ the scalar field
is automatically conformally (minimally) coupled in $2d$ (because $\xi=0$,
see Appendix C). Thus I can use (only in the dilaton model)
the trace anomaly to calculate $\E{T}_2$. All other basic
components (in $4d$ and $2d$) must be derived by more involved
computations.
Finally, I mention that the scalar field is non-minimally
coupled to gravity in a different sense: namely, its
Lagrangian is multiplied by the dilaton field $X$ which
by itself is coupled to the scalar curvature\footnote{If the
scalar field is minimally coupled in $2d$ the dilaton model
becomes intrinsically two-dimensional and the dilaton field
can be removed completely.}. This type
of non-minimal coupling has nothing to do with the coupling
of the scalar field to the scalar curvature discussed above
and it does not interfere with it: the scalar field
remains conformally coupled in $2d$, in whatever way it is coupled
to the dilaton field\footnote{It is possible to spherically reduce 
a four-dimensional scalar-tensor theory \cite{ghk00a}. The resulting
two-dimensional model is characterised by non-minimal coupling
to the dilaton \emph{and} to the scalar curvature -- the latter clearly
cannot be conformal in $2d$!}. The presence of the dilaton
changes the form of the conformal anomaly but not the fact that
it is an anomaly.

\subsection{Boundary Conditions}

The constants $K,Q$ that appear in the CF 
approach are related to the boundary conditions of the expectation
value of the EM tensor (of those components obtained from
the conservation equation). The boundary conditions in the
asymptotic region are connected to the quantum state
of the system as they determine the occupied states on the
boundary of the manifold. A crucial point is to show
that the basic components are independent of the boundary
conditions and hence of the quantum state.

They can be used to set initial
conditions of the scalar field on some time-slice
in the past. The evolution of the mean ``scalar field'' $\St{S}$,
that contains sufficiently many particles such as to be considered
as a statistical mixture with some average energy and momentum,
obeys classical EOM according to the Ehrenfest theorem.
Clearly one cannot make predictions on the trajectory
of a single particle that scatters on the BH,
but the total flux of all particles evolves in a deterministic
way. Hence, we have to consider the state of the system as
a multi-particle state whose actual energy, as measured
at a certain instant of time, is rather insensitive to
the state of single particles.
The \emph{vacuum state is then defined as the state in which the
``average'' energy  of a particle, i.e. the energy density,
becomes minimal}.

If the system is non-static a quantum state can only be defined
on a certain time-slice. In the case of a realistic BH
spacetime such a time-slice is characterised by the actual
mass of the BH. The quantum state then evolves simultaneously
with the BH evolution. In particular, the vacuum state
of an evolving BH differs on each time-slice and is therefore
unstable -- this instability of the vacuum is responsible for
the particle creation and vice versa.

By its construction the CF method is
restricted to the description of matter fields on a
static Schwarzschild spacetime. If one fixes the initial
conditions such that a steady flux of incoming particles
scatters on the BH, sooner or later it will affect
the geometry by increasing the BH mass, no matter how
small the flux is. Nevertheless, if the BH is large
(compared to the matter contribution of the scalar field) and
evolves slowly enough that it can be considered static
the CF method can be applied to calculate the
local flux. Further, the outgoing flux is independent
of the incoming one (as long as its gravitational effect is
very small compared to that of the BH), it solely
depends on the BH mass; the BH steadily produces
some amount of Hawking radiation, while the (spherically symmetric)
incoming flux is swallowed by it. This means that the (arbitrary) fixation
of the asymptotic outgoing radiation is not a physically sensible
(though mathematically permissible) boundary condition.
For instance, if the outgoing flux is set to zero this causes
divergences of the EM tensor at the horizon in global coordinates
(the Hawking temperature goes to infinity because the heat cannot be carried off).
Vice versa, the regularity condition of the EM tensor on the
future horizon completely determines the outgoing Hawking flux
which is the quantity of primary interest. What remains is to fix
the incoming flux which now uniquely characterises the quantum state
(if we insist on the regularity at the horizon). If it does not balance
exactly the outgoing flux the BH grows or shrinks.
The first scenario is in fact realized by heavy BHs
whose Hawking temperature (\ref{Hawking temp.}) lies below
the temperature of the universe (background radiation).
Small BHs radiate at very high temperatures and
shrink rapidly, thereby leaving the quasi-static phase
which is accessible to the CF method.
The physically most interesting state is the one, where the
incoming flux is zero. I will show below that this is indeed
the state of lowest energy density and I will therefore
call it the vacuum state of the system.

\subsubsection{Quantum States}

I define a \emph{quantum state} as a
\emph{complete set of boundary conditions} that fixes
uniquely the expectation value of the EM tensor.
On the one hand these conditions
are imposed explicitly on the fields (or Green functions),
on the other hand they are introduced via the constants
$K,Q$ of the CF representation.
The former kind concerns the basic components, while
the latter determines (through the energy-momentum conservation)
the dynamical components of the EM tensor that possess
radiative terms of the order $r^{-2}$. In the following I
discuss the connection between the constants $K,Q$ and the
quantum states.

In principle $K$ and $Q$ may have arbitrary values, each
combination corresponding to different quantum states.
There are \emph{three states} that are of particular interest
in the examination of BH radiation.

The Boulware state $\left|B\right>$ can be defined by demanding that the 
incoming as well as outgoing flux is zero (and consequently the total flux). 
These boundary conditions guarantee that there are no occupied real
particle states in the asymptotic region. Further, the spacetime is static like
the classical Schwarzschild spacetime because the total asymptotic flux
is zero and hence the BH mass does neither increase nor decrease.
Unfortunately, the EM tensor in $\St{B}$ inevitably becomes (quadratically)
\emph{divergent on the horizon} in global coordinates.
This feature is rather unphysical since there is no principal obstruction
to measure the flux at (or even behind) the horizon.
Therefore I will not consider $\left|B\right>$ as a physical
quantum state of the system, although it is admissible with respect
to energy-momentum conservation. This peculiarity is shared by all
states that are characterised (among other things) by a non-vanishing
constant $Q$. Despite this physical drawback, the Boulware state
is of conceptual interest because it turns out to be the
\emph{natural state of the effective action}.

In the Unruh and Hartle-Hawking states $\left|U\right>,\left|H\right>$
one demands from the beginning that the flux remains finite at the 
(future) horizon in global coordinates (which are regular at the horizon).
This already implies that there is some positive outgoing flux that
is fixed alone by this boundary condition which is realized by setting
$Q=0$, see below.
The second condition handles the incoming flux by fixing $K$.
It can be chosen arbitrarily
between zero and infinite (but not smaller than zero to maintain
the weak energy condition).

In the $\left|H\right>$-state the incoming flux is set to
the same value as the outgoing flux, thereby keeping the
spacetime static. Thus the total flux is zero which corresponds to $K=0$.
One says that the BH is in \emph{thermal equilibrium
with some heat-bath} at infinity. This situation is given e.g. by a BH
whose Hawking temperature equals the temperature of the universe
(corresponding to the background radiation). Such an equilibrium is
unstable because a BH whose temperature lies minimally beyond
the one of the universe radiates away its mass while its temperature
increases. Nevertheless, this state is of some interest because
it can be described geometrically by a \emph{static spacetime with the
appropriate asymptotic behaviour} (which differs from the asymptotically
flat Schwarzschild spacetime).

In the $\left|U\right>$-state the incoming flux is set to
zero. As there is always some outgoing flux the BH looses
mass until it finally disappears. This state describes a BH
that is surrounded by vacuum or a background radiation whose temperature
is much lower than the Hawking temperature.
Clearly, the latter must be low enough so that the BH is
still in the quasi-static phase, otherwise the CF
method cannot be applied. Because I can show that \emph{the $\left|U\right>$
state is the physical state of lowest asymptotic energy density}
(thereby I exclude unphysical states like $\left|B\right>$),
I identify it with the vacuum state of the system: $\left|U\right>
\equiv\left|0\right>$. The Unruh state is most appropriate to describe
an evaporating BH and backreaction effects because in the
interesting region the outgoing flux is assumed to be much
higher than the actual background radiation (this does certainly not apply for
BHs which are surrounded by matter sources). In the quasi-static
phase such effects are negligible and the calculated Hawking flux
is identical to the one in the $\left|H\right>$-state.

In the following I examine the boundary conditions in the four-dimensional
theory. The results can be adopted directly in the dilaton model.

First, I will consider the regularity condition at the future horizon.
A sufficient condition\footnote{A necessary condition would be that
the flux is integrable at the horizon. This allows poles at the horizon
that behave like $(r-2M)^{-\alpha},\alpha<1$, as well as logarithmic divergences.}
is that all components of the EM tensor in Kruskal coordinates
$\E{T_{UU},T_{VV},T_{UV},T_{\theta\theta}}$ are finite for $U=0$. 
I start with the assumption that the basic components \emph{$\E{T},\E{\Th}$
are finite at the horizon}; this is sensible since both quantities are independent
of the coordinate systems considered. Then it follows that also
$\E{T_{UV}}\propto \E{T}$ (\ref{EM-light1},\ref{EM-Krus3}) and
$\E{T_{VV}}$ (\ref{EM-light3},\ref{EM-Krus2}) are finite. The last
component can be expressed as (\ref{sol1},\ref{sol2})
\begin{multline}
\E{T_{UU}}\propto\frac{1}{(r-2M)^2}\biggm\{\left(\E{T}-2\E{\Th}
\right)(r-2M)r-2Q\\
-2\int_{2M}^r\left[M\E{T}+2(r'-3M)\E{\Th}\right]dr'\biggm\}\\
\stackrel{r\to2M}{\approx}\frac{1}{(r-2M)^2}\biggm\{\left(\E{T}-
2\E{\Th}\right)(r-2M)r-2Q\\
-2(r-2M)\left[M\E{T}+2(r-3M)\E{\Th}\right]\biggm\}\\
=-\frac{2Q}{(r-2M)^2}+O\left((r-2M)^0\right).
\end{multline}
Thus \emph{the choice $Q=0$ guarantees the regularity of the remaining 
components $\E{T_{UU},T_{VV},T_{UV}}$ at the future horizon}.

Second, I want to examine the meaning of the constant $K$.
Obviously $K$ determines the total flux into the r-direction (\ref{sol1})
which can be written as the difference of the outgoing and incoming fluxes
(\ref{total flux}).
Further, we see from (\ref{sol1},\ref{sol2},\ref{EM-light2},\ref{EM-light3})
and $\E{T^t{}_t}=\E{T}-\E{T^r{}_r}-2\E{\Th}$ that
\begin{eqnarray}
\E{T_{--}}&\stackrel{r\to\infty}{\approx}&\frac{1}{r^2}
\left\{-\frac{2Q}{M^2}-2f(r)\right\}+O\left(r^{-3}\right)\label{T-out}\\
\E{T_{++}}&\stackrel{r\to\infty}{\approx}&\frac{1}{r^2}
\left\{\frac{4K-2Q}{M^2}-2f(r)\right\}+O(r^{-3}),\label{T-in}
\end{eqnarray}
where I have defined the function
\begin{multline}
f(r):=\int_{2M}^r\left[M\E{T}+2(r'-3M)\E{\Th}\right]dr'\\
\stackrel{dilaton}{\to}\int_{2M}^r\left[\frac{M\E{T}_2}{(r')^2}
+2(r'-2M)\E{\Th}_2\right]dr'.
\label{f}
\end{multline}
The value of $f$ at spacelike infinity has to be finite: $f(\infty)<\infty$.
The necessary conditions are that $\E{T}$ and $\E{\Th}$ are regular
on the horizon and that they go \emph{faster} to zero than $r^{-1}$, respectively
$r^{-2}$ for large $r$. The explicit calculations in the main Chapters
of this thesis will show that the basic components fall off even faster
than demanded here (\ref{T.massive},\ref{Th.massive}).

The outgoing flux $\E{T_{--}}$ only depends on $Q$,
while the incoming flux $\E{T_{++}}$ also depends on $K$.
Thus, one can obtain an arbitrarily large incoming flux
by adjusting $K$. Clearly this increases the number of occupied
asymptotic particle states until one cannot speak of a vacuum
spacetime anymore (the geometry close to the horizon shall be
dominated by the BH mass and not by the incoming particles).

I demand that \emph{the weak energy condition is fulfilled asymptotically:
there are no negative energies or fluxes in the asymptotic region}.
This means that $\E{T_{--}},\E{T_{++}}$ must become positive for $r\to\infty$.
Hence the constants $K$ and $Q$ have to obey the inequalities
\begin{equation}
Q\le-M^2f(\infty)\,\,\,,\,\,\,K\ge\frac{Q}{2}+\frac{M^2f(\infty)}{2}.
\end{equation}
This already implies that $f(\infty)\le0$ if I want
regularity of the EM tensor on the horizon $Q=0$. The flux condition for
$K$ then becomes $K\ge\frac{1}{2} M^2f(\infty)$. The asymptotic energy
density is given by
\begin{equation}
\E{T_{tt}}\stackrel{r\to\infty}{\approx}\frac{1}{r^2}\left\{
\frac{K-Q}{M^2}-f(\infty)\right\}.
\end{equation}
For $Q=0$ the weak energy condition demands $K\ge M^2f(\infty)$ 
(which is less restrictive than the flux condition). We observe that
the limiting value $K=\frac{1}{2}M^2f(\infty)$ then corresponds to the state of
lowest energy density (everywhere) and accordingly to the smallest number
of occupied states. It coincides with the Unruh state 
$\left|U\right>$ for which the incoming asymptotic flux (\ref{T-in})
is zero: $\E{T_{++}}\stackrel{r\to\infty}{=}0$. Therefore, \emph{the
Unruh state is indeed the vacuum state of the system}.

In the Hartle-Hawking state an incoming flux equal to the outgoing one
is achieved by setting $K=0$. Since there is no total flux
this state is static. This is also the case for the Boulware state
where additionally the asymptotic energy density
vanishes by fixing $Q=-M^2f(\infty)$. Thereby one looses the regularity
of the flux on the horizon. In Table 2 these results are collected
and Figure 4 shows the weak energy condition for the fluxes
and the energy density.
\\
\\
\hspace*{.3cm}\begin{tabular}{||c|c|c|c||}
\hline\hline
State & Q & K & Description \\ \hline
$\left|U\right>$ & 0 & $\frac{1}{2}M^2f(\infty)$ & regular, non-static,
lowest energy\\
$\left|H\right>$ & 0 & 0 & regular, static, thermal equilibrium\\
$\left|B\right>$ & $-M^2f(\infty)$ & 0 & singular at horizon, static, zero flux\\
\hline\hline
\end{tabular}\\ \\ \small\hspace*{6cm}Table 2\normalsize
\\
\\

\begin{figure}[h]

\hspace{2cm}\epsfig{file=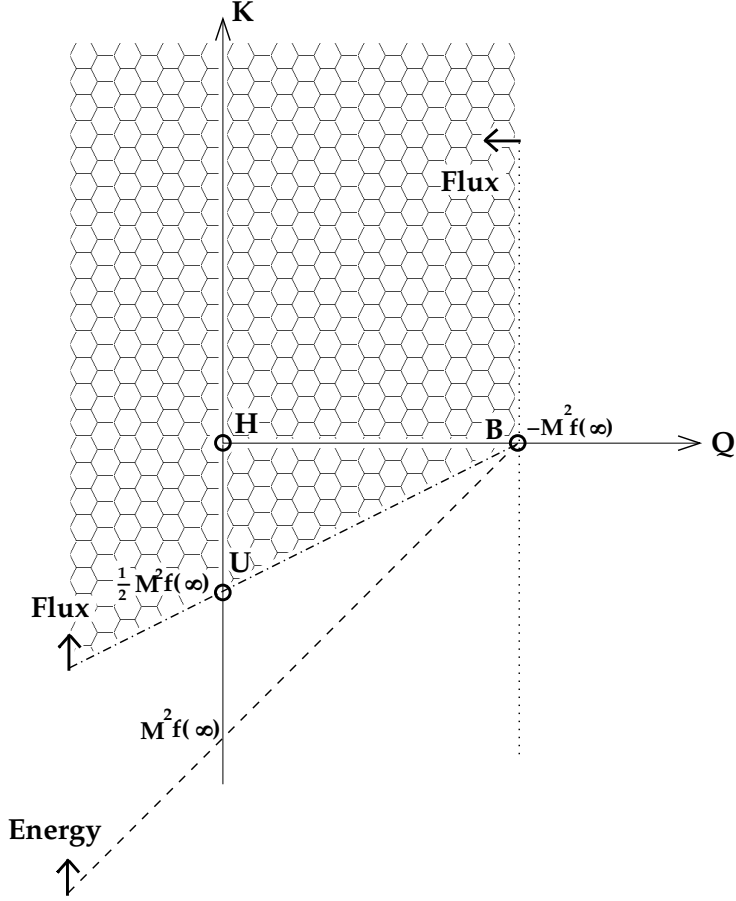,height=12cm}

\caption[fig2]{Weak Energy Condition and Quantum States}
\end{figure}

\subsubsection{State-Independence and Boundary Conditions of
the Basic Components}

In the last Section I have demonstrated how different choices of
the constants $K,Q$ lead to different expectations values of the
EM tensor. I could show how particular values are related to
specific boundary conditions that determine the local
quantum state of the system.
Thereby I have assumed that the basic components,
being the input of the conservation equation, are independent
of these boundary conditions and exhibit a sufficiently
nice behaviour near the boundary of the manifold. 

I have noted already in the last Section that $\Th$ must fall off
at least like $r^{-3}$ (and $T$ like $r^{-2}$) so that $f(\infty)$
is a finite number. This can be traced back to the fact that the
propagation of the particles (showing an $r^{-2}$ behaviour) occurs radially
as I only consider s-waves. This manifests itself also by the
free constants $K,Q$ which determine the radiational
part of those components of the EM tensor that represent
the physical degrees of freedom of the system.

The state-independence of the components $\Th$ and $T$ can be
demonstrated by some simple physical considerations.
If they were state-dependent the function $f$ (\ref{f}) would
be different in the $\left|U\right>$,$\left|H\right>$ and
$\left|B\right>$-state and I would mark it then by a corresponding
index $f_U,f_H,f_B$. In particular, its value at spatial infinity
would depend on the choice of quantum state. The asymptotic
energy density in the $\left|U\right>$  and $\left|H\right>$
state would become
\begin{equation}
T_{tt}\stackrel{r\to\infty}{\propto}\left(-\frac{1}{2r^2}
+O\left(r^{-3}\right)\right)\cdot\left\{\begin{array}{c}2f_H(\infty)
\hspace{2cm},\left|H\right>
\\1f_U(\infty)\hspace{2cm},\left|U\right>\end{array}\right. .
\end{equation}
In the static approximation the asymptotic energy density of the
$\left|H\right>$-state is exactly twice
the one in the $\left|U\right>$-state. This is due to the
balancing incoming flux that doubles the number of particles
with a certain energy in the asymptotic region.
Thus I conclude that $f_H(\infty)=f_U(\infty)$ and, because
$f$ is a linear combination of $T$ and $\Th$, it
is likely that $\U{T}=\HH{T}$ and $\U{\Th}=\HH{\Th}$
on the whole spacetime (in fact the weaker result
$f_H(\infty)=f_U(\infty)$ is at least sufficient to calculate
the correct Hawking flux).

It is not possible to show the equivalence of the basic
components in the Boulware state by similar physical arguments
since the state itself exhibits unphysical properties.
However, the essential difference to the other two states
is that the asymptotic states (incoming and outgoing) are
eliminated completely. The state-independence of the
basic components just shows that they are independent
of the asymptotic particle states but influenced only
by the spacetime geometry. Thus, it
holds for all states in which the total flux is negligible
as compared to the spacetime curvature near the horizon:
\begin{eqnarray}
&\U{T}=\HH{T}=\B{T}&\nonumber\\
&\U{\Th}=\HH{\Th}=\B{\Th}.&
\end{eqnarray}
The state-independence of the basic components presumably
breaks down when backreaction effects become important because
then the scalar field contributes as much to the spacetime
curvature as the BH.

\newpage
\section{The Effective Action}

The effective action contains the whole information
on a quantum system and is therefore the basis to
compute expectation values of the physical observables.
The classical scalar action (\ref{scalar action})
formally describes free particles as it does not contain
a self-interaction term (e.g. $\lambda\cdot S^4$) or any
interaction with other particles. In this thesis the gravitational
metric field $g$ is treated as a classical field,
hence it does not increase the order in $\hbar$ through interaction with $S$.
The only possible Feynman graphs are isolated lines
(corresponding to classical propagation) and single scalar loops
(a line that is bent to a circle and connected at the ends).
The dynamics of the system therefore is purely classical, the
quantum theory enters only by the vacuum energy.
The full interaction is thus described by a one-loop effective
action and can be expressed by the functional determinant
of the Laplace operator which determines the classical
dynamics of the scalar field (I define the Laplace operator
as the total quadratic term in the scalar action).

The main problem is that the classical EOM of
a scalar field on a Schwarz\-schild spacetime cannot
be solved exactly, i.e. the Green function cannot
be given in a compact form. This means that the interaction
of the scalar particle with the gravitational field
needs to be solved perturbatively. The starting point
of the perturbational analysis in my approach is the so-called \emph{heat kernel}
which essentially is the exponentiated Laplacian times a proper time.
At this point one has to distinguish between massless and massive
scalar particles.

Massless particles (or even very light particles)
may cross very large distances during a finite period
of proper time (measured in their rest frame).
Their interaction with the gravitational field therefore
cannot be considered as being local, merely they seem
to interact with the whole manifold at once -- the scalar
loop, corresponding to a measurement of the EM tensor,
is extremely large and extends over the whole spacetime.
If the particle has a very small though finite mass $m$,
the gravitational field, proportional to the BH
mass $M$, slows down the particle and thereby reduces the
size of the loop -- one can say that the mass term
$m^2S^2$ has the effect of a localising potential
in the gravitational field. A particle can be considered
as being localised if its mass times the BH
mass is much larger than one: $mM\gg1$.
This condition determines the form of the effective action:
if it is fulfilled, the effective action can be given
in a local form and all expectation values measured at a
certain spacetime point can be calculated from the geometry
at this point. If the particle is too light, on a given
BH spacetime, the effective action becomes
a non-local expression that can be expressed as multiple
integrals over the manifold.

In the case of massless particles the perturbation theory
consists of counting the number of (non-local) interactions with the
gravitational field.
It is introduced by a formal separation of the spacetime
metric into a flat background and a perturbing part
representing the gravitational interaction.
The result is a series of multiple integrals over
the manifold where the integrand is a curvature term
and a Green function which gives the probability
of the particle to travel from the point of measurement to
the point of interaction. Each integration corresponds
to an interaction of the scalar particle with
the gravitational field and increases the order in the
curvature (the order in $\hbar$
is always $1$ as all interactions occur on a single loop).
The convergence of the series thus depends on the strength
of the curvature (as compared to the Planck curvature).

\begin{figure}[h]
\hspace{3.5cm}\epsfig{file=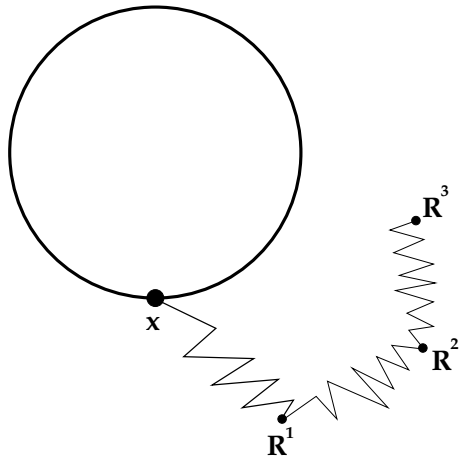,height=6cm}
\caption[fig1]{Non-local Interaction of Massless Particles}
\end{figure}

In Figure 5 I have represented the scalar particle by
a smooth line and the gravitational field by jagged lines.
$x$ is some fixed point of measurement, while each point
of interaction is averaged over the whole manifold.
The perturbation series converges, because the zeroth order
vanishes (i.e. the flat spacetime value) and the main
contribution comes from the first non-local order.
The difficulty lies in finding the Green functions which
are generally not given in a closed analytical form.
The two-dimensional case is particular in this respect,
because there one can find a local form of the effective
action even in the massless case.

For sufficiently massive particles $m\gg\frac{1}{M}$ the
loop is contracted to the point of measurement and the
interaction can be considered as to happen at this single point.
The effective action is
then given by a local expansion, where by each order
increases again the order in the curvature and simultaneously
the order in the parameter $\alpha=\frac{1}{mM}$.
The convergence condition then becomes $\alpha\ll1$.
The perturbation is performed by expanding the heat kernel
for small values of the proper time, corresponding to
small lengths of the scalar loop. For increasing parameter
$\alpha$ the approximation to consider the measurement process
as being local gets worse and contributions from larger
proper times spoil the convergence of the series.
\\
\\
To deal with functional determinants of operators requires
mathematical methods which are restricted to elliptic operators
\cite{gil95}. It is therefore necessary to
Euclideanise the Lorentzian Laplace operator by the introduction
of a Riemannian spacetime metric (I will speak of Euclideanisation,
as it is common in particle physics, though a curved spacetime
with Euclidean signature $(1,1,1,1)$ clearly has a Riemannian metric).
This is realized by defining an imaginary time-variable $\tau=i\cdot t$
and multiplication of the whole metric by a factor $-1$, so that 
the complete transformation becomes $(1,-1,-1,-1)\to(1,1,1,1)$
(the overall minus sign is the greatest disadvantage of my sign-convention
as one has to be particularly careful with signs when switching from
the Lorentzian to the Riemannian spacetime).
Accordingly the measure changes as $\sqrt{-g}\to\sqrt{g}$. 
The Lorentzian Laplace (d'Alembert) operator $\square$ is
replaced by minus the Euclidean Laplace operator\footnote{I denote
the Lorentzian Laplace operator with $\square$ and the Euclidean one
with $\triangle$, not to be confused with the flat Laplacian $\p^2$.} 
$\triangle$. For the transformation of the action see Appendix E
(\ref{Euclid.action.funct.}).
Whenever confusion is possible I mark Euclidean
quantities by an index ${\cal E}$ and Lorentzian quantities by an
index $\cal M$. 

% Thus, the matter action in the path integral transforms as $iL_m\to-L^{\cal E}_m$, 
% see Appendix D. In the same way I introduce the Euclidean effective
% action by $iW\to-W_{\cal E}$, being related to the Euclidean matter
% action by $e^{-W_{\cal E}}=\int{\cal D}\bar{S}e^{-L^{\cal E}_m}$.
% The relation between Lorentzian and Euclidean expectation values,
% defined in (\ref{Euclid.exp.value}), is given by (\ref{relation.exp.value}).
% Whenever confusion with Lorentzian quantities is possible, I will
% denote Euclidean quantities by an index ${\cal E}$.

The effective action (\ref{eff.action}) is defined as a sum over
all possible configurations of the scalar field on the background
spacetime. 
In principle one could approximate numerically the eigenfunctions 
$\phi_n$ and eigenvalues $-\lambda_n$ of the Laplace operator 
$\triangle$ and calculate the effective action explicitly.
For analytical calculations, however, the form (\ref{eff.action}) is
rather inconvenient.
As the eigenfunctions and eigenvalues are determined
by the geometrical and topological (boundary terms) properties of the 
manifold one can describe the effective action in a purely geometrical
and topological way. I start with formally integrating
out the path integral by expanding the scalar field into eigenfunctions
$S=\sum_nc_n\phi_n$:
\begin{multline}
W_{\cal E}[g]=-\ln\int{\cal D}S\cdot e^{-L^{\cal E}_m[g]}
=-\ln\int{\cal D}S\cdot e^{\frac{1}{2}
\int_M S\triangle S\sqrt{g}d^4x}\\
=-\ln\int\prod_n\left\{dc_n\cdot e^{-\frac{\lambda_n}{2}c^2_n}\right\}=
-\ln\prod_n\sqrt{\frac{2\pi}{\lambda_n}}\\
=:\frac{1}{2}\ln\det(-\triangle)+\text{const.}=\frac{1}{2}\textrm{tr}
\ln(-\triangle)+\text{const.}\,.\label{eff.action2}
\end{multline}
The integrations over the coefficients $c_n$ are simply Gaussian integrals
\begin{equation}
\int_{-\infty}^{\infty}e^{-\alpha x^2}dx=\sqrt{\frac{\pi}{\alpha}}.
\end{equation}
In the exponent I have used the orthogonality condition of the eigenfunctions:
\begin{equation}
(\phi_n,\phi_m)=\int_M\phi_n\cdot\phi_m\sqrt{g}d^4x=\delta_{mn}.
\end{equation}
The path integral measure has been derived by $||\delta S||^2
=(\delta S,\delta S)=\sum_n(\delta c_n)^2$ and hence ${\cal D}S=\prod_ndc_n$.

By equation (\ref{eff.action2}) I have eliminated the scalar field
from the effective action and reduced it to a purely geometrical
expression. This simple relation holds for any (free) scalar field
action that is quadratic in the fields. Hence I can generalise
the Laplace operator by adding an \emph{endomorphism} $E$ to the geometric
Laplacian ${\cal O}=-\triangle\1-E$. The sign is chosen such that the
eigenvalues of ${\cal O}$ are positive.
\emph{$E$ is defined as some bounded, linear map from the space of fields
into itself}.
If I consider commuting scalar fields that do not possess inner 
degrees of freedom an endomorphism is simply given by the multiplication 
by some function. $\1$ denotes the identity endomorphism which in
the case of scalar fields corresponds to a multiplication by $1$.

I mention another important point:
by performing a partial integration in the first line of (\ref{eff.action2})
I have dropped a boundary term
\begin{equation}
\frac{1}{2}\int_Mg^{\rho\sigma}\n_{\rho}(S\n_{\sigma}S)\sqrt{-g}d^4x
\end{equation}
by assuming natural boundary conditions
(i.e. sufficiently rapidly vanishing fields). As discussed in
Section 2.4 the boundary conditions are closely related to the
quantum state of the effective action, and hence all expectation values
derived from it. Natural boundary conditions generally imply
that there are no occupied real particle states in the asymptotic
region. This means that there is no incoming or outgoing flux
which, according to the examinations of Section 2.4, corresponds to
the Boulware state $\left|B\right>$. For the moment I will
forget about the boundary conditions and concentrate on the
local properties of the functional determinant.
When considering expectation values it will turn out
that general boundary conditions can be restored
at the level of expectation values by use of the CF
method.

The effective action associated to the Laplacian ${\cal O}$
is related to its eigenvalues, up to a constant that I discard
in the following, by
\begin{equation}
W[g]=\frac{1}{2}\ln\det {\cal O}=\frac{1}{2}\ln\prod_n\lambda_n
=\frac{1}{2}\textrm{tr}\left(\ln {\cal O}\right)=\frac{1}{2}
\sum_n\ln\lambda_n.\label{eff.action3}
\end{equation}
Expression (\ref{eff.action3}) in general is IR ($\ln0$)
and UV divergent ($\ln\infty$). To extract information
from it I must therefore employ some regularisation.
For the moment I postpone the IR problem by assuming that
the particles are massive (this can be managed by adding
an $m^2$-term to the endomorphism $E$).

\subsection{Zeta-Function and Heat Kernel}

With regard to the UV divergence I will use the
``zeta-function regularisation''.
This divergence appears if there are infinitely many large eigenvalues
and (or) the eigenvalues increase without bound. In this
regularisation every term in the sum of the $\ln\lambda_n$'s is multiplied
by a factor $\lambda_n^{-s},s>0$. The contribution of large eigenvalues
may then be neglected: $\lim_{\lambda\to\infty}\lambda^{-s}\ln\lambda=0$.
The regularised effective action is given by
\begin{equation}
W[g,s]^{reg}=\frac{1}{2}\sum_n\lambda_n^{-s}\ln\lambda_n
=-\frac{1}{2}\frac{d}{ds}\sum_n\lambda_n^{-s}.
\label{eff.action-reg}
\end{equation}
The regularisation is removed by taking the limit $s\to0$. By analogy
to the Riemann zeta-function one introduces the zeta-function
of the operator
${\cal O}$:
\begin{equation}
\zeta_{\cal O}[s]:=\sum_n\lambda_n^{-s}=\textrm{tr}\left({\cal O}^{-s}\right).
\label{zeta-function}
\end{equation}
The effective action can then be written as
\begin{equation}
W[g]=-\frac{1}{2}\frac{d}{ds}\zeta_{\cal O}[s]\biggm|_{s=0}.
\label{eff.action4}
\end{equation}
Now I bring the zeta-function into a form that is particularly appropriate
for perturbational analysis. The $\Gamma$-function for
complex arguments is defined as
\begin{equation}
\Gamma(s)=\int_0^{\infty}d\tau\tau^{s-1}e^{-\tau}.
\end{equation}
A shift in the integration variable $\tau\to\lambda\tau$ leads to
\begin{equation}
\Gamma(s)=\lambda^s\int_0^{\infty}d\tau\tau^{s-1}e^{-\lambda\tau}.
\end{equation}
Thus, the zeta-function can be written as
\begin{equation}
\zeta_{\cal O}[s]=\frac{1}{\Gamma(s)}\int_0^{\infty}d\tau\tau^{s-1}
\sum_n e^{-\lambda_n\tau}=\frac{1}{\Gamma(s)}\int_0^{\infty}d\tau\tau^{s-1}
\textrm{tr}\left(e^{-{\cal O}\tau}\right).\label{zeta-function2}
\end{equation}
$e^{-{\cal O}\tau}$ is the so-called heat kernel of the operator ${\cal O}$.
In a coordinate basis it reads:
\begin{multline}
G_{\cal O}(x,y;\tau)=\left<x\left|e^{-{\cal O}\tau}\right|y\right>
=\sum_n\sum_m\E{ x|n}\E{ n|e^{-{\cal O}\tau}|m}\E{ m|y}\label{heatkernel}\\
=\sum_n\phi_n(x)\phi_n(y)e^{-\lambda_n\tau}.
\end{multline}
The coordinate basis functions obey the orthonormality condition
$\left<x|y\right>=\frac{1}{\sqrt{g}}\delta^4(x-y):=\delta^4(x,y)$.
The heat kernel fulfils the heat equation
\begin{equation}
\left(\frac{\partial}{\partial \tau}+{\cal O}_x\right)G_{\cal O}
(x,y;\tau)=0\label{heat eq.}
\end{equation}
with initial condition $G_{\cal O}(x,y;0)=\delta^4(x,y)$.
It can be interpreted as the probability of a particle
to be found at a spacetime point $y$ after a time $\tau$ has passed in
its rest frame when starting from point $x$. The diagonal
heat kernel $G_{\cal O}(x,x;\tau)$ describes the closed loop of a
test particle in the measurement process.
A mass term in the action leads
to an $m^2$-term in the Laplacian and hence causes an exponential
damping of big loops with large values of the proper time $\tau$.
If it is sufficiently strong the heat kernel can be expanded
into a local series for small values of $\tau$.
For massless particles even infinite proper
times $\tau$ contribute to the same order and the heat kernel
inevitably produces non-localities.

The relation between the zeta-function and the heat kernel is given by
\begin{equation}
\zeta_{\cal O}[s]=\frac{1}{\Gamma(s)}\int_0^{\infty}d\tau\tau^{s-1}
\int_M G_{\cal O}(x,x;\tau)\sqrt{g}d^4x.
\end{equation}

\subsubsection{Seeley-DeWitt Expansion}

I start with discussing the local expansion of the heat kernel.
Note that the methods of this Section can only be applied
if a sufficient damping for large proper times $\tau$ is
provided (see Section 3.1.3).
On a $d$-dimensional manifold the diagonal heat kernel
can be expanded around $\tau=0$ as \cite{gil95}
\begin{equation}
G_{\cal O}(x,x;\tau)\stackrel{\tau\to 0}{=}
\frac{1}{(4\pi\tau)^{\frac{d}{2}}}\sum_{n=0}^{\infty}
a_{n}(x,x)\tau^{\frac{n}{2}}.\label{heat.exp.}
\end{equation}
The $a_n$ are called Seeley-DeWitt coefficients.
The odd coefficients $a_{2n+1}$ are zero if the fields
fulfil natural boundary conditions. From now
on I assume that $a_{2n+1}=0$ for all $n$.

The coefficients with even index can be calculated from
the heat equation (\ref{heat eq.}). Following \cite{sch51,dew63}
I make an ansatz for the heat kernel with different spacetime
point arguments:
\begin{equation}
G_{\cal O}(x,y;\tau)=\frac{\sqrt{\Delta(x,y)}}{(4\pi\tau)^{\frac{d}{2}}}
e^{-\frac{\sigma(x,y)}{2\tau}}\sum_{n=0}^{\infty}a_{2n}(x,y)\tau^{n}.
\label{heat expansion}
\end{equation}
The \emph{bi-tensors} $\sigma(x,y)$ (\ref{world-function})
and $\Delta(x,y)$ (\ref{Delta}) which appear
in this expression are explained in Appendix F.2.
$\sigma$ is related to the geodesic distance of two points
and leads to an exponential damping for large point-separations.
$\Delta$ is needed for coordinate invariance and becomes $1$ for $y=x$.
First I show that the Ansatz
(\ref{heat expansion}) fulfils the
initial condition:
\begin{multline}
\lim_{\tau\to 0}\int_MG_{\cal O}(x,y;\tau)\sqrt{g}d^dx\\
\approx\lim_{\tau\to 0}\frac{1}{(4\pi\tau)^{\frac{d}{2}}}
\int_{-\infty}^{\infty} e^{-\frac{g_{00}(x^0-y^0)^2+g_{11}(x^1-y^1)^2+\dots}
{4\tau}}\sqrt{g}dx^0dx^1\dots=1.
\end{multline}
Thus $G_{\cal O}(x,y;\tau)$ becomes a delta-function for $\tau\to0$.
If we put (\ref{heat expansion}) into (\ref{heat eq.}) we obtain
recurrence relations for the $a_n(x,y)$:
\begin{multline}
\left[-\frac{d}{2\tau}+\frac{\sigma}{2\tau^2}\right]
\sum_n a_{2n}\tau^{n}+\sum_nn\,a_{2n}\tau^{n-1}\\
=\left[\frac{\triangle\sqrt{\Delta}}{\sqrt{\Delta}}
-\left(\frac{\n\sqrt{\Delta}\n\sigma}{\sqrt{\Delta}}
+\frac{\Delta\sigma}{2}\right)\frac{1}{\tau}
+\frac{(\n\sigma)^2}{4\tau^2}\right]\sum_n a_{2n}\tau^{n}\\
+2\sum_n\frac{\n\sqrt{\Delta}\n a_{2n}}{\sqrt{\Delta}}\tau^{n}
-\sum_n\n\sigma\n a_{2n}\tau^{n-1}+\sum_n(\triangle+E)a_{2n}\tau^{n}.
\end{multline}
Using the identities $(\n\sigma)^2=2\sigma\,,\,D^{-1}\n_{\mu}(D\n^{\mu}
\sigma)=d$ (\ref{world-vector},\ref{Morette-identity}) this can be
simplified to
\begin{eqnarray}
\n\sigma\n a_0&=&0\\
(n+2)a_{2n+2}+\n\sigma\n a_{2n+2}&=&\frac{\triangle(\sqrt{\Delta} a_{2n})}
{\sqrt{\Delta}}+Ea_{2n},\,n>0.\label{recursion}
\end{eqnarray}
By taking the coincidence limit $y\to x$ one can calculate the 
diagonal coefficients $a_{2n}(x,x)$. For this we need the coincidence
limits of the involved bi-tensors and their derivatives:
\\
\\ \large
\hspace*{1cm}\begin{tabular}{|c|c|c|}
\hline
\normalsize $\sigma\to0$ & \normalsize $\sqrt{\Delta}\to1$ 
& \normalsize $a_0\to1$ \\ \hline
\normalsize $\sigma_{;\mu}\to0$ & \normalsize $\sqrt{\Delta}_{;\mu}\to0$ 
& \normalsize$ {a_0}_{;\mu}\to0$ \\ \hline
\normalsize $\sigma_{;\mu\nu}\to g_{\mu\nu}$ 
& \normalsize $\sqrt{\Delta}_{;\mu\nu}\to
\frac{R_{\mu\nu}}{6}$ 
& \normalsize$ {a_0}_{;\mu\nu}\to\frac{\Omega_{\mu\nu}}{2}$ \\ \hline
\normalsize $\sigma_{;\mu\nu\rho\sigma}\to-\frac{1}{3}
(R_{\mu\nu\rho\sigma}+R_{\mu\sigma\rho\nu})$ & \normalsize & \normalsize \\ \hline
\end{tabular}\\ \\ \small\hspace*{6cm}Table 3
\\ \normalsize
Some of these relations are derived in Appendix F.2. 
$\Omega_{\mu\nu}=-\Omega_{\mu\nu}$ is the gauge curvature associated 
to the gauge connection that might be present, see Appendix F.1
(\ref{gauge-curvature}) --
if there are no gauge degrees of freedom it vanishes.
Starting from $a_0(x,x)=1$ I can calculate all higher diagonal
Seeley-DeWitt coefficients. For instance, for $n=0$ I obtain the
relation
\begin{equation}
a_2+\n\sigma\n a_2=\frac{\triangle\sqrt{\Delta}}{\sqrt{\Delta}}a_0
+2\frac{\n\sqrt{\Delta}\n a_0}{\sqrt{\Delta}}+\triangle a_0+Ea_0
\end{equation}
which in the coincidence limit yields $a_2=\frac{R}{6}\1+E$
(note that $g^{\mu\nu}\Omega_{\mu\nu}=0$).
The first three Seeley-DeWitt coefficients are given by:
\\
\\ \large
\hspace*{0cm}\begin{tabular}{||l||}
\hline\hline
\normalsize $a_0=\1$ \\
\normalsize $a_2=\frac{1}{6}(R\1+6E)$ \\
\normalsize $a_4=\frac{1}{360}[60\triangle E+60RE+180E^2
+30\Omega_{mn}\Omega^{mn}$\\
\normalsize\hspace{1cm}$
+(12\triangle R+5R^2-2R_{mn}R^{mn}+2R_{mnop}R^{mnop})\1]$ \\
\normalsize $a_6=\frac{1}{360}[6\triangle\triangle E+
30E\triangle E+30(\triangle E) E+30E_{;m}E^{;m}+60E^3
+12E\Omega_{mn}\Omega^{mn}$\\
\normalsize\hspace{1cm}$+6\Omega_{mn}E\Omega^{mn}+12\Omega_{mn}\Omega^{mn}E
+10R\triangle E+4R_{mn}E^{;mn}+12R_{;m}E^{;m}$\\
\normalsize\hspace{1cm}$-6E_{;m}\Omega^{mn}{}_{;n}+6\Omega^{mn}{}_{;n}E_{;m}
+30E^2R+12E\triangle R+5ER^2-2ER_{mn}R^{mn}$\\
\normalsize\hspace{1cm}$+2ER_{mnop}R^{mnop}]+\frac{1}{9\cdot 7!}
[81R_{mnop,q}R^{mnop,q}+108R_{mnop}\triangle R^{mnop}$\\
\normalsize\hspace{1cm}$-44R_{mnop}R^{mn}{}_{qr}R^{opqr}
-80R_{mnop}R^{m\,\,o}_{\,\,\,\,\,q\,\,r}R^{nqpr}$\\
\normalsize\hspace{1cm}$+42RR_{mnop}R^{mnop}-48R_{mn}R^m{}_{opq}R^{nopq}]\1+\dots$
\\ \hline\hline
\end{tabular}\\ \\ \small\hspace*{6cm}Table 4
\\ \normalsize
They are taken from \cite{gil95} where $a_6$ is listed
completely -- here I have just quoted the terms I will need
in this work. Note that all geometric objects like the Ricci
tensor $R_{mn}$ belong to the Euclideanised manifold. For
convenience I have written all tensors in vielbein components.
Terms that differ only by the position of its elements, like
$E\Omega_{mn}\Omega^{mn}$ and $\Omega_{mn}E\Omega^{mn}$, become
identical if the fields commute.

\subsubsection{General Form of the Zeta-Function}

The zeta-function (\ref{zeta-function}) of an operator ${\cal O}$ 
is related to the effective action of the corresponding quantum field
(\ref{eff.action4}).
When calculating vacuum expectation values by taking
functional derivatives of the effective action this relation
may be modified by the appearance of new operators in $\zeta_{\cal O}[s]$.
In this respect it proves useful to
define a general zeta-function \cite{gil95}
\begin{equation}
\zeta_{\cal O}[s;{\cal Q}]=\textrm{tr}_{L^2}({\cal Q\cdot O}^{-s})=
\sum_n\lambda^{-s}_n\cdot\textrm{tr}(\pi(\lambda_n;{\cal O}){\cal Q}),
\end{equation}
where ${\cal Q}$ is some other operator. $\pi[\lambda_n;{\cal O}]$
is the projection operator that projects onto the eigenspace
of ${\cal O}$ belonging to the eigenvalue $\lambda_n$.
By performing the same steps as before I can introduce the
general heat kernel
\begin{equation}
G_{\cal O}(x,y;\tau)=\left<x\left|{\cal Q}e^{-{\cal O}\tau}\right|y\right>.
\end{equation}
Again the diagonal heat kernel can be expanded around $\tau=0$:
\begin{equation}
G_{-\triangle}(x,x;\tau)=\frac{1}{(4\pi)^{\frac{d}{2}}}\sum_{n=0}^{\infty}
a_n[{\cal Q;O}]\tau^{\frac{n-d}{2}}.
\end{equation}
The original Seeley-DeWitt coefficients are given by the special
case that ${\cal Q}$ is the identity endomorphism: $a_n=a_n[\1;{\cal O}]$.
The simplest case of a non-trivial general zeta-function is that
${\cal Q}$ is a function\footnote{In the following I omit
the identity endomorphism $\1$ as I am only interested
in commuting scalar fields, hence $\1$ means just multiplication by $1$.}
${\cal Q}=f\1,f\in C^{\infty}(M)$.
The Seeley-DeWitt coefficients of the corresponding heat kernel
expansion can then be obtained easily from the original ones
(still ${\cal O}=-\triangle-E$). 
Defining an operator ${\cal O}'=-\triangle-E-\ve f$ yields the relation
\begin{equation}
\frac{d}{d\epsilon}\textrm{tr}\left[e^{(\triangle+E+
\ve f)\tau}\right]\biggm|_{\ve=0}=\tau\cdot\textrm{tr}
\left[fe^{(\triangle+E)\tau}\right].
\end{equation}
If I insert the corresponding Seeley-DeWitt expansions into this 
relation I get a formula for the general coefficients
\begin{equation}
a_n[f;{\cal O}]=\frac{d}{d\ve}
a_{n+2}[{\cal O}']\biggm|_{\ve=0},\label{gen.Seeley}
\end{equation}
where on the r.h.s. we have the original coefficient $a_{n+2}$ of the
operator ${\cal O}'$. Some general $a_n[f;{\cal O}]$ are collected in Table 5.
\\
\\ \large
\hspace*{0cm}\begin{tabular}{||l||}
\hline\hline
\normalsize $a_0[f;{\cal O}]=f$ \\
\normalsize $a_2[f;{\cal O}]=\frac{1}{6}[Rf+\triangle f+6Ef]$ \\
\normalsize $a_4[f;{\cal O}]=[\frac{1}{60}\triangle\triangle f
+\frac{1}{6}(E\triangle f+f\triangle E+E_{;m}f^{;m})+\frac{1}{2}E^2f
+\frac{1}{6}EfR$\\
\normalsize\hspace{1.8cm} $+\frac{5}{60}f\Omega_{mn}\Omega^{mn}+\frac{1}{72}
(2R\triangle f+fR^2)+\frac{1}{90}R_{mn}f^{;mn}$ \\ 
\normalsize\hspace{1.8cm} $+\frac{1}{30}(R_{;m}f^{;m}
+f\triangle R)-\frac{1}{180}fR_{mn}R^{mn}
+\frac{1}{180}fR_{mnop}R^{mnop}]$\\ \hline\hline
\end{tabular}\\ \\ \small\hspace*{6cm}Table 5
\normalsize

\subsubsection{Application of the Asymptotic Expansion}

In the following I specify for a four-dimensional spacetime.
The same arguments apply in the two-dimensional case.
There are two important cases when the asymptotic heat kernel
expansion can be applied successfully. First, I can calculate
the trace anomaly in arbitrary even dimensions directly by
$\zeta[0]$ which is always a finite expression, see Appendix C.2.
As it only consists of a finite number of terms I need not
worry about convergence conditions.
Second, if there exists a sufficiently strong damping
for big values of the proper time $\tau$ the Seeley-DeWitt
expansion converges and I can calculate the effective action.
Note that the latter always contains UV divergent terms that
must be renormalised -- the convergence of the series only
guarantees IR finiteness.

In the classical Euclidean action a damping term by definition
has the form
\begin{equation}
L^{\cal E}_D=\int_M\frac{SDS}{2}\sqrt{g}d^4x.
\end{equation}
$D$ might be a constant like $m^2$ or an arbitrary
analytical function. Such a term modifies the Laplacian
as ${\cal O}\to{\cal O}+D$. I can pull the damping term
out of the modified heat kernel, whereby commutator terms with
the original Laplacian have to be taken into account\footnote{This
works similar to the Baker-Campell-Hausdorff formula.
The commutator terms can be determined by expanding the involved
exponentials in a Taylor series.}:
\begin{equation}
G_{\cal O}(x,y;\tau)=e^{-D\tau}\left<x\left|\left(1+
\frac{\tau^2}{2}[D,{\cal O}]+O(\tau^3)\right)e^{{\cal O}\tau}
\right|y\right>.
\end{equation}
The commutators can be lifted back into the 
heat kernel to the right, thereby producing new Laplacians ${\cal O}_n$
with new endomorphisms\footnote{The whole argument
is in fact a bit trickier, but for an estimate of order
this assumption is sufficient.}$E_n$:
\begin{equation}
G_{\cal O}(x,y;\tau)=e^{-D\tau}\left<x\left|e^{{\cal O}\tau}+
c_1\,e^{{\cal O}_1\tau}\tau+c_2\,e^{{\cal O}_2\tau}\tau^2+O(\tau^3)
\right|y\right>.
\end{equation}
$c_1,c_2\dots$ are constants. Now I insert the Seeley-DeWitt
expansions of the new heat kernels. Clearly, the Seeley-DeWitt coefficients 
depend on the endomorphisms, being different for each term in general. 
The zeta-function is then given by an infinite sum of integrals of the type
\begin{equation}
\int_0^{\infty}\tau^{s+k}e^{-D\tau}=\Gamma[1+s+k]\frac{1}{D^{(1+s+k)}},
\end{equation}
where $k$ goes from $-3$ to $+\infty$:
\begin{multline}
\zeta_{\cal O}[s]=\frac{1}{(4\pi)^2\Gamma[s]}\int_M
\left\{\Gamma[s-2]D^{2-s}a_0^{(0)}+\Gamma[s-1]D^{1-s}(a_2^{(0)}
+c_1a_0^{(1)})\right.\\
\left.+\Gamma[s]D^{-s}(a_4^{(0)}+c_1a_2^{(1)}+c_2a_0^{(2)})
+\cdots\right\}\sqrt{g}d^4x.\label{damped-expansion}
\end{multline}
$a_n^{(i)}$ are the coefficients belonging to ${\cal O}_i$,
${\cal O}_0={\cal O}$. 
One can see immediately that $\zeta[0]$ is \emph{always finite}
if $D$ is a regular function. Namely, because of $\Gamma[s]^{-1}=s+O(s^2)$
only the first three terms contribute. There is no IR problem for
large values of $\tau$.

To the effective action contribute all terms in the series as it
is obtained by differentiation for $s$. Therefore, I must
investigate the convergence of the whole series, the basic assumption 
being that $a_{2n+2}<a_{2n}$. There are three different cases: 
\begin{itemize}
\item If all commutators vanish (which is true if $[D,{\cal O}]=0$)
one gets the original series with coefficients $a_{2n}$. 
A necessary condition is that
$a_{2n+4}/D^n$ is finite at all spacetime points $x\in M$ for all $n$.
The series converges if $\lim_{n\to\infty}\text{sup}_{x\in M}(a_{2n+2}
/Da_{2n})<1$.
\item If there is a finite number of nonvanishing commutators, one has to 
consider the convergence criterion for all series separately, 
where it is to be expected that the highest series (corresponding 
to ${\cal O}_{max}$) is the most problematic.
\item If there are infinitely many commutators the condition $D>1$ must be 
fulfilled.
\end{itemize}

\newpage
\subsection{Covariant Perturbation Theory}

In this Section I present a short introduction to the
covariant perturbation theory developed by Barvinsky and
Vilkovisky \cite{bav87,bav90}.
I will only give a guideline how to construct the non-local
effective action, starting from the heat kernel, and demonstrate this
for the simplest case. The formalism only works in even dimensions,
hence I introduce the notation $d=2\omega,\omega=1,2,\dots$.

The Seeley-DeWitt expansion cannot be applied
to massless particles in the general case. As I have shown
in the last Section,
the local perturbation series diverges if the damping factor $D$
(e.g. the squared mass of the particle) goes to zero.
In the absence of any self-interaction that ``localises''
the particle in some finite spacetime volume the scalar
particle can travel large spacetime regions within a small
period of proper time. Hence the effective action, containing the
information on the quantum mechanical expectation values,
is supposed to consist of non-local
terms in the form of multiple integrals over the whole
manifold. 

The appropriate formalism to derive this non-local effective action
directly is the covariant perturbation theory.
Just like the Seeley-DeWitt expansion it is based on the
heat kernel, thus all quantities considered in this Section
are Euclidean (or, more precisely, Riemannian).
As already mentioned, the perturbational treatment is necessary
because the classical EOM of the scalar field cannot be 
solved exactly. Like in QFT, the order
of the perturbation counts the number of interactions,
thereby increasing the order in the curvature.
Note that the number of loops remains constantly $1$ as the
gravitational field is considered as a classical field
whose interaction is described by external lines that
cling to the scalar loop (see Figure 5 at the beginning of
this Chapter). If Quantum Gravity would play a role
(e.g. by the exchange of virtual gravitons) an internal graviton
line would increase the loop order by one, see Figure 3 in Section 1.2.1.

One begins by formally considering
the gravitational field of the BH as some small
perturbation of flat spacetime. Accordingly, the
expectation value of the EM tensor in flat spacetime
is the zeroth order of the perturbation theory.
It is obvious that the first perturbative order
cannot be small as compared to the zeroth order, but
as the latter is renormalised to zero (in the vacuum state)
the first order is the effective starting point of the series.
The metric is divided into two terms\footnote{In this Section
I will mainly use the notation of Barvinsky and Vilkovisky.}
\begin{equation}
g_{\mu\nu}=\tilde{g}_{\mu\nu}+h_{\mu\nu},
\end{equation}
where the metric $\tilde{g}$ represents the zeroth order.
Its associated Riemann tensor thus shall vanish:
\begin{equation}
R^{\mu}{}_{\nu\rho\sigma}[\tilde{g}]=0.
\end{equation}
Hence $\tilde{g}$ is in fact the metric of a flat
Euclidean space and in a Cartesian coordinate system it becomes
the flat metric $\delta$. The split of the metric
into a ``background part'' $\tilde{g}$ and a perturbing
part $h$ seems to contradict the covariance of this approach.
However, this is just an intermediate step in the development
of the perturbation series. The final form of the latter turns out
to be (at each order) a covariant expression in the geometric objects
describing the original manifold, depending solely on $g$.
The physical picture is that the scalar particle is
scattered zero, one, two etc. times on the \emph{full}
manifold (which in fact \emph{is} a covariant
object), whereas e.g. in the case of linearised gravity
waves the manifold is indeed split into a background part $\tilde{g}$
and the part describing the propagating gravitational waves $h$
and hence this separation is manifest.

The effective action of Barvinsky and Vilkovisky covers the most general
case, where the Laplacian acting on the scalar field consists
of a geometric Laplacian, including a ``gauge part'', and
an endomorphism. With gauge part I mean the term of the connection
that is added to the Levi-Civit\'a connection and is associated
to some external gauge transformation applied to the scalar
field. Barvinsky and Vilkovisky use the notation
\begin{equation}
\n_{\mu}S=\tilde{\n}_{\mu}S+\tilde{\Gamma}_{\mu}S.
\end{equation}
$\tilde{\n}$ is the derivative operator associated to the
metric $\tilde{g}$ -- it becomes a partial derivative
$\p$ in a Cartesian coordinate system.
$\tilde{\Gamma}$ meanwhile denotes the ``rest'' of the
connection and hence must include the missing Christoffels
of the full metric $g$ as well as the gauge part.
In the general case the connection $\n$ produces
both a geometric curvature $R^{\mu}{}_{\nu\rho\sigma}$
and a gauge curvature ${\cal R}_{\mu\nu}$ (in my notation
the gauge curvature was $\Omega_{\mu\nu}$, see Appendix F.1
(\ref{gauge-curvature})).
In this work I will only have to deal with the case ${\cal R}_{\mu\nu}=0$.

On the whole we have three types of curvatures contributing
to the same order in the perturbation series (as can
be seen from their equal dimensions), namely the
geometric and gauge curvature, and the endomorphism
that Barvinsky and Vilkovisky name $\tilde{P}$. It is related to the
one used by me by $E=\tilde{P}-\frac{R}{6}$; his notation
is especially convenient for the conformally coupled case in $4d$
where $\tilde{P}=0$.

By expanding the ``curvatures'' in $g$ around $\tilde{g}$,
one obtains relations between $h$ and $\tilde{\Gamma}$ and 
these curvatures, e.g. (see (\ref{var.Riemann}))
\begin{multline}
R^{\mu\nu\rho\sigma}[\tilde{g}+h]=R^{\mu\nu\rho\sigma}[g]+O(h)\\
=0+\frac{1}{2}\left(\tilde{\n}^{\rho}\tilde{\n}^{\mu}h^{\nu\sigma}
-\tilde{\n}^{\rho}\tilde{\n}^{\nu}h^{\mu\sigma}
-\tilde{\n}^{\sigma}\tilde{\n}^{\mu}h^{\nu\rho}
+\tilde{\n}^{\sigma}\tilde{\n}^{\nu}h^{\mu\rho}\right)+O(h^2).
\end{multline}
Vice versa one can consider this relation as expressing
the perturbative metric $h$ in terms of the Riemann tensor
$R^{\mu}{}_{\nu\rho\sigma}$ to the first perturbative order.
By means of the Bianchi identity one can integrate this relation and in
a convenient gauge one obtains (I will specify the meaning of
such non-local expressions like $\triangle^{-1}$ as soon as I use them)
\begin{equation}
h^{\mu\nu}=2\frac{\delta^{\mu\nu}_{\rho\sigma}}
{\triangle}R^{\rho\sigma}[g]+O(R^2).\label{cov.pert.metric}
\end{equation}
An analogous relation can be obtained for the perturbative
connection $\tilde{\Gamma}$ (where on the r.h.s. appear
${\cal R}_{\mu\nu}$ and $\tilde{P}$). These expressions
only contain geometric objects depending on the metric $g$ of the
full manifold. Furthermore, Barvinsky and Vilkovisky show that the full
Laplacian (in this gauge) can be decomposed as
\begin{equation}
\triangle+E=g^{\mu\nu}\n_{\mu}\n_{\nu}
=\triangle_0+h^{\mu\nu}\tilde{\n}_{\mu}\tilde{\n}_{\nu}
+2(\tilde{g}^{\mu\nu}+h^{\mu\nu})\tilde{\Gamma}_{\mu}
\tilde{\n}_{\nu}.\label{cov.pert.Laplace}
\end{equation}
Note that the endomorphism $E$ has been absorbed in the
gauge part of the connection (in the next Sections I will
make use of this trick to shift terms between connection and
endomorphism). $\triangle_0$ is simply the flat
Laplacian for a particle with no gauge group acting on it.

The perturbation of the effective action emerges expanding
the heat kernel by the decomposition of the Laplacian
(\ref{cov.pert.Laplace}) and inserting (\ref{cov.pert.metric}).
Then one can write the heat kernel as a sum of terms with
increasing orders in the curvature factors\footnote{Here I use
my own notation to avoid confusion of the proper time $\tau$ with
the parameter $s$ of the zeta-function.}
\begin{equation}
G_{\cal O}(\tau)=e^{-{\cal O}\tau}
=\sum_{n=0}^{\infty}G^n_{\cal O}(\tau),\label{cov.pert.series}
\end{equation}
where $G^n_{\cal O}(\tau)$ has a power $n$ in the curvatures.
As an illustration I will only consider the simplest
case, where I have an endomorphism $E$ but no curvature
neither gauge curvature, i.e. ${\cal O}=-\triangle_0-E$.
In this case the perturbation of the heat kernel has the form
\begin{eqnarray}
G^0_{\cal O}(\tau)&=&e^{\triangle_0\tau}\\
G^n_{\cal O}(\tau)&=&e^{\triangle_0\tau}
\int_0^{\tau}e^{-\triangle_0\upsilon}E\cdot
G^{n-1}_{\cal O}(\upsilon)d\upsilon,\,n\ge1.
\end{eqnarray}
In a coordinate basis $G^0_{\cal O}(\tau)$ reads (\ref{heat expansion})
\begin{equation}
G^0_{\cal O}(x,y;\tau)=\frac{\sqrt{\Delta(x,y)}}{(4\pi\tau)^{\omega}}
e^{-\frac{\sigma(x,y)}{2\tau}}a_0(x,y).
\end{equation}
Note that $\sqrt{\Delta(x,y)}$, though defined via the metric
$\tilde{g}$ of a flat space, only becomes $1$ in a Cartesian
coordinate system. $a_0(x,y)$ is the parallel displacement
operator (bi-unit) if acting on a scalar field, see Appendix F.2
(\ref{bi-unit}). All other Seeley-DeWitt coefficients are zero due
to the flatness of the space described by $\tilde{g}$.
To construct the effective action I will need the trace of the heat kernel:
\begin{equation}
\text{tr}G^0_{\cal O}(\tau)=\int_MG^0_{\cal O}(x,x;\tau)
\sqrt{g}d^{2\omega}x=\frac{1}{(4\pi\tau)^{\omega}}\int_M
a_0(x,x)\sqrt{g}d^{2\omega}x.
\end{equation}
This integral is divergent if the manifold has infinite size
($a_0(x,x)=1$ for scalar fields without external degrees of freedom).
It produces the zero-point energy of the quantised EM tensor
on a flat spacetime which is renormalised to zero -- if some
small finite value would remain after renormalisation (a cosmological
constant) we would have a non-vanishing zeroth order and the
general covariance would be broken (a background radiation selects out a
preferred reference system).
The term of first order in the curvature is still local:
\begin{multline}
\text{tr}G^1_{\cal O}(\tau)\\
=\int^x_M\int^z_M
\int^{z'}_M\left<x\right|e^{\triangle\tau}\left|z\right>
\int_0^{\tau}\left<z\right|e^{-\triangle\upsilon}\left|z'\right>
E(z')\left<z'\right|e^{\triangle\upsilon}\left|x\right>d\upsilon
\sqrt{g}d^{2\omega}x\cdots\\
=\int^x_M\int^z_M
\int^{z'}_M\left<x\right|e^{\triangle\tau}\left|z\right>
E(z)\int_0^{\tau}\left<z\right|e^{-\triangle\upsilon}\left|z'\right>
\left<z'\right|e^{\triangle\upsilon}\left|x\right>d\upsilon
\sqrt{g}d^{2\omega}x\cdots\\
=\int_M\left<x\right|e^{\triangle\tau}\left|x\right>E(x)
\tau\sqrt{g}d^{2\omega}x=\frac{\tau}{(4\pi\tau)^{\omega}}\int_M
E(x)\sqrt{g}d^{2\omega}x.
\end{multline}
In two dimensions ($\omega=1$) this local term is IR divergent,
in higher dimensions UV divergent. In any case it may contribute
to expectation values and must be regularised.
Next I consider the term of second order in the curvature:
\begin{multline}
\text{tr}G^2_{\cal O}(\tau)\\
=\int^x_M\int^z_M\int_0^{\tau}
\left<x\right|e^{\triangle(\tau-\upsilon)}\left|z\right>E(z)
\upsilon\left<z\right|e^{\triangle\upsilon}\left|x\right>
E(x)d\upsilon\sqrt{g}d^{2\omega}x\sqrt{g}d^{2\omega}z\\
=\int^x_M\int^z_M\int_0^{\tau}\upsilon
\frac{\Delta(x,z)}{(4\pi\tau)^{\omega}}\left(\frac{\tau}
{4\pi(\tau-\upsilon)\upsilon}\right)^{\omega}
e^{-\frac{\sigma(x,z)\tau}{2(\tau-\upsilon)\upsilon}}\\
\cdot a_0(x,z)E(z)a_0(z,x)E(x)d\upsilon\sqrt{g}d^{2\omega}x\sqrt{g}d^{2\omega}z\\
=\frac{1}{(4\pi\tau)^{\omega}}\int^x_M\int_0^{\tau}\upsilon
E(x)G^0_{\cal O}\left(x,x;\frac{(\tau-\upsilon)\upsilon}{\tau}\right)
E(x)d\upsilon\sqrt{g}d^{2\omega}x\\
=\frac{1}{(4\pi\tau)^{\omega}}\int^x_M
E(x)\int_0^{\tau}\upsilon\cdot e^{\triangle\frac{(\tau-\upsilon)
\upsilon}{\tau}}d\upsilon E(x)\sqrt{g}d^{2\omega}x\\
=\frac{\tau^2}{(4\pi\tau)^{\omega}}\int^x_M
E(x)\int_0^1a\cdot e^{\triangle a(1-a)\tau}
da E(x)\sqrt{g}d^{2\omega}x\\
=\frac{\tau^2}{(4\pi\tau)^{\omega}}\int^x_M
E(x)\int_0^1(1-a)e^{\triangle a(1-a)\tau}
da E(x)\sqrt{g}d^{2\omega}x\\
=\frac{\tau^2}{(4\pi\tau)^{\omega}}\int^x_M
E(x)\int_0^1\frac{e^{\triangle a(1-a)\tau}}{2}
da E(x)\sqrt{g}d^{2\omega}x\\
:=\frac{\tau^2}{(4\pi\tau)^{\omega}}\int^x_M
E(x)\frac{f(-\triangle\tau)}{2}E(x)\sqrt{g}d^{2\omega}x,
\end{multline}
where I have defined the function
\begin{equation}
f(x)=\int_0^1e^{-a(1-a)x}da.\label{f2}
\end{equation}
This function produces the non-localities in the second order
of the covariant perturbation theory (as will be seen
when doing explicit calculations).

The geometric curvature and the gauge curvature can be
treated in the same way as the endomorphism by using the
decomposition (\ref{cov.pert.Laplace}) of the full Laplacian.
I will restrict myself to the second order of the perturbation
theory, although in principle all orders are accessible by this
formalism. In \cite{bav90}, Equation 2.1, the most general form of the
trace of the heat kernel in general dimensions to the second
order is given. Note that the Riemann tensor does not appear
because it has been eliminated by the Bianchi identities.
I adapt this formula according to my purpose, namely to
study Hawking radiation of massless scalar fields
from Schwarzschild BHs in the quasi-static phase:
\begin{itemize}
\item
I substitute the endomorphism by $E=\tilde{P}-\frac{R}{6}$
because I consider \emph{minimally coupled scalars}
(in the $4d$ action).
\item
\emph{The scalar fields shall commute}, hence I can drop
the gauge curvature by setting ${\cal R}_{\mu\nu}=0$.
\item
The BH spacetime is almost perfectly a \emph{vacuum spacetime
in the quasi-static phase}, thus I can eliminate
the Ricci tensor by the vacuum Einstein equations\footnote{In two dimensions
this relation is identically fulfilled for \emph{any} geometry. Namely,
the Euler number of a two-dimensional manifold $L$ is given by
$\chi(L)=\int_LR\sqrt{g}d^2x$. As a topological invariant
it is invariant under variations of the geometry, i.e. $\delta_g\chi(L)\propto
R_{\mu\nu}-\frac{g_{\mu\nu}}{2}R=0$.}
$R_{\mu\nu}=\frac{g_{\mu\nu}}{2}R$.
\end{itemize}
The trace of the heat kernel up to the second order
of the non-covariant perturbation theory in even dimensions
$d=2\omega$ then reads
\begin{multline}
\text{tr}\,e^{-{\cal O}\tau}=\frac{1}{(4\pi\tau)^{\omega}}
\int_M\text{tr}\biggm\{\1+\tau\left(\frac{R}{6}+E\right)\\
+\tau^2\biggm[R\left(\frac{1}{16(-\triangle\tau)}
+\frac{f(-\triangle\tau)}{32}
+\frac{f(-\triangle\tau)-1}{8(-\triangle\tau)}
+\frac{3[f(-\triangle\tau)-1]}{8(\triangle\tau)^2}\right)R\\
+E\left(\frac{f(-\triangle\tau)}{6}+\frac{f(-\triangle\tau)-1}
{2(-\triangle\tau)}\right)R+R\frac{f(-\triangle\tau)}{12}E
+E\frac{f(-\triangle\tau)}{2}E\biggm]\biggm\}\sqrt{g}d^{2\omega}x.
\label{nonlocal.effective.action}
\end{multline}
It is now manifest that the perturbation theory is indeed
covariant which means that it only depends on tensorial objects of the
\emph{full} manifold described by $g$. The crucial point
is that all geometric objects are in fact associated to the
perturbing part of the metric $h$, but as the starting point
of the series is a flat spacetime (given by $\tilde{g}$)
one can replace (to the linear order) all objects by those
associated to $g=\tilde{g}+h$. What remains is the zeroth
order contribution (in (\ref{nonlocal.effective.action})
represented by the unity term $\1$). But as the latter does not
contribute to the expectation values (apart from a renormalisation
constant related to the cosmological constant which is
here assumed to be zero) it does not spoil the general covariance.

The non-local form of the trace of the heat kernel can be
regarded as the more comprising case -- in the limit
of sufficiently strong damping the Seeley-DeWitt expansion is recovered
(it is neither difficult nor illuminating to show this
explicitly, hence I refer to the papers of Barvinsky and Vilkovisky
\cite{bav87,bav90}).
\\
\\
Finally, I will mention some points discussed by Barvinsky and Vilkovisky
that will be important for this work. First they state that
the conformally coupled scalar field in two dimensions (where $E=0$)
is the only case where the covariant perturbation theory is
applicable in $2d$.

I do not agree to this point
and demonstrate in Chapter 5 that a non-local effective action
(derived from the trace of the heat kernel) can be established
in all cases and for all endomorphisms and metrics if an
IR regularisation is introduced.

Barvinsky and Vilkovisky argue that the non-local form of the heat kernel
allows to recover all boundary conditions of Lorentzian
manifolds (which are not directly accessible in the Euclidean formalism)
by simply representing the non-local terms as integrals
over Green functions, thereby choosing the appropriate
Lorentzian Green functions. Further, they state that
expectation values in the common sense, the
initial and final state both being in-states, are obtained
by the use of the retarded Green function; the Feynman Green function
merely leads to scattering probabilities from in- to
out-states. With regard to these ideas I investigate the
role of the Green functions by the Green function perturbation
series in Section 5.4. Thereby I will come to the conclusion that
the effective action is independent of the choice of the Green function
in the static approximation. The quantum state merely seems
to be determined by the spacetime geometry alone and in the
Schwarzschild approximation it can be identified with the
Boulware state $\left|B\right>$.

The way Barvinsky and Vilkovisky derive their effective action from
the heat kernel differs from mine, see (\ref{eff.action4}).
This fact is not related to any of the considerations
above, but it has some relevance for the renormalisation.

\newpage
\section{Hawking Radiation of Massive Scalars}

The contribution of massive particles to the Hawking flux
is assumed to be much smaller than that of massless particles
if the BH is in the quasi-static phase.
The BH temperature is then much lower than the rest
mass of all elementary particles (except possibly neutrinos!)
and the probability that massive particles are produced is
therefore very small as compared to massless particles.
The physical picture is that particle-antiparticle states
of heavy particles self-interact on smaller distances and within
smaller periods of time and thus decrease
the probability that the BH catches the constituent
with negative energy before pair-annihilation occurs.
With decreasing BH mass, and hence increasing
surface temperature, the particle production rate
becomes more and more mass-independent.

In any case, there exists a lower limit for what I call
``massive particles'' imposed by the convergence condition
of the Seeley-DeWitt expansion. It will turn out that
this limiting value depends on the inverse of the BH
mass which means that the strength of surface gravity
determines if a particle is heavy enough as to be considered
as massive. In this Chapter I will only treat massive particles
in this sense, lighter particles are object of the non-local
formalism developed for the massless case.
The great advantage of (sufficiently) massive particles
is that one can establish a \emph{local covariant perturbation
series} of the effective action via the Seeley-DeWitt expansion.
The particle dynamics enters only by the perturbational
expansion of the Euclidean Feynman Green function.
This means that it is not necessary to know explicitly
the Green functions of the massive scalar particles to compute
the expectation values. Hence, the effective action
is directly available for all even spacetime dimensions
(for massless particles the two-dimensional case is somewhat
particular because only there one can still find a local form
of the effective action).

According to the above considerations, the effective
action is given in a particular quantum state (namely
the Boulware state) which is not suitable for the
calculation of Hawking radiation. Therefore, I will only
use it to compute the basic
components which have been shown to be state-independent
(see Section 2.4.2). The remaining components are then
calculated by the conservation equation a la CF.

The advantage of massive particles is that one can work
directly in four spacetime dimensions and avoids
the additional difficulties and uncertainties of the dilaton
model. Besides, one can still go over to two dimensions
and compare the results of the dilaton model with the
``correct'' results of the $4d$-theory. This provides
a check of the dilaton model at the quantum level that
may give hints for the treatment of massless particles.

\subsection{$4d$ Theory}

The procedure in four dimensions is quite straightforward.
I establish an effective action, whereby the mass-term
guarantees IR regularity. Then I compute the basic
components by variation for the metric to the first order
of the perturbation theory; this part of the calculation is
the most tedious one. Finally, I complete the EM tensor
by the CF method and calculate the Hawking-flux
in the vacuum state.

\subsubsection{Effective Action}

I consider the action (\ref{scalar action}) of a massive
particle without internal degrees of freedom and
without (further) self-interaction. The Euclidean Laplacian
thus is given by $-\triangle+m^2$ (note that the mass
term is invariant under Euclideanisation). To cover also the 
general case I add an (Euclidean) endomorphism $E$
to the Laplacian ${\cal O}+m^2=-\triangle-E+m^2$.
${\cal O}$ is the Laplacian of a massless scalar field with
an arbitrary endomorphism. The corresponding heat kernel is given by
\begin{equation}
e^{(-{\cal O}-m^2)\tau}=e^{-{\cal O}\tau}\cdot e^{-m^2\tau}.
\end{equation}
The mass-term can be separated without producing further
terms because it commutes with the remaining Laplacian.
In the following I will assume that the mass-term
provides a sufficient damping such that the heat kernel
of ${\cal O}$ can be expanded in a Seeley-DeWitt series
(\ref{heat.exp.}). The zeta-function of the full Laplacian
then reads (\ref{zeta-function2}):
\begin{multline}
\zeta_{{\cal O}+m^2}[s]=\frac{1}{\Gamma(s)}\int_0^{\infty}d\tau\tau^{s-1}
\int_M G_{{\cal O}+m^2}(x,x;\tau)\sqrt{g}d^4x\\
=\frac{1}{\Gamma(s)}\int_0^{\infty}d\tau\tau^{s-1}
\int_M \left<x\left|e^{(-{\cal O}-m^2)\tau}\right|x\right>\sqrt{g}d^4x\\
=\frac{1}{\Gamma(s)}\int_0^{\infty}d\tau\tau^{s-1}e^{-m^2\tau}
\int_M G_{\cal O}(x,x;\tau)\sqrt{g}d^4x\\
=\frac{1}{\Gamma(s)}\int_0^{\infty}d\tau\tau^{s-1}\frac{e^{-m^2\tau}}
{(4\pi\tau)^2}\int_M\sum_{n=0}^{\infty}a_{2n}\tau^n\sqrt{g}d^4x\\
=\frac{1}{(4\pi)^2\Gamma(s)}\int_0^{\infty}d\tau e^{-m^2\tau}\int_M
\Bigm[\tau^{s-3}a_0+\tau^{s-2}a_2+\tau^{s-1}a_4
+\tau^{s}a_6+\dots\Bigm]\sqrt{g}d^4x.\label{zeta-massive}
\end{multline}
Note that the $\tau$-integrations are divergent in the limit $m\to0$.
The integrals over the proper time $\tau$ are simply 
representations of the Gamma-function.
Thus I can formally write:
\begin{multline}
\zeta_{{\cal O}+m^2}[s]=\frac{1}{(4\pi)^2}\int_M\frac{1}{\Gamma(s)}
\Bigm[\Gamma[s-2](m^2)^{2-s}a_0+\Gamma[s-1](m^2)^{1-s}a_2\\
+\Gamma[s](m^2)^{-s}a_4+\frac{a_6}{m^2}+\frac{a_8}{m^4}+\dots\Bigm]
\sqrt{g}d^4x.
\end{multline}
The Gamma-functions have poles of first order at $s=0$ and at
all negative integer values of $s$. By expanding them around $s=0$ I get
\begin{eqnarray}
\Gamma[s-2]&=&\frac{1}{2s}+\left(\frac{3}{4}-
\frac{\gamma_E}{2}\right)+O(s)^1
\label{Gamma-exp.1}\\
\Gamma[s-1]&=&-\frac{1}{s}+(\gamma_E-1)+O(s)^1\\
\Gamma[s]&=&\frac{1}{s}-\gamma_E+O(s)^1\\
\frac{1}{\Gamma(s)}&=&s+s^2\cdot \gamma_E+O(s)^3\\
\left(m^2\right)^{(k-s)}&=&
\left(m^2\right)^k
-s\cdot\left(m^2\right)^k
\ln\left(m^2\right)+O(s),\label{R^(k-s)-exp.}
\end{eqnarray}
where $\gamma_E$ is the Euler-Mascheroni constant. Plugging
this into the zeta-function gives
\begin{multline}
\zeta_{{\cal O}+m^2}[s]\\
=\frac{1}{(4\pi)^2}\int_M\biggm\{\frac{a_0m^4}{2}\left[1+
s\left(3/4-\ln m^2\right)\right]
-a_2m^2\left[1+s\left(1-\ln m^2\right)\right]\\
+[1-s\ln m^2]a_4+s\left(\frac{a_6}{m^2}+\frac{a_8}{m^4}
+\frac{2!a_{10}}{m^6}\right)+\dots+O(s)^2
\biggm\}\sqrt{g}d^4x.
\end{multline}
The Euclidean effective action according to (\ref{eff.action4}) is obtained
by differentiation for $s$ and then setting $s=0$:
\begin{multline}
W[g,m^2]=-\frac{1}{32\pi^2}\int_M\biggm[
\frac{a_0m^4}{8}(3-2\ln m^2)+a_2m^2(\ln m^2-1)-a_4\ln m^2\\
+\frac{a_6}{m^2}+\frac{a_8}{m^4}+\frac{2!a_{10}}{m^6}+
\dots\biggm]\sqrt{g}d^4x.\label{eff.action-mass}
\end{multline}
For $m\to\infty$ the terms in the first line diverge, while
the remaining terms tend to zero. In the last line of (\ref{zeta-massive})
one can see that these terms further exhibit bad UV behaviour,
namely their $\tau$-integrals have poles at $\tau=0$.
The first (divergent) terms in the Seeley-DeWitt expansion
are connected to the renormalisation of the vacuum energy
and gravitational radiation. The first coefficient $a_0$, which is
simply constant, has to be removed to obtain zero vacuum
energy in flat spacetime. In other words, the vacuum energy density
in curved spacetime is renormalised by substracting the flat
spacetime value which is assumed to be zero. The coefficients $a_2,a_4$ cannot
be interpreted this simply. $a_2$, as it only contains the
scalar curvature, does not contribute to the EM tensor on a vacuum
spacetime because there $R=0,R_{\mu\nu}=0$ (if the backreaction is
neglected). Hence only $a_4$ causes problems. The factor $\ln m^2$
leads to diverging vacuum energy density for infinitely heavy particles,
though it is expected that such particles do not contribute at all
(at least in the stationary phase).
After gravitational radiation has been renormalised a term $\propto a_4$
may still appear in the effective action (I come back to this
point in Section 4.3). For the moment, and
following \cite{dew63}, I drop these coefficients
and define the renormalised Euclidean effective action as
\begin{equation}
W_{ren}[g,m^2]:=-\frac{1}{32\pi^2}\int_M\biggm[
\frac{a_6}{m^2}+\frac{a_8}{m^4}+\frac{2!a_{10}}{m^6}+
\dots\biggm]\sqrt{g}d^4x.\label{eff.action-mass.ren}
\end{equation}
The Lorentzian effective action is related to the Euclidean one
by $W_{\cal M}=iW_{\cal E}$, see Appendix E. If each term in the
action is switched back by its own I must add
a minus sign: $W_{\cal M}=-iW_{\cal E}$!. The Euclidean time-coordinate
transforms as $d\tau=idt$. Further, one has to replace the
geometric objects in the Seeley-DeWitt coefficients
according to the rules given in Appendix E.
The renormalised Lorentzian effective action reads
\begin{equation}
W^{ren}_{\cal M}[g,m^2]:=-\frac{1}{32\pi^2}\int_M\biggm[
\frac{a_6^{\cal M}}{m^2}+\frac{a_8^{\cal M}}{m^4}
+\frac{2!a_{10}^{\cal M}}{m^6}+\dots\biggm]\sqrt{-g}d^4x_{\cal M}.
\label{eff.action-mass-lorentz}
\end{equation}
In the following I will omit the index ${\cal M}$ as long as
I work with the Lorentzian effective action.

Until now I have assumed that the damping provided by the mass
term is sufficient to guarantee the convergence of the Seeley-DeWitt
expansion (otherwise I cannot apply it). Now I will examine when
this is indeed the case.
According to the considerations in Section 3.1.3 the whole series converges
if $\lim_{n\to\infty}\text{sup}_{x\in M}(a_{2n+2}/m^2a_{2n})<1$.
Because I only know the first coefficients explicitly I can
only make a crude estimate to see how the convergence condition
can be fulfilled. For the known coefficients one observes that
(in the present case of a Schwarzschild spacetime with vanishing
endomorphism $E$) their ratio is of the order of a curvature
\begin{equation}
\frac{a_{2n+2}}{a_{2n}}=c|R|\approx c\,\frac{2M}{r^3}\,,\,c<1.
\end{equation}
The supremum of these values is found on the horizon $r=2M$
(the series need not converge inside the horizon because the
expectation values which are calculated from it correspond
to local measurements outside the horizon).
Thus the condition
\begin{equation}
m\gg\frac{1}{M}
\end{equation}
seems to be crucial to the convergence of the series.
This means that the Compton wave length of the massive
particle has to be much smaller than the radius
of the event horizon.
The masses are given in Planck units 
($m_{Pl}=1$). For instance the mass
of the sun in Planck units is $M_{\odot}=6\cdot10^{38}$.
Near a BH with the mass of the sun the scalar
particles would need to be heavier than $10^{-38}$.
For comparison, the mass of an electron is $m_{e^-}=2.9\cdot10^{-22}$.

Thus it seems that for a quasi-static BH the series
converges nicely for all known fundamental massive particles,
except probably neutrinos -- the latter can be considered
massless and must be treated respectively. Once again I
mention that, although the existence of fundamental scalar
particles is not yet established (the Higgs still awaits observation),
the treatment of particles with higher spin will not be fundamentally
different from that of scalar particles. Hence, the Hawking flux
calculated in this Chapter will give a good approximation
for the flux of electrons and positrons in a realistic BH
scenario. In the quasi-static region the total contribution
of massive particles will be negligible. Namely, the Hawking
temperature $T_H=\frac{1}{8\pi M}$ is much smaller than the rest mass
of the produced particles. The conclusion is that the local
expansion can only be applied for particles whose mass
is so large that their contribution to the flux (for the
given temperature) can be neglected.

\subsubsection{Computation of the Basic Components}

The most direct way to calculate the expectation value of the
EM tensor is to vary the effective action for the metric,
according to its definition (\ref{EM tensor}). Thereby
one immediately obtains \emph{all} components of the EM
tensor. There is the problem that the effective action does
not deliver expectation values in arbitrary quantum states,
see Section 2.4. Therefore I restrict myself to the
state-independent basic observables $\E{T}$ and $\E{\Th}$.
The Lorentzian expectation value of the EM tensor
is given by the variation of the effective action
(\ref{eff.action-mass-lorentz})
\begin{equation}
\E{T_{\mu\nu}}=\frac{2}{\sqrt{-g}}\frac{\delta}{\delta g^{\mu\nu}}
W_{\cal M}[g,m^2],
\end{equation}
where the Lorentzian Seeley-DeWitt coefficients correspond to
the geometric Laplacian on a Schwarzschild spacetime and $E=0$.
I consider only the coefficient $a_6$, see Table 4 in Section 3.1.1:
\begin{multline}
a_6^{\cal M}=-a_6^{\cal E}=-\frac{1}{9\cdot 7!}
\Bigm[81R_{mnop,q}R^{mnop,q}+108R_{mnop}\square R^{mnop}\\
-44R_{mnop}R^{mn}{}_{qr}R^{opqr}-80R_{mnop}R^{m}{}_{q}{}^{o}{}_{r}R^{nqpr}\\
+42R\cdot R_{mnop}R^{mnop}-48R_{mn}R^m{}_{opq}R^{nopq}\Bigm].
\end{multline}
The minus sign appears because each term contains an odd
number of metrics. Before performing the variation I can do some
simplifications.
The terms in the first line can be brought to a single expression
by a partial integration (in the effective action). By use of the vacuum Einstein
equations $R_{\mu\nu}=\frac{g_{\mu\nu}}{2}R$ the terms in the
last line become form-equivalent. There remain four different
types of terms:
\begin{multline}
\frac{1}{9\cdot 7!}\Bigm[27R_{mnop,q}R^{mnop,q}-18R\cdot R_{mnop}R^{mnop}\\
+44R_{mnop}R^{mn}{}_{qr}R^{opqr}+80R_{mnop}R^{m}{}_{q}{}^{o}{}_{r}R^{nqpr}\Bigm].
\end{multline}
On a Schwarzschild spacetime $a_6^{\cal M}$ becomes (for a purely geometric
scalar field Laplacian with vanishing endomorphism) (see Appendix B.3):
\begin{equation}
a_6^{\cal M}=\frac{1}{105}\left(-\frac{27M^2}{r^8}+\frac{46M^3}{r^9}\right).
\end{equation}
In the following I will compute the variation of each of these
terms for the metric. Another term contributes to the EM tensor,
namely the one where the measure is varied for the metric (\ref{var.measure});
this term can be written down immediately:
\begin{equation}
\E{T_{\mu\nu}}=\frac{g_{\mu\nu}}{32m^2\pi^2}\cdot a_6^{\cal M}
+\text{terms}\propto\frac{\delta a_6^{\cal M}}{\delta g^{\mu\nu}}.
\end{equation}
Now I come to the remaining four terms which are a bit trickier
to handle.
Because it will be necessary to perform partial integrations I will
write down the action integral including the spacetime measure, but I
will not vary the measure as this term already has been considered.
For the variations of the geometric objects see Appendix B.3.
I start with the most simple one, whereby I set $R=0,R_{\mu\nu}=0$
after the variation:
\begin{multline}
\frac{\delta}{\delta g^{\mu\nu}}\int_MR\cdot R_{\rho\sigma\tau\upsilon}
R^{\rho\sigma\tau\upsilon}\sqrt{-g}d^4x
=\int_M\frac{\delta R}{\delta g^{\mu\nu}}\cdot R_{\rho\sigma\tau\upsilon}
R^{\rho\sigma\tau\upsilon}\sqrt{-g}d^4x\\
=\int_Mg^{\rho\sigma}\frac{\delta R_{\rho\sigma}}{\delta g^{\mu\nu}}
\cdot R_{\rho\sigma\tau\upsilon}R^{\rho\sigma\tau\upsilon}\sqrt{-g}d^4x
=\sqrt{-g}[\square g_{\mu\nu}-\n_{\mu}\n_{\nu}]R_{\rho\sigma\tau\upsilon}
R^{\rho\sigma\tau\upsilon}.
\end{multline}
Next I consider the terms with each three Riemann tensors.
I write them in the form
\begin{eqnarray}
\Bigm(g^{\gamma\kappa}g^{\delta\lambda}\Bigm)\cdot\Bigm(g^{\rho\tau}
g^{\sigma\upsilon}\Bigm)\cdot\Bigm(g^{\omega\alpha}g^{\eta\beta}\Bigm)
\cdot R_{\alpha\beta\gamma\delta}R_{\kappa\lambda\rho\sigma}
R_{\tau\upsilon\omega\eta}\label{var.expr1}\\
\Bigm(g^{\beta\kappa}g^{\delta\rho}\Bigm)\cdot\Bigm(g^{\lambda\tau}
g^{\sigma\omega}\Bigm)\cdot\Bigm(g^{\upsilon\alpha}g^{\eta\gamma}\Bigm)
\cdot R_{\alpha\beta\gamma\delta}R_{\kappa\lambda\rho\sigma}
R_{\tau\upsilon\omega\eta}\label{var.expr2}
\end{eqnarray}
to take advantage of the existing symmetries. It is obvious that
each pair of metrics in brackets can be exchanged with each other
pair without changing the expressions. Additionally, in the first
expression (\ref{var.expr1}) one can exchange the two metrics within a certain pair
by use of the symmetry $R_{\alpha\beta\gamma\delta}R_{\kappa\lambda\rho\sigma}
=R_{\alpha\beta\delta\gamma}R_{\kappa\lambda\sigma\rho}$. Hence, the variations
for the metric $\delta g^{\mu\nu}$ (neglecting for the moment the metrics
in the Riemann tensors) lead to
\begin{eqnarray}
\frac{\delta}{\delta g^{\mu\nu}}(\ref{var.expr1})\to6\Bigm(g^{\delta\lambda}\Bigm)
\cdot\Bigm(g^{\rho\tau}
g^{\sigma\upsilon}\Bigm)\cdot\Bigm(g^{\omega\alpha}g^{\eta\beta}\Bigm)
\cdot R_{\alpha\beta(\mu)\delta}R_{(\nu)\lambda\rho\sigma}
R_{\tau\upsilon\omega\eta}\\
=6R_{(\mu)\delta\alpha\beta}R_{\nu}{}^{\delta}{}_{\rho\sigma}
R^{\alpha\beta\rho\sigma}\\
\frac{\delta}{\delta g^{\mu\nu}}(\ref{var.expr2})\to3\Bigm(g^{\delta\rho}\Bigm)
\cdot\Bigm(g^{\lambda\tau}
g^{\sigma\omega}\Bigm)\cdot\Bigm(g^{\upsilon\alpha}g^{\eta\gamma}\Bigm)
\cdot R_{\alpha(\mu)\gamma\delta}R_{(\nu)\lambda\rho\sigma}
R_{\tau\upsilon\omega\eta}\nonumber\\
+3\Bigm(g^{\beta\kappa}\Bigm)\cdot\Bigm(g^{\lambda\tau}
g^{\sigma\omega}\Bigm)\cdot\Bigm(g^{\upsilon\alpha}g^{\eta\gamma}\Bigm)
\cdot R_{\alpha\beta\gamma(\mu)}R_{\kappa\lambda(\nu)\sigma}
R_{\tau\upsilon\omega\eta}\\
=6R_{(\mu)\alpha\beta\delta}R_{(\nu)\rho\sigma}{}^{\delta}
R^{\alpha\rho\beta\sigma}.
\end{eqnarray}
For visibility I have set the free indices under brackets.
Now I vary the Riemann tensors in the two expressions.
By the obvious symmetry it suffices to vary the first of the three Riemann
tensors (\ref{var.Riemann}):
\begin{multline}
\delta R_{\alpha\beta\gamma\delta}=\frac{1}{2}[\n_{\delta}\n_{\beta}
\delta g_{\alpha\gamma}+\n_{\delta}\n_{\gamma}\delta g_{\alpha\beta}
-\n_{\delta}\n_{\alpha}\delta g_{\beta\gamma}\\
-\n_{\gamma}\n_{\beta}\delta g_{\alpha\delta}-\n_{\gamma}\n_{\delta}
\delta g_{\alpha\beta}+\n_{\gamma}\n_{\alpha}\delta g_{\beta\delta}].
\end{multline}
The contribution of the first expression (\ref{var.expr1}) is:
\begin{multline}
\int_M3(\delta R_{\alpha\beta\gamma\delta})R^{\gamma\beta}{}_{\mu\nu}
R^{\alpha\beta\mu\nu}\sqrt{-g}d^4x\\
=\int_M\frac{3}{2}\delta g_{\alpha\beta}\cdot g_{\mu\kappa}\,g_{\nu\lambda}
\Bigm[\n_{\gamma}\n_{\delta}(R^{\beta\delta\kappa\lambda}R^{\alpha\gamma\mu\nu}
+R^{\gamma\delta\kappa\lambda}R^{\alpha\beta\mu\nu}-R^{\alpha\delta\kappa\lambda}
R^{\gamma\beta\mu\nu})\\
-\n_{\delta}\n_{\gamma}(R^{\gamma\beta\kappa\lambda}R^{\alpha\delta\mu\nu}
+R^{\gamma\delta\kappa\lambda}R^{\alpha\beta\mu\nu}-R^{\gamma\alpha\kappa\lambda}
R^{\delta\beta\mu\nu})\Bigm]\sqrt{-g}d^4x\\
=\int_M3\,\delta g_{\alpha\beta}\n_{\gamma}\n_{\delta}
(R^{\beta\delta}{}_{\mu\nu}R^{\alpha\gamma\mu\nu}
+R^{\alpha\delta}{}_{\mu\nu}R^{\beta\gamma\mu\nu})\sqrt{-g}d^4x\\
=-\int_M3\,\delta g^{\alpha\beta}\n_{\gamma}\n_{\delta}
(R_{\beta}{}^{\delta}{}_{\mu\nu}R_{\alpha}{}^{\gamma\mu\nu}
+R_{\alpha}{}^{\delta}{}_{\mu\nu}R_{\beta}{}^{\gamma\mu\nu})\sqrt{-g}d^4x.
\end{multline}
Note that the antisymmetric part has been eliminated by the symmetry of the
varied metric $\delta g_{\alpha\beta}=\delta g_{\beta\alpha}$.
The second expression (\ref{var.expr2}) leads to a contribution
\begin{multline}
\int_M3(\delta R_{\alpha\beta\gamma\delta})R^{\beta}{}_{\mu}{}^{\delta}{}_{\nu}
R^{\mu\alpha\nu\gamma}\sqrt{-g}d^4x\\
=\int_M\frac{3}{2}\delta g_{\alpha\beta}\cdot g_{\mu\kappa}\,g_{\nu\lambda}
\Bigm[\n_{\gamma}\n_{\delta}(R^{\gamma\kappa\delta\lambda}R^{\mu\alpha\nu\beta}
+R^{\beta\kappa\delta\lambda}R^{\mu\alpha\nu\gamma}-
R^{\beta\kappa\delta\lambda}R^{\mu\gamma\nu\alpha})\\
-\n_{\delta}\n_{\gamma}(R^{\delta\kappa\beta\lambda}R^{\mu\alpha\nu\gamma}
+R^{\beta\kappa\delta\lambda}R^{\mu\alpha\nu\gamma}-
R^{\beta\kappa\alpha\lambda}R^{\mu\delta\nu\gamma})\Bigm]\sqrt{-g}d^4x\\
=\int_M\frac{3}{2}\delta g_{\alpha\beta}\cdot g_{\mu\kappa}\,g_{\nu\lambda}
\Bigm\{\Bigm[\n_{\gamma}\n_{\delta}-\n_{\delta}\n_{\gamma}\Bigm]
(R^{\gamma\kappa\delta\lambda}R^{\mu\alpha\nu\beta}
+R^{\beta\kappa\delta\lambda}R^{\mu\alpha\nu\gamma})\\
-\Bigm[\n_{\gamma}\n_{\delta}+\n_{\delta}\n_{\gamma}\Bigm]
R^{\beta\kappa\delta\lambda}R^{\mu\gamma\nu\alpha}\Bigm\}\sqrt{-g}d^4x\\
=-\int_M3\delta g_{\alpha\beta}\n_{\gamma}\n_{\delta}
(R^{\beta}{}_{\mu}{}^{\delta}{}_{\nu}R^{\mu\gamma\nu\alpha})\sqrt{-g}d^4x\\
=\int_M\frac{3}{2}\delta g^{\alpha\beta}\n_{\gamma}\n_{\delta}
(R_{\beta}{}_{\mu}{}^{\delta}{}_{\nu}R^{\mu\gamma\nu}{}_{\alpha}
+R_{\alpha}{}_{\mu}{}^{\delta}{}_{\nu}R^{\mu\gamma\nu}{}_{\beta})\sqrt{-g}d^4x.
\end{multline}
The symmetry in the indices $\gamma$ and $\delta$ is induced
by the symmetry in $\alpha$ and $\beta$, e.g.:
\begin{equation}
\delta g_{\alpha\beta}R^{\beta}{}_{\mu}{}^{\delta}{}_{\nu}R^{\mu\gamma\nu\alpha}
=\delta g_{\alpha\beta}R^{\alpha}{}_{\mu}{}^{\delta}{}_{\nu}R^{\mu\gamma\nu\beta}
=\delta g_{\alpha\beta}R^{\beta\nu\gamma\mu}R^{\alpha}{}_{\mu}{}^{\delta}{}_{\nu}
=\delta g_{\alpha\beta}R^{\beta}{}_{\nu}{}^{\gamma}{}_{\mu}R^{\nu\delta\mu\alpha}.
\end{equation}
Now I come to the final expression:
\begin{equation}
g^{\mu\nu}\cdot g^{\alpha\rho}g^{\beta\sigma}g^{\gamma\tau}g^{\delta\upsilon}
(\n_{\mu}R_{\alpha\beta\gamma\delta})(\n_{\nu}R_{\rho\sigma\tau\upsilon}).
\end{equation}
The variation for the free metrics yields
\begin{equation}
\frac{\delta}{\delta g^{\mu\nu}}\to(\n_{\mu}R_{\alpha\beta\gamma\delta})
(\n_{\nu}R^{\alpha\beta\gamma\delta})+4(\n_{\kappa}
R_{\mu\beta\gamma\delta})(\n^{\kappa}R_{\nu}{}^{\beta\gamma\delta}).
\end{equation}
The variation of the curvature term I compute again by use of the
auxiliary metric $\tilde{g}$, see Appendix B.2:
\begin{multline}
\tilde{\n}_{\mu}\tilde{R}_{\alpha\beta\gamma\delta}=
\n_{\mu}\tilde{R}_{\alpha\beta\gamma\delta}-C^{\kappa}_{\mu\alpha}
\tilde{R}_{\kappa\beta\gamma\delta}-C^{\kappa}_{\mu\beta}
\tilde{R}_{\alpha\kappa\gamma\delta}-C^{\kappa}_{\mu\gamma}
\tilde{R}_{\alpha\beta\kappa\delta}-C^{\kappa}_{\mu\delta}
\tilde{R}_{\alpha\beta\gamma\kappa}\\
=\n_{\mu}\delta R_{\alpha\beta\gamma\delta}-C^{\kappa}_{\mu\alpha}
R_{\kappa\beta\gamma\delta}-C^{\kappa}_{\mu\beta}
R_{\alpha\kappa\gamma\delta}-C^{\kappa}_{\mu\gamma}
R_{\alpha\beta\kappa\delta}-C^{\kappa}_{\mu\delta}
R_{\alpha\beta\gamma\kappa}+O(\delta g^2).
\end{multline}
By symmetry considerations one can see that all terms in $C$
lead to the same contribution:
\begin{multline}
(-C^{\kappa}_{\mu\alpha}R_{\kappa\beta\gamma\delta}
-C^{\kappa}_{\mu\beta}R_{\alpha\kappa\gamma\delta}
-C^{\kappa}_{\mu\gamma}R_{\alpha\beta\kappa\delta}-C^{\kappa}_{\mu\delta}
R_{\alpha\beta\gamma\kappa})\n^{\mu}R^{\alpha\beta\gamma\delta}\\
=-4C^{\kappa}_{\mu\beta}R_{\alpha\kappa\gamma\delta}\n^{\mu}
R^{\alpha\beta\gamma\delta}.
\end{multline}
Hence, the whole variation of the curvature term gives
\begin{multline}
\int_M2(\n_{\mu}\delta R_{\alpha\beta\gamma\delta}-4C^{\kappa}_{\mu\alpha}
R_{\kappa\beta\gamma\delta})(\n^{\mu}R^{\alpha\beta\gamma\delta})\sqrt{-g}d^4x\\
=\int_M\Bigm[\n_{\mu}\bigm(\n_{\gamma}\n_{\beta}\delta g_{\alpha\delta}
+\n_{\gamma}\n_{\delta}
\delta g_{\alpha\beta}-\n_{\gamma}\n_{\alpha}\delta g_{\beta\delta}
-\n_{\delta}\n_{\beta}\delta g_{\alpha\gamma}\\
-\n_{\delta}\n_{\gamma}\delta g_{\alpha\beta}
+\n_{\delta}\n_{\alpha}\delta g_{\beta\gamma}\bigm)\\
-4\bigm(\n_{\mu}\delta g_{\nu\alpha}+\n_{\alpha}\delta g_{\nu\mu}
-\n_{\nu}\delta g_{\mu\alpha}\bigm)R^{\nu}{}_{\beta\gamma\delta}\Bigm]
\n^{\mu}R^{\alpha\beta\gamma\delta}\sqrt{-g}d^4x\\
=\int_M\Bigm[4\delta g_{\beta\delta}\n_{\alpha}\n_{\gamma}\square
R^{\alpha\beta\gamma\delta}\\
+2\bigm(\delta g_{\nu\alpha}\n_{\mu}+\delta g_{\nu\mu}\n_{\alpha}
-\delta g_{\mu\alpha}\n_{\nu}\bigm)R^{\nu}{}_{\beta\gamma\delta}
(\n^{\mu}R^{\alpha\beta\gamma\delta})\Bigm]\sqrt{-g}d^4x\\
=-\int_M\delta g^{\mu\nu}\Bigm[4\n_{\kappa}\n_{\lambda}\square
R^{\kappa}{}_{\mu}{}^{\lambda}{}_{\nu}\\
+2\n_{\kappa}\bigm(
R_{\mu\lambda\gamma\delta}\n^{\kappa}R_{\nu}{}^{\lambda\gamma\delta}
+R_{\mu\lambda\gamma\delta}\n_{\nu}R^{\kappa\lambda\gamma\delta}
-R^{\kappa}{}_{\lambda\gamma\delta}\n_{\mu}R_{\nu}{}^{\lambda\gamma\delta}
\bigm)\Bigm]\sqrt{-g}d^4x.
\end{multline}
Note that all terms in this expression need to be symmetrised
in the indices $\mu,\nu$ of the EM tensor. Now I have
computed all types of contributions to the EM tensor from the
Seeley-DeWitt coefficient $a_6$. In this approximation the EM
tensor thus reads
\begin{multline}
\E{T_{\mu\nu}}=\frac{1}{16m^2\pi^2}\biggm\{\frac{g_{\mu\nu}}{2}\cdot a_6^{\cal M}
+\frac{1}{7!}(2g_{\mu\nu}\square-\n_{\mu}\n_{\nu}-\n_{\nu}\n_{\mu})
R_{\alpha\beta\gamma\delta}R^{\alpha\beta\gamma\delta}\\
-\frac{24}{9\cdot 7!}\bigm(11R_{(\mu)\kappa\alpha\beta}R_{(\nu)}{}^{\kappa}
{}_{\gamma\delta}R^{\alpha\beta\gamma\delta}+20R_{(\mu)\alpha\beta\kappa}
R_{(\nu)\gamma\delta}{}^{\kappa}R^{\alpha\gamma\beta\delta}\bigm)\\
+\frac{12}{9\cdot 7!}\n_{\kappa}\n_{\lambda}\Bigm[11\bigm(R_{\mu}{}^{\lambda}
{}_{\alpha\beta}R_{\nu}{}^{\kappa\alpha\beta}+R_{\nu}{}^{\lambda}
{}_{\alpha\beta}R_{\mu}{}^{\kappa\alpha\beta}\bigm)\\
+10\bigm(R_{\mu\alpha}{}^{\lambda}{}_{\beta}R_{\nu}{}^{\beta\kappa\alpha}
+R_{\nu\alpha}{}^{\lambda}{}_{\beta}R_{\mu}{}^{\beta\kappa\alpha}\bigm)\Bigm]\\
+\frac{3}{7!}\Bigm[\bigm(\n_{\mu}R_{\alpha\beta\gamma\delta}\bigm)
\bigm(\n_{\nu}R^{\alpha\beta\gamma\delta}\bigm)+4\bigm(\n_{\kappa}
R_{\mu\alpha\beta\gamma}\bigm)\bigm(\n^{\kappa}R_{\nu}{}^{\alpha\beta\gamma}\bigm)\\
-2\n_{\kappa}\n_{\lambda}\square\bigm(R^{\kappa}{}_{\mu}{}^{\lambda}{}_{\nu}
+R^{\kappa}{}_{\nu}{}^{\lambda}{}_{\mu}\bigm)
-\n_{\kappa}\bigm(R_{\mu\alpha\beta\gamma}\n^{\kappa}R_{\nu}{}^{\alpha\beta\gamma}
+R_{\nu\alpha\beta\gamma}\n^{\kappa}R_{\mu}{}^{\alpha\beta\gamma}\\
+R_{\mu\alpha\beta\gamma}\n_{\nu}R^{\kappa\alpha\beta\gamma}
+R_{\nu\alpha\beta\gamma}\n_{\mu}R^{\kappa\alpha\beta\gamma}
-R^{\kappa}{}_{\alpha\beta\gamma}\n_{\mu}R_{\nu}{}^{\alpha\beta\gamma}
-R^{\kappa}{}_{\alpha\beta\gamma}\n_{\nu}R_{\mu}{}^{\alpha\beta\gamma}
\bigm)\Bigm]\biggm\}.
\end{multline}
From this formula I can calculate the desired basic components
(to the first perturbational order) by inserting the geometric
objects of a Schwarzschild spacetime. The explicit computation
of the various terms is done in Appendix B.3.
The quantum trace of the EM tensor is now given by
\begin{multline}
\E{T}=g^{\mu\nu}\E{T_{\mu\nu}}=\frac{1}{16\cdot 7!\cdot m^2\pi^2}
\Bigm\{21R_{\alpha\beta\gamma\delta;\kappa}R^{\alpha\beta\gamma\delta;\kappa}
+3\square(R_{\alpha\beta\gamma\delta}R^{\alpha\beta\gamma\delta})\\
-\frac{176}{9}R_{\alpha\beta\gamma\delta}R^{\alpha\beta}{}_{\kappa\lambda}
R^{\gamma\delta\kappa\delta}-\frac{320}{9}R_{\alpha\beta\gamma\delta}
R^{\alpha}{}_{\kappa}{}^{\gamma}{}_{\lambda}R^{\beta\kappa\delta\lambda}\\
+6\n_{\kappa}\bigm(R^{\kappa}{}_{\alpha\beta\gamma}\n^{\mu}R_{\mu}
{}^{\alpha\beta\gamma}-R_{\mu\alpha\beta\gamma}\n^{\mu}R^{\kappa\alpha\beta\gamma}
\bigm)\Bigm\}\\
=\frac{1}{80640\cdot m^2\pi^2}\left(-\frac{13392M^2}{r^8}
+\frac{30048M^3}{r^9}\right).\label{T.massive}
\end{multline}
The other component that I calculate directly from the effective
action is
\begin{multline}
\E{\Th}=g^{\theta\theta}\E{T_{\theta\theta}}
=\frac{a_6^{\cal M}}{32m^2\pi^2}\\
+\frac{1}{16\cdot 7!\cdot m^2\pi^2}\Bigm\{2\square(
R_{\alpha\beta\gamma\delta}R^{\alpha\beta\gamma\delta})
-2\eta^{22}\omega^1{}_2(E_2)E_1(R_{\alpha\beta\gamma\delta}
R^{\alpha\beta\gamma\delta})\\
-\frac{88}{3}4\sum_aR_{2a2a}R^{2a}{}_{2a}R^{2a2a}
-\frac{160}{3}2\sum_{a,b}R_{2a2a}R^2{}_{b}{}^{2}{}_{b}R^{abab}\\
+\frac{176}{3}\sum_a\n_a\n^a(R_{2a2a}R_{2a2a})
+\frac{160}{3}\sum_a\n_2\n^2(R_{2a2a}R_{2a2a})\\
+3(\n_2R_{abcd})\n^2R_{abcd}+12(\n_kR_{2bcd})\n^kR^{2bcd}
-12\eta_{22}\n_k\n_l\square R^{k2l2}\\
-6(\n_kR_{2abc})\n^kR^{2abc}-6R_{2abc}\square R^{2abc}\\
+6\eta^{22}\n_k(R_{2abc}\n_2R^{kabc})-6\eta^{22}
\n_k(R^{kabc}\n_2R^{2abc})\Bigm\}\\
=\frac{1}{3360m^2\pi^2}\left(-\frac{27M^2}{r^8}+\frac{46M^3}{r^9}\right)\\
+\frac{1}{16\cdot 7!\cdot m^2\pi^2}\biggm\{-\frac{2880M^2}{r^8}
+\frac{6912M^3}{r^9}+\frac{576M^2}{r^8}-\frac{1152M^3}{r^9}\\
+\frac{704M^3}{r^9}+\frac{1280M^3}{r^9}-\frac{10912M^2}{r^8}+\frac{23936M^3}{r^9}
+\frac{1920M^2}{r^8}-\frac{3840M^3}{r^9}\\
+3\cdot0-\frac{648M^2}{r^8}+\frac{1296M^3}{r^9}
-\frac{2880M}{r^7}+\frac{18144M^2}{r^8}-\frac{25920M^3}{r^9}\\
+\frac{432M}{r^8}-\frac{1296M^3}{r^9}\biggm\}
=\frac{1}{80640\cdot m^2\pi^2}\left(-\frac{2880M}{r^7}+\frac{5984M^2}{r^8}
+\frac{3024M^3}{r^9}\right).\label{Th.massive}
\end{multline}

\subsection{Flux and Energy Density}

The basic components calculated in the last
Section, $\E{T}$ and $\E{\Th}$, have been shown to be state-independent
(see Section 2.4.2). As I want to study the vacuum expectation value
of the EM tensor which has been identified with the expectation
value in the $\left|U\right>$-state, I must rely on the
CF method to compute the missing components.
All that has to be done is to calculate the flux-determining constant $K$ which
is related to the value $f(\infty)$ of the function (\ref{f}).
In the $\left|U\right>$-state this relation is given by
$K_U=\frac{1}{2}M^2f(\infty)$, see Table 2 at the end of Section 2.4.1.
With (\ref{T.massive},\ref{Th.massive}),
the results of the last Section, I obtain
\begin{multline}
K_U=\frac{M^2f(\infty)}{2}=M^2\int_{2M}^{\infty}
\left[\frac{M\E{T}}{2}+(r-3M)\E{\Th}\right]dr\\
=-\frac{769}{54190080\cdot\pi^2m^2M^2}.
\end{multline}
Now all components of the vacuum expectation value of the EM tensor
are determined (remember that $Q=0$). From (\ref{sol1})
the flux is given by
\begin{equation}
\U{T^r{}_t}=\frac{769}{54190080\cdot\pi^2m^2M^4r^2}.
\label{flux.massive}
\end{equation}
The radial stress and the energy density can be calculated
by the formulas (\ref{sol2}) and $T^t{}_t=T-T^r{}_r-2\Th$.
I will present here only the analytical form of the asymptotic energy density:
\begin{equation}
\U{T_{tt}}\stackrel{r\to\infty}{\approx}\frac{1}{r^2}\left(\frac{K_U}{M^2}
-f(\infty)\right)=\frac{769}{54190080\cdot\pi^2m^2M^4r^2}.
\label{asymp.energy.dens.massive}
\end{equation}
The energy density near the event horizon is shown in Figure 6.

\begin{figure}[h]

\hspace{2cm}\epsfig{file=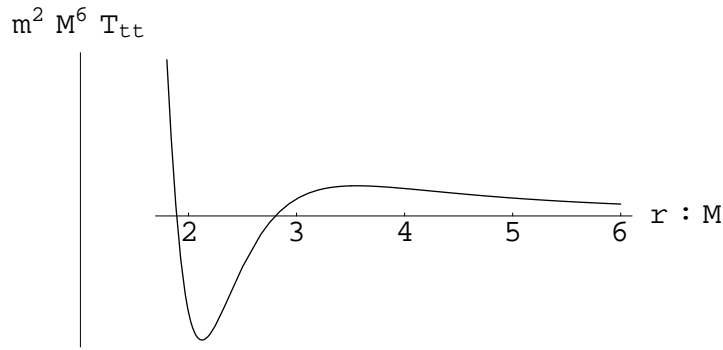,height=6cm}

\caption[fig1]{Energy Density for Massive Particles}
\end{figure}

Note that the result is only valid until the horizon.
The graph shows that the energy density becomes negative outside
the horizon and reverses its sign at a radius $r\approx2.749M$
and then falls off like (\ref{asymp.energy.dens.massive}).
The point where it becomes zero is almost mass-independent
if $m\gg\frac{1}{M}$ in the quasi-static region $M\gg m_{Pl}$=1.
In the interesting range, where the mass of the scalar particle
is small $m\approx\frac{1}{M}$, one must take into account
higher order terms of the perturbation series that change the radius
of vanishing energy density. Presumably, it is shifted closer to
the horizon with decreasing mass. Namely, within the area of negative
energy density the massive particles cannot escape to infinity
but fall back into the BH. Another consideration leads to the
same result: a BH in the quasi-static phase with a certain
mass $M\gg1$ provides a characteristic energy (given by the Hawking temperature)
to produce particles. If the major part of this energy is consumed
by the rest mass of some produced particle, a smaller amount of energy
remains as kinetic energy. Hence, such particles with higher mass
must start further away from the BH to reach infinity
and for them the zone of negative energy density increases.
I emphasize that this idea is not a consequence of my results and could
only be verified by a calculation that covers the interesting region
$m\approx\frac{1}{M}$.

\emph{In the region of negative energy density the weak
energy condition is violated} and the spacetime curvature is
effectively reduced from its vacuum value (produced by the BH).
This can be explained by the presence of \emph{virtual particles
with negative energy that flow into the BH}
and thereby decrease its mass -- clearly these particles
(being virtual) cannot be measured! This flux of particles with
negative energy cannot be observed by the flux component of the
EM tensor alone (\ref{flux.massive}) -- a flux of negative energy
into the BH contributes equally as a flux of positive
energy out of the BH.

Physically this process is best described by particles
that have to tunnel through the barrier of negative energy density.
With increasing particle mass the barrier becomes larger and
the tunneling probability decreases.

\subsection{Renormalisation}

At this point I will shortly come back to the renormalisation problem.
From the zeta-function (\ref{zeta-massive}) we have seen that
the coefficients $a_0,a_2,a_4$ appear in divergent integrals
over the proper time $\tau$, having poles at zero proper time,
and lead to divergences in the effective
action (\ref{eff.action-mass}) in the limit $m\to\infty$.
The poles correspond to the well-known UV-divergences
that already appear in QFT in flat spacetime. More precisely
(as I only consider the one-loop order),
they are divergent contributions to the
vacuum energy. The coefficient $a_0$ is simply $1$ in the present
case, hence its contribution to the EM tensor reads
\begin{equation}
\E{T_{\mu\nu}}_{a_0}=\frac{g_{\mu\nu}}{256\pi^2}m^4(3-2\ln m^2).
\end{equation}
It is independent of the spacetime curvature. In particular,
on a Schwarzschild spacetime one has in particular an energy
density
\begin{equation}
\E{T_{tt}}_{a_0}=\frac{\Sss}{256\pi^2}m^4(3-2\ln m^2).
\end{equation}
The limit $M\to0$ corresponds to the flat spacetime value.
Note that these expectation values are in the $\left|B\right>$-state
(and not in the vacuum state)
as computed directly by the effective action. The state-independent
components $\E{T}$ and $\E{\Th}$ have exactly the same form
on a Schwarzschild spacetime as on a flat spacetime (because of
$g^{\mu\nu}g_{\mu\nu}=\eta^{\mu\nu}\eta_{\mu\nu}=d,g_{\theta\theta}
=\eta_{\theta\theta}$).
This suggests to introduce a \emph{zero-point renormalisation
only for the state-independent basic components} by simply substracting the
flat spacetime value
\begin{equation}
\E{T_{\mu\nu}}_{ren;g}:=\E{T_{\mu\nu}}_g-\E{T_{\mu\nu}}_{flat}.
\end{equation}
The remaining components need not be renormalised as they are
calculated from the latter a la CF.

The coefficient $a_2=\frac{R}{6}+E$ does not contribute
at all to the zero-point energy on a vacuum spacetime
(since there we have $R=0,R_{\mu\nu}=0,E=0$).

Unfortunately, there remains a non-vanishing contribution to the EM
tensor from the coefficient $a_4$ by the term $\frac{R_{\mu\nu\sigma\tau}
R^{\mu\nu\sigma\tau}}{180}$. It is well-known that such terms
may arise in the renormalisation of the EM tensor and lead to
ambiguities because the renormalisation parameters cannot be
absorbed by parameters already present in the theory. This is
the renormalisation problem of gravity theory and the dilemma
is that one must rely on physical observables to determine
these parameters at each order in $\hbar$. Generally one expects
a contribution 
\begin{equation}
\E{T_{\mu\nu}}_{a_4}=c_{ren}\frac{\ln m^2}{16\pi^2\sqrt{-g}}
\frac{\delta\int_Ma_4\sqrt{-g}d^4x}{\delta g^{\mu\nu}},
\end{equation}
where $c_R$ is an arbitrary renormalisation constant.
The physical condition to determine this constant is that
the EM tensor shall vanish for $m\to\infty$ -- very massive
particles have zero probability to be produced spontaneously
by the vacuum. This can only be satisfied by setting $c_{ren}=0$.
It could be in principle that the perturbation series \emph{locally}
sums up to a function in $m^2$ that vanishes in this limit
though the first term is finite (one could set
$c_{ren}\propto\frac{1}{\ln m^2}$). Beside the fact that I do not
know any analytic function with the appropriate behaviour, this
could hardly be satisfied \emph{globally} for both basic
components.
\\
\\
To sum up, I renormalise the vacuum expectation value of the
EM tensor by discarding the first three terms in the Seeley-DeWitt
expansion of the heat kernel $a_0,a_2,a_4$. For the first two of
them this can be interpreted as substracting the flat spacetime
value of the EM tensor. For $a_4$ I must argue by the
necessity that particles with large rest-mass produce a negligible
contribution to the vacuum energy.
Following these ideas one may agree that (\ref{eff.action-mass-lorentz})
is indeed the physically sensible effective action and that
the expectation values derived from it are indeed correct.

\subsection{Phenomenology}

In this Section I will discuss the relevance of the results
obtained so far. In view of the obvious lack of experimental data I can
only compare them with existing results, and estimates
derived from the Black Body hypothesis.

The crucial condition in the derivation of the quantum EM tensor
for massive particles was the convergence condition of the
Seeley-DeWitt expansion: $m\gg\frac{1}{M}$. Particles heavier
than this lower limit can be considered as being localised on
a spacetime with mass $M$. On the other hand, this is the
region where the Hawking flux is exponentially damped if
one follows semi-classical arguments, see below.
This would mean that the results were obtained in a region
where they are of no importance. Surprisingly, the results
obtained here by direct quantisation of the scalar particle
do not agree with the simple estimate based on the Black Body
hypothesis.

In the following I will examine how the radiation law of a Black Body
is modified if the emitted particles have a non-zero rest mass
$m$. The relation between energy and momentum is now given by
$E^2=p^2+m^2$. The density of states thus is changed as
$p^2dp=E\sqrt{E^2-m^2}dE$. Further, the mass $m$ introduces a
lower boundary in the integral of (\ref{Flux.tot}). For
large parameter $\alpha=(mM)^{-1}$ and using $T_H=\frac{1}{8\pi M}$
the total flux can be approximated as
\begin{multline}
\text{Flux}_{tot-mass}=\frac{2 M^2}{\pi}\int_m^{\infty}
\frac{E^2\sqrt{E^2-m^2}}{e^{\frac{E}{T_H}}-1}dE\\
\stackrel{\alpha<1}{\approx}\frac{2}{\pi M^2\alpha^4}
\int_0^{\infty}x^{\frac{5}{2}}\sqrt{x-1}e^{-\frac{8\pi x}{\alpha}}dx\\
=\frac{1}{128\pi^3M^2\alpha^2}\,e^{-\frac{4\pi}{\alpha}}
\biggm[\frac{8\pi}{\alpha}K_1\left(\frac{4\pi}{\alpha}\right)
+\left(3+\frac{8\pi}{\alpha}\right)
K_2\left(\frac{4\pi}{\alpha}\right)\biggm].
\label{Flux-mass.tot}
\end{multline}
$K_{1,2}(x)$ are modified Bessel-functions of the first kind.
For $\alpha<1$ the behaviour of the Hawking flux is dominated
by the exponential damping. Figure 7 shows the total flux
calculated by (\ref{Flux-mass.tot}) as a function
of the Hawking temperature $T_H$ and the mass of the scalar particle $m$
(both given in Planck units).

\begin{figure}[h]

\hspace{2cm}\epsfig{file=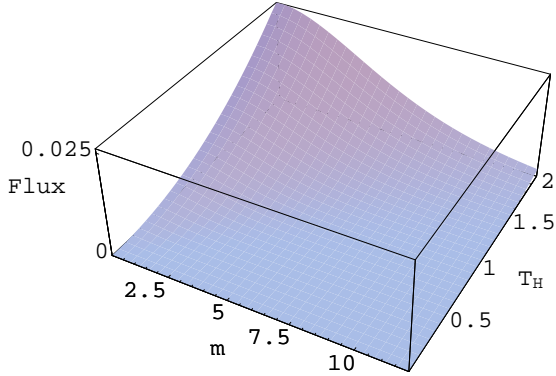,height=6cm}

\caption[fig1]{Total Flux for Massive Particles, calculated by the
Black Body Hypothesis}
\end{figure}

For computational reasons I have plotted the flux in a range
that is far beyond the interesting physical range. Nevertheless,
it shows nicely its qualitative behaviour. For fixed BH
mass the flux is almost constant for $m\ll T_H$, it falls off
exponentially for $m\approx T_H$ and it is negligible for
$m\gg T_H$. Clearly, for a BH in the quasi-static region
$M\gg1$, e.g. $M=M_{\odot}\approx10^{40}$, the Hawking temperature
is extremely small $T_H\approx10^{-40}$ and the interesting region
lies at $m\approx10^{-40}$. Table 6 compares the
result of the full quantum calculation (\ref{flux.massive}) with that
of the Black Body calculation.
\\
\\ \large
\hspace*{2.8cm}\begin{tabular}{|c|c|c|}
\hline
\normalsize $m$ &\normalsize Black Body & \normalsize Field Quantisation\\ \hline
\normalsize $0$ & \normalsize $1.036\cdot10^{-5}M^{-2}$ &
\normalsize -\\ 
\normalsize $\frac{1}{10M}$ & \normalsize $5.5\cdot10^{-6}M^{-2}$ &
\normalsize $1.8\cdot10^{-3}M^{-2}$\\ 
\normalsize $\frac{1}{M}$ & \normalsize $8.8\cdot10^{-14}M^{-2}$ &
\normalsize $1.8\cdot10^{-5}M^{-2}$\\ 
\normalsize $\frac{1.7}{M}$ & \normalsize $7.2\cdot10^{-21}M^{-2}$ &
\normalsize $6.2\cdot10^{-6}M^{-2}$\\
\normalsize $\frac{2}{M}$ & \normalsize $5.7\cdot10^{-24}M^{-2}$ &
\normalsize $4.5\cdot10^{-6}M^{-2}$\\ 
\normalsize $\frac{10^{10}}{M}$ & \normalsize $0$ &
\normalsize $1.8\cdot10^{-26}M^{-2}$\\ \hline
\end{tabular}\\ \\ \small\hspace*{6cm}Table 6
\\ \normalsize
In the first line I have listed once more the flux for massless particles
(\ref{flux.massless.blackbody}). The values in the column ``Field Quantisation''
may be changed crucially if one adds the next perturbative order in $\alpha$.
Especially the value at the point $\alpha=1$, where the perturbation
series breaks down,
might differ drastically from the correct value. Beyond this point
the Black Body result goes exponentially fast to zero while the quantum result
remains at finite values. Note that the first order of the perturbation series
should approach the exact quantum value as $\alpha\to0$, while its deviation
from the Black Body result increases. This disagreement clearly cannot be traced
back to the use of the conservation equation because it already concerns
the basic components which are derived directly from the effective action
(the conservation equation is not changed for massive particles as long
as we are in the quasi-static phase $M\gg1$).

Apart from the missing accordance with the BH radiation law with respect
to the qualitative behaviour the result (\ref{Flux-mass.tot}) fits nicely
to the quantum calculation for massless particles (\ref{Hawking.flux.massless}).
Namely, at the
critical point $mM=1$ the flux for massive particles (in this approximation) lies
just one order below that of massless particles. This means that, compared to the
massless case, the absolute flux is not too high but still contributes with
a considerable amount, as expected.
\\
\\
The reason why the behaviour of the expectation value of the EM tensor
deviates in such a drastic way from the one suggested by the
Black Body law may be found in the very basis of the approach
which is the \emph{local} expansion of the heat kernel.
I have shown that this is only possible if
the condition $mM\gg1$ is fulfilled. On the other hand, the estimate
(\ref{Flux-mass.tot}) demonstrates that this is exactly the range
where the flux is effectively zero. Indeed, the point where
the exponential damping sets in, namely for $mM\approx1$ (see
Table 6 on the last page), 
is already out of the scope of the perturbation series which
converges for $mM\gg1$. By this circumstance one cannot describe the
physically interesting range, where the BH is still in the
quasi-static phase $M>1$ (e.g. $M=10^{10}$) and the geometry can be described
accurately by a Schwarzschild spacetime, and where the mass of the
scalar particle $m$ lies between $0$ and $\frac{1}{M}$, such that
$mM<1$.

I conclude that \emph{the local expansion of the heat kernel does not seem
appropriate to describe the characteristic features of the radiation
of massive particles}. Merely, I suppose that the use of the
more comprising non-local expansion, see Section 3.2, may reveal
the exact behaviour around the critical point $mM\approx1$.
This idea is discussed in the Outlook of my thesis.
Nevertheless, the Seeley-DeWitt expansion is the most simple
way to derive one-loop expectation values on a curved spacetime
in arbitrary dimensions as it does not afford the knowledge
of the Green function. In particular, one might use this
advantage to compare the two-dimensional dilaton model with
the $4d$ theory at the quantum level.

Frolov and Zel'nikov used a similar approach to obtain
some of the components of the EM tensor \cite{frz82}. Their result
for the trace and the $\Th$-component is of the same
order as mine. Because they derived all components directly
from the effective action they could not give an estimate
on the radiation component $T^r{}_t$. If I calculate the
radiation constant in the Unruh state by their basic
components I obtain $K=-\frac{3.6}{10^6\cdot\pi^2m^2M^2}$.
Note that they use the inverse sign convention, thus I have
multiplied their results by $-1$!

\newpage
\section{Hawking Radiation of Massless Scalars}

The main contribution to the Hawking flux of a BH
in the quasi-static phase comes from massless particles, namely
photons, because their Hawking temperature lies far below
the rest mass of the known fundamental massive particles
(except probably neutrinos). In the absence of a mass term
or in the limit of very low masses the local Seeley-DeWitt
expansion of the heat kernel breaks down ($\alpha=\frac{1}{mM}>1$)
and the effective action becomes a non-local expression
that can be derived by the covariant perturbation theory
(Section 3.2). This approach is rather new in the present context
and it is more direct and unambiguous as compared to
the methods found in the literature \cite{kuv99}. Further, it turns out
that the so-called trace anomaly induced effective action,
obtained by functional integration of the trace anomaly,
only corresponds to (some part of) the first order of the covariant perturbation
theory (apart from the fact that this method is not
applicable in four dimensions if the scalar field is coupled minimally).
The examinations of the last Chapter have shown
that the local expansion does not reproduce the significant
qualitative behaviour of the Hawking flux for massive particles
as it was expected from the Black Body hypothesis.
The non-local expansion can therefore be seen as the more
comprising case which presumably yields interesting
results also for massive scalar fields (I come back to this
point in the Outlook). 

The non-local effective action, once written down
in a compact form, is still difficult to handle.
Namely, the non-localities come in by negative
powers of Laplacians that can be transformed into
multiple integrals over Green functions. Because of their
fundamental importance for the effective action
I will call these integrals \emph{basic integrals}.
However, the Green functions of a scalar field
(massive or massless) on a Schwarzschild spacetime
cannot be given in a closed analytical form.

Here emerges the main advantage of the dilaton model:
a two-dimensional manifold has only one gravitational
degree of freedom, i.e. there is only one independent
component of the Riemann tensor. As a consequence the
EOM describing the gravitational dynamics of a two-dimensional
manifold become integrable and the dynamical evolution is
completely determined by the boundary conditions.
As a consequence the basic integrals, forming the effective action,
can be reduced to boundary terms that are fixed by the
boundary conditions of the Green functions.
Thereby the IR renormalisation plays a crucial role.
The whole Chapter is devoted to the investigation of
the dilaton model.

Although the \emph{explicit} knowledge of the Green functions
is not necessary in the dilaton model, I will study
the two-point Green functions of massless scalar fields
on a two-dimensional Schwarzschild spacetime
by a \emph{perturbation series}, starting from
the exactly known Green functions of a flat spacetime.
Most importantly, I can show that the perturbation
series converges if the Green functions are applied
to basic integrals and if \emph{the BH interior
is excluded} from the domain of the Green functions.
This is justified by the fact that no classical particles
(even those produced in the Hawking effect) can leave the
region inside of the horizon -- all particles measured
as Hawking flux have been produced at or outside the horizon.
There are several reasons why it is worthwhile
investigating the Green functions: by explicitly calculating
the basic integrals one can check consistency of the
boundary conditions. Further, I can examine whether
the choice of Green function affects the quantum state,
as proposed by Barvinsky and Vilkovisky (see the discussion at
the end of Section 3.2). Finally, it is likely that these
results can be adopted in the four-dimensional theory,
where the explicit usage of Green functions is inevitable.

\subsection{Dilaton Model}

In Section 2.2 I have presented the dilaton model
and shown that it is classically equivalent to
the s-modes of a scalar field on a four-dimensional Schwarzschild
spacetime. In the present Chapter I go one step further
and apply the formalism of QFT to the
dilaton model. By the methods developed in the Chapters
2 and 3 it will be possible to carry out the whole
task of quantising a free scalar field and to obtain
the expectation value of the EM tensor to the first order
of the perturbation theory.

Before going into details I sketch the procedure:
\begin{itemize}
\item First, I establish the non-local effective
action of the dilaton model by the covariant perturbation
theory to the first perturbational order.
\item I discuss the boundary conditions of the Green function
and how they affect the basic integrals.
\item I derive and discuss the flat retarded and Feynman
Green functions and develop a perturbation series for
the corresponding Schwarzschild Green functions.
\item I calculate the expectation value of the EM tensor and
the Hawking flux and discuss the result.
\item Finally, I reconsider the effective action with respect
to its quantum state and show that an arbitrary quantum state
can be fixed by adjusting the constants of the CF
approach.
\end{itemize}

\subsection{Effective Action and Expectation Values}

In Section 3.2 I have presented the general form of the
non-local effective action introduced by Barvinsky and Vilkovisky.
Now I want to fix the ``parameters'' as to
describe the dilaton model.

The ``parameters'' are the spacetime dimension, which from now
on is always two $d=2\omega=2$, and the characteristic
Euclidean Laplacian ${\cal O}=-\triangle-E$. The dimension
of the spacetime always causes a characteristic number, type,
and form of IR and UV divergences. As already mentioned,
IR divergences are a typical feature of massless particles.

I use the zeta-function regularisation to handle the
UV-divergences of the heat kernel. Remember that
in this regularisation scheme the effective action was related
to the heat kernel as
\begin{equation}
W[g]=-\frac{1}{2}\frac{d\zeta[s]}{ds}\biggm|_{s=0}=
-\frac{1}{2}\frac{d}{ds}\frac{1}{\Gamma(s)}
\int_0^{\infty}\frac{d\tau}{\tau^{1-s}}
\text{tr}\,e^{-{\cal O}\tau}\biggm|_{s=0}.
\end{equation}
In the zeta-function the IR-divergent terms are those that behave like
$\tau^{-n},\newline n\le1$. In any case they demand some additional
regularisation (beside the UV-regularisation naturally provided
by the zeta-function).

To the second order of the covariant perturbation theory
there are five types of terms in the non-local
effective action, see (\ref{nonlocal.effective.action}), with
different UV and IR behaviour:
\begin{eqnarray}
\int_0^{\infty}\tau^{s-2}d\tau\,,\,\int_0^{\infty}
\tau^{s-1}d\tau\hspace{8.5cm}\\
\int_0^{\infty}\tau^{s}f(-\triangle\tau)d\tau\,,\,
\int_0^{\infty}\tau^{s-1}\frac{f(-\triangle\tau)-1}{-\triangle}d\tau\,,\,
\int_0^{\infty}\tau^{s-2}\frac{f(-\triangle\tau)-1}{\triangle^2}d\tau.
\hspace{.5cm}
\label{nonloc.int.}\end{eqnarray}
In the integrals in the first line I must introduce
an additional IR regularisation by cutting off the upper
bound at some value $T$, the other integrals are regularised by the
exponential function in $f(-\triangle\tau)$ (\ref{f2}).
The first integral does not contribute to the effective action:
\begin{equation}
\frac{d}{ds}\frac{1}{\Gamma(s)}\int_0^T\tau^{s-2}d\tau\biggm|_{s=0}
=-\frac{1}{T}\stackrel{T\to\infty}{\to}0.
\end{equation}
The second integral has a finite and a divergent term:
\begin{equation}
\frac{d}{ds}\frac{1}{\Gamma(s)}\int_0^T\tau^{s-1}d\tau\biggm|_{s=0}
=\gamma_E+\ln T.
\end{equation}
The remaining integrals (\ref{nonloc.int.}) only lead to finite contributions.
This can be seen if I change the order of the $\tau$ and the $a$ integration
in the regularised integrals and finally differentiate
for $s$ and carry out the limit $s\to0$. Note that the
$1$ in $f(-\triangle\tau)-1$ can be pulled into the
integral over $a$ since $1=\int_0^a1\cdot da$! The first term in (\ref{nonloc.int.})
after the $\tau$-integration reads
\begin{multline}
\frac{\Gamma(1+s)}{(-\triangle)^{1+s}}
\int_0^1\frac{1}{[a(1-a)]^{1+s}}da\\
=\frac{\Gamma(1+s)}{(-\triangle)^{1+s}}\int_{-1/2}^{1/2}
\frac{(-1)^{1+s}}{\left[u^2-\frac{1}{4}\right]^{1+s}}du
=\frac{2^{1+2s}\sqrt{\pi}\cdot\Gamma(1+s)\Gamma(-s)}
{(-\triangle)^{1+s}\Gamma(\frac{1}{2}-s)}.
\end{multline}
The other two integrals in (\ref{nonloc.int.}) can be treated
analogously. The contribution of these three terms to the
effective action reads
\begin{eqnarray}
\frac{d}{ds}\frac{1}{\Gamma(s)}\int_0^{\infty}\tau^{s}
f(-\triangle\tau)d\tau\biggm|_{s=0}&=&
\frac{2\cdot\ln(-\triangle)}{-\triangle}\\
\frac{d}{ds}\frac{1}{\Gamma(s)}\int_0^{\infty}\tau^{s-1}
\frac{f(-\triangle\tau)-1}{-\triangle}d\tau\biggm|_{s=0}&=&
\frac{2-\ln(-\triangle)}{-\triangle}\\
\frac{d}{ds}\frac{1}{\Gamma(s)}\int_0^{\infty}\tau^{s-2}
\frac{f(-\triangle\tau)-1}{\triangle^2}d\tau\biggm|_{s=0}&=&
\frac{\ln(-\triangle)-\frac{8}{3}}{6(-\triangle)}.
\end{eqnarray}
Now I have all pieces at hand to construct the effective
action for a general two-dimensional scalar model. 
Putting together these results, the formula for the heat kernel
(\ref{nonlocal.effective.action}), and the defining equation of
the effective action (\ref{eff.action4}) I obtain\footnote{I denote
the two-dimensional manifold by $L$.}
\begin{multline}
W_{\cal E}[g]=\frac{1}{96\pi}\int_L\biggm[-(2R+12E)(\gamma_E+\ln T)\\
+(R+12E)\frac{1}{\triangle}R
+2R\frac{\ln(-\triangle)}{\triangle}E-2E\frac{\ln(-\triangle)}{\triangle}R
\biggm]\sqrt{g}d^2x.\label{non-local.eff.action}
\end{multline}
The last two terms vanish if I consider commuting fields as
in this case also the scalar curvature and the endomorphism commute,
i.e. $RE=ER$.
The logarithm in the last term can then be expanded in a power series and
by partial integration the Laplacians of this series can be made acting
on $E$ without a change of sign (the surface terms vanish if I assume
that $M$ is flat in the infinite future and past).

The very first term in the effective action must be renormalised
in any case. I define a renormalisation constant $c_R:=\gamma_E+\ln T$
that I fix later by the physical requirements of the EM tensor.

Now I go over to the Lorentzian effective action:
\begin{equation}
W_{\cal M}[g]=\frac{1}{96\pi}\int_L\left[-c_R(2R-12\ve_{\cal E}E_{\cal M})
+(R-12\ve_{\cal E}E_{\cal M})
\frac{1}{\square}R\right]\sqrt{-g}d^2x.
\end{equation}
Note that beside $R\to-R,\triangle\to-\square$ we have
$W_M=iW_E$ and $d\tau=i dt$. This action is in fact of the
most general form for all \emph{commuting scalar
fields}. It covers all kinds of couplings and potentials
that are specified by the Laplacian and the associated
endomorphism.

Now I will examine how the dilaton model (\ref{2daction}) can be
described by this effective action (at the quantum level).
The Lorentzian Laplacian reads
\begin{equation}
{\cal O}_{\cal M}=X\square+(\n X)\n.
\end{equation}
To bring it into the standard form I must
define a conformally related metric
$\hat{g}^{\alpha\beta}=Xg^{\alpha\beta}$:
\begin{equation}
{\cal O}_{\cal M}=\left(\hat{\n}+\frac{\hat{\n}X}{2X}\right)
\left(\hat{\n}+\frac{\hat{\n}X}{2X}\right)+\frac{(\hat{\n}X)^2}{4X^2}
-\frac{\hat{\square}X}{2X}=\hat{\square}_{tot}+\frac{(\n X)^2}{4X}
-\frac{\square X}{2}\label{Lapl.dil}
\end{equation}
$\hat{\square}_{tot}$ is the total geometric Laplacian, hence
I must replace all geometric objects in the effective action
by those associated to the connection $\hat{\n}_{tot}=\hat{\n}
+\frac{\hat{\n}X}{2X}$, whereby
I can omit the ``gauge part'' $\frac{\hat{\n}X}{2X}$ because
of its commutativity (see Appendix F.1 (\ref{gauge-curvature})):
\begin{equation}
\Omega_{ab}S=[\hat{\n}^{tot}_a,\hat{\n}^{tot}_b]S=[\hat{\n}_a,\hat{\n}_b]S
+\left(\hat{\n}_{[a}\hat{\n}_{b]}\ln X\right)S=0.
\end{equation}
The Lorentzian endomorphism from (\ref{Lapl.dil}) reads
\begin{equation}
E_{\cal M}=\frac{(\n X)^2}{4X}-\frac{\square X}{2}
\end{equation}
and exhibits an Euclidean sign $\ve_{\cal E}(E)=-1$.
Further, I have to replace the scalar curvature by
$\hat{R}=XR-\frac{(\n X)^2}{X}+\square X$ and the volume
element by $\sqrt{-\hat{g}}=X^{-1}\sqrt{-g}$, see Appendix C
(\ref{conf.R},\ref{conf.g-1}).
Hence the non-local part of the effective action
(\ref{non-local.eff.action}) has the form
\begin{multline}
W_{nl}[g]=\frac{1}{96\pi}\int_L(X^{-1}\sqrt{-g})
\left[XR-\frac{(\n X)^2}{X}+\square X+12E_{\cal M}\right]\\
\cdot\frac{1}{X\square}\left[XR-\frac{(\n X)^2}{X}+\square X
\right]d^2x\\
=-\frac{1}{96\pi}\int_L\left[R+2\frac{(\n X)^2}{X^2}-5\frac{\square X}
{X}\right]\\
\cdot\int'_LG(x,x')\left[R'-\frac{(\n X')^2}{(X')^2}+\frac{\square'X'}
{X'}\right]\sqrt{-g'}d^2x'\sqrt{-g}d^2x.
\end{multline}
In the second line I have again replaced the negative power
of the geometric Laplacian by an integral over a Green
function: $\frac{1}{\square}=-\int'_L G(x,x')\sqrt{-g'}d^2x'$.

The local part of the effective action reads
\begin{equation}
W_{l}[g]=\frac{1}{96\pi}\int_L(-c_R)\left[2R+\frac{(\n X)^2}{X^2}
-\frac{\square X}{X}\right]\sqrt{-g}d^2x.
\end{equation}
Now it is convenient to introduce the field $\phi$
by $X=e^{-2\phi}$. This gives the identities
\begin{equation}
\frac{(\n X)^2}{X^2}=4(\n\phi)^2\,\,,\,\,\frac{\square X}{X}
=4(\n\phi)^2-2\square\phi.
\end{equation}
The non-local part of the effective action then becomes
\begin{multline}
W_{nl}[g]=-\frac{1}{96\pi}\int_L[R-12(\n\phi)^2+10\square\phi]\\
\cdot\int'_LG(x,x')[R'-2\square'\phi']\sqrt{-g'}d^2x'\sqrt{-g}d^2x.
\label{nonloc.eff.action}
\end{multline}
In this form the effective action, together with the local part,
allows to derive all one-loop expectation values of the
scalar field $S$ in the dilaton model\footnote{Clearly, I
have not yet discussed the boundary conditions. Below I
discuss how to specify the quantum state.}. The non-local part
represents the second order of the covariant perturbation series
(\ref{cov.pert.series})
-- higher orders (corresponding to further curvature lines
in Figure 5) would appear in the form of multiply
staggered integrals. The information on the scalar field
provided by the effective action is hidden in the Green
functions. Their knowledge, beside the geometric data of the manifold
given by $R$ and $\phi$, is the key to the expectation values.

In two dimensions the situation is very particular. Namely,
the geometric variables only possess one dynamical field degree
of freedom. This implies that all geometric objects can be described
by just one variable. So the scalar curvature $R$
already determines the complete Riemann tensor. Further, one
can find a representation of the metric where its component matrix is a
multiple of the flat metric $\eta_{\alpha\beta}=\text{diag}(1,-1)$:
$g_{\alpha\beta}=e^{2\rho}\eta_{\alpha\beta}$. This
representation is called the \emph{conformal gauge}, see
Appendix A.3. In this gauge the scalar curvature can be written
as $R=-2\square\rho$. If I substitute this into the effective
action I can perform the $x'$-integration and obtain
a completely local effective action\footnote{I have not written
down terms $\propto\square\rho$ and $\propto\square\phi$ in the
(originally) local part of the effective action because they are just
vanishing surface terms.}:
\begin{equation}
W[g]=\frac{1}{24\pi}\int_L\Bigm\{-3c_R(\n\phi)^2
+[\square\rho+6(\n\phi)^2-5\square\phi]
[\rho+\phi]\Bigm\}\sqrt{-g}d^2x.\label{dilaton.eff.action}
\end{equation}
This result is obtained by setting $\square^{-1}\square=1$
in the non-local action. Thereby I have dropped a homogeneous
solution $\chi$ of the Laplace equation $\square\chi=0$. Generally,
for a function $f(x)$, we would have
\begin{equation}
\int_LG(x,x')\square f(x)\sqrt{-g}d^2x=
-\frac{1}{\square}\square f=-\frac{1}{\square}\square(f+\chi)
=-(f+\chi).
\end{equation}
By looking to the l.h.s. and the r.h.s. of this equation we
see that $\chi$ (as a function of $x$) is determined
uniquely by the Green function in the integral.
It will be the subject of the next Section to show
that $\chi$ is indeed zero for the considered integrals
and therefore (\ref{dilaton.eff.action}) is the
appropriate effective action for my purposes.

It will turn out that the \emph{Hawking flux is independent}
of the (originally) local term in the action and hence \emph{of
the renormalisation constant $c_R$}. Nevertheless, it cannot
be chosen arbitrarily because it affects other components
of the EM tensor like the energy density. In the CF
approach it appears in the $\Th$-component of the EM tensor
in the same place like the homogeneous solution $\chi$, hence,
if the latter is only a constant (as it will be the case), it can
be removed by the renormalisation term. For the moment I will
disregard this term by setting $c_R=0$.
\\
\\
\textbf{A final Remark:} The effective action (\ref{dilaton.eff.action})
differs from the one derived in \cite{kuv99}. The difference can
be traced back to a different choice of path integral
measure and a field redefinition in the four-dimensional
effective action (\ref{eff.action}).
In a first step a factor $\sqrt[4]{-g}$ is introduced into
the path integral measure\footnote{Such a factor may come
from the normalisation constant ${\cal N}$.} $Z[g]={\cal N}'
\int{\cal D}(\sqrt[4]{-g}S)\cdot e^{iL_m[g,S]}$ \cite{fln88}.
This does clearly not affect the expectation values because
there the normalisation drops out. Then a new field is
defined by $S':=S\sqrt[4]{-g}$ and the classical action
becomes $L'=-\int S'\square S'd^2x$. Note that neither the
new field $S'$ nor the volume element $d^4x$ are diffeomorphism
invariant anymore! The action is spherically reduced
by introducing the spherically reduced Laplacian $\square_4=\square_2
+\frac{\n X}{X}\n$ (see Appendix D (\ref{Laplace-sph.red.}))
and by integration over the angular coordinates $\theta,\vp$.
One can construct the corresponding
non-local effective action by the same steps as before.
Things are a bit easier now because the geometric Laplacian
is simply $\triangle$, the metric must not be conformally transformed,
and the endomorphism becomes $E_{\cal M}=\frac{(\n X)^2}{4X^2}
-\frac{\square X}{2X}=-(\n\phi)^2+\square\phi$. In its local form the effective
action reads $W'[g]=\frac{1}{24\pi}\int_L\left\{-3c_R(\n\phi)^2
+\rho\square\rho+6\rho(\n\phi)^2-6\rho\square\phi\right\}\sqrt{-g}d^2x$
which is identical to the effective action derived in \cite{kuv99}.
$c_R$ is again a renormalisation constant that does not affect the
Hawking flux. Therefore, the basic components and hence the Hawking
flux agree with those of \cite{kuv99}: $\E{T}_2=\frac{M}{3\pi r^3}\,,
\E{\Th}_2=-\frac{1}{8\pi r^5}\left\{4M+(4M-r)\left[\ln\Sss-c_R\right]\right\},
\E{T^r{}_t}_2=\frac{1}{768\pi M^2}$.
Interestingly, the flux agrees with (\ref{Hawking.flux.massless}),
the one calculated in this thesis
from (\ref{dilaton.eff.action}), although
all other components of the EM tensor differ.

\subsection{Boundary Conditions}

In the local form of the effective action (\ref{dilaton.eff.action})
the boundary conditions have already been fixed implicitly
by the assumption $\square^{-1}\square=1$.
As already mentioned above one might have an additional contribution
by a homogeneous Green function. This contribution corresponds to
a boundary term that can be fixed by the boundary conditions
imposed on the Green functions. This can be studied best
by Green's theorem in curved spacetime
\begin{multline}
\int_V\Bigm[f(x')\square^{x'}G(x,x')-G(x,x')\square^{x'}f(x')
\Bigm]\sqrt{-g'}d^2x'\\
=\oint_{\p V}\Bigm[f(x')\n_{\alpha}^{x'}G(x,x')-G(x,x')\n_{\alpha}^{x'}
f(x')\Bigm]\sqrt{-g'}\cdot g^{\alpha\beta}\ve_{\beta\gamma}(dx')^{\gamma}.
\label{Green's-theorem}
\end{multline}
If I apply it to the basic integrals in the effective action, where
$F=\square f$, I can express the non-localities by the sum of
a local term and a boundary term:
\begin{multline}
-\frac{1}{\square}F=\int_LG(x,x')F(x')\sqrt{-g'}d^2x'\\
=-f(x)-\oint_{\p L}\Bigm[f\n_{\alpha}'G-G\n_{\alpha}'
f\Bigm]\sqrt{-g'}\ve^{\alpha}{}_{\beta}(dx')^{\beta}\\
=-f(x)-\int_{2M}^{\infty}f(x')\p_{t'}G\frac{dr'}
{\left(1-\frac{2M}{r'}\right)}\biggm|_{t'=\infty}
+\,{\dots}\biggm|_{t'=-\infty}\\
-\int_{-\infty}^{\infty}\Big[f(x')\p_{r'}G-G\p_{r'}f(x')\Big]
\left(1-\frac{2M}{r'}\right)dt'\biggm|_{r'=2M}
+\,{\dots}\biggm|_{r'=\infty}.\label{bound.basic.int}
\end{multline}
One immediately observes that the boundary terms in the
third line are not well-defined if $G$ does not vanish
at least linearly at $r'=2M$ (for arbitrary $x$).
If this is provided, the function $f$ may still have a logarithmic 
divergence on the horizon (as it is the case with the
conformal metric factor $\rho$). This trouble can be traced
back to the fact that the Schwarzschild coordinates break down
at the horizon. In fact, from the point of view of this
gauge the spacetime must be considered as to end at the horizon,
and the two-dimensional spacetime spanned by $t,r$ is thus
isomorphic to a rectangle $\{-\infty,\infty\}\times\{2M,\infty\}$
in coordinate space. Therefore one is forced to fix boundary
conditions on the horizon, namely by demanding that the
eigenfunctions of the Laplace operator, and hence the Green
functions, vanish there. According to the reasoning in the introduction
of this Chapter the usage of Schwarzschild coordinates seems
to be appropriate to calculate the Hawking flux.

In (\ref{bound.basic.int}) one also can see the possible
IR problems, caused by the fact that the Green functions
of massless particles do not have a sufficient fall-off behaviour for
large point-separations $|x-x'|$. The physical picture is the
following: the measurement of the Hawking flux takes place
somewhere at a finite distance from the BH and at
some instant of time during the quasi-static phase, i.e.
when the BH mass is much larger than the Planck mass
$M\gg1$. Hence, only particles with effectively zero energy
can contribute to this expectation value by travelling from
infinity until the point of measurement -- such particles
clearly are a mathematical idealization as physical particles
always possess a minimum energy, therefore
their (formally infinite) contribution must be renormalised to zero.
This can be achieved by several methods. The most obvious
way is to introduce a mass of the scalar particle such
that the Green functions become exponentially damped.
The boundary terms then vanish and afterwards the mass can
be set to zero; the elimination of the boundary terms
in (\ref{bound.basic.int}) represents the only difference to the
unrenormalised case. Or one can work with a finite
spacetime volume, by limiting the range of the coordinates to finite values,
and allows only wave lengths that fit into
this rectangular of finite size. Thereby one singles out
those eigenfunctions that vanish (at least linearly) on the boundary and hence
the Green functions inherit this boundary condition.
Note that these Green functions naturally vanish also
on the horizon as it belongs to the boundary.
The same happens by the introduction of a mass term because
particles with a finite mass need an infinite time to reach
or leave the horizon!

By these considerations I conclude that any possible contribution
from the boundaries of the manifold (including the event horizon)
are unphysical and must be removed by some \emph{infrared
renormalisation}. The easiest way to realize this in
practical calculations is to \emph{drop all boundary terms} if the
integrand contains a Green function and the remaining
integrand is at most logarithmically divergent on the
boundary. This procedure will be one of the basic ingredients
in the Green function perturbation theory of the next Section.

By applying this renormalisation prescription to (\ref{bound.basic.int})
all boundary terms cancel (the spacelike terms in the third
line are finite on the horizon and vanish because they are
in the remote past, respectively future) if $f$ has at most
logarithmic divergences. For such functions the relation
\begin{equation}
\square^{-1}\square=1
\end{equation}
is indeed valid. In the effective action one has instead of $f$
the functions $\rho=\frac{1}{2}\ln\Ss$ and $\vp=-\ln r$ which
fulfil the necessary regularity condition. Thus,
(\ref{dilaton.eff.action}) is the unique local form of the effective
action that is compatible with the required IR
renormalisation. Further, as the Green function only appears
in the boundary terms, the effective
action, and accordingly the expectation values, is independent
of the \emph{type} of Green function.

\subsection{Green Function Perturbation}

In Section 5.2 I have used the conformal gauge to render
the effective action local. Thereby I have thrown away
a homogeneous solution $\chi$ of the geometric Laplace
equation. In the last Section I have shown that the
IR renormalisation justifies this step as it removes the
boundary terms that correspond to this homogeneous solution.
In principle, this would be enough to guarantee the existence
of a unique local form of the effective action and hence the uniqueness
of the expectation values derived from it.

Apart from the fact that the Hawking flux of massless
particles in the dilaton model can be calculated \emph{without}
knowing explicitly the Green function of the two-dimensional Schwarzschild
Laplacian, there are several reasons to examine them.
Basically, it would be interesting to know if the
considerations on the boundary terms can be affirmed by
a direct calculation. Further, by explicitly dealing with the
Green functions one can choose between different types
among them and investigate their effect (or not existing effect)
on the basic integrals. Finally, the method developed
in this Section may be a guiding line for the investigations
in four dimensions.

The procedure presented here is strongly
adapted as to examine the specific types of basic integrals
that appear in the non-local effective action.
Beside the IR renormalisation discussed in the last Section,
also the fact that the perturbative Green function
is applied to the basic integrals will be crucial for its
convergence.

The objects of my investigations will be integrals of the type
\begin{equation}
\int G(x,x')F(r')\sqrt{-g'}d^2x'=\int G(x,x')\square'f(r')
\sqrt{-g'}d^2x'=-[f(r)+\chi_f(r)],\label{int.basic}
\end{equation}
that I call \emph{basic integrals}. The functions
$F,f,\chi$, and hence the whole equation shall be
\emph{time-independent}. For $F$ and $f$ this is just the
static approximation, i.e. I approximate the
BH spacetime in the slowly evolving region by
a Schwarzschild BH. As a result the basic integrals
themselves will be time-independent (as it can be seen from the explicit
calculations in Section 5.6.2). Thus, also $\chi$ must be
time-independent. The general time-independent solutions of the
homogeneous Laplace equation $\square\chi$ 
(on a Schwarzschild background) can be written as
\begin{equation}
\chi_f=C_1^f+C_2^f\frac{r_{\ast}}{2M}
=C_1^f+C_2^f\left[\frac{r}{2M}+\ln\left(\frac{r}{2M}-1\right)\right],
\label{chi}
\end{equation}
where $r_{\ast}$ is the Tortoise coordinate, see Appendix A.3.
In Section 5.6.2 I will show that the Hawking flux and
all other components of the EM tensor are not affected by
the constant $C_1^f$. Nevertheless it is
formally useful because it connects my results to the
ones in the literature by an infinite constant that appears
in the perturbation series of the Green functions -- I
have always the freedom to add such a constant without
changing measurable quantities.

The Green functions, even of the two-dimensional
Laplace operator on a Schwarzschild spacetime fulfilling
the equation
\begin{equation}
\square G(x,x')=-\delta(x-x'),
\end{equation}
are not known in a closed analytical form. Also the eigenfunctions
of the Laplace operator are not known exactly, therefore
one cannot proceed as usual and construct the Green
functions by their decomposition into eigenfunctions. For this reason I
cannot solve the basic integrals analytically. 

From equations (\ref{int.basic},\ref{chi}) one can see
that the asymptotic behaviour (i.e. for large values of the
radius $r$) of the basic integrals
is sufficient to determine the constant $C_2^f$.
This encourages me to approximate the exact Green
function by the one from flat spacetime $G_0(x,x')$ and introduce
a perturbation series to calculate the next order
approximations.

The perturbation series has the form
\begin{multline}
G(x,x')=G_0(x,x')+\int''_LG_0(x,x'')\delta\square''G_0(x'',x')d^2x''\\
+\int''_L\int'''_LG_0(x,x'')\delta\square''G_0(x'',x''')
\delta\square'''G_0(x''',x')d^2x''d^2x'''+\dots,\label{pert.Green}
\end{multline}
where $\delta\square=\square-\square_0$ is the perturbing
Laplacian and $\square_0=\p_t^2-\p_r^2$ is the flat Laplacian.
When computing more complicated expressions I will use the notation
$G_0(x,x')=G^x_{'}$ (only for the flat Laplacian!) and
$f(x)=f^x,f(x'')=f''$. Further, I will omit the volume
element $d^2x$ and the index $L$ of the integrals. That (\ref{pert.Green})
is a Green function of the full Laplacian $\square$
can be seen immediately:
\begin{multline}
\square^xG(x,x')=\square^xG^x_{'}-\delta\square^xG^x_{'}
+\int'\delta\square^xG^x_{''}\delta\square''G''_{'}\\
-\int'''\delta\square^xG^x_{'''}\delta\square'''G'''_{'}
+\int''\int'''\delta\square^xG^x_{''}\delta\square''G''_{'''}
\delta\square'''G'''_{'}+\dots\\
=\square_0G^x_{'}=-\delta(x-x').
\end{multline}
The series converges if the condition
\begin{equation}
\left|\int''\delta\square''G_0(x'',x')\right|<1\label{conv.condition}
\end{equation}
is fulfilled. In the following I will discuss when
this is the case. I can write the perturbed Laplacian
in the form (remember that $\square=\Sss^{-1}\p_t^2-\p_r[\Sss\p_r]$)
\begin{equation}
\delta\square=\frac{2M}{r-2M}\p_t^2+\p_r\left(\frac{2M}{r}
\p_r\right).\label{pert.Laplace}
\end{equation}
Before starting with the estimate I emphasize some critical
points. From (\ref{pert.Laplace}) it is obvious that
condition (\ref{conv.condition}) cannot be fulfilled in the
general case. But as I will only apply the Green functions
in the basic integrals I can make use of the time-independence
of the latter. Second, I will restrict the range of the $r$-integration
to the horizon, i.e. the radius-coordinate $r$ will only be
integrated from $2M$ to $\infty$, and employ the boundary
conditions as discussed in the last Section. 

I will put some function $f(r)$ into the integral of
(\ref{conv.condition}). This makes the calculation more
transparent and corresponds to the application of the
perturbation series to a basic integral. If I write a partial
derivative without coordinate index this will always mean differentiation
for $r$, e.g. $\p''=\p_{r''}$.
\begin{multline}
\int'f'\frac{2M}{r'-2M}\p_{t'}^2G'_x+\int'f'\p'\left(
\frac{2M}{r'}\p'G'_x\right)\\
=\p_t^2\int'f'\frac{2M}{r'-2M}G'_x+\int'\p'\left(f'\frac{2M}{r'}
\p'G'_x\right)-\int'(\p'f')\frac{2M}{r'}\p'G'_x\\
=-\int'(\p'f')\frac{2M}{r'}\p'G'_x.\label{conv.proof}
\end{multline}
I have used the relations $\p_tG_0(x,x')=-\p_{t'}G_0(x,x'),
\p_t^2G_0(x,x')=\p_{t'}^2G_0(x,x')$ and $\p_rG_0(x,x')=\p_{r'}G_0(x,x')$
(see the Fourier-transform of the flat Green functions
in the next Sections (\ref{flat.Green})).
Hence I can pull the time-derivative in the first line out
of the integral and it becomes zero because of its time-independence
(see below). The factor $(r'-2M)^{-1}$ does not cause problems
because $G_0$ vanishes on the horizon. The surface term in the second
line vanishes by the boundary conditions (the $r'$-derivative
of $G_0$ can be transformed into an $r$-derivative).
Now I consider the identity
\begin{multline}
f(r)=-\int'f'\square_0'G'_x=\int'f'(\p_{r'}^2-\p_{t'}^2)G'_x\\
=-\p_t^2\int'f'G'_x+\int'\p'(f'\p'G'_x)-\int'(\p'f')\p'G'_x
=-\int'(\p'f')\p'G'_x.
\end{multline}
Obviously the absolute value of this integral is larger than
that of (\ref{conv.proof}). Thus I have the inequality
\begin{equation}
\left|\int''f(r'')\delta\square''G_0(x'',x')\right|<f(r')
\end{equation}
which proves the convergence of the perturbation series
if applied to time-independent functions in basic integrals.

\subsubsection{Second and Third Order}

In this Section I will bring the second and third order
of the perturbation series in a simple form, such that
the basic integrals can be solved by simple integrations
if the flat Green functions are known.
Again I will assume that the basic integrals are time-independent,
hence all terms with time-derivatives do not contribute.
Further, I will use the boundary condition that $G$ vanishes
on the boundary of the manifold (that begins at $r=2M$).

The second order of the perturbation then reads
\begin{equation}
-\int''[\p_{r''}G_0(x,x'')]\frac{2M}{r''}
\p_{r''}G_0(x'',x')d^2x'',
\end{equation}
where I have partially integrated one $r''$-derivative.
I define the function
\begin{equation}
g(r):=\frac{2M}{r}.
\end{equation}
Note that $\p_r^2g(r)=-R(r)$ on a two-dimensional Schwarzschild spacetime.
Now I derive a useful identity, introducing light-cone
derivatives $\p_{\pm}=\p_t\pm\p_r$\,, $\square_0=\p_+\p_-=\p_-\p_+$.
Note that because of the time-independence I can effectively
set $\square_0=-\p_r^2$ and $\p_+=-\p_-=\p_r$. Again I use the
notation $\p'=\p_{r'}$.
\begin{multline}
0=\int''\square_0''(G^x_{''}G''_{'}g'')\\
=\int''\left\{g''\square_0''(G^x_{''}G''_{'})-\p''(G^x_{''}G''_{'})
\p''g''-\p''(G^x_{''}G''_{'})\p''g''-(G^x_{''}G''_{'})
(\p'')^2g''\right\}\\
=\int''\left\{g''[-\delta(x,x'')G''_{'}-\delta(x'',x')G^x_{''}
-2(\p''G^x_{''})\p''G''_{'}]+(G^x_{''}G''_{'})(\p'')^2g''\right\}\\
=-gG^x_{'}-g'G^x_{'}-\int''\left\{(G^x_{''}G''_{'})R''
+2g''(\p''G^x_{''})\p''G''_{'}\right\}\label{identity}
\end{multline}
By this identity the second order of the perturbation
series can be written in the compact form
\begin{equation}
\frac{1}{2}\left\{[g(r)+g(r')]G_0(x,x')+\int''G_0(x,x'')
G_0(x'',x')R(r'')d^2x''\right\}.\label{pert.sec.order}
\end{equation}
In a similar manner I can compute the third order:
\begin{multline}
\int''\int'''G^x_{''}\p''(g''\p''G''_{'''})\p'''(g'''\p'''G'''_{'})
=\int''(\p''G^x_{''})g''\p''\int'''(\p'''G''_{'''})g'''\p'''
G'''_{'}\\
=-\frac{1}{2}\int''(\p''G^x_{''})g''\p''\left\{(g''+g')G''_{'}+\int'''
G''_{'''}G'''_{'}R'''\right\}\\
=\frac{1}{4}\left\{[g^2+(g')^2]G^x_{'}-\int''G^x_{''}G''_{'}(\p'')^2
(g'')^2-\int''(\p''G^x_{''})G''_{'}\p''(g'')^2\right\}\\
-\frac{g'}{2}\int''(\p''G^x_{''})g''\p''G''_{'}
-\frac{1}{2}\int'''G'''_{'}R'''\int''(\p''G^x_{''})g''\p''G''_{'''}\\
=\frac{1}{4}\biggm\{[g^2+(g')^2]G^x_{'}+\int''G^x_{''}[\p''(g'')^2]
\p''G''_{'}
+[g'g+(g')^2]G^x_{'}+g'\int''G^x_{''}G''_{'}R''\\
+\int'''G'''_{'}R'''\left([g+g''']G^x_{'''}+\int''G^x_{''}G''_{'''}R''
\right)\biggm\}.
\end{multline}
Here I have used again identity (\ref{identity}).
Finally, I write down the third order of the perturbation
series in a compact form, replacing $\p_rg^2(r)=2R(r)$:
\begin{multline}
\frac{1}{4}\biggm\{[g^2(r)+g(r)g(r')+2g^2(r')]G_0(x,x')\\
+\int''G_0(x,x'')R(r'')[2\cdot\p_{r''}+g(r')]G_0(x'',x')d^2x''\\
+\int'''G_0(x''',x')R(r''')\\
\cdot\left([g(r)+g(r''')]G_0(x,x''')+\int''G_0(x,x'')G_0(x'',x''')
R(r'')d^2x''\right)d^2x'''\biggm\}.\label{pert.thi.order}
\end{multline}

\subsection{Flat Green Functions}

The flat Green functions on the half-plane
are the basis of the perturbation series.
I will derive the Feynman Green function $G_F$ and the
retarded Green function $G_{ret}$
by choosing different integration paths in the Fourier
decomposition of the Green functions.
At the end of Section 3.2 I have mentioned the suggested
relation between $G_F$ and the $\St{H}$-state and between $G_{ret}$ and the
$\St{U}$-state. From the physical point of view
$\St{U}$ is more interesting as it is the vacuum state
and describes a radiating BH surrounded by empty space.
However, the Euclidean origin of the effective action
suggests to use the Feynman Green function. As I have
proofed in Section 2.4.2, the expectation values of the basic components
$\E{T},\E{\Th}$ which
I compute by the effective action are state-independent, hence
I could use any Green function corresponding to a sensible
physical quantum state. Nevertheless, I consider $G_F$ \emph{and}
$G_{ret}$ as to investigate the proposed correspondence
between quantum states and Green functions.

\subsubsection{Retarded and Feynman Green Functions on the
Half-Plane}

All Green functions can be expressed by their Fourier
decomposition into eigenfunctions $e^{ikx}$ of the flat
Laplacian $\square_0$. I will first calculate the Green
functions on the half-plane $0<r<\infty,-\infty<t<\infty$
and then shift the left border from $0$ to $2M$.
At $r=0$ (or finally $r=2M$) I will demand that the
eigenfunctions vanish -- formally the r-coordinate still
ranges from $\{-\infty,\infty\}$, but the line $r=0$ now
divides the physically interesting part from the rest of the
manifold\footnote{Therefore also the range of the momentum-coordinates
$k^{\alpha}$ is still $\{-\infty,\infty\}$.}. This can be accomplished
by selecting out the $\sin r$-modes; the complete
eigenfunctions that fulfil this boundary condition are $\phi_{hp}
\propto\sin k^1r\cdot e^{ik^0t}$. The Green functions
on the half-plane can be written as
\begin{multline}
G_0(x,x')\propto\frac{1}{2\pi^2}\int\frac{\phi_{hp}(x)
\phi^{\ast}_{hp}(x')}{k^2}d^2k\\
=\frac{1}{2\pi^2}\int\frac{(\sin k^1r)
(\sin k^1r')e^{ik^0(t-t')}}{(k^0)^2-(k^1)^2}d^2k.\label{flat.Green}
\end{multline}
At this point I have chosen an arbitrary normalisation constant.
It will be fixed later by the condition $\square_0G_0(x,x')
=-\delta(x-x')$. The integrand has single poles at $k^0=\pm k^1$.
By choosing a particular path in the complex $k^0$-plane
I select the type of Green function. If I pass
both poles \emph{above the real axis} I obtain the
retarded Green function (in the following I omit the index $0$
of the flat Green functions):
\begin{multline}
G_{ret}\propto\frac{i}{2\pi}\theta(t-t')
\int(\sin k^1r)(\sin k^1r')\left[\frac{e^{ik^1(t-t')}}{k^1}
-\frac{e^{-ik^1(t-t')}}{k^1}\right]dk^1\\
=-\frac{1}{2\pi}\theta(t-t')
\int\frac{[\cos k^1(r-r')-\cos k^1(r+r')]\sin k^1(t-t')}{k^1}dk^1\\
=\frac{\theta(t-t')}{2}[\theta(r-r'+t'-t)-\theta(r-r'+t-t')
-\theta(r+r'+t'-t)+\theta(r+r'+t-t')].
\end{multline}
Now I check the normalisation of the Green function by
comparing its action on a test-function $\vp$ with that of the
delta-distribution:
\small
\begin{multline}
\int^x\vp(x)\square_0^xG^{ret}(x,x')=\frac{1}{2}
\int^x\vp(x)(\p_0^2-\p_1^2)^xG^{ret}(x,x')\\
=\frac{1}{2}\int^x\Bigm\{-(\p_0G)\p_0\vp\\
-\theta(t-t')\Bigm[\delta'(r-r'+t'-t)+\delta'(r+r'+t-t')
-\delta'(r-r'+t-t')-\delta'(r+r'+t'-t)\Big]\vp\Bigm\}\\
=\int^x\Bigm\{-\frac{\delta(t-t')}{2}
\Bigm[\theta(r-r'+t'-t)+\theta(r+r'+t-t')
-\theta(r-r'+t-t')-\theta(r+r'+t'-t)\Big]\p_0\vp\\
-\theta(t-t')\Bigm[-\delta(r-r'+t'-t)+\delta(r+r'+t-t')
-\delta(r-r'+t-t')+\delta(r+r'+t'-t)\Big]\p_0\vp-{\dots}\Bigm\}\\
=-\frac{1}{2}\int_0^{\infty}0\cdot\p_0\vp dr\\
+\frac{1}{2}\int^x\Bigm\{\delta(t-t')
\Bigm[-\delta(r-r'+t'-t)+\delta(r+r'+t-t')
-\delta(r-r'+t-t')+\delta(r+r'+t'-t)\Big]\\
+\theta(t-t')\Bigm[\delta'(r-r'+t'-t)+\delta'(r+r'+t-t')
-\delta'(r-r'+t-t')-\delta'(r+r'+t'-t)\Big]-{\dots}\Bigm\}\vp\\
=\frac{1}{2}\int_0^{\infty}\Bigm[-\delta(r-r')+\delta(r+r')
-\delta(r-r')+\delta(r+r')\Big]\vp(r,t')dr=-\vp(r',t').
\end{multline}
\normalsize
Obviously the normalisation factor already has been chosen correctly.
The next step is the shift of the left border of the manifold.
This can be achieved by simply replacing $r\to r-2M\,,\,
r'\to r'-2M$. It can be easily checked that the
retarded Green function
\begin{multline}
G_{ret}(x,x')=\frac{1}{2}\theta(t-t')\\
\cdot\Big[\theta(r-r'+t'-t)-\theta(r-r'+t-t')
+\theta(r+r'-4M+t-t')-\theta(r+r'-4M+t'-t)\Big]\label{retarded.Green}
\end{multline}
indeed vanishes for $r=2M$ or $r'=2M$.

Next I calculate the Feynman Green function starting from
the general expression (\ref{flat.Green}). This time
I bypass the pole at $k^0=-k^1$ below the real axis and
the pole at $k^0=+k^1$ above the real axis\footnote{In the present
case of chargeless particles the causal and the acausal Green
functions coincide, hence a mirroring of the integration path
on the real $k^0$-axis would lead to the same result.}:
\begin{multline}
G_F=\frac{i}{\pi}\int(\sin k^1r)(\sin k^1r')
\frac{e^{ik^1|t-t'|}}{k^1}dk^1\\
=-\frac{1}{\pi}\int\frac{(\sin k^1r)(\sin k^1r')
(\sin k^1|t-t'|)}{k^1}dk^1\\
=\frac{1}{4}\Bigm[\theta(r-r'-|t-t'|)-\theta(r-r'+|t-t'|)
+\theta(r+r'+|t-t'|)-\theta(r+r'-|t-t'|)\Bigm].
\end{multline}
From $e^{ik^1|t-t'|}$ the $\cos$-part drops out because
it gives an odd integrand in $k^1$.
Again I can shift the left border to the horizon and then
show by a test-function that the normalisation
factor is correct:
\small
\begin{multline}
\int^x\vp(x)\square_0G_F(x,x')dx\\
=-\frac{1}{4}\int^x\Bigm[\delta'(r-r'+|t-t'|)
+2\delta(r-r'+|t-t'|)\delta(t-t')\\
-\delta'(r-r'-|t-t'|)+2\delta(r-r'-|t-t'|)\delta(t-t')\\
+\delta'(r+r'-4M-|t-t'|)-2\delta(r+r'-4M-|t-t'|)\delta(t-t')\\
-\delta'(r+r'-4M+|t-t'|)-2\delta(r+r'-4M+|t-t'|)\delta(t-t')\\
-\delta'(r-r'+|t-t'|)+\delta'(r-r'-|t-t'|)\\
-\delta'(r+r'-4M-|t-t'|)+\delta'(r+r'-4M+|t-t'|)\Bigm]\vp(x)dx\\
=-\frac{1}{2}\int^x\Bigm[\delta(r-r'+|t-t'|)+\delta(r-r'-|t-t'|)\\
-\delta(r+r'-4M-|t-t'|)-\delta(r+r'-4M+|t-t'|)\Bigm]\delta(t-t')\vp(x)dx\\
=-\int_{2M}^{\infty}\Bigm[\delta(r-r')-\delta(r+r'-4M)\Bigm]\vp(r,t')dr
=-\vp(r',t').
\end{multline}
\normalsize
Thus, the Feynman Green function I will use is
\begin{multline}
G_F(x,x')=\frac{1}{4}\Bigm[\theta(r-r'-|t-t'|)-\theta(r-r'+|t-t'|)\\
+\theta(r+r'-4M+|t-t'|)-\theta(r+r'-4M-|t-t'|)\Bigm].\label{Feynman.Green}
\end{multline}

\subsubsection{Euclidean Feynman Green Function}

In the last Section both Green functions have been derived without
the need of any regularisation prescription. This is somewhat
astonishing if one considers the Feynman Green function
normally used on the full plane which I will discuss here.
It requires an IR renormalisation already
during its derivation and by its logarithmic dependence
on the point separation $x-x'$ it must be regularised
in a complicated way whenever applying it in calculations \cite{fms97}.
Also the Green functions of the last Section exhibit IR divergences,
but they can always be removed by simply dropping boundary
terms. In the following I will call the Feynman Green
function, if it is derived by Euclideanisation of spacetime,
the \emph{Euclidean Feynman Green function}, even if
its arguments are Minkowskian coordinates (or Lorentzian
if the flat Green function is object of the perturbation theory)
-- the name ``Euclidean'' only emphasizes the way it has been computed.

The problem on the full plane becomes apparent
when considering its Fourier decomposition
\begin{equation}
G_F\propto\frac{1}{2\pi^2}
\int\frac{e^{-i[k^1(r-r')-k^0(t-t')]}}{(k^0)^2-(k^1)^2}d^2k
=\frac{i}{\pi}\int\frac{e^{-ik^1[(r-r')-|t-t'|]}}{k^1}dk^1.
\end{equation}
Because of the presence of the $\cos$-term,
the integral over $k^1$ is now divergent.
This problem is normally resolved by first Euclideanizing
spacetime by introducing an imaginary time-coordinate $\tau=it$
and an imaginary frequency $k^0_{\cal E}=-ik^0$ (note that
$k^0t=k^0_{\cal E}\tau$) and then separating the finite
part of the integral. The divergent part goes linearly
to $\infty$ with some regularisation parameter $\lambda$.
Hence the finite part of $G_F$ can be obtained by picking
out the constant term (respectively to $\lambda$)
of the regularised Green function and it has the form
\begin{equation}
G^{ren}_F(x,x')=\frac{d}{d\lambda}\left[\lambda\cdot G_F^{reg}(x,x')
\right]\Bigm|_{\lambda=0}
=\frac{i}{4\pi}\ln[(t-t')^2-(r-r')^2].\label{ren.Feynman}
\end{equation}
In calculations one has to use a regularised Green
function, e.g. \cite{bkp92}
\begin{equation}
G^{reg}_F(x,x')=\frac{-i}{2\pi}\frac{\Gamma[-\frac{\lambda}{2}]}
{\Gamma[1+\frac{\lambda}{2}]}\frac{[(t-t')^2-(r-r')^2]^{\lambda/2}}
{2^{1+\lambda}},\label{reg.Feynman}
\end{equation}
and finally pick out the finite part by the rule (\ref{ren.Feynman}).
I check the normalisation by the Euclidean Green function
$G_F^{\cal E}(x,x')=\frac{i}{4\pi}\ln[(\tau-\tau')^2+(r-r')^2]
=\frac{1}{4\pi}\ln\rho^2$, where I introduce polar coordinates
$\vp,r$ by $x^{\alpha}_{\cal E}=\rho(\cos\vp,\sin\vp)$ and choose $x'=(0,0)$
on the origin of the manifold. I define a volume $V:=L-B_{\ve}(0)$
as the whole manifold minus a ball of radius $\ve$ around the origin
(note that we are on the full plane $-\infty<r<\infty$).
Then $\triangle_0G_F^{\cal E}(x,0)\equiv0$ for $x\in V$.
Now I use Green's theorem (\ref{Green's-theorem}), where $\phi(\rho)$ shall
be an angular-independent test-function:
\begin{multline}
\int_VG_F(x,0)\square_0\phi\sqrt{-g}d^2x
=i\int_VG_F^{\cal E}(x,0)\triangle_0\phi\sqrt{g}d^2x_{\cal E}\\
=i\oint_{\p V}\Big[G_F^{\cal E}(x,0)\n_{\alpha}\phi
-\phi\n_{\alpha}G_F^{\cal E}(x,0)\Big]\sqrt{g}
\ve^{\alpha}{}_{\beta}dx^{\beta}_{\cal E}\\
=\frac{1}{4\pi}\int_0^{2\pi}\Big[\ln\ve^2\p_{\ve}\phi
-\frac{2\phi}{\ve}\Big]\ve d\phi\stackrel{\ve\to0}{\to}
-\phi(0).
\end{multline}
The orientation of the boundary is such that the normal vector
points outwards of $V$ (note that $dt\wedge dr$ has the
same orientation as $d\vp\wedge dr$ if $\vp$ goes anti-clockwise
from $0$ to $2\pi$). I have dropped the outer boundary
because the testfunction, as well as $G_F^{\cal E}$, are
assumed to have compact support (see IR problem).
The first line is just the distributional action
of the Laplacian on the Green functions. Bearing in
mind that $\square_0=-\triangle_0$ I conclude that
the normalisation is correct, i.e.
 $\square_0G^{ren}_F(x,x')=-\delta(x-x')$.

The method of Euclideanizing spacetime (and momentum
space) has some peculiarities that (beside other problems)
reduce the freedom one has in adjusting a Green function.
The basic trick thereby, called ``Wick rotation'', is to
treat the imaginary frequency $k^0_{\cal E}$ as being real,
i.e. integrating it from $\{-\infty,\infty\}$ instead
of ${i\infty,-i\infty}$ as it should be. This corresponds
to a $90$ degree rotation of the integration path in the
complex plane. Note that the two auxiliary paths cannot be
harmless at the same time.
To neglect them can already be seen as the first step of
renormalisation. The poles of the denominator now
lie at the points $k^0_{\cal E}=\pm i k^1$ and are no
more crossed by the integration path. By rotating back
one recognises the Feynman contour -- in the Euclidean
formalism no other Green function is accessible.

It is convenient to introduce polar coordinates $k:=
(k^0_{\cal E})^2+(k^1)^2,k^a_{\cal E}=k(\cos\vp,\sin\vp)$
in the Euclidean plane to perform the integrals over the
momentum coordinates:
\begin{equation}
G_F\propto\frac{1}{2\pi^2}\int_{-\frac{\pi}{2}}^{\frac{\pi}{2}}
\int_{-\infty}^{\infty}\frac{e^{ik\cos\vp(\tau-\tau')}e^{-ik\sin\vp(r-r')}}
{k}dkd\vp.
\end{equation}
Instead of the usual parametrisation $0<\vp<2\pi\,,\,0<k<\infty$
of the manifold I use $-\frac{\pi}{2}<\vp<\frac{\pi}{2}\,,\,-\infty<k<\infty$
because it fits to the boundary condition.
Now I introduce the boundary condition of the half-plane
by selecting out the appropriate eigenfunctions. Again the
integrals then become convergent and lead to the Euclidean
Feynman Green function on the half-plane:
\begin{multline}
G_F\propto\frac{1}{2\pi^2}\int_{-\frac{\pi}{2}}^{\frac{\pi}{2}}
\int_{-\infty}^{\infty}\frac{e^{ik\cos\vp(\tau-\tau')}\sin[k(\sin\vp)r]
\sin[k(\sin\vp)r']}{k}dkd\vp\\
=\frac{1}{2\pi^2}\int_{-\frac{\pi}{2}}^{\frac{\pi}{2}}
\int_{-\infty}^{\infty}\frac{\sin[k\cos\vp(\tau-\tau')]\sin[k(\sin\vp)r]
\sin[k(\sin\vp)r']}{k}dkd\vp\\
=-\frac{1}{4\pi}\int_{-\frac{\pi}{2}}^{\frac{\pi}{2}}
\Bigm\{\theta[\sin\vp(r+r')+\cos\vp(\tau-\tau')]
-\theta[\sin\vp(r+r')-\cos\vp(\tau-\tau')]\\
+\theta[\sin\vp(r-r')-\cos\vp(\tau-\tau')]
-\theta[\sin\vp(r-r')+\cos\vp(\tau-\tau')]\Bigm\}d\vp\\
=-\frac{1}{2\pi}\left[\arctan\frac{\tau-\tau'}{r+r'}-
\arctan\frac{\tau-\tau'}{|r-r'|}\right]\\
\to-\frac{1}{2\pi}\left[\text{Artanh}\frac{t-t'}{r+r'}-
\text{Artanh}\frac{t-t'}{|r-r'|}\right]\\
=-\frac{i}{4\pi}\ln\frac{[|r-r'|+(t-t')][(r+r')-(t-t')]}
{[|r-r'|-(t-t')][(r+r')+(t-t')]}.\label{Eukl.Feynman.Green.calc}
\end{multline}
In the last two lines I have made the transition back to
the Lorentzian manifold.
By shifting the left border of the manifold from $0$ to $2M$
I obtain the ``Euclidean'' Feynman Green function on the
half-plane:
\begin{equation}
G_F^{hp}=-\frac{i}{4\pi}\ln\frac{[|r-r'|+(t-t')]
[(r+r'-4M)-(t-t')]}{[|r-r'|-(t-t')][(r+r'-4M)+(t-t')]}.
\label{Eukl.Feynman.Green}
\end{equation}
Obviously this Green function is not correct!
Namely, it is not symmetric in its arguments as it should
be, whereas it fulfils $G_F^{hp}(x,x')=-G_F^{hp}(x',x)$.
This can already be seen from the second line in
(\ref{Eukl.Feynman.Green.calc}). Note that 
(\ref{Eukl.Feynman.Green}) is not unique. If I would
have set $k^a_{\cal E}=k(\sin\vp,-\cos\vp)$ I would have
obtained a similar expression, where instead of the
radius-coordinates the time-coordinates would have
appeared as absolute values $|t-t'|$. It seems as if
the Wick-rotation which is based on the symmetry between
time and radius coordinate is not compatible with
the boundary condition on the half-plane.

The correct Euclidean Feynman Green function
\begin{equation}
G_{F}^{mod}=\frac{i}{4\pi}\ln\frac{(t-t')^2-(r-r')^2}
{(t-t')^2-(r+r'-4M)^2}.\label{Eukl.Feynman.Green.mod}
\end{equation}
can be constructed by substracting from  (\ref{ren.Feynman})
the ``mirrored'' Green function on the horizon.
Indeed (\ref{Eukl.Feynman.Green.mod}) is symmetric and
vanishes at the horizon.
Because of $\square_0\frac{1}{4\pi}\ln[(t-t')^2-(r+r')^2]
=-\delta(t-t')\delta(r+r'-4M)=0$ for $r,r'>0$ the mirrored
term is just a homogeneous Green function with respect to
the positive half-plane. In calculations (\ref{Eukl.Feynman.Green.mod})
can simply be regularised by using the
regularised Feynman Green function (\ref{reg.Feynman}),
where a similar term $\propto[(t-t')^2-(r+r'-4M)^2]^{\lambda/2}$
must be substracted.

The problem in ``calculating'' the correct Feynman Green
function on the half-plane in the Euclidean approach once more
demonstrates the limitations of this formalism.
Although I think that the Green functions
(\ref{Feynman.Green},\ref{retarded.Green}) are the correct ones,
it will be interesting to estimate which result for the Hawking
flux is obtained when using the ``Euclidean'' Green function 
(\ref{Eukl.Feynman.Green.mod}).

\subsubsection{Flat Green Functions (Summary)}

In the last Sections I have derived the retarded and the
Feynman Green functions of a scalar field on a flat
two-dimensional spacetime with boundary. Thereby I have
chosen as boundary condition that the eigenfunctions,
and hence the Green functions,
vanish at the horizon $r=2M$, representing the ``left'' boundary
of the manifold. As a consequence also the Green functions
on the curved Schwarzschild spacetime $L$ fulfil these
boundary conditions, as can be seen from the perturbation
series (\ref{pert.Green}).

In Figures 8,9 I present the flat Green functions
(\ref{retarded.Green},\ref{Feynman.Green}).

\begin{figure}[h]

\hspace{2cm}\epsfig{file=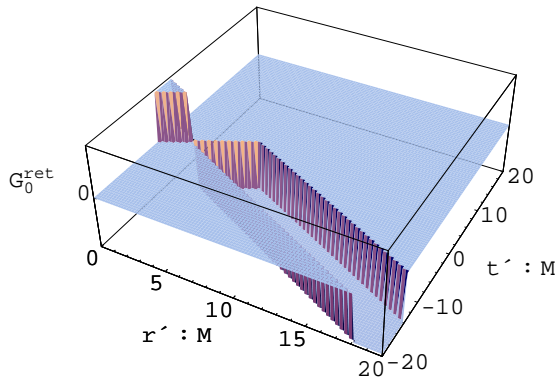,height=6cm}

\caption[fig1]{Flat retarded Green Function}
\end{figure}

The plots show $G_0(x,x')$ for fixed $x$, where I have set
$t=0,r=6M$. The values for $x'$ range over $-20M<t'<20M$
and $0<r'<20M$. I have included the strip $0<r'<2M$
in the plots to show the mathematical continuation, although
the physical manifold ends at $r'=2M$.
The retarded Green function (Figure 8) is zero everywhere except
on the past light cone. Interestingly it also vanishes
below the past-directed light-ray that starts from $x$ and
is reflected on the horizon.

The Feynman Green function (Figure 9) looks similar but further
has non-vanishing support on the future light-cone
(except the reflection zone).

\begin{figure}[h]

\hspace{2cm}\epsfig{file=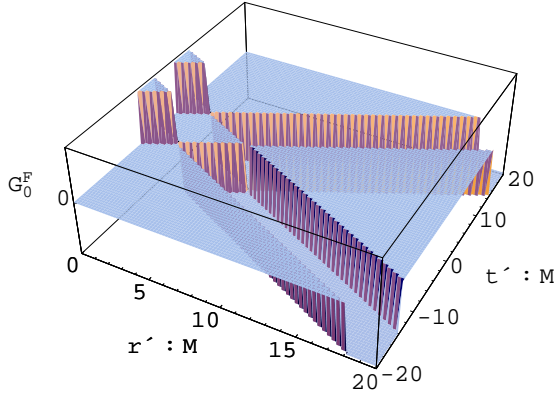,height=6cm}

\caption[fig1]{Flat Feynman Green Function}
\end{figure}

The support of both Green functions reaches until
lightlike infinity, their amplitude remains constant, and hence
causes IR divergences that are removed by the IR renormalisation.

Next I consider the symmetries of the Green functions:
the Feynman Green function is symmetric in
its arguments $G^F_0(x,x')=G^F_0(x',x)$, while the retarded
Green function obeys the relation
\begin{multline}
G^{ret}_0(x',x)=G^{ret}_0(x,x')\\
-\frac{1}{2}\Big[\theta(r-r'+t'-t)-\theta(r-r'+t-t')
+\theta(r+r'-4M+t-t')-\theta(r+r'-4M+t'-t)\Big].
\end{multline}
At first sight one might conclude that this difference between
$G^F_0$ and $G^{ret}_0$ might lead to different results
if employed in basic integrals, as suggested. 
However, an explicit calculation shows that the contribution
is actually \emph{the same}:
\begin{multline}
\int^x_{L}G^F_0(x,x')F(r')\sqrt{-g'}dx'\\
=\frac{1}{4}\int_{-\infty}^t\int_{2M}^{\infty}
\Big[\theta(r-r'-t+t')-\theta(r-r'+t-t')\\
+\theta(r+r'-4M+t-t')-\theta(r+r'-4M-t+t')\Big]F(r')dr'dt'\\
+\frac{1}{4}\int^{\infty}_t\int_{2M}^{\infty}
\Big[\theta(r-r'+t-t')-\theta(r-r'-t+t')\\
+\theta(r+r'-4M-t+t')-\theta(r+r'-4M+t-t')\Big]F(r')dr'dt'\\
=\frac{1}{2}\int_{-\infty}^0\int_{2M}^{\infty}
\Big[\theta(r-r'+z)-\theta(r-r'-z)\\
+\theta(r+r'-4M-z)-\theta(r+r'-4M+z)\Big]F(r')dr'dz\\
=\int^x_{L}G^{ret}_0(x,x')F(r')\sqrt{-g'}dx'.\label{equiv.Green}
\end{multline}
Indeed, if applied to basic integrals, the retarded Green function
reveals the same symmetry properties as the Feynman Green function,
as can be seen from (\ref{equiv.Green})
\begin{equation}
\int_{L}G^{ret}_0(x,x')F(r')\sqrt{-g'}d^2x'
=\int_{L}G^{ret}_0(x',x)F(r')\sqrt{-g'}d^2x'.
\end{equation}
Further, because the perturbation series is nothing but a multiple
of basic integrals (\ref{pert.sec.order},\ref{pert.thi.order})
this equivalence extends to the exact Green function of the
full Schwarzschild Laplacian. This means that \emph{the effective
action of the dilaton model is independent of the type
of Green function used}. Clearly, this property already has been
implied by the existence of a unique local effective action
in $2d$, as discussed in Section 5.3, but by the examinations
of the last Sections it can now be traced back all the way to the
explicit selection of a particular Green function.

\newpage
Finally I discuss the Euclidean Feynman Green function
(\ref{Eukl.Feynman.Green.mod}). The real part (Figure 10) takes exactly
the same values as the original Feynman Green functions
on the physical manifold (below $r=2M$ the latter is inverted).

\begin{figure}[h]

\hspace{2cm}\epsfig{file=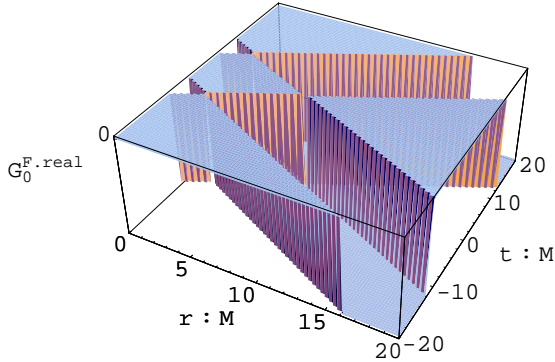,height=6cm}

\caption[fig1]{Euclidean Feynman Green Function, real part}
\end{figure}

This is rather surprising as it reveals some relationship
between the two Feynman Green functions that is not
apparent from their mathematical form. The Euclidean Feynman
Green function further has an imaginary part that shows
the logarithmic divergences on the light-cones which require
a complicated regularisation in all explicit calculations.

The fact that the two Feynman Green functions
(\ref{Feynman.Green},\ref{Eukl.Feynman.Green.mod}) only differ
by a divergent imaginary part suggests once more
that probably something has gone wrong during the procedure
of Euclideanisation.
In the context of the basic integrals the imaginary part
of the Euclidean Feynman Green function cannot contribute because
the result shall be a real value, while the real part
should give the same result as the original Feynman Green function.
On the other hand one must introduce a complicated
regularisation prescription, such as (\ref{ren.Feynman},\ref{reg.Feynman})
(which is not unique), to handle the logarithmic divergences
of the imaginary part. It is possible, and shall be shown explicitly
in one case, that the renormalisation destroys the correlation
with the original Green function and finally yields completely
different results.

\subsection{Hawking Radiation}

Now all pieces are collected to calculate the complete
EM tensor and, most importantly, the Hawking flux.
I use the local form of the effective action (\ref{dilaton.eff.action})
to calculate the basic components $\E{T}_2,\E{\Th}_2$.
By means of the Green function perturbation series, developed
in the last Sections, I show that the basic integrals over
$\square\rho\,,\,\square\phi$ indeed do not produce homogeneous
solutions and therefore (\ref{dilaton.eff.action}) is correct.
Then I calculate the remaining components and the flux by
the CF method, whereby the integration constants
are chosen according to the Unruh state $\St{U}$.
Finally I discuss how an incorrect choice of homogeneous solution
leads to unphysical expectation values.

\subsubsection{Expectation Values}

In Section 2.4.2 I have argued that $\E{T}_2,\E{\Th}_2$ are
independent of the quantum state. This allows to calculate
these components from the effective action which by construction
leads to expectation values in the $\St{B}$-state.
Further, they can be obtained easily
by variation from (\ref{dilaton.eff.action}) for $\rho$, respectively
$\phi$.

In the conformal gauge the Schwarzschild metric reads $g_{\bar{\alpha}
\bar{\beta}}=a_{\bar{\alpha}\bar{\beta}}e^{2\rho}$, where
\begin{equation}
a_{\bar{\alpha}\bar{\beta}}=\left(\begin{array}{cc}0&\frac{1}{2}
\\\frac{1}{2}&0\end{array}\right)\,\,,\,\,\left(a^{-1}\right)
^{\bar{\alpha}\bar{\beta}}=\left(\begin{array}{cc}0&2
\\2&0\end{array}\right),
\end{equation}
and its inverse is $g^{\bar{\alpha}\bar{\beta}}=\left(a^{-1}\right)
^{\bar{\alpha}\bar{\beta}}e^{-2\rho}$. Thus, the trace anomaly
can be calculated as
\begin{multline}
\E{T}_2=g^{\bar{\alpha}\bar{\beta}}\frac{2}{\sqrt{-g}}
\frac{\delta W}{\delta g^{\bar{\alpha}\bar{\beta}}}
=\frac{2\left(a^{-1}\right)^{\bar{\alpha}\bar{\beta}}
e^{-2\rho}}{\sqrt{-g}}\frac{\delta W}{\left(a^{-1}\right)
^{\bar{\alpha}\bar{\beta}}(-2)e^{-2\rho}\delta\rho}\\
=-\frac{1}{\sqrt{-g}}\frac{\delta W}{\delta\rho}=
-\frac{1}{12\pi}\Big[\square\rho+3(\n\phi)^2-2\square\phi\Big].
\label{E(T)}
\end{multline}
Note that one has to vary not only for $\rho$ where it appears
explicitly, but also where it is hidden in the metric that
depends on $\rho$, as for instance in $g^{\alpha\beta}(\rho)
\p_{\alpha}\phi\p_{\beta}\phi$.
As the latter variations are traceless in $2d$, these terms do
not contribute to the trace anomaly.

$\E{\Th}_2$ is given by the variation of the effective action for
the dilaton field, see Section 2.2 (\ref{Th-2}).
Remembering that $X=e^{-2\phi}$ we get
\begin{multline}
\E{\Th}_2=\frac{1}{2X\sqrt{-g}}\frac{\delta W}{\delta\phi}\\
=-\frac{1}{24\pi X}\Big\{2\square\rho+3(\n\phi)^2
+6\n\rho\n\phi+\square\phi\bigm[5+6(\phi+\rho+\chi_{\phi}+\chi_{\rho})
-3c_R\bigm]\Big\}.\label{E(Th)}
\end{multline}
I have given here the homogeneous solutions $\chi_{\phi},\chi_{\rho}$
of the basic integrals and the renormalisation constant $c_R$
to show that they appear in the same place in the
expectation values. This means that if the homogeneous solutions
are only constants (as I will show in the next Section), they can be
absorbed by the renormalisation constant, which is then the
only remaining ambiguity of the model.

In the Schwarzschild gauge the geometric fields
and their derivatives are given by
\begin{eqnarray}
\rho=\frac{1}{2}\ln\Sss,&\phi=-\ln r,&X=r^2=e^{-2\phi}\nonumber\\
(\n\rho)^2=-\frac{M^2}{r^4\Sss},&(\n\phi)^2=-\frac{\Sss}{r^2},
&(\n X)^2=-4r^2\Sss\nonumber\\
\square\rho=\frac{2M}{r^3},&\square\phi=\frac{4M-r}{r^3},&
\square X=-2.\nonumber\\
\n\rho\n\phi=\frac{M}{r^3}.&&
\label{2d-formulas}
\end{eqnarray}
Inserting this into the general expressions (\ref{E(T)},\ref{E(Th)})
we obtain the expectation values in the static approximation:
\begin{eqnarray}
\E{T}_2&=&\frac{1}{12\pi r^2}\label{expl.E(T)}\\
\E{\Th}_2&=&\frac{1}{2\pi}\biggm\{\frac{2}{3r^4}-\frac{3M}{r^5}\nonumber\\
&&+\frac{r-4M}{4r^5}\left[\ln\left(\frac{r-2M}{r^3}\right)
+2(\chi_{\rho}+\chi_{\phi})-c_R\right]\biggm\}.
\label{expl.E(Th)}
\end{eqnarray}
In Section 2.4.2 I have proven the state-independence of these
two expectation values. Accordingly I can simply say that
the system is in the $\left|U\right>$-state
which is the vacuum state $\left|0\right>$.
Because of the state-independence I omit the quantum state symbols.

The component $\E{\Th}_2$ shows a logarithmic divergence on the
horizon. Because $\E{\Th}_2$ is invariant if we consider global
coordinates this singularity is not simply a coordinate singularity.
All other components of the EM tensor are regular on the horizon
(in global coordinates).

\subsubsection{Basic Integrals}

In this Section I will show that the results obtained so far,
namely the expectation values (\ref{expl.E(T)},\ref{expl.E(Th)}),
are correct. They are based on the assumption that
the homogeneous solutions $\chi_f$ that appear on the r.h.s.
of the basic integrals (\ref{int.basic}) are only constants
that do not contribute to the flux.

I start with the basic integral
\begin{multline}
\int_{L}G(x,x')R(r')\sqrt{-g'}d^2x'\\
=-\int_{L}G(x,x')2\square'\rho(r')\sqrt{-g'}d^2x'
=2\rho(r)+2\chi_{\rho}(r).
\end{multline}
The first order approximation is obtained by replacing the full Green
function by the flat one. I have already shown that the result
of the perturbational analysis is independent from the choice of
flat Green functions (\ref{equiv.Green}); in the following I will
only work with the retarded Green function:
\begin{multline}
\int_{L}G^{ret}_0(x,x')R(x')d^2x'=-2M\int_{-\infty}^t
\biggm[\int_{2M}^{r-(t-t')}\theta(r-2M-(t-t'))\\
+\int_{2M}^{\infty}-\int_{2M}^{r+t-t'}-\int_{4M+t-t'-r}^{\infty}
\theta(2M+t-t'-r)-\int_{2M}^{\infty}\theta(r-2M-(t-t'))\biggm]
\frac{dr'dt'}{(r')^3}\\
=2M\biggm[\int_{t-(r-2M)}^{t}\int_{r-(t-t')}^{\infty}
-\int_{-\infty}^t\int_{r+t-t'}^{\infty}+\int_{-\infty}^{t-(r-2M)}
\int_{4M+t-t'-r}^{\infty}\biggm]
\frac{dr'dt'}{(r')^3}\\
=\Sss.\label{first.order}
\end{multline}
The first order already reveals the expected properties: it is indeed
independent of the time $t$ and there are no IR divergences.
Clearly it vanishes for $r=2M$. The second order of the
perturbation series is given by the integral $\int_{L}
G^{2nd}(x,x')R(x')\sqrt{-g'}d^2x'$, where $G^{2nd}(x,x')$
is (\ref{pert.sec.order}). The first two terms can be integrated
out directly like (\ref{first.order}):
\begin{multline}
\frac{1}{2}\int_{L}[g(r)+g(r')]G^{ret}_0(x,x')R(x')d^2x'\\
=\frac{M}{r}\Sss+\frac{1}{2}\int_{L}G^{ret}_0(x,x')
R(x')\frac{2M}{r'}d^2x'=\frac{1}{6}+\frac{M}{r}-\frac{8M^2}{3r^2}.
\end{multline}
The third term in (\ref{pert.sec.order}) is computed by
first integrating over $x'$:
\begin{multline}
\frac{1}{2}\int''G^{ret}_0(x,x'')R(r'')\int'
G^{ret}_0(x'',x')R(r')d^2x'd^2x''\\
=\frac{1}{2}\int''G^{ret}_0(x,x'')R(r'')\left(1-\frac{2M}{r''}
\right)d^2x''=\frac{4M^2-6Mr+2r^2}{6r^2}.
\end{multline}
The complete second order therefore reads
\begin{equation}
\frac{1}{2}-\frac{2M^2}{r^2}.
\end{equation}
Generally, all terms in the perturbation series can be reduced
to multiple integrals of the type $\int'G(x,x')K(r')d^2x'$,
where $K(r)$ is an analytical function on $L$.
In the following I summarise the integrals that appear
up to the third order of the perturbation series:
\begin{eqnarray}
\int'G^x_{'}R'&=&1-\frac{2M}{r}\nonumber\\
\int'G^x_{'}R'g'&=&\frac{1}{3}-\frac{4M^2}{3r^2}\nonumber\\
\int' G^x_{'}R'(g')^2&=&\frac{1}{6}-\frac{4M^3}{3r^3}\nonumber\\
\int'G^x_{'}R'\Sss&=&\frac{4M^2-6Mr+2r^2}{3r^2}\nonumber\\
\int'G^x_{'}R'g'\Sss&=&\frac{8M^3-8M^2r+r^3}{6r^3}\nonumber\\
\int'G^x_{'}R'\left(\frac{1}{3}-\frac{4M^2}{3(r')^2}\right)
&=&\frac{8M^3-12Mr^2+5r^3}{18r^3}\nonumber\\
\int'G^x_{'}R'\frac{4M^2-6Mr'+2(r')^2}{3(r')^2}&=&
\frac{-8M^3+24M^2r-24Mr^2+7r^3}{18r^3}\hspace{0.5cm}\nonumber\\
\int'G^x_{'}[\p'(g')^2]\p'\left(1-\frac{2M}{r'}\right)
&=&\frac{1}{6}-\frac{4M^3}{3r^3}.
\end{eqnarray}
I have used the condensed notation $G^x_{'}=G_0(x,x')$,
$G_0$ being either the flat retarded or the flat Feynman
Green function. By (\ref{pert.thi.order}) and this table
of integrals I can calculate the third order of the
perturbation series. To this approximation the considered
basic integral becomes
\begin{equation}
\int_{L}G(x,x')R(x')\sqrt{-g'}d^2x'=1+\frac{1}{2}
+\frac{1}{3}+\dots-\frac{2M}{r}-\frac{2M^2}{r^2}
-\frac{8M^3}{3r^3}-\dots.
\end{equation}
The r-dependent part equals the first terms in the power expansion
of\\$\ln\Sss$,
while the constants seem to sum up to $-\ln0$ because: $-\ln0=\int_0^1
\frac{1}{1-x}dx=\int_0^1(1+x+x^2+\dots)dx=1+\frac{1}{2}
+\frac{1}{3}+\dots$\,. Thus, I conjecture that the perturbation
series of this basic integral ``converges'' to
\begin{equation}
\int_{L}G(x,x')R(x')\sqrt{-g'}d^2x'=\ln\frac{\Sss}{0}.
\end{equation}
This result shows that the homogeneous solution of this
basic integral is just an infinite constant $\chi_{\rho}=-\frac{1}{2}\ln0$,
because $\rho=\frac{1}{2}\ln\Sss$, and the integral
over the full Green function acts as $-\square^{-1}$.
I come back to this constant at the end of this Section.

Now I consider the second basic integral that appears in
the effective action:
\begin{equation}
\int_{L}G(x,x')\square'\phi(r')\sqrt{-g'}d^2x'
=-\phi(r)-\chi_{\phi}(r).
\end{equation}
$\phi=-\frac{1}{2}\ln X=-\ln r$ was the ``re-defined''
dilaton field and $\square\phi=\frac{4M-r}{r^3}$.
The first order of the perturbation series now gives
\begin{equation}
\int_{L}G_0(x,x')\frac{4M-r'}{(r')^3}d^2x'
\approx-\Sss+\ln\left(\frac{r}{2M}\right).
\end{equation}
The first two terms of the second order (\ref{pert.sec.order}) are
\begin{equation}
\frac{M}{r}\left[-\Sss+\ln\left(\frac{r}{2M}\right)\right]
+\frac{1}{12}\Sss\left(1-\frac{4M}{r}\right),
\end{equation}
while the third term results in
\begin{equation}
\frac{3}{4}\Sss-\frac{M}{r}\ln\left(\frac{r}{2M}\right).
\end{equation}
Hence, the considered basic integral up to the second
order perturbation theory reads
\begin{equation}
\int_{L}G_0(x,x')\frac{4M-r'}{(r')^3}d^2x'
=\ln\left(\frac{r}{2M}\right)-\Sss\left(\frac{1}{6}+\frac{4M}{3r}
\right).
\end{equation}
This shows that the leading order, up to a constant $-\ln 2M$,
is $\ln r$. Hence, again the homogeneous solution can only
be a constant $\chi_{\phi}=\ln 2M$. I conjecture that
the remaining orders of the perturbation series slowly decrease
the superfluous higher orders in $r$. But, as already concluded
before in Section 5.3, the leading order is sufficient to determine
the homogeneous solution, therefore the latter \emph{must} be
$r$-independent.
\\
\\
I have shown now that for both basic integrals appearing
in the non-local form of the effective action (\ref{nonloc.eff.action}),
the homogeneous solutions are just constants:
\begin{eqnarray}
\int_{L}G(x,x')R(r')d^2x'&=&2\rho(r)-\ln0\\
\int_{L}G(x,x')\square\phi(r')d^2x'&=&-\phi(r)-\ln2M.
\end{eqnarray}
These results differ only by the constants from the ones
obtained by setting $\square^{-1}\square=1$, see Section 5.3.
They are a direct consequence of the
boundary condition that the Green functions vanish
on the horizon and guarantee that the basic integral
vanish for $r=2M$. Hence, by including these boundary
conditions, one has
\begin{equation}
\square^{-1}\square f(r)=f(r)-f(2M).
\end{equation}
From equation (\ref{bound.basic.int}) the appearance of these
constants was not apparent, at first sight. They are absent, because at that
point I had only adapted the Green functions to the half-plane.
Consequently one must also adapt the \emph{delta-function to the half-plane}.
If it is decomposed into eigenfunctions
it adopts their boundary conditions, namely that they vanish
at the horizon. An easier way is to guess the correct delta-function.
Just like the Green functions it can be constructed by
substracting from the original one the mirrored delta-function on the
horizon:
\begin{equation}
\delta_{hp}(x-x'):=\delta(r-r')\delta(t-t')-\delta(r+r'-4M)\delta(t-t').
\label{delta-hp.}
\end{equation}
For $r>2M\vee r'>2M$ the mirrored part does not contribute, because
then the condition $r=4M-r'$ cannot be fulfilled. For $r=2M$
or $r'=2M$ the delta-function vanishes. Going back to Green's theorem
(\ref{Green's-theorem}) and inserting $\square^{x'}G(x,x')=-\delta_{hp}(x-x')$
instead of $\square^{x'}G(x,x')=-\delta(x-x')$ we obtain the
same relation as before (\ref{bound.basic.int}) where $f(x)$
is replaced by $f(r)-f(2M)$.
Note that throughout the derivation of the Green function perturbation
theory I have used the ordinary delta-function of the full-plane.
The delta-function only appears in the perturbation series where
a flat Laplacian $\square_0$ is separated from the perturbing
Laplacian $\delta\square$. As $\delta\square$ is always accompanied
by an integration over the variable on which it acts and the integrand
contains a flat Green function with the same argument, the mirrored
part of the delta-function cannot lead to contributions because
the Green function vanishes at the horizon!

Finally, by the modification of the delta-function according
to (\ref{delta-hp.}) the picture is complete: the
Green function perturbation theory on the half-plane
equipped with the boundary conditions of Section 5.3 is now
\emph{fully self-consistent and yields the expected results},
up to constants.
\\
It remains to examine if and how these constants
affect the expectation values. The effective action is
simply changed by replacing $\rho\to\rho+\chi_{\rho}$
and $\phi\to\phi+\chi_{\phi}$. From (\ref{dilaton.eff.action})
and the expectation values (\ref{E(T)},\ref{E(Th)})
we see that only $\E{\Th}_2$ acquires an additional term
\begin{equation}
\E{\Th}_2=\text{old}+\frac{r-4M}{4\pi r^5}(\chi_{\rho}+\chi_{\phi}).
\end{equation}
Interestingly, the Hawking flux, given by the component
$\E{T^r{}_t}$ of the EM tensor, remains the same.
This can be seen by the integral over $\E{\Th}_2$ that
determines the constant $K$ in the CF
approach, see Section 2.4.1:
\begin{multline}
K\propto\int_{2M}^{\infty}(r'-2M)\E{\Th}_2dr'
=\text{old}+(\chi_{\rho}+\chi_{\phi})\int_{2M}^{\infty}
\frac{(r'-2M)(r'-4M)}{4\pi(r')^5}dr'\\
=\text{old}+0.\label{flux-ind.}
\end{multline}
Accordingly, the \emph{asymptotic energy density and asymptotic
radial stress are unaffected by the constants}.
Note that \emph{the flux is unaffected everywhere},
not only asymptotically, because it only contains an $r^{-2}$-term
(in the static approximation).
However, energy density and stresses for finite radius $r$ do
depend on them.

The actual values of the constants $\chi_{\rho},\chi_{\phi}$
are a direct consequence of the boundary condition at
$r=2M$ as they only shift the absolute value of the basic
integrals to the correct position. It is rather obvious that
a different boundary condition on the horizon only changes
the values of the constants in a way that they shift
the r.h.s. of (\ref{bound.basic.int}) in agreement with
the l.h.s., i.e. the value of the basic integral on the horizon.
If this assumption is correct, the \emph{Hawking flux is independent
of the boundary conditions imposed at the horizon}.
A crucial role is certainly played by the IR renormalisation
that in my approach enters as a boundary condition at infinity.

I have already observed that (\ref{expl.E(Th)}) and thus all expectation
values of the EM tensor have the same dependence on the homogeneous
solutions as on the renormalisation constant $c_R$.
The remaining ambiguity of the EM tensor (if the quantum state
has already been fixed) can therefore be put into this single
constant.
\\
\\
\textbf{Remark:} Until now I have only considered the flat Green functions
derived directly in Minkowski spacetime to evaluate the basic
integrals via the perturbation theory.
Because the Euclidean Feynman Green function (\ref{ren.Feynman})
is often employed in two-dimensional problems, it is
interesting whether it reproduces the obtained results 
of the perturbational analysis. Clearly it must fulfil
the required boundary conditions. The Euclidean Green
function on the half-plane (\ref{Eukl.Feynman.Green.mod}) vanishes
at the horizon, i.e. for $r=2M$ or $r'=2M$.
The IR renormalisation is no more carried out by dropping
boundary terms, but by using the regularised version
of the Euclidean Feynman Green function (\ref{reg.Feynman}).
On the half-plane again the mirrored term will be added.
I assume that the series converges and that the first
order term gives the leading order in $r$ for $r\to\infty$.
The basic integral with the scalar curvature $R$ is then
approximated by
\begin{multline}
\lim_{\lambda\to0}\frac{d}{d\lambda}\biggm\{\lambda\frac{2M\,i}{\pi}
\frac{\Gamma\left(-\frac{\lambda}{2}\right)}
{\Gamma\left(1+\frac{\lambda}{2}\right)2^{1+\lambda}}\\
\int_L\frac{[(t-t')^2-(r-r')^2]^{\frac{\lambda}{2}}
-[(t-t')^2-(r+r'-4M)^2]^{\frac{\lambda}{2}}}{(r')^3}d^2x'\biggm\}.
\end{multline}
The integration over $t'$ gives
\begin{multline}
\int_{-\infty}^{\infty}\left\{[(t-t')^2-(r-r')^2]^{\frac{\lambda}{2}}
-[(t-t')^2-(r+r'-4M)^2]^{\frac{\lambda}{2}}\right\}dt'\\
=\sqrt{\pi}(-1)^{\frac{\lambda+1}{2}}\frac{\Gamma\left(-\frac{\lambda+1}{2}
\right)}{\Gamma\left(-\frac{\lambda}{2}\right)}
\left\{\left[(r-r')^2\right]^{\frac{\lambda+1}{2}}
-\left[(r+r'-4M)^2\right]^{\frac{\lambda+1}{2}}\right\}.
\end{multline}
As I am interested in the asymptotic behaviour ($r\to\infty$) of the basic
integral I can facilitate the integration over $r'$ by assuming
that $r>r'$ and therefore $\left[(r-r')^2\right]^{\frac{\lambda+1}{2}}
=(r-r')^{\lambda+1}$ and $\left[(r+r'-4M)^2\right]^{\frac{\lambda+1}{2}}
=(r+r'-4M)^{\lambda+1}$. The integral over $r'$ from $2M$ to $\infty$
can then be performed analytically and yields a hypergeometric function:
\begin{multline}
\int_{2M}^{\infty}\frac{(r-r')^{\lambda+1}-(r+r'-4M)^{\lambda+1}}{(r')^3}dr'\\
=\frac{(2M)^{\lambda-1}}{\lambda-1}\biggm\{(-1)^{\lambda}
F\left[-1-\lambda,1-\lambda,2-\lambda,\frac{r}{2M}\right]\\
+F\left[-1-\lambda,1-\lambda,2-\lambda,2-\frac{r}{2M}\right]\biggm\}.
\end{multline}
Now it is a simple task to carry out the renormalisation
procedure to obtain the first order of the basic integral:
\begin{multline}
\int_LG_F^{mod}(x,x')R(x')\sqrt{-g'}d^2x'\stackrel{r\to\infty}{\to}
\lim_{\lambda\to0}\frac{d}{d\lambda}\biggm\{\frac{\lambda}{\lambda-1}
\frac{i\,M^{\lambda}(-1)^{\frac{\lambda+1}{2}}}{2\sqrt{\pi}}
\frac{\Gamma\left(-\frac{\lambda+1}{2}\right)}
{\Gamma\left(1+\frac{\lambda}{2}\right)}\\
\biggm((-1)^{\lambda}
F\left[-1-\lambda,1-\lambda,2-\lambda,\frac{r}{2M}\right]
+F\left[-1-\lambda,1-\lambda,2-\lambda,2-\frac{r}{2M}\right]\biggm)\biggm\}\\
=\frac{r}{2M}-1.
\end{multline}
This result is obviously not in agreement with the one obtained by the
original Feynman Green function. By (\ref{int.basic},\ref{chi})
and $R=-2\square\rho$ I see that the homogeneous solution associated
to $\rho$ must be
\begin{equation}
\chi_{\rho}=\frac{1}{2}\left(\frac{r_{\ast}}{2M}-1-\ln0\right)
,\label{wrong-hom.sol.}
\end{equation}
i.e. $C_1^{\rho}=-\frac{1}{2}-\ln0\,,\,C_2^{\rho}=\frac{1}{2}$,
and the exact basic integral is
\begin{multline}
\int_LG_F^{mod}(x,x')\square'\rho(x')\sqrt{-g'}d^2x'=-(\rho+\chi_{\rho})\\
=\frac{1}{2}+\ln0-\frac{r}{4M}-\frac{1}{2}\ln\left[\biggm(1-\frac{2M}{r}\biggm)
\biggm(\frac{r}{2M}-1\biggm)\right].
\end{multline}
If the perturbation theory works correctly with the Euclidean
Feynman Green function it should produce the asymptotic expansion
of the logarithmic term at the higher orders.
Anyway, this result is in conflict with the one obtained by the
original Green functions and also with the considerations on the
boundary conditions in Section 5.3. There I have shown that the
homogeneous solution only can be a constant if the boundary
terms in (\ref{bound.basic.int}) vanish through the elimination
of IR divergences.

That the boundary terms vanish, in particular those at infinity,
was one of my basic assumptions, inspired by physical arguments.
If this is correct, the usage of the Euclidean Feynman Green
function leads to an inconsistency, namely the r.h.s. of
(\ref{bound.basic.int}) is still $-f$, while the
l.h.s. produces an additional homogeneous solution like
$\chi_{\rho}$ above.

\subsubsection{Hawking Flux and Energy Density}

With the computation of the components $\E{T}_2$ and
$\E{\Th}_2$ most of the work is done to determine
the vacuum expectation value of the complete EM tensor.
It remains to calculate the constant $K$ of the
CF approach for the Unruh state $\left|U\right>$
which has been identified with the vacuum state of the model
(see Section 2.4.1). Remember that the other constant $Q$
was set to zero for all physical states. In the Unruh state
we have the relation $K_U=\frac{M^2f(\infty)}{2}$ (see Table 2
at the end of Section 2.4.1) where $f(r)$ is the
state-independent function (\ref{f}):
\begin{equation}
K_U=\frac{M^2}{2}\int_{2M}^{\infty}\left[\frac{\E{T}_2M}{(r')^2}
+2(r'-2M)\E{\Th}_2\right]dr'=-\frac{1}{768\pi}.
\end{equation}
By the CF equations (\ref{2dsol1},\ref{2dsol2}) the total
flux through a spherical shell surrounding the BH is then given by
\cite{kuv99}
\begin{equation}
\text{Flux}_{tot}=\U{T^r{}_t}_2=-\frac{K}{M^2}=\frac{1}{768\pi M^2},
\end{equation}
and the measurable four-dimensional flux is (see (\ref{EM-tensor.4d-2d}))
\begin{equation}
\U{T^r{}_t}=\frac{1}{3072\pi^2M^2r^2}.\label{Hawking.flux.massless}
\end{equation}
Hence it is by a factor $40$ larger than the total flux calculated
by the Black Body hypothesis (\ref{flux.massless.blackbody})
which for a ``minimal'' effective area $A=16\pi M^2$ is the widely
accepted result. It might well be
that a quantum calculation directly in four dimensions could reproduce
the correct flux for the Black Body law; in this case the deviation between
(\ref{Hawking.flux.massless}) and (\ref{flux.massless.blackbody})
would indicate the failure of the dilaton model at the quantum
level to describe a spherically symmetric four-dimensional theory.
On the other hand, it is questionable if the Black Body hypothesis
alone is sufficient to determine the Hawking flux as it is
based on a semi-classical\footnote{Not to be confused with
semi-classical Quantum Gravity, where matter is described by
a QFT, while the geometry remains classical.}
calculation and an assumption concerning the effective area;
the cross section of particles without self-interaction on a
Schwarzschild background involves only free propagators and no loop-graphs.
In contrast to
that the current approach includes the full quantum theory of free
scalar particles which is provided by the one-loop order.
Such a calculation clearly describes the processes near
the horizon that occur in QFT. By this reasoning (\ref{Hawking.flux.massless})
could instead be closer to the correct value of the Hawking flux.
\enlargethispage{0.5cm}

Although the results obtained here may not
be the final answer, they reveal some nice qualitative
features which are in agreement with the expected properties
of BH radiation. First of all, the flux as well as
the energy density (see below) do not violate the weak energy
condition in the asymptotic Minkowski region of spacetime.
I emphasize this seemingly trivial point, because there has
been lots of confusion during the last years in the literature,
where exactly this happened by wrong calculations and it was
interpreted as a principle failure of the dilaton model.
Interestingly, (\ref{Hawking.flux.massless}) is identical
to the result for the intrinsic two-dimensional model
\cite{kuv99}.

Finally, I present the vacuum expectation values of the
remaining components of the EM tensor in the Schwarzschild gauge.
The energy density is given by
\begin{multline}
\E{T_{tt}}_2=\frac{1}{768\pi M^2}+\frac{7}{24\pi r^2}
-\frac{7 M}{6\pi r^3}+\frac{9 M^2}{8\pi r^4}\\
+\frac{\Sss^2\left[\ln\frac{r-2M}{r^3}+2(\chi_{\rho}+\chi_{\phi})-c_R
\right]}{8\pi r^2}\label{energy-density-dil.}
\end{multline}
and the radiational stress can be calculated by $\E{T_{rr}}_2=
g_{rr}(\E{T}_2-\E{T^t{}_t}_2)$.
Again, the four-dimensional components can be reconstructed
by (\ref{EM-tensor.4d-2d}). Note that the logarithmic term
in the energy density brings in a further mass-dependence
as compared to the case of massive particles (\ref{asymp.energy.dens.massive}).
In particular, this means that the point where the energy
density changes sign now depends on the BH mass. Figures 12,13,14
show the energy density for two BHs
whose mass is one time much larger than the Planck mass
and the other time much smaller. For a very small BH
$M=10^{-20}$ the point of sign-reversal lies exactly on the horizon.

\begin{figure}[h]
\hspace{2cm}\epsfig{file=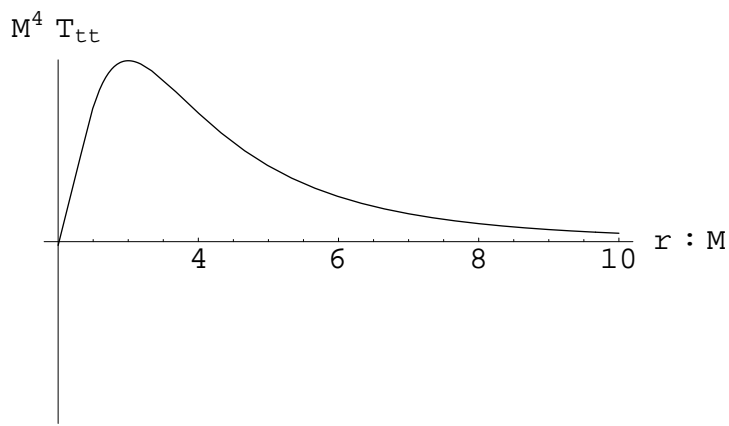,height=6cm}
\caption[fig1]{Small Black Hole, $M=10^{-20}$}
\end{figure}

For huge BHs $M=10^{40}$ this point is shifted far away
from the horizon (Figure 13).

\begin{figure}[h]
\hspace{2cm}\epsfig{file=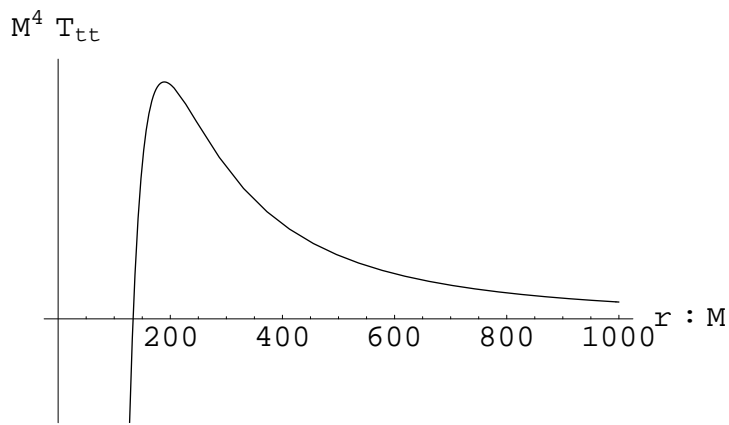,height=6cm}
\caption[fig1]{Large Black Hole, $M=10^{40}$}
\end{figure}

Note that for masses smaller than the Planck mass $M<1$
the results loose their significance because the BH
is then in a rapidly evolving state where backreaction
effects play an important role. With respect to this,
Figure 12 cannot be taken too seriously (although
it shows that the zone of negative energy density
decreases with the BH mass).
Figure 14 shows the energy density for a BH of the same
size but close to the horizon.

\begin{figure}[h]
\hspace{2cm}\epsfig{file=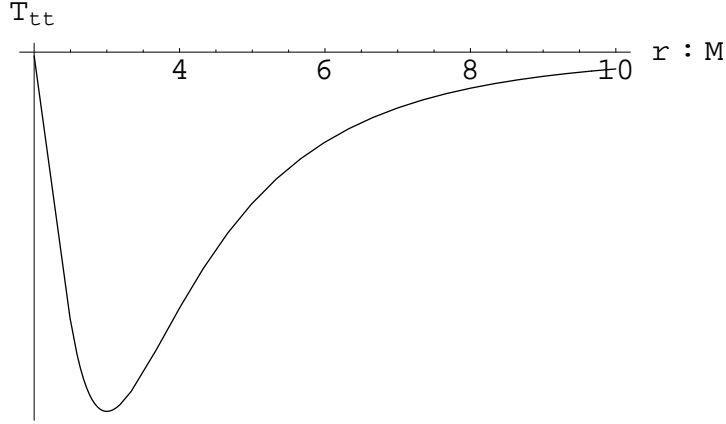,height=6cm}
\caption[fig1]{Large Black Hole, $M=10^{40}$}
\end{figure}

Surprisingly, the event horizon is surrounded by a large zone
of negative energy density, caused by the virtual particles
that are swallowed by the BH and thereby decrease its mass.
Such an effect already has been observed for massive particles, see Figure 6.
There I argued that for fixed BH mass this zone decreases with
the rest mass of the emitted particles.
This idea is in agreement with the current results, because
it states that particles with higher kinetic energy can be produced
closer to the horizon, as they have a higher probability to
leave the BH. As the Hawking temperature, and hence
the energy of massless scalar particles, becomes higher with
decreasing BH mass, the point of production
accordingly approaches the horizon.
\\
\\
\textbf{Remark:}
It is now a trivial task to calculate the components of
the EM tensor in the Hartle-Hawking state $\left|H\right>$. The total
flux is just zero, as $K$ is set to zero, and the asymptotic
energy density is by a factor $2$ larger than in the
$\left|U\right>$-state. All higher orders in $r$ of the energy density
are identical in the two states.
\\
\\
\textbf{Further Remark:} In the last Section I have shown that the homogeneous
solutions $\chi_{\rho},\chi_{\phi}$ are constants
and therefore do not affect the Hawking flux.
If, for some reason, one uses an r-dependent homogeneous
solution (like e.g. the one obtained for the Euclidean Feynman Green
function (\ref{wrong-hom.sol.})) the flux and
other components of the EM tensor are changed.
For instance, I will assume that
\begin{equation}
\chi_{\rho}^{new}=\frac{1}{2}\left[\frac{r}{2M}
+\ln\left(\frac{r}{2M}-1\right)-1-\ln0\right],
\end{equation}
and $\chi_{\phi}=\ln2M$ as before. This means that the
first basic integral would give the result
\begin{equation}
\int_{L}G(x,x')R(r')d^2x'=\frac{r}{2M}+\ln\left[\biggm(\frac{r}{2M}-1\biggm)
\biggm(1-\frac{2M}{r}\biggm)\right]-1-2\ln0.
\end{equation}
This choice is compatible with the boundary condition
for the Green function at the horizon (the basic integral vanishes
for $r=2M$), but not with the ones at infinity.
Namely, as discussed in Section 5.3 and Section 5.6.2, if the Green
functions vanish asymptotically (or merely are set to zero by an IR
renormalisation) the homogeneous solutions are
\emph{constants} that are unambiguously fixed by the boundary
condition at the horizon. If for some reason they are nontrivial
functions in $r$ (e.g. by using the Euclidean Feynman Green
function) one must reconsider the problem of IR renormalisation
to compute the correct boundary terms that restore
consistency in (\ref{bound.basic.int}).

In the $\left|U\right>$-state the flux-determining constant $K$
acquires an additional contribution, such that
\begin{equation}
K_U^{new}=\text{old}+M^2\int_{2M}^{\infty}\frac{(r'-2M)(r'-4M)}{4\pi(r')^5}
\chi_{\rho}^{new}dr'=\text{old}+\frac{1}{128\pi}=\frac{5}{768\pi}.
\label{BalbFabb}
\end{equation}
As a consequence, the flux as well as the asymptotic energy density
become negative which means that \emph{the weak energy condition is
violated} in the flat (asymptotic) region!
From the physical point of view this result is nonsense and
demonstrates the importance of a consistent IR renormalisation.
The same result (\ref{BalbFabb}) was obtained by Balbinot-Fabbri
\cite{baf99}, Equation 16.

\subsection{Quantum States and the Effective Action}

Concerning the relation between the effective action and
the choice of quantum state, there have been two basic
assumptions which are the very basis of the computations
in my thesis:
\begin{itemize}
\item First of all, I have assumed that \emph{the
effective action does not produce quantum mechanical
expectation values in the $\left|U\right>$-state}
(which is the vacuum state of the system), hence the
components of the EM tensor cannot be derived directly
from the effective action.
\item Second, I have assumed (and proven by heuristic arguments
in Section 2.4.2) that the basic
components of the CF approach, $\E{T}$
and $\E{\Th}$, are \emph{independent} of the quantum
state and therefore nevertheless \emph{can} be derived
directly from the effective action.
\end{itemize}
In the present Section I will reconsider these assumptions
by explicitly calculating the remaining components
of the EM tensor (apart from the basic ones) by variation
of the effective action (\ref{dilaton.eff.action}):
\begin{multline}
\E{T_{\alpha\beta}}_2=\frac{2}{\sqrt{-g}}\frac{\delta W}{\delta g^{\mu\nu}}\\
=\frac{1}{12\pi}\biggm\{\Bigm[\p_{\alpha}\phi\p_{\beta}\phi
-\frac{g_{\alpha\beta}}{2}(\p\phi)^2\Bigm]\Bigm[5+6(\phi+\chi_{\phi}
+\rho+\chi_{\rho})-3c_R\Bigm]\\
-\Bigm[\p_{\alpha}\rho\p_{\beta}\rho-\frac{g_{\alpha\beta}}{2}
(\p\rho)^2\Bigm]+4\Bigm[\p_{\alpha}\phi\p_{\beta}\rho-\frac{g_{\alpha\beta}}{2}
(\p_{\gamma}\phi)(\p^{\gamma}\rho)\Bigm]\biggm\}
+\frac{2}{\sqrt{-g}}\frac{\delta W}{\delta\rho}\frac{d\rho}
{d g^{\alpha\beta}}.\label{EMtensor.dilaton}
\end{multline}
To compute the last term I need the definition of the conformal
metric, see Appendix A.3. Its determinant
is related to $\rho$ by $g=\det(g_{\bar{\alpha}\bar{\beta}})
=-\frac{e^{4\rho}}{4}$ or $\sqrt{-g}=\frac{e^{2\rho}}{2}$,
hence $\frac{d\sqrt{-g}}{d\rho}=2\sqrt{-g}$.
Using further the relation $\frac{d\sqrt{-g}}{d g^{\alpha\beta}}
=-\frac{\sqrt{-g}}{2}\cdot g_{\alpha\beta}$
I can calculate this term to
\begin{equation}
\frac{2}{\sqrt{-g}}\frac{\delta W}{\delta\rho}\frac{d\rho}
{d g^{\alpha\beta}}=\frac{g_{\alpha\beta}}{24\pi r^2}.
\end{equation}
If contracted by the metric it gives the correct trace anomaly,
all other terms in (\ref{EMtensor.dilaton}) are traceless.

The first thing I observe is that the flux component of the
EM tensor now indeed vanishes:
\begin{equation}
\E{T^r{}_t}_2=0\Bigm(=\HH{T^r{}_t}_2=\B{T^r{}_t}_2\Bigm).
\end{equation}
This is a direct consequence of the static approximation
because the Schwarzschild metric has no off-diagonal entries $g_{rt}=0$.
This proves the assumption that the expectation values derived
from the effective action are not the ones in the $\left|U\right>$-state
because there we have a finite flux.
Next I consider the energy density which becomes
\begin{multline}
\E{T_{tt}}_2=\frac{1}{4\pi r^2}-\frac{13M}{12\pi r^3}
+\frac{9M^2}{8\pi r^4}\\
+\frac{\Sss^2\left[\ln\frac{r-2M}{r^3}
+2(\chi_{\rho}+\chi_{\phi})-c_R\right]}{8\pi r^2}.\label{energy-dens.eff.action}
\end{multline}
If I compare this result with the energy density in the
Unruh state (\ref{energy-density-dil.}) I see that it differs
from it by $\frac{1}{768\pi M^2}+\frac{1}{24\pi r^2}\Sss$.
The leading order (which is constant in $2d$) corresponds to the
occupied real particle states. Thus, the state of the effective action
can be identified with the Boulware state $\left|B\right>$ (see Section 2.4.1)
as it is the only state which contains no radiative components
of the EM tensor, i.e. the spacetime is asymptotically vacuum
(in the Hartle-Hawking state $\left|H\right>$ the asymptotic energy
density is twice the one in the Unruh state).
In the CF approach the Boulware state is reached by fixing the
constants $K_B=0$ as $Q_B=\frac{1}{384\pi}$ (remember that
the choice $Q\neq0$ produces a quadratic divergence of the EM
tensor on the horizon in global coordinates, hence $\St{B}$ has
no physical significance); note that the energy density calculated
a la CF still differs from (\ref{energy-dens.eff.action})
by a term $\frac{1}{24\pi r^2}\Sss$!

This result can be understood
by reconsidering the static approximation. If any
radiation components are present (let's say in the stationary
$\left|H\right>$-state) one must observe inevitably occupied
states in the asymptotic region and a non-zero spacetime
curvature at some time-slice. If the spacetime is assumed
to be static this asymptotic spacetime curvature is always
present. Hence, the asymptotically flat Schwarzschild geometry
can only describe correctly a quantum state in the static
approximation with no occupied particle states in the asymptotic region,
i.e. $\left|B\right>$. If one wanted to construct an effective
action that yields expectation values e.g. in the $\left|H\right>$-state
one had to construct a complicated geometry capable to
describe the asymptotic region. Clearly, this problem is created
by the static approximation as it continues the
actual spacetime geometry into the remote past and future.
In the quasi-static phase a BH spacetime (or better space)
is indeed best described by the Schwarzschild geometry, but
by infinitely extending it one must either take into account
the occupied asymptotic states or creates the divergence on the horizon
of the Boulware state.

This result is not at all disappointing as it has been proven
by Wald \cite{wal77,wal78} that the expectation value of the EM tensor
is unique up to a generally conserved expression (in the dilaton
model this expression fulfils the non-conservation equation).
The CF approach shows explicitly
how one can switch between representations of the EM
tensor (corresponding to different quantum states) by simply adding a
covariantly conserved expression of the form
\begin{equation}
\Delta\E{T^{\mu}{}_{\nu}}=\large\frac{1}{M^2}\left(\begin{array}{cc}
\frac{\Delta K_2-\Delta Q_2}{\Sss} & \frac{\Delta K_2}{\Sss^2}\\
-\Delta K_2 & \frac{\Delta Q_2-\Delta K_2}{\Sss}
\end{array}\right).\normalsize\label{EM-tensor,diff}
\end{equation}
By these considerations I can summarise and answer the problem
of the effective action and its expectation values by some unexpected but
nice statements:
\begin{itemize}
\item \emph{The effective action yields only expectation values
in the} unphysical, but the conservation equation
satisfying, \emph{Boulware state}.
\item \emph{The Unruh state} (which is the vacuum state of the
model) as well as the Hartle-Hawking state, leading both
to the same outgoing Hawking flux, \emph{can be
adjusted by} adding an expression (\ref{EM-tensor,diff})
with the appropriate choice of \emph{the constants $\Delta K_2,\Delta Q_2$}.
\item The desired \emph{quantum state cannot be fixed} directly
in the effective action \emph{by the choice of Green function},
as supposed, neither by the fixation of the renormalisation
ambiguity. The \emph{quantum state} of the effective action
\emph{originates from the geometry} of the static approximation.
\end{itemize}
These points have often caused confusion in the literature.
It has been stated \cite{baf99} that the logarithmic divergence of the
$\Th$-component (\ref{expl.E(Th)}) shows that it corresponds
to the Boulware state, and it was claimed that the same component
in the Unruh state would be finite at the horizon.

In Section 2.4 I have demonstrated how a \emph{quadratic}
divergence in the radiation components of the EM tensor
(not in $\Th$) can be removed by adjusting the constant $Q$.
Any other divergences that enter by the basic functions
cannot be removed and are thus state-independent.
Further, I have shown in the current Chapter that the
logarithmic divergence found in $\Th$ is a direct consequence
of the IR renormalisation that fixes the homogeneous
solutions in the basic integrals (\ref{bound.basic.int}).
Because of these arguments I have very good reasons to contradict
the authors of \cite{baf99}.

\newpage
\section{Conclusions}

The aim of my thesis was to calculate quantum mechanical
expectation values of a scalar field on a Schwarzschild
spacetime whose mass $M$ was assumed to be sufficiently large so that
the spacetime could be considered as static.
I could show, developing further previous work, that this can
be done for massive particles directly in $4d$, and
for massless particles in the two-dimensional dilaton model.
In the following, I will sketch the considerations
and calculations which led to these results in a chronological
order.

One of the basic tools of this work was the CF
method \cite{chf77}. It allows to calculate the complete EM tensor
on a Schwarzschild spacetime from two ``basic components'' $T,\Th$ and
two integration constants $K,Q$ by the conservation equation
(\ref{cons.equation}). The EM tensor is thereby
restricted by the spherical symmetry conditions which
means singling out the s-modes of the scalar field, and
thus has the form (\ref{s-wave-1}) at the classical and quantum
level. Because I applied the method to expectation values
I had to show that the conservation equation still holds for
the renormalised expectation value of the EM tensor (\ref{EM.cons.quant}).
Further, I could verify that also the non-conservation
equation of the dilaton model survives the quantisation
procedure (\ref{EM.cons.quant.dil}).

The problem of the boundary conditions and the quantum
state appeared throughout this thesis. The first observation
was that the classical solution of the scalar field $S_0$,
which is always added to the quantum fluctuations and
enters the path integral, is most conveniently set to zero
(see Section 1.2.1). In particular, this choice was in agreement
with the representation of the effective action as a functional
determinant (\ref{eff.action2}) which demanded the vanishing of
a boundary term. In Section 2.4 I considered three different
quantum states that correspond to different choices of the
constants $K,Q$ (see Table 2 at the end of Section 2.4.1).
Following \cite{chf77} I could verify that
the condition $Q=0$ guarantees the regularity of the EM tensor
in global coordinates and hence singles out the physical
states. The remaining constant $K$ was shown to regulate the incoming
flux and thus the number of occupied states at past null infinity
${\cal I}^-$ (Figure 1). The outgoing flux turned out to be independent
of $K$. I identified the Unruh state $\left|U\right>$
with the vacuum state of the theory as it led to the minimal
energy density of the physical states (see Figure 4).
Further, I considered the Boulware state $\left|B\right>$:
because of $Q_B\neq0$
it does not belong to the physical states. However, as
it is characterised by vanishing asymptotic fluxes I
could identify it with the state of the effective action.
In Section 5.7 I demonstrated by direct calculation that
the effective action indeed leads to expectation values
in the $\left|B\right>$-state. I interpreted this result
in the following way: the effective action yields expectation
values in the form of geometric objects, such as the scalar curvature.
In the static approximation the asymptotic state must be the
vacuum state, i.e. there cannot be occupied particle states in the asymptotic
region. This is only fulfilled by the $\left|B\right>$-state
which is therefore the only state in accord with the
static approximation. Nevertheless, it was still possible to compute
expectation values in arbitrary quantum states. Namely,
I used the effective action only for the computation of the
basic components, whereby the crucial point was to show the state-independence
of the latter in Section 2.4.2. The physical argument was that
the basic components did not correspond to radiational degrees
of freedom. The remaining components
were calculated a la CF anyway, so the
state was finally fixed by the constants $K,Q$ alone.

The next task was to establish the effective action.
In a first step I used the zeta-function regularisation
(\ref{zeta-function},\ref{eff.action4}) to express the 
effective action in terms of the heat kernel
(\ref{zeta-function2},\ref{heatkernel}).
Then I distinguished between massive and massless scalar
fields: in the massive case I could find a local form
of the heat kernel by the Seeley-DeWitt expansion (\ref{heat.exp.})
whose convergence was guaranteed by the mass term (for sufficiently
large particle mass $m$).
In Section 3.1.3 I examined the general conditions for the
convergence of the local expansion in the presence of a damping
term in the classical action. 
In the massless case I derived a non-local effective action
(\ref{nonlocal.effective.action}) by the covariant perturbation
theory invented by Barvinsky and Vilkovisky \cite{bav87,bav90}.

The treatment of massive particles was straightforward:
I calculated the basic components by the effective action
and used the CF approach to obtain
the remaining components like the energy density and the flux.
I renormalised the effective action by substracting the
flat spacetime values of the expectation values and by
setting the remaining renormalisation constant $c_{ren}$ to zero so that
the expectation value of the EM tensor vanishes (in the Unruh state)
if the particle mass $m$ goes to infinity. 
For fixed $m$ the obtained energy density revealed the expected
behaviour (Figure 6): near the horizon it is negative by the presence
of virtual particles with negative flux that fall into the BH.
Then it becomes positive for $r\approx 2.75 M$ and falls
off like $r^{-2}$ in the asymptotic region (\ref{asymp.energy.dens.massive})
(i.e. a spherical shell has constant energy, corresponding to the
occupied states). The result for the local flux was (\ref{flux.massive}).
Near the critical point of the Seeley-DeWitt expansion $\alpha=(mM)^{-1}\approx1$
the agreement with the Black Body law for massive particles
(\ref{Flux-mass.tot}) and the result for massless particles
(\ref{Hawking.flux.massless}) (which should be some orders beyond) 
was reasonable (though at this point the perturbation series
actually diverges). However, in the convergence region $\alpha\ll1$
the quantum calculation could not reproduce the expected exponential
damping as illustrated in Table 6. The reason was that the
convergence condition $\alpha\ll1$ of the perturbation series
automatically restricted the scope of the analysis
to the region where the exponential damping already had set in
and thereby practically eliminated the contribution of the
massive particles. For a correct description in the interesting
range $\alpha\approx1$ the usage of a non-local effective
action therefore seems to be unavoidable (see in the Outlook).
From the phenomenological point of view such an investigation
could be of interest as there is sufficient freedom
to adjust the BH mass $1\ll M<10^{20}$ such that the BH
is still in the quasi-static phase and the particle
mass is within the range of the fundamental particles $m\approx M^{-1}$.

The examinations of massless particles in the two-dimensional
dilaton model has been the main concern of my thesis.
First of all, I could establish a non-local effective action
by the covariant perturbation theory including a non-trivial
endomorphism. This was a nice
result, even more as Barvinsky and Vilkovisky \cite{bav87,bav90}
themselves claimed that the case of a conformally coupled
scalar field would represent the only possible application
of their formalism in two dimensions!
Furthermore, the effective action up to now had only been
derived by integration of expectation values \cite{kuv99}, such as the
trace anomaly, and not directly from the Euclidean path integral.
Therefore, the construction of the effective action
(\ref{dilaton.eff.action}) by the method of Barvinsky and Vilkovisky
can be seen as a missing link in the current literature
which I have supplemented in my thesis.
Another point which was widely discussed have been the boundary
conditions and their connection to the quantum state.
My first observation was that the quantum state of the effective
action is fixed by the homogeneous solution that emerges
through the non-local expression $\square^{-1}$.
Because of the integrability of two-dimensional gravity
the explicit knowledge of the Green functions of the
two-dimensional Schwarzschild
Laplacian was not necessary to compute the effective action.
Therefore, the behaviour of the Green functions on the
boundary, including the event horizon $r=2M$, was sufficient
to determine uniquely the quantum state of the effective action
which could then be given in a local form.
In Section 5.3 I argued that the IR renormalisation
suggests a vanishing of the Green functions at the boundary
which fixes the homogeneous solutions (as constants)
by the prescription $\square^{-1}\square=1$ (\ref{bound.basic.int}).

Beyond that, I investigated the two-dimensional Green functions
by a perturbation series (\ref{pert.Green}). I could show
that the series converges if
\begin{itemize}
\item The analysis is restricted to the half-plane $2M<r<\infty$
(which is sensible as the flux is produced outside the horizon).
\item The boundary conditions are imposed according to the
IR renormalisation.
\item The Green functions are applied to basic integrals (\ref{int.basic}).
\end{itemize}
As starting point of the perturbation series I chose the flat
retarded (\ref{retarded.Green}) and Feynman (\ref{Feynman.Green})
Green function on the half-plane which I calculated by Fourier
transformation without Euclideanizing spacetime.
The explicit evaluations with the Green functions 
proved consistency with the considerations on the boundary terms
in Section 5.3, whereby the boundary condition at the horizon
fixed the constant homogeneous solutions. The latter could be
shown to leave the Hawking flux invariant (\ref{flux-ind.}).

The result for the Hawking flux agreed with the one already obtained
in \cite{kuv99}. The progresses achieved in my thesis were
on the one hand the \emph{direct derivation of the effective action 
from the heat kernel} and on the other hand the \emph{explicit
examination of boundary terms and their effect on the
expectation values} of the EM tensor.
The energy density (\ref{energy-density-dil.}) which
I calculated a la CF (like the flux)
revealed a similar behaviour like the one in the massive case.
It is characterised by the existence of a zone of negative
energy density which in the massless case turned out to
be scale-dependent (Figures 12,13,14).

Finally, I concluded that the quantum state of the effective
action is indeed the Boulware state as it produces an EM
tensor that in the CF approach corresponds to the
choice $Q=K=0$. Nevertheless, all quantum states of the
expectation values are accessible by adding a covariantly
conserved EM tensor of the type (\ref{EM-tensor,diff}),
the basis of this method being the state-independence of
the basic components $\E{T},\E{\Th}$.

\newpage
\section{Outlook}

The examinations of my thesis can be continued
and supplemented into various directions.
I give a list of some of them, whereby I quote those
first which are a direct extension of this work.

The investigations of the massive particles by
the local Seeley-DeWitt expansion of the heat kernel
in four dimensions have shown that the interesting
range of the parameters in the static
approximation $M>1,m\approx M^{-1}$ lies beyond the
scope of the local expansion.
This suggests to consider massive particles in the
two-dimensional dilaton model, where one can work
easily with the non-local effective action of
the covariant perturbation theory (\ref{nonlocal.effective.action}).
The mass term modifies
the two-dimensional Laplacian by a term $X\cdot m^2$,
$X$ being the dilaton field. Now one can either keep
this term as an endomorphism $E$ in the Laplacian and work
with an effective action of the type (\ref{non-local.eff.action}).
As the corresponding part of the endomorphism is now
proportional to $m^2$, the expectation values might now
have the form of a power series in the mass that sums
up to an exponential function (the local expansion led
to an inverse series which was not interpretable as a
power series). Alternatively the mass term can be separated
from the Laplacian at the level of the heat kernel by
the use of some Baker-Campell-Hausdorff-like formula.
Because of the spacetime dependence of the dilaton field $X(x)$
one has then contributions from commutator terms like $[X,\triangle]$.
By this procedure the exponential damping possibly
might be obtained more directly.

Another, less physical, application of the non-local
effective action (\ref{non-local.eff.action}) is the investigation
of general dilaton models which have not been produced by spherical
reduction. As it includes arbitrary couplings of the dilaton
field to gravity, the covariant perturbation theory allows
the derivation of an effective action for all known dilaton
models. In particular one could consider spherically
reduced scalar-tensor theories, where the scalar field
is non-minimally coupled to the scalar curvature.

The current approach does not seem to be appropriate to
examine the backreaction of the quantum field on the spacetime.
In principle, the first order in $\hbar$ of the metric can be
calculated by the differential equations (\ref{backreaction}). 
However, the next order of the EM tensor cannot be obtained
by the CF method as the latter is based on the static
Schwarzschild metric. A more promising approach would be
to integrate out the geometric variables in the path integral
as in \cite{klv97a}. In $2d$, as there is no dynamical degree
of freedom of gravity, this might provide a method to treat
the classical geometry non-perturbatively which seems to
be unavoidable for the investigation of the final phase
of a BH.

The methods I used in Chapter 5 to examine massless scalar fields in the
two-dimensional dilaton model could be applied to some extent
in the four-dimensional theory. A non-local effective action
can be established straightforwardly in the same manner
as in Section 5.2. To extract expectation values
from this effective action will be more involved because
there is no conformal gauge in $4d$. Probably, one can
employ a Green function perturbation theory, similar
to (\ref{pert.Green}), to evaluate the basic integrals
of the four-dimensional effective action in the static
approximation. This may be sufficient to obtain the
expectation values of at least the basic components
of the EM tensor. Such calculations are necessary
to verify the reliability of the dilaton model at the quantum
level and to determine a possible spherical reduction anomaly.

\begin{appendix}

\newpage
\section{Conventions and Notations}

\subsection{Signs}

I use the metric sign convention $(+,-,-,-)$. In gravity theory
often the inverse sign convention $(-,+,+,+)$ is used because spacelike
distances are supposed to be measured by positive numbers.
Further sign differences
concern the EM tensor and the definition of geometric objects,
such as the Riemann tensor. I compare my convention with the
one from Wald which is the most common in the literature.
I mark the quantities by an index $I$, respectively $W$.
The metrics are related by
\begin{equation}
g_{\mu\nu}^I=-g_{\mu\nu}^W.
\end{equation}
The Ricci tensor is therefore invariant but the
scalar curvature transforms by a sign
\begin{equation}
R_{\mu\nu}^I=R_{\mu\nu}^W\,\,,\,\,R_I=-R_W.
\end{equation}
Because the Ricci tensor in both conventions is related to
the Riemann tensor by $R_{\mu\nu}^{I,W}=g_{I,W}^{\kappa\lambda}
R^{I,W}_{\kappa\mu\lambda\nu}$ the latter is also invariant
\begin{equation}
(R^{\mu}{}_{\nu\kappa\lambda})_I=(R^{\mu}{}_{\nu\kappa\lambda})_W.
\end{equation}
The Lagrangian and the EM tensor are defined as
\begin{eqnarray}
L_I=R_I-\frac{g_I^{\mu\nu}\p_{\mu}S\p_{\nu}S}{2}-\frac{m^2S^2}{2}&,&
T_{\mu\nu}^I=\frac{2}{\sqrt{-g_I}}\frac{\delta S^I_m}{\delta g^{\mu\nu}_I}\\
L_W=R_W-\frac{g_W^{\mu\nu}\p_{\mu}S\p_{\nu}S}{2}-\frac{m^2S^2}{2}&,&
T_{\mu\nu}^W=-\frac{2}{\sqrt{-g_W}}\frac{\delta S^W_m}{\delta g^{\mu\nu}_W}.
\end{eqnarray}
This entails that the EM tensor is invariant
\begin{equation}
T_{\mu\nu}^I=T_{\mu\nu}^W
\end{equation}
which is desired to guarantee the positivity of the energy density
$T_{tt}$ in both conventions.

\subsection{Indices}

Sometimes it will be necessary to distinguish between the tensor
indices associated to different coordinate systems. As I work
in spacetimes with different dimensions it is also useful
to denote indices belonging to different (sub)manifolds
by different symbols. If confusion cannot occur I will use
arbitrary indices.

I use Greek letters to denote tensor indices when I work
in a coordinate basis:
\begin{equation}
T=T^{\mu}{}_{\nu\kappa}\cdot\p_{\mu}\otimes dx^{\nu}\otimes dx^{\kappa}.
\end{equation}
In a vielbein basis (see Appendix A.4 and Appendix B.3 for Schwarzschild)
indices are denoted by Latin letters $A_me^m$.

Indices associated to a light-cone coordinate system
are marked by a bar on top, e.g. $T_{\bar{\mu}\bar{\nu}},e^{\bar{m}}$.
In Eddington-Finkelstein coordinates I use primed indices
$A^{\mu'}$ and in Kruskal coordinates I use capital Latin letters
$T_{UU}$.

The indices of tensors that live on a $d$-dimensional
Lorentz spacetime $M$ ($d>2$) are taken from the end of the alphabet,
e.g. $T_{\rho\sigma},e^{r}$. If a tensor lives on the
Lorentz sub-manifold $L$ which is obtained by dimensional reduction
(spherical reduction see Appendix D) of a $d$-dimensional Lorentz
spacetime its indices are taken from the beginning
of the alphabet $T_{\alpha\beta},e^{a}$ and take the values
$0,1$. If a tensor lives on the $(d-2)$-sphere $S^{d-2}$ its indices
are taken from the middle of the alphabet $T_{\kappa\lambda},e^k$
and run from $2$ to $d$.

If confusion is possible I mark objects with an index $M,L,S$,
according to their associated manifold. Sometimes I also
use the numbers $4,2$ instead of $M,L$ in the case of spherical
reduction from four to two dimensions.

\subsection{Coordinate Systems}

In \emph{Schwarzschild coordinates} $t,r,\theta,\vp$, the metric
of a spherically symmetric four-dimensional BH reads
\begin{equation}
ds^2=\Sss dt^2-\frac{1}{\Sss}dr^2-r^2(d\theta^2+\sin^2\theta\,d\vp^2),
\label{SS-metric-4d}
\end{equation}
where $2M$ is the radius of the event horizon in Planck units.
By introducing the \emph{Regge-Wheeler (Tortoise) coordinate}
\begin{equation}
r_{\ast}=r+2M\ln\left(\frac{r}{2M}-1\right)\,
,\,\frac{dr_{\ast}}{dr}=\left(1-\frac{2M}{r}\right)^{-1}\label{Tortoise}
\end{equation}
the metric becomes
\begin{equation}
ds^2=\Sss(dt^2-dr_{\ast}^2)-r^2(d\theta^2+\sin^2\theta\,d\vp^2).
\end{equation}
\emph{Light-cone coordinates} are defined by
\begin{equation}
x^-=t-r_{\ast}\,\,,\,\,x^+=t+r_{\ast}.\label{light-cone}
\end{equation}
The transformation rules for vector and one-form components are given by
\begin{eqnarray}
A^{\bar{\mu}}&=&\left(\begin{array}{c}A^- \\ A^+ \\ A^{\theta}
\\ A^{\vp} \end{array}\right)
={\cal T}^{\bar{\mu}}{}_{\nu}A^{\nu}
=\left(\begin{array}{cccc}1 & -1 &&\\ 1 & 1 &&\\ && 1 & 0\\ && 0 & 1
\end{array}\right)\left(\begin{array}{c}A^t \\ A^{r_{\ast}}
\\ A^{\theta} \\ A^{\vp} \end{array}\right)\\
\omega_{\bar{\mu}}&=&\left(\begin{array}{c}\omega_- \\ \omega_+
\\ \omega_{\theta} \\ \omega_{\vp}\end{array}\right)
={\cal T}_{\bar{\mu}}{}^{\nu}\omega_{\nu}
=\frac{1}{2}\left(\begin{array}{cccc}1 & -1 &&\\ 1 & 1 && \\
&& 1 & 0\\ && 0 & 1\end{array}\right)
\left(\begin{array}{c}\omega_t \\ \omega_{r_{\ast}} \\
\omega_{\theta}\\ \omega_{\vp}\end{array}\right).
\end{eqnarray}
In the following I consider the two-dimensional metric $g_{\alpha\beta}^L$
of the submanifold $L$ which is spanned by the $t,r$-coordinates.
The metric of $M$ (\ref{SS-metric-4d}) can be obtained from $g_{\mu\nu}^L$
by simply adding the remaining three blocks in the matrix-representation.
The same holds for the transformation matrices between different coordinate
systems.

Often I use the \emph{conformal gauge}\footnote{In this gauge one
can see that a two-dimensional metric only differs by a local factor
from the flat metric $\eta_{\alpha\beta}$. The conformal factor
$\rho$ represents the single gravitational degree of freedom
of a two-dimensional spacetime.} for the two-dimensional metric:
\begin{equation}
ds^2_{2d}=e^{2\rho}dx^-dx^+=\Sss dx^-dx^+.
\end{equation}
In this gauge only the off-diagonal elements are non-vanishing: $g_{-+}=g_{+-}
=\frac{e^{2\rho}}{2}$. Further, all Christoffel symbols vanish but
$\Gamma^-_{--}=2(\p_-\rho)\,,\,\Gamma^+_{++}=2(\p_+\rho)$. 
Of particular use for the investigation of the future horizon are
the hybrid \emph{Eddington-Finkelstein coordinates} of the type
$x^-,r$ with line-element $ds^2=\Sss (dx^-)^2+2dx^-dr'$ and (inverse)
metric $g_{x^-x^-}=\Sss\,,\,g_{x^-r'}=g_{r'x^-}=1\,,\,g_{r'r'}=0\,;\,
g^{x^-x^-}=0\,,\,g^{x^-r'}=g^{r'x^-}=1\,,\,g^{r'r'}=-\Sss$.
The transformation rule for tensors (in the first quadrant)
is given by\footnote{I denote the radius coordinate by $r':=r$.}
\begin{eqnarray}
A^{\alpha'}=\left(\begin{array}{c}A^{x^-} \\ A^{r'} \end{array}\right)
=\left(\begin{array}{cc}1 & -\Sss^{-1} \\ 0 & 1 \end{array}\right)
\left(\begin{array}{c}A^t \\ A^{r} \end{array}\right)\\
\omega_{\alpha'}=\left(\begin{array}{c}\omega_{x^-} \\ \omega_{r'}\end{array}\right)
=\left(\begin{array}{cc}1 & 0 \\ \Sss^{-1} & 1 \end{array}\right)
\left(\begin{array}{c}\omega_t \\ \omega_{r} \end{array}\right).
\end{eqnarray}
For examinations close to the horizon $r=2M$ one must employ
a global coordinate system defined by
\begin{equation}
U=-e^{-\frac{x^-}{4M}}\,\,,\,\,V=e^{\frac{x^+}{4M}}.\label{global}
\end{equation} 
In these coordinates the metric has the form
\begin{equation}
ds^2_{2d}=\frac{32M^3e^{-\frac{r}{2M}}}{r}dUdV.
\end{equation}
Thereby I have used the relation $r_{\ast}=(x^+-x^-)/2$. These coordinates
are the light-cone version of \emph{Kruskal coordinates}.
The unphysical singularity of the metric at the horizon $U=V=0$ is no
more present. The transformation between light-cone and global coordinates
is given by
\begin{equation}
\left(\begin{array}{c}\omega_U \\ \omega_V\end{array}\right)
=4M\left(\begin{array}{cc}-\frac{1}{U} & 0 \\ 0 & \frac{1}{V}\end{array}\right)
\left(\begin{array}{c}\omega_- \\ \omega_+\end{array}\right),
\left(\begin{array}{c}A^U \\ A^V\end{array}\right)
=\frac{1}{4M}\left(\begin{array}{cc}-U & 0 \\ 0 & V\end{array}\right)
\left(\begin{array}{c}A^- \\ A^+\end{array}\right).\label{Kruskal}
\end{equation}

\subsubsection{Schwarzschild Metric in $2d$}

The only non-vanishing Christoffel symbols of the two-dimensional
Schwarz\-schild metric $ds^2=\Sss dt^2-\Sss^{-1}dr^2$ are:
\begin{eqnarray}
\Gamma^r_{tt}&=&\frac{M\Sss}{r^2}\,\,,\,\,\Gamma^r_{rr}
=-\frac{M}{r^2\Sss}\\
\Gamma^t_{rt}&=&\Gamma^t_{tr}=\frac{M}{r^2\Sss}.
\end{eqnarray}

The Laplace operator in two dimensions, acting on a scalar field
$f(x)$, is thus given by
\begin{multline}
\square f=g^{tt}(\p_t)^2f+g^{rr}(\p_r)^2f-g^{tt}\Gamma_{tt}^r\p_rf
-g^{rr}\Gamma_{rr}^r\p_rf\\
=\frac{1}{\Sss}(\p_t)^2f-\Sss(\p_r)^2f-\frac{2M}{r^2}\p_rf.
\label{Laplace-2d}
\end{multline}

\subsection{Cartan Variables}

A vielbein basis defines a flat metric $\eta_{mn}=\text{diag}(1,-1,-1,-1)$
at each spacetime point by establishing a freely falling system
(which is unique up to local Lorentz transformations):
\begin{equation}
ds^2=g_{\mu\nu}dx^{\mu}dx^{\nu}=\eta_{mn}e^me^n.
\end{equation}
The $e^m$ provide a basis of the one-forms on $M$ and are
related to the coordinate differentials in a particular
coordinate system by
\begin{equation}
e^m=e^m{}_{\mu}dx^{\mu}.
\end{equation}
The $e^m{}_{\mu}$ are matrices which are called the \emph{inverse vielbeine}.
A basis of the tangent space is given by
\begin{equation}
E_m=e_m{}^{\mu}\p_{\mu},
\end{equation}
where the matrices $e_m{}^{\mu}$ are called the \emph{vielbeine} and
fulfil the condition $e_m{}^{\mu}e^n{}_{\mu}=\delta_m^n$.
Further, they obey $e_m{}^{\mu}e^m{}_{\nu}=\delta^{\mu}_{\nu}$.
Sloppily I will call the $E_m$ vielbeine and the $e^m$ inverse vielbeine.
A connection $\n_m$ in a vielbein basis is defined by
its action on the basis forms
\begin{equation}
\n_me^n=-\omega^n{}_k(E_m)e^k.
\end{equation}
The coefficients $\omega^n{}_m(E_k)$ are called spin-connection
and define a matrix-valued one-form $\omega^n{}_m(E_k)e^k$.
The action of the connection on a general tensor field
is given by
\begin{equation}
\n_mT^k{}_l=E_mT^k{}_l+\omega^k{}_n(E_m)T^n{}_l-\omega^n{}_l(E_m)T^k{}_n.
\end{equation}
The Levi-Civit\'a connection is, as usual, the unique connection
which is torsion-free and metric compatible. The torsion
condition in a vielbein frame reads
\begin{equation}
T^m=de^m+\omega^m{}_ne^n=0,\label{torsion}
\end{equation}
where $d$ is the exterior derivative. The metric compatibility
is fulfilled if the matrix-indices of the spin-connection
are anti-symmetric $\omega_{mn}=-\omega_{nm}$.
The Riemann tensor is represented by the two-form
\begin{equation}
R^m{}_n=d\omega^m{}_n+\omega^m{}_o\wedge\omega^o{}_n.\label{curvature}
\end{equation}
The collection $e^m,\omega^m{}_n,T^m,R^m{}_n$ are called the \emph{Cartan
variables} and (\ref{torsion},\ref{curvature}) are called
the \emph{Cartan equations}. Note that $e^m,\omega^m{}_n$ are one-forms
and $T^m,R^m{}_n$ are two-forms. The relation between
the Riemann tensor in a vielbein basis and in a coordinate basis
is given by
\begin{equation}
R^m{}_n(E_o,E_p)=e^m{}_{\mu}e_n{}^{\nu}e_o{}^{\kappa}e_p{}^{\lambda}
R^{\mu}{}_{\nu\kappa\lambda}.
\end{equation}
In Appendix B.3 I give a vielbein basis and the corresponding
Levi-Civit\'a spin-connection on a Schwarzschild spacetime.

\subsection{Energy Momentum Tensor}

The EM tensor describes the energy density, fluxes, and stresses
of some physical field and can be calculated by variation
of the classical (or quantum) matter action for the metric
(\ref{EM tensor}). In Section 1.1.3 I discuss the general
meaning of its components and some of its properties on a
curved spacetime.

Here I will only show how the components of the EM tensor
are related in different coordinate systems and then consider
a non-minimal coupling of the scalar field $S$ to the
scalar curvature and its effect on the EM tensor.

\subsubsection{Coordinate Systems}

In light-cone coordinates (\ref{light-cone}) the components of the 
EM tensor are given by
\begin{eqnarray}
T_{-+}&=&\frac{\Sss}{4}(T-2\Th)\label{EM-light1}\\
T_{--}&=&\frac{\Sss}{4}(T^t{}_t-
T^{r_{\ast}}{}_{r_{\ast}}+2T^{r_{\ast}}{}_{t})\label{EM-light2}\\
T_{++}&=&\frac{\Sss}{4}(T^t{}_t-
T^{r_{\ast}}{}_{r_{\ast}}-2T^{r_{\ast}}{}_{t}).\label{EM-light3}
\end{eqnarray}
Note that asymptotically one has $T^{r_{\ast}}{}_{r_{\ast}}\to
T^r{}_r$ and $T^{r_{\ast}}{}_{t}\to T^r{}_{t}$
because of $\frac{dr_{\ast}}{dr}\to1$ as $r\to\infty$ (\ref{Tortoise}).
The relation between Eddington-Finkelstein gauge and Schwarzschild gauge reads
\begin{eqnarray}
T_{x^-x^-}&=&T_{tt}\\
T_{x^-r'}&=&\Sss^{-1}T_{tt}+T_{rt}\\
T_{r'r'}&=&\Sss^{-2}T_{tt}+2\Sss^{-1}T_{tr}+T_{rr}.
\end{eqnarray}
Finally, I give the transformation from global coordinates (\ref{global})
to light-cone coordinates:
\begin{eqnarray}
T_{UU}&=&\frac{16M^2}{U^2}T_{--}=\frac{64M^4e^{\frac{-r}{M}}}
{(r-2M)^2}T_{--}\label{EM-Krus1}\\
T_{VV}&=&\frac{16M^2}{V^2}T_{++}\label{EM-Krus2}\\
T_{UV}&=&-\frac{16M^2}{UV}T_{-+}=\frac{32M^3e^{\frac{-2r}{M}}}
{(r-2M)}T_{-+}\label{EM-Krus3}.
\end{eqnarray}

\subsubsection{Non-Minimal Coupling}

I consider a non-minimally coupled, massless scalar field
on a $d$-dimensional Schwarzschild spacetime $M$ described by the action
\begin{equation}
L_{nm}=\int_M\left[R+\xi S^2R+\frac{(\p S)^2}{2}\right]
\sqrt{-g}d^dx\hspace{0.2cm};\,\xi\in\mathbb{R}.\label{non-min.action}
\end{equation}
I do not add a mass term with respect to conformal coupling
as it would destroy the tracelessness of the EM tensor.
Now I want to compute the EM tensor in the case of general coupling.
By (\ref{var.contr.Ricci}) I can calculate the
variation of the matter action for the metric:
\begin{multline}
\frac{\delta L_{nm}}{\delta g^{\mu\nu}(x')}=\int_M(1+\xi S^2)\frac{\delta}
{\delta g^{\mu\nu}}\left(g^{\kappa\lambda}R_{\kappa\lambda}
\sqrt{-g}\right)d^dx+\frac{\sqrt{-g}}{2}T^{old}_{\mu\nu}\\
=\sqrt{-g}\left\{(1+\xi S^2)\left(R_{\mu\nu}-\frac{g_{\mu\nu}}{2}R\right)+
\frac{1}{2}T^{old}_{\mu\nu}\right\}\\
+\int_M\xi S^2\left(g_{\mu\nu}\square-\n_{\mu}\n_{\nu}\right)
\delta(x-x')\sqrt{-g}d^dx\\
=\sqrt{-g}\left\{(1+\xi S^2)\left(R_{\mu\nu}-\frac{g_{\mu\nu}}{2}R\right)+
\frac{1}{2}T^{old}_{\mu\nu}+\xi\left(g_{\mu\nu}\square S^2-\n_{\mu}\n_{\nu}S^2
\right)\right\}=0.
\end{multline}
$T^{old}_{\mu\nu}=\p_{\mu}S\p_{\nu}S-\frac{g_{\mu\nu}}{2}(\p S)^2$
is just the EM tensor of a minimally coupled
massless scalar. The Einstein equations now read
\begin{multline}
R_{\mu\nu}-\frac{g_{\mu\nu}}{2}R=-\frac{1}{2}\Bigm\{
2\xi S^2\left(R_{\mu\nu}-\frac{g_{\mu\nu}}{2}R\right)+T^{old}_{\mu\nu}\\
+4\xi\Bigm[g_{\mu\nu}(S\square S+(\p S)^2)-S\n_{\mu}\p_{\nu}S
-\p_{\mu}S\p_{\nu}S\Bigm]\Bigm\}.
\end{multline}
The EM tensor and its trace for general coupling are thus given by
\begin{eqnarray}
T^{\xi}_{\mu\nu}&=&\left(4\xi-\frac{1}{2}\right)g_{\mu\nu}(\p S)^2+(1-4\xi)\p_{\mu}S
\p_{\nu}S+2\xi S^2\left(R_{\mu\nu}-\frac{g_{\mu\nu}}{2}R\right)\nonumber\\
&&+\,\,4\xi(g_{\mu\nu}S\square S-S\n_{\mu}\p_{\nu}S)\label{EMtensor.nm}\\
T^{\xi}&=&\left[4\xi(d-1)+\frac{2-d}{2}\right](\p S)^2
+4\xi(d-1)S\square S\nonumber\\
&&+(2-d)\xi S^2R.
\end{eqnarray}
In $d=4$ and $d=2$ the trace becomes
\begin{equation}
T^{\xi}_4=(12\xi-1)(\p S)^2+12\xi S\square S-2\xi S^2R\,\,,\,\,
T^{\xi}_2=4\xi\left[S\square S+(\p S)^2\right].
\end{equation}
The choices $\xi_4=\frac{1}{12}$ and $\xi_2=0$ lead on-shell
to a vanishing trace of the EM tensor in four, respectively two dimensions.
In $4d$ one must use the EOM of the scalar field $\square S=2\xi RS$,
as derived from (\ref{non-min.action}) by variation for $S$.
One can choose in arbitrary even dimensions a constant $\xi_d$
such that the trace of the EM tensor vanishes on-shell. This type
of coupling is called \emph{conformal coupling} as it implies
the invariance of the action functional under conformal transformations,
see Appendix C.1. In general this property may be destroyed at
the quantum level -- in this case one speaks of a \emph{trace anomaly}
(or conformal anomaly).

\newpage
\section{Differential Geometry}

In this Appendix I collect some useful formulas and derivations
that are frequently used in differential geometry and the
variational calculus on manifolds.

\subsection{Notations and Basics}

I fix the sign between Riemann tensor and Ricci tensor by
\begin{equation}
R_{\mu\nu}=R^{\rho}{}_{\mu\rho\nu}.
\end{equation}
The ``trace'' of the Levi-Civit\'a connection can be written as
\begin{equation}
\Gamma^{\mu}_{\mu\nu}=
\frac{g^{\mu\sigma}}{2}[\p_{\mu}g_{\sigma\nu}
+\p_{\nu}g_{\sigma\mu}-\p_{\sigma}g_{\mu\nu}]
=\frac{g^{\mu\sigma}}{2}\p_{\nu}g_{\mu\sigma}
=\frac{1}{2}\p_{\nu}\ln g,
\end{equation}
where I have used $dg=g\cdot g^{\mu\nu}dg_{\mu\nu}$,
because
\begin{equation}
d\,\ln\det(g_{\mu\nu})=\frac{dg}{g}=d\,\text{tr}(\ln g_{\mu\nu})
=d\sum_n\ln\lambda_n=\sum_n\frac{d\lambda_n}{\lambda_n}=g^{\mu\nu}dg_{\mu\nu}
\end{equation}
in a diagonal gauge of the metric.
From this one can derive a useful relation:
\begin{equation}
\p_{\mu}(\sqrt{-g}\p^{\mu}f)=\sqrt{-g}\left(\p_{\mu}\p^{\mu}f
+\frac{\p_{\mu}\ln g}{2}\p^{\mu}f\right)=\sqrt{-g}\cdot\square f
=\sqrt{-g}\n_{\mu}\p^{\mu}f,
\end{equation}
i.e. the covariant derivative becomes a partial derivative if
pulled through the measure (in this special case).

\subsection{Variations of Geometric Objects}

In this Section I show how geometric objects
transform if the metric of the manifold $M$ is varied
infinitesimally. The manifold shall be equipped with
some metric $g$. I define another metric $\tilde{g}:=g+\delta g$
that differs by the infinitesimal quantity $\delta g$ from $g$.
In components we have the relations
\begin{eqnarray}
\tilde{g}_{\mu\nu}&=&g_{\mu\nu}+\delta g_{\mu\nu}\\
\tilde{g}^{\mu\nu}&=&g^{\mu\nu}-\delta g^{\mu\nu}+O(\delta g^2).
\end{eqnarray}
Indices are raised (and lowered) by the metric $g$.
Both metrics provide a Levi-Civit\'a connection on $M$,
namely $\n$ and $\tilde{\n}$, respectively. I define
a tensorial object $C$ by the difference connection
\begin{equation}
(\tilde{\n}_{\mu}-\n_{\mu})v^{\rho}=C^{\rho}_{\mu\nu}v^{\nu},
\end{equation}
where $v$ is some vector field on $M$. Analogously, for a one-form
$\omega$ holds the relation $(\tilde{\n}_{\mu}-\n_{\mu})\omega_{\rho}
=-C^{\nu}_{\mu\rho}\omega_{\nu}$. The difference connection
obviously is torsion free, i.e. $C^{\rho}_{\mu\nu}=C^{\rho}_{\nu\mu}$.
By the metric compatibility condition for $\tilde{g}$ (both connections
are compatible with their associated metrics as they are Levi-Civit\'a)
with permuted indices
\begin{equation}
\tilde{\n}_{\kappa}\tilde{g}_{\mu\nu}=0=\n_{\kappa}\tilde{g}_{\mu\nu}
-C^{\rho}_{\kappa\mu}\tilde{g}_{\nu\rho}-C^{\rho}_{\kappa\nu}\tilde{g}_{\rho\mu}
\end{equation}
one can show that
\begin{multline}
C^{\rho}_{\mu\nu}=\frac{\tilde{g}^{\rho\tau}}{2}[\n_{\mu}\tilde{g}_{\tau\nu}
+\n_{\nu}\tilde{g}_{\tau\mu}-\n_{\tau}\tilde{g}_{\mu\nu}]
=\frac{\tilde{g}^{\rho\tau}}{2}[\n_{\mu}\delta g_{\tau\nu}
+\n_{\nu}\delta g_{\tau\mu}-\n_{\tau}\delta g_{\mu\nu}]\\
=\frac{g^{\rho\tau}}{2}[\n_{\mu}\delta g_{\tau\nu}
+\n_{\nu}\delta g_{\tau\mu}-\n_{\tau}\delta g_{\mu\nu}]+O(\delta g^2).
\end{multline}
The Riemann tensors associated to the metrics $g$ and $\tilde{g}$
are related by
\begin{equation}
\tilde{R}^{\mu}{}_{\nu\rho\sigma}=R^{\mu}{}_{\nu\rho\sigma}+
\n_{\rho}C^{\mu}_{\nu\sigma}-\n_{\sigma}C^{\mu}_{\nu\rho}
+C^{\tau}_{\nu\sigma}C^{\mu}_{\tau\rho}-C^{\tau}_{\nu\rho}C^{\mu}_{\tau\sigma}.
\end{equation}
Hence, to the first order the variation of the Riemann tensor
is given by
\begin{multline}
\delta R^{\mu}{}_{\nu\rho\sigma}=\tilde{R}^{\mu}{}_{\nu\rho\sigma}
-R^{\mu}{}_{\nu\rho\sigma}=\n_{\rho}C^{\mu}_{\nu\sigma}
-\n_{\sigma}C^{\mu}_{\nu\rho}+O(\delta g^2)\\
=\frac{g^{\mu\tau}}{2}[\n_{\rho}\n_{\nu}\delta g_{\tau\sigma}
+\n_{\rho}\n_{\sigma}\delta g_{\tau\nu}-\n_{\rho}\n_{\tau}\delta g_{\nu\sigma}\\
-\n_{\sigma}\n_{\nu}\delta g_{\tau\rho}-\n_{\sigma}\n_{\rho}\delta g_{\tau\nu}
+\n_{\sigma}\n_{\tau}\delta g_{\nu\rho}]+O(\delta g^2)\label{var.Riemann}.
\end{multline}
The variation of the Ricci tensor reads
\begin{equation}
\delta R_{\nu\sigma}=\frac{1}{2}[\n^{\tau}\n_{\nu}\delta g_{\tau\sigma}
+\n^{\tau}\n_{\sigma}\delta g_{\tau\nu}-\square\delta g_{\nu\sigma}
-\n_{\sigma}\n_{\nu}g^{\tau\upsilon}\delta g_{\tau\upsilon}]+O(\delta g^2)
\label{var.Ricci}.
\end{equation}
The contraction of this formula by the metric yields the relation
\begin{equation}
g^{\nu\sigma}\delta R_{\nu\sigma}=[\n^{\nu}\n^{\sigma}
-\square g^{\nu\sigma}]\delta g_{\nu\sigma}
=[\square g_{\nu\sigma}-\n_{\nu}\n_{\sigma}]\delta g^{\nu\sigma}.
\label{var.contr.Ricci}
\end{equation}
In the last step I have used the fact that
\begin{equation}
\delta(g^{\mu\nu}g_{\nu\rho})=\delta\delta^{\mu}_{\rho}=0
=(\delta g^{\mu\nu})g_{\nu\rho}+g^{\mu\nu}\delta g_{\nu\rho}.
\end{equation}
By this method one can also calculate the variation of a
Laplace operator, acting on a metric-independent scalar field $f$
(if it depends on $g$ one has to compute additionally the inner
variation):
\begin{multline}
\delta\square f=\delta g^{\alpha\beta}\n_{\alpha}\p_{\beta}f
-g^{\alpha\beta}C_{\alpha\beta}^{\gamma}\p_{\gamma}f\\
=\delta g^{\alpha\beta}\n_{\alpha}\p_{\beta}f
+(\n_{\alpha}\delta g^{\alpha\beta})\p_{\beta}f-\frac{g_{\alpha\beta}}{2}
(\n^{\gamma}\delta g^{\alpha\beta})\p_{\gamma}f.
\label{var.Laplace}
\end{multline}
Finally, I show how the spacetime measure is varied for the metric.
The variation of the determinant of the metric works analogously to
its differentiation:
\begin{equation}
\delta g=g\cdot g^{\mu\nu}\delta g_{\mu\nu}=-g\cdot g_{\mu\nu}\delta g^{\mu\nu}
\end{equation}
The variation of the measure thus reads
\begin{equation}
\delta\sqrt{-g}=-\frac{\sqrt{-g}}{2}g_{\mu\nu}\delta g^{\mu\nu}\label{var.measure}.
\end{equation}

\subsection{Computation of Geometric Objects on a
Four-Dimensional Schwarzschild Spacetime}

The quantum mechanical expectation values of a scalar field on
a Schwarz\-schild spacetime are expressed by geometric tensor fields
and contractions of higher tensorial objects by the metric. In the quasi-static
phase of a BH the spacetime curvature caused by the Hawking
flux (known as the backreaction) is negligible compared to that
caused by the BH. Thus, the Schwarzschild solution
describes accurately the spacetime curvature and hence the Ricci tensor,
as well as the scalar curvature, vanish almost perfectly outside the
horizon $R_{\mu\nu}=0,R=0$. Therefore, the only nonvanishing geometric
objects are the Riemann tensor $R^{\mu}{}_{\nu\kappa\lambda}$
and the metric $g_{\mu\nu}$. By taking covariant derivatives and
then contracting with the metric one can construct arbitrarily complicated
geometric objects that contribute to higher perturbational
orders in the expansion of the effective action.
The metric in this Section is \emph{Lorentzian}, the corresponding
Euclidean (Riemannian) expressions can be obtained easily by inserting
the appropriate minus signs in the basic expression.
I will perform the calculations by use of Cartan variables
introduced in Appendix A.4.
A (inverse) vielbein basis for a Schwarzschild metric is given by\footnote{Other
vielbein bases are connected to the current one by local Lorentz rotations.}
\begin{equation}
e^0=\Ss dt\,\,,\,\,e^1=\frac{1}{\Ss}dr\,\,,\,\,e^2=r\,d\theta
\,\,,\,\,e^3=r\sin\theta d\vp.
\end{equation}
The line-element can thus be written as $ds^2=\eta_{mn}e^me^n$, where
$\eta_{mn}=\text{diag}(1,-1,-1,-1)$ is the metric of Minkowski
spacetime. The vielbeine and their tensor products form a basis of
arbitrary covariant tensor fields and differential forms. Note,
that I only use Latin letters to denote vielbein indices.
The vielbeine
\begin{equation}
E_0=\frac{1}{\Ss}\p_t\,\,,\,\,E_1=\Ss\p_r\,\,,\,\,E_2=\frac{1}{r}\p_{\theta}
\,\,,\,\,E_3=\frac{1}{r\sin\theta}\p_{\vp}
\end{equation}
and their tensor products form a basis of arbitrary contravariant tensor
fields. The only non-vanishing coefficients of the spin-connection
on a Schwarzschild spacetime are given by
\begin{eqnarray}
&\omega^0{}_1=\frac{M}{r^2\Ss}\,e^0\,\,,\,\,\omega^2{}_1=\frac{\Ss}{r}\,e^2&
\nonumber\\
&\omega^3{}_1=\frac{\Ss}{r}\,e^3\,\,,\,\,\omega^3{}_2=\frac{\cot{\theta}}{r}\,e^3.&
\label{con.4d}
\end{eqnarray}
The metric compatibility can be expressed as
\begin{equation}
\omega^0{}_i=\omega^i{}_0\,,\,\omega^i{}_j=-\omega^j{}_i\hspace{0.2cm}
;\,\,i,j\in\left\{1,2,3\right\}.
\end{equation}
For later convenience, and to demonstrate this formalism, I derive
the Schwarz\-schild Laplacian, acting on a function $f(r)$ that \emph{only depends
on the radial coordinate $r$}:
\begin{multline}
\square f(r)=\bigm[\eta^{11}E_1E_1-\eta^{mn}\omega^1{}_n(E_m)E_1\bigm]f(r)\\
=\bigm[-E_1E_1-\eta^{00}\omega^1{}_0(E_0)E_1-\eta^{22}\omega^1{}_2(E_2)E_1
-\eta^{33}\omega^1{}_3(E_3)E_1\bigm]f(r)\\
=\left[-\Ss\p_r\left(\Ss\p_r\right)-\left(\frac{M}{r^2}
+\frac{2\Sss}{r}\right)\p_r\right]f(r)\\
=\left[-\Sss\p_r^2-\left(\frac{2}{r}-\frac{2M}{r^2}\right)\p_r\right]f(r).
\label{Laplace.SS}
\end{multline}
\\
\\
Now I come to the computations. The non-vanishing covariant components of the
Riemann tensor on a Schwarzschild spacetime with ADM mass $M$ are
\begin{eqnarray}
&R_{0101}=-R_{1001}=-R_{0110}=R_{1010}=\frac{2M}{r^3}&\label{Riemann.SS}\\
&R_{0202}=R_{0303}=-\frac{M}{r^3}\,\,,\,\,R_{1212}=R_{1313}=\frac{M}{r^3}
\,\,,\,\,R_{2323}=-\frac{2M}{r^3}.&\nonumber
\end{eqnarray}
From this one can already observe the following: because a certain
index always appears twice, the indices of the
Riemann tensor can be lowered or raised without change of sign
if this is done for \emph{all four indices} simultaneously: $R^{mnop}=
\eta^{mq}\eta^{nr}\eta^{os}\eta^{pt}R_{qrst}=R_{mnop}$.
The metrics always multiply to a total factor $1$. 
I start with calculating pure contractions to scalars of the Riemann
tensor:
\begin{multline}
R_{mnop}R^{mnop}=R_{mnop}R_{mnop}=\sum_{a<b}R_{abab}R_{abab}\\
=4\Bigm(R_{0101}^2+R_{0202}^2+R_{0303}^2+R_{1212}^2+R_{1313}^2+
R_{2323}^2\Bigm)\\
=4\Bigm(\frac{4M^2}{r^6}+4\frac{M^2}{r^6}+\frac{4M^2}{r^6}\Bigm)=48\frac{M^2}{r^6}.
\label{RR}
\end{multline}
To the third order I will need two different types of contractions
that cannot be transformed into each other by simple symmetry
considerations. The first one is
\begin{multline}
R_{mnop}R^{mn}{}_{qr}R^{opqr}=8\sum_{a<b}R_{abab}R^{ab}{}_{ab}R^{abab}\\
=8\sum_{a<b}(\eta^{aa})^3(\eta^{bb})^3R_{abab}R_{abab}R_{abab}
=8\sum_{a<b}(\eta^{aa}\eta^{bb}R_{abab})^3\\
=8\bigm[(-R_{0101})^3+(-R_{0202})^3+(-R_{0303})^3+(R_{1212})^3+(R_{1313})^3+
(R_{2323})^3\bigm]\\
=\frac{96M^3}{r^9}.
\end{multline}
The second kind of term is
\begin{multline}
R_{mnop}R^{m}{}_{q}{}^{o}{}_{r}R^{nqpr}=4\sum_{a<b}\sum_{c\neq a,b}
R_{abab}R^{a}{}_{c}{}^{a}{}_{c}R^{bcbc}\\
=4\sum_{a<b}\sum_{c\neq a,b}(\eta^{aa}\eta^{bb}\eta^{cc})^2
R_{abab}R_{acac}R_{bcbc}\\
=4\bigm[R_{0101}R_{0202}R_{1212}+R_{0101}R_{0303}R_{1313}
+R_{0202}R_{0101}R_{2121}+R_{0202}R_{0303}R_{2323}\\
+R_{0303}R_{0101}R_{3131}+R_{0303}R_{0202}R_{3232}
+R_{1212}R_{1010}R_{2020}+R_{1212}R_{1313}R_{2323}\\
+R_{1313}R_{1010}R_{3030}+R_{1313}R_{1212}R_{3232}
+R_{2323}R_{2020}R_{3030}+R_{2323}R_{2121}R_{3131}\bigm]\\
=12\bigm[R_{0101}R_{0202}R_{1212}+R_{0101}R_{0303}R_{1313}
+R_{0202}R_{0303}R_{2323}+R_{1212}R_{1313}R_{2323}\bigm]\\
=-\frac{96M^3}{r^9}.
\end{multline}
Next I will consider terms in two curvatures on which two
covariant derivatives act. The first term I consider is
\begin{multline}
(\n_qR_{mnop})\n^qR^{mnop}\\
=\eta^{qr}\bigm[E_qR_{mnop}-\omega^s{}_m(E_q)R_{snop}
-\dots\bigm]\bigm[E_rR^{mnop}+\omega^m{}_s(E_r)R^{snop}+\dots\bigm]\\
=4\sum_{a<b}\eta^{qr}\bigm[E_qR_{abab}-\omega^s{}_a(E_q)R_{sbab}
-\dots\bigm]\bigm[E_rR^{abab}+\omega^a{}_s(E_r)R^{sbab}+\dots\bigm]\\
=-4\sum_{a<b}(E_1R_{abab})(E_1R^{abab})=-4\sum_{a<b}(E_1R_{abab})(E_1R_{abab})\\
=-4\bigm[(E_1R_{0101})^2+(E_1R_{0202})^2+(E_1R_{0303})^2\\
+(E_1R_{1212})^2+(E_1R_{1313})^2+(E_1R_{2323})^2\bigm]\\
=-4\Sss\left(2\cdot\frac{36M^2}{r^8}+4\cdot\frac{9M^2}{r^8}\right)\\
=-\Sss\frac{432M^2}{r^8}=-\frac{432M^2}{r^8}+\frac{864M^3}{r^9}.
\end{multline}
Note that the connection terms cannot contribute because the free
indices always coincide and $\omega^a{}_a=0$. This is due to the
fact that the Riemann tensor on a Schwarzschild manifold has only
components of the form $R_{abab},R_{abba}$, see (\ref{Riemann.SS}).
As a consequence, a covariant derivative always acts on the Riemann
tensor as if it were a scalar field:
\begin{multline}
\n_kR_{abab}=E_kR_{abab}-\omega^s{}_a(E_k)R_{sbab}-\omega^s{}_b(E_k)R_{asab}
-\dots\\
=E_kR_{abab}-\omega^a{}_a(E_k)R_{abab}-\omega^b{}_b(E_k)R_{abab}
-\dots=E_kR_{abab}.
\end{multline}
There are two similar types
of terms that, by the current symmetries of the Riemann tensor, yield
the same contribution, namely:
\begin{multline}
(\n^kR_{kmno})\n_lR^{lmno}=2\sum_a\sum_b(\n^kR_{kaba})\n_lR^{laba}\\
=2\sum_a\sum_b(\n^bR_{baba})\n_bR^{baba}
=2\eta^{11}\sum_a(E_1R_{1a1a})E_1R^{1a1a}\\
=-2\bigm[(E_1R_{1010})^2+(E_1R_{1212})^2+(E_1R_{1313})^2\bigm]
=-\frac{432M^2}{r^8}+\frac{864M^3}{r^9},
\end{multline}
and analogously
\begin{equation}
(\n_kR_{lmno})\n^lR^{kmno}=2\sum_a\sum_b(\n_bR_{baba})\n^bR^{baba}
=-\frac{432M^2}{r^8}+\frac{864M^3}{r^9}.
\end{equation}
Next I consider terms with two covariant derivatives in a row,
the most simple being a Laplacian acting on a scalar (\ref{Laplace.SS})
\begin{multline}
\square(R_{mnop}R^{mnop})=\square\frac{48M^2}{r^6}
=\left[-\Sss\p_r^2-\left(\frac{2}{r}-\frac{2M}{r^2}\right)\p_r\right]
\frac{48M^2}{r^6}\\
=-\frac{1440M^2}{r^8}+\frac{3456M^3}{r^9}.
\end{multline}
Further, there are terms where the covariant derivatives are contracted
with the Riemann tensor:
\begin{multline}
R_{kmno}\n^k\n_lR^{lmno}=2\sum_{a,b}R_{kaba}\n^k\n_lR^{laba}\\
=2\sum_{a,b}R_{baba}\n^b\n_bR^{baba}
=2\sum_{a,b}R_{baba}[E^bE_b-\omega^s{}_b(E^b)E_s]R^{baba}\\
=2\eta^{11}\sum_aR_{a1a1}E_1E_1R_{a1a1}-2\sum_{a,b}R_{abab}\eta^{bb}
\omega^1{}_b(E_b)E_1R_{abab}\\
=-2\sum_aR_{a1a1}\left[\Sss\p_r^2+\frac{M}{r^2}\p_r\right]R_{a1a1}
-2\sum_{a,b}R_{abab}\eta^{bb}\omega^1{}_b(E_b)E_1R_{abab}\\
=-2R_{0101}\left[\Sss\p_r^2+\frac{M}{r^2}\p_r\right]R_{0101}
-4R_{2121}\left[\Sss\p_r^2+\frac{M}{r^2}\p_r\right]R_{2121}\\
-2\frac{M}{r^2}\bigm(R_{1010}\p_rR_{1010}+R_{2020}\p_rR_{2020}+R_{3030}
\p_rR_{3030}\bigm)\\
-4\frac{\Sss}{r}\bigm(R_{0202}\p_rR_{0202}+R_{1212}\p_rR_{1212}+R_{3232}
\p_rR_{3232}\bigm)\\
=-\frac{144M^2}{r^8}+\frac{324M^3}{r^9}+\frac{36M^3}{r^9}+\frac{72M^2}{r^8}
-\frac{144M^3}{r^9}=-\frac{72M^2}{r^8}+\frac{216M^3}{r^9}.
\end{multline}
The same expression with exchanged covariant derivatives
yields the same result:
\begin{equation}
R_{kmno}\n_l\n^kR^{lmno}=2\sum_{a,b}R_{baba}\n_b\n^bR^{baba}
=-\frac{72M^2}{r^8}+\frac{216M^3}{r^9}.
\end{equation}

\newpage
\section{Conformal Transformations}

An active conformal transformation maps the metric $g_{\mu\nu}$ of a 
$d$-dimensional manifold $M$ onto a new metric $\hat{g}_{\mu\nu}$ through the 
multiplication by a (local) factor:
\begin{equation}
\hat{g}_{\mu\nu}=\Omega^2(x)g_{\mu\nu}.\label{conf.g}
\end{equation}
The dual metric and the measure transform as 
\begin{equation}
\hat{g}^{\mu\nu}=\Omega^{-2}g^{\mu\nu}\,\,
,\,\,\sqrt{-g}=\Omega^{-d}\sqrt{-\hat{g}}.\label{conf.g-1}
\end{equation}
Whenever I speak of a conformal transformation in this work I mean
an active transformation that results in a real change of the metric
$g=g_{\mu\nu}dx^{\mu}\otimes dx^{\nu}$, as opposed to a coordinate 
transformation (the invariance with respect to the latter is a trivial
consequence of general diffeomorphism invariance).
Because a Riemannian manifold is defined by its metric, a
conformal transformation maps one spacetime $M$ onto another 
spacetime $\hat{M}$.
Accordingly, all geometric objects describing the manifold, such as the 
Levi-Civit\'a connection and the scalar curvature, change under such
a transformation\footnote{Under a coordinate transformation the scalar 
curvature clearly would remain unchanged.}. 
I will compute these changes by using Cartan variables.
The vielbeine and their inverse obey the transformation rules
\begin{equation}
\hat{E}_m=\Omega^{-1}E_m\,\,,\,\,\hat{e}^m=\Omega e^m.
\end{equation}
The torsion condition on the new manifold
\begin{equation}
0=\hat{T}^m=(d\Omega)e^m+\Omega de^m+\Omega\hat{\omega}^m{}_ne^n
=(E_n\Omega)e^n\wedge e^m-\Omega\omega^m{}_ne^n+\Omega\hat{\omega}^m{}_ne^n
\end{equation}
constrains the new connection one-form to be $\hat{\omega}^m{}_n=
\omega^m{}_n+\frac{(E_n\Omega)}{\Omega}e^m+\propto e^n$. The metric compatibility
$\omega_{mn}=-\omega_{nm}$ fixes the last term and we have
\begin{equation}
\hat{\omega}^m{}_n=\omega^m{}_n+\frac{(E_n\Omega)}{\Omega}e^m
-\frac{(E^m\Omega)}{\Omega}e_n.
\end{equation}
The scalar curvature transforms as
\begin{equation}
\Omega^2\hat{R}=R+(d-1)\left[(4-d)\left(\frac{\p\Omega}{\Omega}\right)^2
-2\frac{\square\Omega}{\Omega}\right].\label{conf.R}
\end{equation}
The Laplacians of the two manifolds are related by
\begin{equation}
\Omega^2\hat{\square}S=\square S+(d-2)\frac{\p\Omega\p S}{\Omega}.
\label{conf.square}
\end{equation}
A Laplace operator or a scalar product of partial derivatives 
without hat mean contraction with the original metric.
The Riemann tensor transforms as
\begin{multline}
\hat{R}^m{}_{nop}=\frac{1}{\Omega^2}R^m{}_{nop}
+\frac{2}{\Omega}\left\{\eta^m_{\,\,[p}\n_{o]}\left(\frac{\n_n\Omega}
{\Omega^2}\right)
+\eta_{n[o}\n_{p]}\left(\frac{\n^m\Omega}{\Omega^2}\right)\right\}\\
+\frac{2(\n\Omega)^2}{\Omega^4}\eta^m_{\,\,[p}\eta_{o]n}.\label{conf.Riemann}
\end{multline}
Attention: the antisymmetrisation bracket of the indices includes
a factor $\frac{1}{2}$ in my notation: $V_{[mn]}=\frac{1}{2}(V_{[mn]}-V_{[nm]})$.
Finally, the Ricci tensor transforms as
\begin{multline}
\hat{R}_{mn}=\frac{1}{\Omega^2}R_{mn}+\frac{1}{\Omega}
\left\{(2-d)\n_n\left(\frac{\n_m\Omega}{\Omega^2}\right)
-\eta_{mn}\n_o\left(\frac{\n^o\Omega}{\Omega^2}\right)\right\}\\
+(1-d)\frac{(\n\Omega)^2}{\Omega^4}\eta_{mn}.
\label{conf.Ricci}
\end{multline}
If the variation of geometric objects for the metric
is in question, the use of a vielbein frame is sometimes not
very convenient (clearly one could vary for the vielbeine instead,
but this requires a completely new formalism).
In this case one has to transform the obtained expressions
back to a coordinate basis:
\begin{eqnarray}
\hat{R}^{\mu}{}_{\nu\kappa\lambda}&=&\hat{e}_m{}^{\mu}\hat{e}^n{}_{\nu}
\hat{e}^o{}_{\kappa}\hat{e}^p{}_{\lambda}\hat{R}^m{}_{nop}
=\Omega^2e_m{}^{\mu}e^n{}_{\nu}e^o{}_{\kappa}e^p{}_{\lambda}\hat{R}^m{}_{nop}
=R^{\mu}{}_{\nu\kappa\lambda}+\dots\nonumber\\
\hat{R}_{\mu\nu}&=&\hat{e}_m{}^{\mu}\hat{e}_n{}^{\nu}\hat{R}_{mn}
=\Omega^2e_m{}^{\mu}e_n{}^{\nu}\hat{R}_{mn}=R_{\mu\nu}+\dots.
\end{eqnarray}
If an index is lowered or raised one clearly gets additional factors in $\Omega$,
e.g.
\begin{equation}
\hat{R}_{\mu\nu\kappa\lambda}=\hat{g}_{\mu\sigma}
\hat{R}^{\sigma}{}_{\nu\kappa\lambda}
=\Omega^2g_{\mu\sigma}\hat{R}^{\sigma}{}_{\nu\kappa\lambda}
=\Omega^2R_{\mu\nu\kappa\lambda}+\dots.
\end{equation}

\subsection{Conformal Invariance}

A classical theory is said to be conformally invariant
if the EOM are \emph{form-invariant} under arbitrary active conformal
transformations. In Appendix A.5.2 I have shown for the
case of two and four dimensions that the trace of the
EM tensor of a massless scalar field vanishes on-shell if a term
$\xi S^2R$ is added to the Lagrangian, where $\xi$
takes the values $\frac{1}{12}$ and $0$ in respectively
$4d$ and $2d$. This result can be extended to arbitrary
even dimensions. Note that a mass term destroys the tracelessness
of the EM tensor, as it contributes by a term $\frac{d}{2}m^2S^2$.
The vanishing of the trace automatically implies conformal
invariance because the trace is proportional to the change of the action
functional under a conformal transformation:
\begin{equation}
\delta_{\Omega}L_m=\delta g^{\mu\nu}\frac{\delta L_m}{\delta g^{\mu\nu}}
=(\Omega^{-2}-1)g^{\mu\nu}\frac{\sqrt{-g}}{2}T_{\mu\nu}
=\frac{(\Omega^{-2}-1)\sqrt{-g}}{2}T.
\end{equation}
In the following I will show explicitly the conformal invariance
of the four-dimensional conformal scalar model. The multiplication of the
metric by some factor can be interpreted as a rescaling of the
units of the theory. For dimensional reason the scalar field
must transform as $\hat{S}=\Omega^{-1}S$, i.e. it has conformal
weight $-1$. This can be seen from the transformation of the
measure and the scalar curvature (\ref{conf.g-1},\ref{conf.R})
which lead to a total rescaling factor $\Omega^{-2}$ of the action.
In four dimensions the scalar curvature transforms like
$R=\Omega^2\hat{R}+6\frac{\square\Omega}{\Omega}$.
The l.h.s of the EOM $\square S=\frac{SR}{6}$ changes as
\begin{multline}
\square S=\Omega^2\hat{\square}S-2\Omega(\hat{\p}\Omega)(\hat{\p} S)\\
=\Omega^3\hat{\square}\hat{S}+\hat{S}\Omega^2\hat{\square}\Omega+2\Omega^2
(\hat{\p}\Omega)(\hat{\p}\hat{S})-2\Omega^2(\hat{\p}\Omega)
(\hat{\p}\hat{S})-2\hat{S}(\hat{\p}\Omega)^2,
\end{multline}
while the r.h.s. changes as 
\begin{equation}
\frac{SR}{6}=\Omega^3\frac{\hat{S}\hat{R}}{6}+\hat{S}\square\Omega
=\Omega^3\frac{\hat{S}\hat{R}}{6}+\hat{S}\Omega^2\hat{\square}\Omega
-2\hat{S}(\hat{\p}\Omega)^2.
\end{equation}
Hence the conformally transformed EOM has the same form as the
original one: $\hat{\square}\hat{S}=\frac{\hat{S}\hat{R}}{6}$.

\subsection{Conformal Anomaly}

At the quantum level the conformal invariance of a classical theory
might be broken. In this case one speaks of a conformal anomaly.
In analogy to the classical theory, the conformal invariance of
a quantised model is accompanied by the tracelessness
of the expectation value of the EM tensor:
\begin{equation}
\delta_{\Omega}W\propto\E{T}.
\end{equation}
$W$ is the effective action of the considered model.
In this Section I will calculate explicitly the trace anomaly
for a scalar field model in two and four dimensions.
I use the representation of the effective action as a
functional determinant; accordingly, the scalar field
exhibits natural boundary conditions and the quantum
state is the Boulware state $\left|B\right>$ (in the static approximation
on a Schwarzschild spacetime $\E{T}$ is state-independent).
By equations (\ref{zeta-function},\ref{eff.action4})
a conformal transformation of the effective action can be
written as
\begin{multline}
\delta_{\Omega}W[g]=-\frac{1}{2}\frac{d}{ds}\delta_{\Omega}\,
\text{tr}({\cal O}^{-s})\biggm|_{s=0}
=-\frac{1}{2}\frac{d}{ds}\delta_{\Omega}\int_M
\left<x\left|{\cal O}^{-s}\right|x\right>\sqrt{g}d^dx\biggm|_{s=0}\\
=\frac{1}{2}\frac{d}{ds}\int_Ms
\left<x\left|{\cal O}^{-s-1}\delta_{\Omega}({\cal O}\sqrt{-g})\right|x\right>
d^dx\biggm|_{s=0}\\
=\frac{1}{2}\text{tr}\left[{\cal O}^{-s-1}\frac{\delta_{\Omega}({\cal O}
\sqrt{g})}{\sqrt{g}}\right]\biggm|_{s=0}.
\end{multline}
Now one can show that the conformal transformation of
${\cal O}\sqrt{g}$ is proportional to ${\cal O}$ if the
scalar action, associated to ${\cal O}=-\triangle+2\xi R$, is conformally
invariant. If this is the case, the conformal transformation of this
action can only be proportional to a term that vanishes on-shell:
\begin{multline}
0\stackrel{EOM}{=}\int_M\delta g^{\mu\nu}_{\Omega}
\frac{\delta L_{\cal E}[{\cal O}]}{\delta g^{\mu\nu}(x)}\sqrt{g}d^dx
=\int_M\delta g^{\mu\nu}_{\Omega}\int_MS\frac{\delta ({\cal O}\sqrt{g})}
{\delta g^{\mu\nu}(x)}Sd^dx'\sqrt{g}d^dx\\
\propto\int_Mf(x)S{\cal O}S\sqrt{g}d^dx
\end{multline}
Note that the EOM can be written as ${\cal O}S=0$. The
arbitrariness of $S$ in the last equation implies the result.
It follows that the conformal transformation
of the effective action is determined by the 
zeta-function of the conformal operator:
\begin{equation}
\delta_{\Omega}W[g]\propto\text{tr}({\cal O}^{s-1}{\cal O})\bigm|_{s=0}
=\text{tr}({\cal O}^{s})\bigm|_{s=0}=\zeta_{\cal O}[0].
\end{equation}
The trace anomaly is given by
\begin{equation}
\E{T}=\frac{2g^{\mu\nu}}{\sqrt{-g}}\frac{\delta W_{\cal M}}{\delta g^{\mu\nu}}
=-\zeta_{\cal O}[0].
\end{equation}
$\zeta_{\cal O}[0]$ can be calculated by the Seeley-DeWitt expansion
of the heat kernel. I use (\ref{damped-expansion}) and assume
that $D$ is a constant which I set to zero finally. Therefore,
the commutator terms $[D,{\cal O}]$ vanish and the constants $c_n$
are all zero. The first two terms in the expansion are finite
in the limit $s\to0$:
\begin{equation}
\frac{1}{(4\pi)^2}\int_M\left(\frac{D^2a_0}{(s-1)(s-2)}+\frac{Da_1}{(s-1)}\right)
\sqrt{g} d^4x
\end{equation}
Because the damping $D$ also goes to zero they simply vanish.
Hence, the only contribution in $4d$ comes from the coefficient $a_4$:
\begin{equation}
\zeta_{\cal O}[0]=\frac{1}{(4\pi)^2}\int_Ma_4\sqrt{g}d^4x.
\end{equation}
The remaining terms are proportional to $s$ and thus vanish.
The explicit form of the Lorentzian trace anomaly in four dimensions
can be found by Table 4 in Section 3.1.1, whereby the Euclidean
endomorphism is $E=-2\xi_4R=-\frac{1}{6}R$
and $\Omega_{mn}=0$:
\begin{equation}
\E{T}_4=\frac{-1}{2880\pi^2}\int_M\left(\square R-R_{\mu\nu}R^{\mu\nu}
+R_{\mu\nu\sigma\tau}R^{\mu\nu\sigma\tau}\right)\sqrt{-g}d^4x.
\label{trace-anomaly4d}
\end{equation}
In two dimensions the heat kernel differs by a factor $4\pi\tau$
from the four-dimensional one. Thus, the zeta-function for $s=0$,
calculated by the $2d$-version of (\ref{damped-expansion}), reads
\begin{equation}
\zeta_{\cal O}[0]=\frac{1}{4\pi}\int_Ma_2\sqrt{g}d^2x.
\end{equation}
In $2d$ the endomorphism (induced by the conformal coupling)
vanishes, i.e. $E=0$, and thus the 
Lorentzian trace anomaly becomes
\begin{equation}
\E{T}_2=\frac{1}{24\pi}\int_MR\sqrt{-g}d^2x.\label{trace-anomaly2d}
\end{equation}
Note that one might add arbitrary traceless terms to the action
which do not destroy the conformal coupling and
thus produce a non-vanishing endomorphism.
The scalar curvature in the last equation is then replaced by 
$R+6E$.

\newpage
\section{Spherical Reduction}

In this Appendix I present the spherical reduction procedure,
which is the basis of the two-dimensional dilaton model of Section 2.2.
The primary aim is to find a two-dimensional Einstein-Hilbert
action that is (at least) classically equivalent to a four-dimensional
Einstein-Hilbert action describing the s-waves of a scalar field
on a spherically symmetric (non-static) spacetime.
The spherical reduction can be generalised to $d$-dimensional
spacetimes, whereby a $(d-2)$-sphere $S^{d-2}$ is integrated out and
the physical spacetime is still two-dimensional. Throughout this
Appendix I use strictly my index-notations given in Appendix A.2.

The coordinates describing the $d$-dimensional manifold $M$ can
be separated in a two-dimensional (Lorentzian) part $x^{\alpha}$ (e.g. $t,r$)
spanning the physical Lorentz manifold $L$,
and a $(d-2)$-dimensional (Riemannian) part $x^{\kappa}=\theta,\vp\dots$
describing the $(d-2)$-sphere $S^{d-2}$. As the $x^{\kappa}$
denote symmetry directions of $M$, i.e. the tangent vectors to these coordinate
directions are Killing fields and $S^{d-2}$ is the corresponding Killing orbit,
all geometric objects (including the metric) on $M$ depend
solely on the coordinates $x^{\alpha}$.
$M$ possesses a general spherically-symmetric metric 
\begin{equation}
ds^{2}=g_{\alpha\beta}dx^{\alpha}dx^{\beta}-\Phi^{2}(x^{\alpha})
g_{\kappa\lambda}dx^{\kappa}dx^{\lambda}.
\end{equation}
$g_{\alpha\beta}$ is the induced metric on the Lorentzian submanifold
$L$ and $g_{\kappa\lambda}$ the one on $S^{d-2}$.
I define the \emph{dilaton field} $\Phi=\sqrt{X}$ (compare with
Section 2.2) which is more convenient when working with Cartan variables,
see Appendix A.4. I will perform the whole calculation in a vielbein basis
in which the line-element can be written as
\begin{equation}
ds^{2}=\eta_{ab}e^{a}e^{b}-\delta_{ij}e^{i}e^{j}.
\end{equation}
The $e^r$ form a vielbein basis on $M$. One can define
a vielbein basis on $L$ and $S^{d-2}$ which I denote
by $\tilde{e}^a$ and $\tilde{e}^i$, respectively\footnote{I mark
geometric objects belonging to $L$ or $S^{d-2}$ by a tilde on top;
between tensorial objects on the submanifolds one can distinguish
easily by the different indices used.}.
They are related to the $e^r$ by
\begin{eqnarray}
e^{a} &=&\tilde{e}^{a}\,\,,\,\,e^{i}=\Phi\tilde{e}^{i}
\label{vielbein transformation} \\
E_{a} &=&\tilde{E}_{a}\,\,,\,\,E_{i}=\Phi^{-1}\tilde{E}_{i}.
\end{eqnarray}
Further, one has a spin-connection $\omega^r{}_s$ on $M$
which induces connections on the submanifolds:
\begin{equation}
\tilde{\omega}^a{}_b=\omega^a{}_b\,\,,\,\,\tilde{\omega}^i{}_j=\omega^i{}_j.
\end{equation}
All three connections are Levi-Civit\'a and thus fulfil the torsion
condition (\ref{torsion}) and the metric compatibility.
I denote the exterior derivative on $M$ by $d$ (not to be confused
with the spacetime dimension). It can be expressed as a sum
of the exterior derivative on $L$ and $S^{d-2}$:
\begin{equation}
d=dx^{\rho}\partial _{\rho}=e^{r}E_{r}=\tilde{e}^{a}\tilde{E}_{a}
+\tilde{e}^{i}\tilde{E}_{i}=d^{L}+d^{S}.
\end{equation}
The torsion conditions on $L$ and $S^{d-2}$ read
\begin{equation}
\tilde{T}^{a}=d^{L}\tilde{e}^{a}+\tilde{\omega}^a{}_b
\tilde{e}^{b}:=0\,\,,\,\,\tilde{T}^{i}=d^{S}\tilde{e}^{i}+\tilde{\omega}^i{}_j
\tilde{e}^{j}:=0.
\end{equation}
In the following I want to express the Riemann tensor and its contractions
on $M$ by geometric objects of $L$. First I must find the
remaining block $\omega^i{}_a$ of the spin-connection on $M$.
The torsion condition gives
\begin{multline}
T^i=de^i+\omega^i{}_a\wedge e^a+\omega^i{}_j\wedge e^j
=(d^{L}+d^{S})\Phi\tilde{e}^i+\omega^i{}_a\wedge \tilde{e}^a+\Phi
\tilde{\omega}^i{}_j\wedge e^j\\
=(d^{L}\Phi)\tilde{e}^i+\omega^i{}_a\wedge \tilde{e}^a+\Phi(d^{S}\tilde{e}^i
+\tilde{\omega}^i{}_j\wedge e^j)=(\tilde{E}_a\Phi)\tilde{e}^a\tilde{e}^i
-\tilde{e}^a\wedge\omega^i{}_a=0.
\end{multline}
This restricts the connection to the form $\omega^i{}_a=(\tilde{E}_a\Phi)
\tilde{e}^i+\Delta^i\tilde{e}_a$. The remaining ambiguity
$\Delta_i$ is fixed by the metric compatibility condition
\begin{equation}
\omega_{ia}=(\tilde{E}_a\Phi)\tilde{e}_i+\Delta_i\tilde{e}_a
=-\omega_{ai}.
\end{equation}
Thus we have $\Delta_i=-(\tilde{E}_i\Phi)=0$. The complete Levi-Civit\'a
connection on $M$ is given by
\begin{equation}
\omega^r{}_s=\left(\begin{array}{cc}\tilde{\omega}^a{}_b &
(\tilde{E}^a\Phi)\tilde{e}_i\\(\tilde{E}_a\Phi)\tilde{e}^i&
\tilde{\omega}^i{}_j\end{array}\right).\label{con.sph.red.}
\end{equation}
Note that the lowering or raising of the vielbein indices
(and those of other geometric objects) on $L$ and $S^{d-2}$ is defined
via the metric on these manifolds, for instance
$\eta_{ij}\tilde{e}^j=-\delta_{ij}\tilde{e}^j=-\tilde{e}_i$
and thus $\tilde{e}^i=\tilde{e}_i$!

By the use of this connection the Riemann tensor on $M$
can be expressed by objects on $L$ and $S^{d-2}$. In the following
calculations I omit the $\wedge$-symbol between differential forms.
\begin{multline}
R^a{}_b=d\omega^a{}_b+\omega^a{}_c\omega^c{}_b+\omega^a{}_i\omega^i{}_b\\
=\left(d^L\tilde{\omega}^a{}_b+\tilde{\omega}^a{}_c\tilde{\omega}^c{}_b\right)
+\left(\tilde{E}^a\Phi\right)\left(\tilde{E}_b\Phi\right)\tilde{e}^i\tilde{e}^i
=\tilde{R}^a{}_b.
\end{multline}
Further, one gets
\begin{eqnarray}
R^i{}_j&=&\tilde{R}^i{}_j+\left(\tilde{E}_c\Phi\right)\left(\tilde{E}^c\Phi\right)
\tilde{e}^i\tilde{e}^j\\
R^a{}_i&=&\left(\tilde{E}_b\tilde{E}^a\Phi\right)\tilde{e}^b\tilde{e}^i
+\left(\tilde{E}^b\Phi\right)\tilde{\omega}^a{}_b\tilde{e}^i\\
R^i{}_a&=&\left(\tilde{E}_b\tilde{E}_a\Phi\right)\tilde{e}^b\tilde{e}^i
-\left(\tilde{E}_b\Phi\right)\tilde{\omega}^b{}_a\tilde{e}^i.
\end{eqnarray}
The Ricci tensor is obtained by contraction of the vector
index with the first of the two-form indices:
\begin{equation}
R_{a}\left(\bullet\right)=R^{b}{}_a\left(\tilde{E}_{b},\bullet\right)
+\frac{1}{\Phi}R^{i}{}_a\left(\tilde{E}_{i},\bullet \right).
\end{equation}
It reads
\begin{eqnarray}
R_{a} &=&\tilde{R}_{a}
+(d-2)\frac{1}{\Phi }\left[ \left( \tilde{E}_{b}\Phi \right) 
\tilde{\omega}^{b}{}_{a} -\left( \tilde{E}_{b}%
\tilde{E}_{a}\Phi \right) \tilde{e}^{b} \right]
\label{Ricci R spherically red.}\\
R_{i} &=&\frac{1}{\Phi }\tilde{R}_{i} +\biggm[\left(\tilde{E}_{b}
\tilde{E}^{b}\Phi \right)+\left( \tilde{E}^{b}\Phi \right)
\tilde{\omega}^{a}{}_b\left( \tilde{E}_{a}\right)\nonumber\\
&&+(d-3)\frac{1}{\Phi }\left( \tilde{E}_{c}\Phi\right)\left(\tilde{E}^{c}\Phi
\right)\biggm]\tilde{e}^{i}
\end{eqnarray}
Further contraction gives the scalar curvature on $M$:
\begin{equation}
R^{M}=\eta^{rs}R_{r}\left(E_{s}\right) =\eta ^{ac}R_{a}\left( 
\tilde{E}_{c}\right)-\frac{1}{\Phi}\delta ^{ij}R_{i}
\left(\tilde{E}_{j}\right).
\end{equation}
It can be expressed by the scalar curvatures on $L$ and $S^{d-2}$
and geometric objects of $L$:
\begin{multline}
R^{M}=R^{L}-\frac{1}{\Phi^{2}}\left[R^{S}
+(d-2)(d-3)\left(\tilde{E}_{b}\Phi\right)\left(\tilde{E}^{b}\Phi\right)\right]\\
-\frac{2}{\Phi}(d-2)\left[ \left( \tilde{E}_{b}\tilde{E}
^{b}\Phi \right) +\left( \tilde{E}^{b}\Phi \right)\tilde{\omega}^{a}{}_b
\left(\tilde{E}_{a}\right) \right]\\
=R^{L}-\frac{\left( d-2\right) \left( d-3\right) }{\Phi ^{2}}
\left[ 1+\tilde{\nabla}_{b}\Phi \tilde{\nabla}^{b}\Phi \right] -2\left( 
\frac{d-2}{\Phi }\right)\tilde{\square}_L\Phi.\label{R-sph.red.}
\end{multline}
In the last line I have inserted the (constant) scalar curvature
of the $(d-2)$-sphere $R^S=\left(d-2\right)\left(d-3\right)$
(the Riemann tensor on the $(d-2)$-sphere is given by $\tilde{R}^i{}_j=
\tilde{e}^{i}\wedge \tilde{e}^{j}$). In this form all quantities
on the r.h.s. live on $L$ as it should be.
\newpage
By reduction from $d=4$ the Ricci tensor and the scalar curvature read
\begin{eqnarray}
R^M_{ab}&=&R_a(E_b)=R^L_{ab}-2\frac{\tilde{\nabla}_{a}
\tilde{\nabla}_{b}\Phi}{\Phi}=R^L_{ab}+\frac{\tilde{\nabla}_{a}X
\tilde{\nabla}_{b}X}{2X^2}-\frac{\tilde{\nabla}_{a}\tilde{\nabla}_{b}X}{X}\\
R^M&=&R^L-2\frac{1+(\tilde{\n}\Phi)^2}{\Phi^2}-4\frac{\tilde{\square}\Phi}{\Phi}
=R^L-\frac{2}{X}+\frac{(\tilde{\n}X)^2}{2X^2}-2\frac{\tilde{\square}X}{X}.
\label{R-sph.red.4d}
\end{eqnarray}
In the last equality I have introduced the dilaton field $X=\Phi^2$
which appears in the action functional (\ref{2daction}).
Note that the $\frac{\tilde{\square}X}{X}$-term in the scalar
curvature becomes a surface term in the action when it is multiplied by the
spherically reduced measure $\sqrt{-g_M}=X\sqrt{-g_L}$.
Finally, I consider the relation between
the Laplacian on the $d$-dimensional manifold $M$ and on the
two-dimensional manifold $L$, whereby I assume that it acts
on a scalar field $S(x^{\alpha})$ which depends only on the
coordinates of $L$:
\begin{multline}
\square S=\eta^{rs}\n_rE_sS=\eta^{ab}\n_aE_bS+\eta^{ij}\n_iE_jS\\
=\tilde{\square}S-\eta^{ij}\omega^a{}_j(E_i)E_aS
=\tilde{\square}S-\eta^{ij}\frac{\tilde{E}^a\Phi}{\Phi}\tilde{E}_aS\\
=\tilde{\square}S+(d-2)\frac{\tilde{\n}^a\Phi\tilde{\n}_aS}{\Phi}
=\tilde{\square}S+\frac{d-2}{2}\frac{\tilde{\n}^aX\tilde{\n}_aS}{X}.
\label{Laplace-sph.red.}
\end{multline}
If $M$ is the four-dimensional Schwarzschild spacetime and
the gauge of the dilaton is fixed as $X=r^2$ one has
\begin{equation}
\square S=\tilde{\square}S-\frac{2}{r}\Sss\p_rS,
\end{equation}
to be compared with (\ref{Laplace-2d},\ref{Laplace.SS}).

\newpage
\section{Euclidean Formalism}

I want to map the Lorentzian manifold $M_{\cal M}$,
spanned by the coordinates $x_{\cal M}^{\mu}=(t,
\vec{x})$ and equipped with a Lorentzian metric $g$ with
signature\newline$(+,-,-,-)$, onto a Riemannian manifold $M_{\cal E}$
with a Riemannian metric $g_{\cal E}$ with signature $(+,+,+,+)$
and coordinates $x_{\cal E}^{\mu}=(\tau,\vec{x}_{\cal E})$.
This is achieved by relating the Euclidean time-coordinate $\tau$
to the Lorentzian one $t$ by $\tau=i\cdot t$ (I use the notion Euclidean
instead of Riemannian because it is common in particle physics),
and by multiplying the Lorentzian metric by an overall factor $-1$, i.e.
$g_{\cal E}=-g_{\cal M}$.

In this Section I will consider how the geometric objects and the
action functional of the manifolds are related.
The basis vectors and basis forms transform like
$\p_0=\frac{\p}{\p t}=i\p_{\tau}$ and $dt=-i\cdot d\tau$.
Thus, because vectors are invariant, the zero-components of
vectors catch a factor $i$: $V^0_{\cal E}=V^0_{\cal M}\cdot i$. 
The scalar product of two vectors transforms by a minus sign:
\begin{equation}
g_{\cal M}^{\mu\nu}V^{\cal M}_{\mu}V^{\cal M}_{\nu}
=-g_{\cal E}^{\mu\nu}V^{\cal E}_{\mu}V^{\cal E}_{\nu}.
\end{equation}
Now I consider the geometric objects. Those that contain an
even number of metrics remain invariant, while those those
that contain an odd number are multiplied by $-1$.
The Christoffel symbols (and the connection in general)
catch no overall sign but change by factors $i$ in front of
time-derivatives. The Laplace operator, as a scalar product of
two connections, transforms like:
\begin{equation}
\square_{\cal M}=-\triangle_{\cal E}.\label{E>M}
\end{equation}
The Ricci tensor contains an even number of metrics and hence is invariant
under Euclideanisation. The Riemann tensor is also invariant,
while the scalar curvature is multiplied by a sign:
\begin{equation}
(R^{\mu}{}_{\nu\sigma\tau})_{\cal M}=(R^{\mu}{}_{\nu\sigma\tau})_{\cal E}
\,\,,\,\,R^{\cal M}_{\mu\nu}=R^{\cal E}_{\mu\nu}\,\,,\,\,R_{\cal M}=-R_{\cal E}.
\end{equation}
It is convenient to introduce a \emph{Euclidean sign} $\ve$
for each geometric object, by which one can write the relation
between Lorentzian and Euclidean expression as
\begin{equation}
E_{\cal M}=\ve_{E}E_{\cal E}.
\end{equation}
This is particularly useful for quantities whose Euclidean sign
is not known in general, such as the endomorphism $E$.

The action functional is changed twofold, first by the volume element
and then by the integrand. I will first consider a trivial action
where the integrand is just the volume element:
\begin{equation}
L_{\cal M}=\int_{-\infty}^{\infty}dt\int d\vec{x}\sqrt{-g_{\cal M}}
=-i\int_{-i\infty}^{i\infty}d\tau\int d\vec{x}_{\cal E}\sqrt{g_{\cal E}}
:=-iL_{\cal E}.
\end{equation}
If the integrand contains geometric objects the relation between
Lorentzian and Euclidean action functional is modified by the
Euclidean sign of this object:
\begin{equation}
L_{\cal M}[O_{\cal M}]=\int O_{\cal M}\sqrt{-g_{\cal M}}d^dx
=\ve_{O}\int O_{\cal E}\sqrt{-g_{\cal M}}d^dx=i\ve_{O}L_{\cal E}[O_{\cal E}].
\label{Euclid.action.funct.}
\end{equation}
If $O_{\cal M}=\frac{1}{2}(\p S)^2$ then the corresponding
Euclidean action functional acquires a minus sign
\begin{equation}
L_{\cal M}\left[\frac{(\p S)^2}{2}\right]=-iL_{\cal E}
\left[\frac{(\p S)^2}{2}\right].
\end{equation}
Note that a mass term does \emph{not} change sign under Euclideanisation.

\newpage
\section{Seeley-DeWitt Expansion}

In this Appendix I collect the notations and some useful formulas
which are rarely explained in the text. All expressions are
Euclidean.

\subsection{Generalised Laplacian}

A general Laplace operator has the form
\begin{equation}
{\cal O}=-g^{mn}\n_m\n_n\1-E.
\end{equation}
$g^{mn}$ is the Euclideanised metric with signature $(+,+,+,+)$,
see Appendix E. $E$ is an endomorphism which is defined
as a bounded, linear map of the vector space (on which it acts)
into itself. For instance, if the Laplacian is supposed to act
on spinor fields, the endomorphism may produce a rotation
in spinor space. If the involved fields are scalar fields
it simply acts by multiplication with some function.

The total connection $\n$ is a sum of the Levi-Civit\'a connection $\n^{LC}$,
responsible for the parallel transport of the fields on the
physical manifold, and a gauge connection $A$ acting on the space
of inner symmetries of the fields. If the fields possess such
inner degrees of freedom (e.g. an $SU(2)$-index) the total
connection has the form
\begin{equation}
(\n_m)^A{}_B=(\n^{LC}_m)\1^A{}_B+(A_m)^A{}_B.
\end{equation}
$m$ denotes the spacetime index in a vielbein basis and capital letters
like $A,B$ the indices in gauge space.
The gauge curvature $\Omega_{\mu\nu}$ is defined by the action of a commutator of
two connections on a field which is a scalar with respect to Lorentz
transformations:
\begin{multline}
{\Omega_{\mu\nu}}^A{}_{B}\psi^B=[{\n_{\mu}}^A{}_C,{\n_{\nu}}^C{}_B]\psi^B\\
=(\n^{LC}_{\mu}\1^A{}_C+{A_{\mu}}^A{}_C)(\p_{\mu}\1^C{}_B+{A_{\mu}}^C{}_B)
\psi^B-(\n^{LC}_{\nu}\1^C{}_B+{A_{\nu}}^C{}_B)(\p_{\nu}\1^A{}_C+{A_{\nu}}^A{}_C)
\psi^B\\
=\left\{\left(\p_{\mu}{A_{\nu}}^A{}_C\right)-
\left(\p_{\nu}{A_{\mu}}^A{}_C\right)+[{A_{\mu}}^A{}_B,
{A_{\nu}}^B{}_C]\right\}\psi^B.\label{gauge-curvature}
\end{multline}
The Levi-Civit\'a connection defines the spacetime curvature
(Riemann tensor) as usual via its action on a Lorentz one-form:
\begin{equation}
[\n_m,\n_n]\omega_o=R_{mno}{}^p\omega_p.\label{connection-comm.}
\end{equation}
In this thesis I consider only commuting scalar fields, hence
the gauge curvature vanishes.

\newpage
\subsection{Bi-Tensors}

Bi-tensors are objects which depend on two spacetime
points $x$ and $y$. For instance, Green functions
and the delta function are bi-tensors. 
In particular, the Seeley-DeWitt coefficients $a_n(x,y)$ in the
local heat kernel expansion (\ref{heat expansion})
are also bi-tensors, just like the world function
$\sigma(x,y)$ and the Van Fleck-Morette determinant
$D(x,y)$.

The heat kernel appears in the effective action in
the diagonal form, which means that one needs the
coincidence limits $a_n(x,x)$ of the Seeley-DeWitt coefficients.
They can be calculated from the recurrence
relation (\ref{recursion}) by inserting the Seeley-DeWitt
coefficients, $\sigma(x,y)$, and $D(x,y)$, carrying out the
differentiations and performing the coincidence limit
$x\to y$.

In the following I discuss the basic bi-tensors that
appear in the heat kernel expansion and calculate
some of the coincidence limits. Most of the content of
this Section has been simply taken over from \cite{herxx}.

The world function is defined via the geodesic distance
between two spacetime points:
\begin{equation}
\sigma(x,y):=[\tau(y)-\tau(x)]\text{min}\biggm\{\int_{\tau(x)}^{\tau(y)}
\frac{g_{\mu\nu}}{2}\frac{dx^{\mu}}{d\tau}\frac{dx^{\nu}}{d\tau}d\tau\biggm\},
\label{world-function}
\end{equation}
i.e. it is one half of the square of the geodesic distance from $x$
to $y$, where $\tau$ is the parameter along the geodesic.
In the limit $x\to y$ it can thus be approximated by
\begin{equation}
\sigma(x,y)\approx\frac{g_{\mu\nu}}{2}(y^{\mu}-x^{\mu})(y^{\nu}-x^{\nu}),
\label{sigma-approx}
\end{equation}
for $x=y$ it clearly vanishes $\sigma(x,x)=0$.
Obviously the world function is symmetric in its arguments
$\sigma(x,y)=\sigma(y,x)$. Its covariant derivative
$\n_{\mu}\sigma=\sigma_{;\mu}$ is a vector whose norm equals the distance
between $x$ and $y$, i.e.
\begin{equation}
\sigma_{;\mu}\sigma^{;\mu}=2\sigma,\label{world-vector}
\end{equation}
which is oriented in the direction $y\to x$ and tangent to the point $x$.

The Van Fleck-Morette determinant $D$ and the bi-tensor $D_{\mu\nu'}$
are defined via the second covariant derivative of $\sigma(x,y)$:
\begin{equation}
D(x,y):=-\det(D_{\mu\nu'})=\det(\sigma_{;\mu\nu'}).
\end{equation}
A prime on an index denotes association with $y$. By repeated
differentiations of equation (\ref{world-vector}) one can
derive relations between $\sigma$ and $D_{\mu\nu'}$ like
\begin{eqnarray}
\sigma^{;\mu}D_{\mu\nu'}&=&-\sigma_{;\nu'}\\
D_{\sigma\nu'}&=&\sigma^{;\mu}{}_{;\sigma}D_{;\mu\nu'}+\sigma^{;\mu}
D_{\sigma\nu';\mu},
\end{eqnarray}
and, using the matrix identities $\text{tr}\ln A_{\mu\nu}=\ln\det A$
and $(A^{-1})^{\mu\nu}\p A_{\mu\nu}\newline
=(\det A)^{-1}\p\det A$, one obtains
\begin{equation}
D^{-1}(D\sigma^{;\mu})_{;\mu}=d.\label{Morette-identity}
\end{equation}
Note that the Van Fleck-Morette determinant appears in the form
\begin{equation}
\Delta(x,y):=g^{-1/2}D(g')^{-1/2}\label{Delta}
\end{equation}
in the heat kernel expansion.

Finally, one needs two bi-vectors which are related to the
spacetime metric. The parallel displacement bi-vector $g_{\mu\nu'}$
is defined by $\lim_{x\to y}g_{\mu\nu'}=g_{\mu\nu}$ and
$g_{\mu\nu;\sigma}\sigma^{;\sigma}=0$ and it satisfies
\begin{equation}
g_{\mu'}{}^{\nu}\sigma_{;\nu}=-\sigma_{;\mu'}.\label{displacement}
\end{equation}
The bi-unit $I(x,y)$ fulfils
\begin{equation}
\lim_{x\to y}I(x,y)=1\,\,,\,\,I_{;\mu}\sigma^{;\mu}=0.\label{bi-unit}
\end{equation}
It can be shown by (\ref{recursion}) that the first Seeley-DeWitt
coefficient $a_0$ exhibits the same properties as the bi-unit.
Namely, by setting $n=-2$, one obtains $\n\sigma\n a_0
=\sigma^{;\mu}(a_0)_{;\mu}=0$. Further, the initial condition
of the heat kernel demands $a_0(x,x)=0$. Thus, one can identify
\begin{equation}
a_0(x,y)=I(x,y).
\end{equation}

Now one can calculate all coincidence limits given in Table 3
in Section 3.1.1 by differentiation of the already existing relations.
I demonstrate this for a few cases and refer to
\cite{herxx} for the remaining expressions.

The coincidence limits of the world function $\sigma$
follow from (\ref{world-vector}) and repeated differentiations
of it. $\lim_{x\to y}\sigma\to0$ implies $\lim_{x\to y}\sigma_{;\mu}\to0$.
If one lets two covariant derivatives $\n_{\lambda}\n_{\kappa}$
act on (\ref{world-vector}), one obtains the relation
\begin{equation}
\lim_{x\to y}2\sigma_{;\kappa\lambda}=\lim_{x\to y}
(\sigma_{;\mu\kappa}\sigma^{;\mu}{}_{;\lambda}
+\sigma_{;\mu\lambda}\sigma^{;\mu}{}_{;\kappa})
\end{equation}
which suggests $\lim_{x\to y}\sigma_{;\kappa\lambda}=g_{\kappa\lambda}$.
This is in agreement with (\ref{sigma-approx}). By further
differentiation of (\ref{world-vector}) by $\n_{\nu}$, and using
(\ref{connection-comm.}) and some symmetry considerations, one gets
\begin{multline}
\lim_{x\to y}\sigma_{;\kappa\lambda\nu}=\lim_{x\to y}(\sigma_{;\kappa\lambda\nu}
+\sigma_{;\lambda\kappa\nu}+\sigma_{;\nu\kappa\lambda})
=\lim_{x\to y}(2\sigma_{;\kappa\lambda\nu}+\sigma_{;\nu\kappa\lambda})\\
=\lim_{x\to y}(2\sigma_{;\kappa\lambda\nu}+\sigma_{;\kappa\nu\lambda})
=\lim_{x\to y}(3\sigma_{;\kappa\lambda\nu}+R_{\lambda\nu\kappa}{}^{\mu}
\sigma_{;\mu})=0
\end{multline}
The coincidence limits of $D$ (and $\sqrt{\Delta}$) and $I$
are obtained by repeated differentiation of (\ref{displacement}),
respectively (\ref{bi-unit}), starting from $\lim_{x\to y}D=g$
and $\lim_{x\to y}\sqrt{\Delta}=1$.
They allow the computation of the coincidence limits of the
Seeley-DeWitt coefficients.

\end{appendix}

% \newpage
% \section*{Lebenslauf}
% \addcontentsline{toc}{section}{Lebenslauf}

% \vspace{0.6cm}

% \Large \textbf{Daniel Hofmann}\vspace{0.9cm}\\ \large
% \textbf{Persönliche Daten}\normalsize\vspace{0.2cm}
% \\
% \textbf{Geburtsdatum und Ort:} 16. Mai 1975 in Wien\\
% \textbf{Eltern:} Irmgard Pesendorfer und Werner Hofmann\\
% \textbf{Staatsbürgerschaft:} Österreich\\
% \textbf{Wohnort:} Wien 2. Bezirk\vspace{1.1cm}\\ \large
% \textbf{Ausbildung}\normalsize\vspace{0.2cm}
% \\
% \textbf{1981-1985:} Volksschule in Perchtoldsdorf\\
% \textbf{1985-1987:} AHS in Perchtoldsdorf\\
% \textbf{1987-1989:} AHS in der Anton-Kriegergasse, Wien\\
% \textbf{1989-1993:} ORG in Perchtoldsdorf\\
% \textbf{1993-1999:} Studium der Technischen Physik an der TU-Wien\\
% \textbf{1999-2002:} Doktoratsstudium der Technischen Physik an der TU-Wien

\newpage
\bibliographystyle{mystyle.bst}

\end{document}